\documentclass[a4paper,11pt]{article}
\usepackage{jheppub}
\usepackage[T1]{fontenc}
\usepackage{amsmath,amssymb,amsthm,mathtools}
\usepackage{booktabs,longtable,array}
\usepackage{mathrsfs}
\usepackage{microtype}
\allowdisplaybreaks[2]
\renewcommand{\afterTocRuleSpace}{\clearpage}
\makeatletter
\renewcommand*\l@subsection{\@dottedtocline{2}{1.5em}{3em}}
\makeatother
\hypersetup{
  pdftitle={Casimir multiplets suffice: strong generation in N=4 SU(N) SYM},pdfauthor={Yifan He}}
\newtheorem{theorem}{Theorem}[section]
\newtheorem{lemma}[theorem]{Lemma}
\newtheorem{proposition}[theorem]{Proposition}

\theoremstyle{definition}

\newcommand{\C}{\mathbb C}

\newcommand{\tr}{\operatorname{tr}}

\title{\boldmath
Casimir multiplets suffice:\\
strong generation in $\mathcal{N}=4$ $SU(N)$ SYM
}
\author{Yifan He}
\affiliation{School of Physical Science and Technology, Lanzhou University, Lanzhou 730000, China}
\emailAdd{heyifan@westlake.edu.cn}
\abstract{The Schur sector of four-dimensional $\mathcal N=4$ super Yang--Mills theory contains infinitely many protected operators, while the invariant polynomials of a single adjoint scalar are organized by finitely many Casimir degrees. I prove that the small $\mathcal N=4$ short multiplets built from these invariants strongly generate the associated chiral algebra, thereby establishing the $SU(N)$ case of the strong-generation conjecture of Beem et al. For every $N\ge2$, all field components associated with degrees $2,\ldots,N$ suffice, in every conformal weight and every degree of the standard relative quantum BRST cohomology. The proof brings together cyclic trace cohomology, quantum lifting, and an arbitrary-degree quartic--odd transport identity. Exact field relations and a finite-rank Newton obstruction exclude every additional lowest space, including non-scalar flavor representations; the required normal-ordering corrections and BRST primitives are derived explicitly. The resulting completeness theorem explains how a finite family of supersymmetric multiplets organizes an unbounded quantum spectrum. Every Schur class is a normal polynomial in their components and derivatives, even when its gauge-theory representative involves fermions and arbitrarily many derivatives. This supplies a uniform set of local inputs for meromorphic operator products, bounds protected state multiplicities before cancellations in an index, and gives an algebraic account of the finite-rank reorganization of protected trace operators.}
\keywords{Supersymmetric Gauge Theory, Conformal and W Symmetry, Extended Supersymmetry, BRST Quantization}
\makeatletter
\gdef\@fpheader{}
\makeatother
\begin{document}
\maketitle
\raggedbottom
\section{Introduction}
\label{sec:introduction}

Local operators provide a common language for the dynamics, symmetries,
and observables of quantum field theory. Their operator product expansion
organizes the effect of short distances into a spectrum of operators and
coefficients describing their interactions \cite{Wilson1969}. Conformal
symmetry makes this description especially powerful. In two dimensions,
the infinite-dimensional enhancement of local conformal transformations
turns holomorphic operator products into a tractable algebraic structure,
as exemplified by the conformal bootstrap of
Belavin, Polyakov, and Zamolodchikov \cite{BPZ1984}. A central question in
higher dimensions is how much of this algebraic control survives in an
interacting theory with infinitely many local observables.

Supersymmetry supplies a systematic answer for suitable protected sectors.
Passing to the cohomology of a supercharge removes operators paired by
supersymmetry and preserves structures that are often accessible throughout
a family of theories. Topological twisting \cite{Witten1988} and the exact
description of low-energy dynamics in four-dimensional gauge theory
\cite{SeibergWitten1994} illustrate distinct uses of this principle.
Superconformal indices provide another: they count protected states with
signs in a way compatible with continuous deformations
\cite{Romelsberger2006,Kinney2007}. The different protected constructions
retain different information. Understanding the operator algebra, rather
than only its graded trace, requires control of the actual cohomology and
its products.

For four-dimensional $\mathcal N=2$ superconformal field theories,
the correspondence of Beem et al.~\cite{Beem2013} assigns a vertex
superalgebra to a sector of local operators on a plane. Its meromorphic
correlation functions are twisted four-dimensional correlators. The
relevant Schur operators obey
\begin{equation}
 E-2R-j_1-j_2=0,\qquad r_{4d}+j_1-j_2=0,
 \qquad h=E-R.
 \label{intro:schur}
\end{equation}
Here $E$ is the four-dimensional scaling dimension, $R$ is the
$SU(2)_R$ weight, $r_{4d}$ is the abelian $R$ charge, and $j_1,j_2$
are the Lorentz spins. The conformal weight $h$ grades the resulting
two-dimensional algebra. This construction makes a protected part of
higher-dimensional dynamics accessible to the methods of chiral symmetry.
I refer to the resulting chiral algebra as the Schur algebra when
emphasizing its four-dimensional origin. A broad account of the
correspondence and its applications is given in ref.~\cite{Lemos2020}.
The $\Omega$-deformed topological--holomorphic twist provides a
complementary field-theoretic construction
\cite{OhYagi2019,Jeong2019}: the cohomological reduction and the
localization of the path integral lead to the same meromorphic sector.

Gauging makes this correspondence particularly rich. In theories of
class $\mathcal S$, duality relates different decompositions of the
underlying surface \cite{Gaiotto2012}; on the chiral-algebra side,
these become gluing and associativity constraints
\cite{BeemClassS2015}. The mathematical construction for genus-zero
surfaces identifies the associated varieties with the corresponding
Moore--Tachikawa symplectic varieties \cite{ArakawaClassS2018}.
Derived gluing \cite{Yanagida2021} and the treatment of twisted
class-$\mathcal S$ theories \cite{BeemNair2023} further develop the
cohomological structure of this operation. These developments make
BRST cohomology a natural language for studying the algebra produced
by gauging, including the question of which local fields survive.

The case of $\mathcal N=4$ super Yang--Mills theory is a particularly
sharp setting in which to ask whether these methods capture a complete
sector. Viewed as an $\mathcal N=2$ theory, it contains an adjoint
hypermultiplet and an $\mathcal N=2$ vector multiplet. The corresponding
free fields are adjoint symplectic bosons and fermionic ghosts, with gauge
invariance implemented by relative quantum BRST cohomology. The extra
supersymmetry gives a small $\mathcal N=4$ superconformal algebra. For
gauge algebra $\mathfrak{sl}_N$, its central charge and affine level are
\begin{equation}
 c_{2d}=-3(N^2-1),\qquad k_{2d}=-\frac{N^2-1}{2},
 \qquad c_{2d}=6k_{2d}=-12c_{4d}.
 \label{intro:central}
\end{equation}
The elementary invariant polynomials have degrees $2,\ldots,N$.
Their images in the Schur sector supply one short multiplet at each of
these degrees, starting with the small $\mathcal N=4$ currents.

This finite list suggests a strong form of completeness. Strong
generation means that every field is a finite linear combination of
normally ordered products of derivatives of the proposed generators.
It therefore excludes new independent generators at arbitrarily high
weight, even though it permits infinitely many composite primary fields.
Conjecture~3 of ref.~\cite{Beem2013} proposes precisely a Casimir-indexed
super $\mathcal W$-algebra structure for general simple gauge groups.
For $SU(N)$, the issue is whether the complete Casimir multiplets exhaust
the quantum cohomology at every weight.

Several complementary developments make this question consequential.
The small $\mathcal N=4$ algebra at $c=-9$ admits an explicit free-field
realization with a detailed representation theory
\cite{Adamovic2016}. Bonetti, Meneghelli, and Rastelli constructed
vertex algebras associated with complex reflection groups and proposed
their Weyl-group cases as the chiral algebras of $\mathcal N=4$ gauge
theories \cite{Bonetti2019}. These constructions expose the role of
invariant degrees and provide an efficient organization of candidate
generators. Free-field realizations motivated by Higgs-branch effective
theories further relate chiral algebras to moduli-space geometry and
the four-dimensional $R$ filtration \cite{BeemHiggs2019}. The connection
between Higgs branches, vertex-algebra null states, and modular
differential equations places relations among generators at the center
of a broader program \cite{BeemRastelli2018}.

For the symmetric-group family, Arakawa, Kuwabara, and M\"oller
give a geometric construction using a sheaf of vertex superalgebras
on a Hilbert scheme \cite{AKM2023}. Their results include the
associated variety, an embedding into $N-1$ free $\beta\gamma$--$bc$
systems, and character formulas. This supplies an all-rank geometric
realization of the family suggested by reflection-group invariants.
The construction uses a framed Hilbert-scheme BRST reduction. The
cohomological problem considered here concerns the standard adjoint
relative quantum complex itself and the exhaustion of its classes by
the Casimir multiplets. The distinction between constructing a vertex
algebra with the expected geometric properties and identifying every
class in this gauge-theory complex is central to the argument.

The operator products of the proposed multiplets have also been
studied directly. Gaberdiel and Li analyze their associativity,
finding no additional parameter beyond the central charge in the
products they compute and testing the expected finite-rank
truncations \cite{GaberdielLi2025}. Bonetti and Meneghelli obtain
bootstrap evidence for a universal small-$\mathcal N=4$
$\mathcal W$-algebra, together with an explicit wedge algebra and
evidence for the filtration inherited from four dimensions
\cite{BonettiUniversal2025}. These calculations reveal how a common
multiplet pattern can accommodate different ranks. Strong generation
of the actual BRST cohomology supplies the complementary assertion
that this pattern omits no independent field at a higher weight.

There is a parallel algebraic description of the stable trace sector.
The relation between cyclic homology and stable matrix Lie algebra
homology goes back to Loday--Quillen and Tsygan
\cite{LodayQuillen1984,Tsygan1983}. Zeng constructs a large-$N$
BRST vertex algebra in a Deligne category and relates its classical
limit to a cyclic complex \cite{Zeng2025}. Retaining the rank as a
parameter separates the stable algebraic structure from the matrix
identities at a specified integer rank. In the present proof, the
cyclic calculation, quantum lifting, and finite-rank descent are
established explicitly in the conventions of the relative complex;
the transition between them is part of the proof.
The invariant-theoretic background is the description of simultaneous
matrix invariants by traces \cite{Procesi1976} and of their trace
identities through Cayley--Hamilton relations \cite{Razmyslov1974}.
These classical results explain the special role of rank in trace
algebras; here the normal-ordering terms and relative BRST primitives
make that role explicit at the level of quantum fields.

Protected counting supplies independent and highly structured data.
The Schur and Macdonald limits of the superconformal index organize
operators according to supersymmetric quantum numbers and connect
four-dimensional theories to two-dimensional topological theories
\cite{Gadde2011,Gadde2013}. The class $\mathcal S$ description makes
duality and different gauge-theory constructions part of this same
framework \cite{Gaiotto2012}. Localization also gives direct access to
Schur correlation functions and their torus interpretation
\cite{PanPeelaers2019,DedushenkoFluder2020}. These tools probe the
candidate algebra through protected observables. A proof of strong
generation adds a statement about individual cohomology classes, before
taking the signs and cancellations of an index.

For $\mathcal N=4$ theories, this counting problem admits remarkably
explicit solutions. The free-fermion representation of the unitary
Schur index yields exact finite-rank expressions and an all-orders
large-rank expansion \cite{Bourdier2015}; its extension to circular
quivers brings a wider class of gauge theories into the same framework
\cite{Bourdier2016}. Flavored $\mathcal N=2^*$ indices reveal further
elliptic and quasi-Jacobi structure at arbitrary rank
\cite{Hatsuda2023}. Refined-character proposals also relate a
vertex-algebra filtration to the Macdonald index \cite{Song2017}.
Each refinement records information that is lost upon taking a less
detailed trace, whereas a cohomological spanning theorem controls
the underlying classes themselves.

The geometric and analytic sides of protected counting meet in the
theory of associated varieties. Filtrations turn vertex algebras
into commutative differential algebras \cite{Li2005}, while
quasi-lisse geometry imposes modular differential equations on
ordinary-module characters \cite{ArakawaKawasetsu2018}. Together
with the Higgs-branch interpretation
\cite{BeemRastelli2018,BeemHiggs2019}, these ideas explain why
relations among local fields carry information about both moduli
spaces and torus observables. More recently, Wilson-line indices and
modular transformations have been used to study irreducible
characters and their relation to mixed Hodge structures of Coulomb
branches \cite{LiPanYan2025}. Strong generation concerns the
local fields on which such representation-theoretic constructions
are based.

Several foundational structures help organize the operator algebra
behind these observables. Zhu's construction turns vertex-algebra
products into an associative algebra acting on lowest spaces
\cite{Zhu1996}. Nonlinear Lie conformal algebras encode singular
products of generating fields together with their normally ordered
composite terms \cite{DeSoleKac2005}. These perspectives clarify why
both generator products and their action on a hypothetical lowest
space are useful in a completeness proof. The precise mode identities
needed here are derived directly in the relative quantum complex.

The relation between local fields and representations also has a
physical extension to defects. Superconformal surface defects give
modules of the Schur algebra, with defect indices recording their
characters \cite{Cordova2017}. In another protected construction
within $\mathcal N=4$ gauge theory, vertex algebras at junctions of
supersymmetric interfaces carry modules labeled by interface line
defects and reflect four-dimensional dualities
\cite{GaiottoRapcak2019}. Together these examples show how the local
operator algebra can organize observables beyond its vacuum sector.

Completeness is also important for the superconformal bootstrap. Protected
meromorphic data enter the crossing equations for $\mathcal N=4$
correlators \cite{Bootstrap2013,Bootstrap2017}. Knowing a complete set
of strong generators identifies the operator data from which this
meromorphic sector can be built. It is also relevant to holography:
$SU(N)$ super Yang--Mills theory is the standard boundary theory in
the $AdS_5/CFT_4$ correspondence \cite{Maldacena1998,WittenAdS1998},
and its protected trace operators provide a natural setting in which
to distinguish finite-rank relations from new elementary protected
building blocks.

The protected holographic sector has its own detailed constructions.
Localization in $AdS_5$ leads to a Chern--Simons description of a
protected subsector and motivates a higher-spin bulk description of
the chiral algebra \cite{BonettiAdS2018}. Twisted holography relates
gauged $\beta\gamma$ systems to a topological string construction
\cite{CostelloTwisted2021}, while a worldsheet approach extracts
Schur operators from the proposed string description of free
$\mathcal N=4$ super Yang--Mills theory \cite{LiWorldsheet2024}.
Chang, Lin, and Wu connect the strong-generation conjecture to
Schur graviton cohomology and study it through index bounds,
modular differential equations, and direct cohomology calculations
\cite{Chang2024}. These perspectives give the completeness question
a physical meaning: whether the protected trace multiplets and their
products account for every local Schur class when the finite-rank
identities are fully imposed.

I determine a complete set of strong generators for the $SU(N)$ Schur
algebra by establishing the following theorem. The relative complex, mode
conventions, and the explicit representatives used in its statement
are defined in section~\ref{sec:foundations}.

\begin{theorem}[Strong generation at every rank and weight]
\label{thm:main}
For every integer $N\ge2$, let
\[
 \mathcal H_N=H^\bullet_{\mathrm{rel}}
 \bigl(\mathrm{Sb}(\mathfrak{sl}_N\oplus\mathfrak{sl}_N)
                  \otimes bc(\mathfrak{sl}_N),Q\bigr)
\]
be the standard relative quantum BRST cohomology. Let $\mathcal S_N$
consist of all field components of the small $\mathcal N=4$ short
multiplets whose highest bottom fields are the Casimir fields of degrees
$d=2,\ldots,N$. Then
\begin{equation}
 \mathcal H_N=\operatorname{span}_{\C}
 \left\{:\partial^{a_1}s_1
       \bigl(:\partial^{a_2}s_2(\cdots\partial^{a_\ell}s_\ell)\cdots:\bigr):
       \ \middle|\ a_i\ge0,\ s_i\in\mathcal S_N\right\},
 \label{intro:strong-span}
\end{equation}
including the empty product. The equality holds in every conformal
weight and every cohomological degree. Either the multiplets built from raw traces or those built from
harmonic superconformal-primary bottoms may be used, with the degree-two
multiplet unchanged.
\end{theorem}

The proof has two logically distinct parts. First, a free cyclic
calculation determines the stable trace cohomology. Quantum lifting over
a Laurent-polynomial rank parameter and the Cayley--Hamilton identity
then establish strong closure at every finite rank. This constructs the
candidate vertex subalgebra before any quotient module is considered.
Second, if a class outside it existed, the first missing conformal
weight would carry a finite-dimensional flavor representation. An
arbitrary-degree quartic--odd identity produces a nonzero common kernel
of all odd zero modes. Exact field relations then exclude this kernel,
including its non-scalar flavor components. The contradiction proves
exhaustion without an upper bound on weight. The induction in the proof
is on tensor degree and on finite filtrations; the rank enters as a
parameter.

The theorem identifies the full strong-generating content of the Schur
algebra and gives completeness a definite physical interpretation.
At fixed rank, the protected local building blocks are carried by the
Casimir short multiplets. Higher protected
complexity is organized by their products, derivatives, and relations.
This identifies what must be specified in an algebraic description of
the protected sector and what can be reconstructed from it. I develop
these consequences, including a multigraded spanning bound and the
interpretation of finite-rank trace truncation, in
section~\ref{sec:physics}.

Section~\ref{sec:foundations} constructs the algebra and proves uniform
strong closure. Section~\ref{sec:odd} proves the arbitrary-degree odd
transport and the common-kernel theorem. Section~\ref{sec:descent}
establishes the exclusion of lowest spaces. Section~\ref{sec:exhaustion}
assembles these results into the proof of Theorem~\ref{thm:main}.
The appendices give the normal-ordering formulas, the full field
identities underlying the descent, and the small-rank endpoints.

\subsection{Proof architecture and reading guide}
\label{sec:proof-guide}

I organize the proof around the dependencies in table~\ref{tab:proof-dependencies}.
The table separates the identities required to construct the candidate
algebra from the module argument that proves its completeness. The
appendices are part of the proof: they supply the field identities,
normal-ordering corrections, and relative BRST primitives used in the
main text.

\begin{table}[htbp]
\centering
\small
\renewcommand{\arraystretch}{1.16}
\begin{tabular}{@{}>{\raggedright\arraybackslash}p{0.23\textwidth}>{\raggedright\arraybackslash}p{0.72\textwidth}@{}}
\toprule
Stage & Result and its inputs \\
\midrule
Strong closure &
The cyclic calculation, quantum lifting, and complete-tower reduction
(Theorem~\ref{fnd:cyclic} and Lemmas~\ref{fnd:lifting},~\ref{fnd:tower})
prove Theorem~\ref{thm:strong-closure}. Only after this step is
$\mathcal H_N/A_N$ used as an $A_N$-module. \\
All-degree odd transport &
The field identities of Appendix~\ref{app:odd-fields}, including the
separate septic identity, supply the starting data for the simultaneous
induction in section~\ref{sec:odd}. Equations~\eqref{odd:solved-recursion}
and~\eqref{odd:bootstrap-result} propagate the return identities and
normalization to every odd trace degree. \\
Invariant odd kernel &
The full transport relations and the principal-tensor argument
(Lemma~\ref{odd:principal-tensor}) yield
Theorem~\ref{odd:common-kernel}: a nonzero lowest space has a nonzero
invariant subspace annihilated by all odd zero modes. \\
Lowest-space exclusion &
For $N\ge5$, Theorem~\ref{thm:lowest-exclusion} uses the whole-input
descent of Appendix~\ref{app:descent-fields}. The even action, Newton
identity, and spectrum are proved in
sections~\ref{df:sec-even-lie},~\ref{df:sec-newton},~\ref{df:sec-spectrum}.
The nontrivial flavor branch and the scalar branch
\eqref{low:scalar-branch} both lead to the terminal contradiction
\eqref{low:final-contradiction}. \\
All weights and matrix sizes &
Lemma~\ref{fnd:extremal} gives $j<r$ at a hypothetical first missing
weight. Section~\ref{sec:exhaustion} combines this bound with the kernel
and exclusion theorems. Appendix~\ref{sec:small-rank} supplies separate
all-weight arguments for $N=2,3,4$. \\
\bottomrule
\end{tabular}
\caption{Dependencies in the proof of Theorem~\ref{thm:main}.}
\label{tab:proof-dependencies}
\end{table}

Two distinct arguments account for the unbounded ranges in the theorem.
The induction in section~\ref{sec:odd} concerns trace degree and proves
identities throughout the raw multiplet tower. The final argument in
section~\ref{sec:exhaustion} concerns conformal weight: a nonempty
complement would have a least weight, and that entire lowest space is
excluded. The parameter $N$ remains unrestricted in the uniform argument
for $N\ge5$; the three smaller matrix sizes are its separately proved
endpoints.

For navigation, $N$ denotes the matrix size, $d$ a Casimir or raw trace
degree, $h$ a conformal weight, and $r$ the lowest weight of the module
under consideration. Flavor spin is denoted by $j$; it is distinct from
both trace degree and conformal weight. The symbol $A_N$ always denotes
the normal differential span of the complete Casimir multiplets.
The physical zero mode $o(U)$ is defined only when $U$ has integral
weight, whereas $U^+$ denotes the mode of a half-integral-weight field
that raises weight by $1/2$; the conventions are fixed in
\eqref{fnd:mode-degree}. In the descent, these raising maps take values
in the actual space $M_{r+1/2}$, with the input and output actions kept
separate.

\section{The relative BRST algebra and uniform strong closure}
\label{sec:foundations}

I first construct the candidate algebra independently of the
exhaustion argument. The cyclic calculation identifies the trace
classes, quantum lifting controls their singular products, and
Cayley--Hamilton reduces the complete tower to the prescribed
Casimir degrees. The resulting strong-closure theorem supplies the
quotient module used in the later sections.

I work over $\mathbb C$ and put $\mathfrak g=\mathfrak{sl}_N$,
$N\ge2$. The parameter $N$ is the matrix size; the Lie algebra has rank $N-1$. All cohomology groups in this article include every BRST
cohomological degree. Matrix traces are in the defining representation,
and the invariant form is $\langle a,b\rangle=\operatorname{tr}(ab)$.
In a trace-orthonormal basis, the nonzero elementary products are
\begin{equation}\label{fnd:free-ope}
 X^a(z)Y^b(w)\sim\frac{\delta^{ab}}{z-w},\qquad
 Y^a(z)X^b(w)\sim-\frac{\delta^{ab}}{z-w},\qquad
 b^a(z)c^b(w)\sim c^a(z)b^b(w)\sim\frac{\delta^{ab}}{z-w}.
\end{equation}
The symplectic bosons $X,Y$ have conformal weight $1/2$; the odd
fields $b,c$ have weights $1,0$ and ghost numbers $-1,+1$.
I write $p=\partial c$. The BRST operator and conformal field are
\begin{equation}\label{fnd:brst}
 Q=\bigl(\operatorname{tr}(c[X,Y]-ccb)\bigr)_{(0)},\qquad
 T=\frac12\operatorname{tr}(X\partial Y-(\partial X)Y)
       -\operatorname{tr}(bp).
\end{equation}
Unless composite factors are explicitly separated by colons, a trace
polynomial denotes simultaneous normal ordering of elementary fields.
The notation $:ab:$ always denotes the vertex product $a_{(-1)}b$.
Nested products retain their displayed parentheses.

The relative complex $C_N$ is the common kernel of all antighost
zero modes and constant total gauge-current zero modes. Its PBW
presentation is the space of $\mathfrak g$-invariants in the
supercommutative creation variables
\begin{equation}\label{fnd:relative-pbw}
 \partial^rX,\ \partial^rY,\ \partial^rb\quad(r\ge0),\qquad
 \partial^rc\quad(r\ge1).
\end{equation}
There is no undifferentiated $c$ in this description. The BRST
operation preserves this space. Its square vanishes: the cubic
terms in the simple pole of $j_{\rm BRST}(z)j_{\rm BRST}(w)$ cancel
by the Jacobi identity, and the remaining anomaly is proportional
to the sum $-2N+2N$ of the adjoint symplectic-boson and ghost
contractions. Both coefficients follow from
\begin{equation}\label{fnd:completeness}
 \sum_a(t_a)_{ij}(t_a)_{kl}
   =\delta_{il}\delta_{jk}-\frac1N\delta_{ij}\delta_{kl},\qquad
 \sum_a t_a Zt_a=\operatorname{tr}(Z)I-\frac1N Z.
\end{equation}
The double-contraction trace is
$\operatorname{tr}_{\mathfrak g}(\operatorname{ad}t_a
\operatorname{ad}t_b)=2N\delta_{ab}$; its sign is negative for
the even first-order system and positive for the odd one.
Thus $\mathcal H_N=H^\bullet(C_N,Q)$ is a conformal vertex
superalgebra, with central charge $-3(N^2-1)$.

\subsection{Modes, flavor, and the complete Casimir multiplets}

I use the mode convention $Y(a,z)=\sum_{n\in\mathbb Z}a_{(n)}z^{-n-1}$,
so that
\begin{equation}\label{fnd:mode-degree}
 [L_0,a_{(n)}]=(\operatorname{wt}a-1-n)a_{(n)},\qquad
 o(a)=a_{(\operatorname{wt}a-1)}
 \quad\text{if }\operatorname{wt}a\in\mathbb Z.
\end{equation}
Half-integral-weight fields have no weight-preserving modes in an
untwisted module. Their least raising mode raises weight by $1/2$;
I denote its action on a lowest space by a superscript $+$.
The flavor currents and supercurrents are
\begin{align}\label{fnd:small-fields}
 e&=\tfrac12\operatorname{tr}X^2,\qquad
 H=-\operatorname{tr}XY,\qquad f=-\tfrac12\operatorname{tr}Y^2,\notag\\
 G_X&=\operatorname{tr}(Xb),\quad G_Y=\operatorname{tr}(Yb),\quad
 \bar G_X=\operatorname{tr}(Xp),\quad\bar G_Y=\operatorname{tr}(Yp).
\end{align}
They are relative cycles. Their zero-mode flavor algebra is
$[H,e]=2e$, $[H,f]=-2f$, $[e,f]=H$; its current level is
$-(N^2-1)/2$. The mixed supercurrent products needed below are
\begin{align}
 [G_X{}_{\lambda}\bar G_X]&=(\partial+2\lambda)e,&
 [G_Y{}_{\lambda}\bar G_Y]&=-(\partial+2\lambda)f,\notag\\
 [G_X{}_{\lambda}\bar G_Y]&=\tfrac{N^2-1}{2}\lambda^2
                     -\lambda H-T-\tfrac12\partial H,\notag\\
 [G_Y{}_{\lambda}\bar G_X]&=-\tfrac{N^2-1}{2}\lambda^2
                     -\lambda H+T-\tfrac12\partial H.
 \label{fnd:small-ope}
\end{align}
Products of two unbarred or two barred supercurrents vanish.
These equations follow by a single matter contraction, a ghost
contraction, and their simultaneous contraction in
\eqref{fnd:free-ope}; the latter sums over $N^2-1$ adjoint indices.
In particular, writing $F=f_{(0)}$ and
$\alpha=G_{X,(0)}$, $\beta=G_{Y,(0)}$,
$\gamma=\bar G_{X,(0)}$, $\delta=\bar G_{Y,(0)}$, I have
\begin{equation}\label{fnd:super-derivations}
 \{\alpha,\delta\}=-\partial,\quad
 \{\beta,\gamma\}=\partial,\qquad
 [F,\alpha]=\beta,\quad[F,\gamma]=\delta,
\end{equation}
with all other odd anticommutators and $[F,\beta],[F,\delta]$ zero.
All these zero modes are superderivations of vertex normal products
and commute with $Q$ and $\partial$.

Put $P_d=\operatorname{tr}X^d$ for $d\ge2$. Its raw short multiplet
consists of the flavor strings generated from
\begin{align}\label{fnd:normalization}
 &P_d,\qquad B_d=-\frac1d\beta P_d=\operatorname{tr}(X^{d-1}b),\qquad
 C_d=-\frac1d\delta P_d=\operatorname{tr}(X^{d-1}p),\notag\\
 &H_d=-\frac1d\delta\beta P_d+\frac1{d^2}\partial FP_d.
\end{align}
Their spins are $d/2,(d-1)/2,(d-1)/2,(d-2)/2$, respectively,
and their weights are $d/2,(d+1)/2,(d+1)/2,(d+2)/2$.
The degree-two multiplet is precisely
$e,H,f,G_X,G_Y,\bar G_X,\bar G_Y,T$, since $H_2=-T$.
For a highest tensor $v$ of spin $s$ I normalize its components by
$v_i=(2s-i)!F^iv/(2s)!$; thus
$[f,v_i]=(2s-i)v_{i+1}$ and $[e,v_i]=iv_{i-1}$.
I denote the four component families by
$W_{d,i},B_{d,i},C_{d,i},H_{d,i}$, with highest components
$W_{d,0}=P_d$, $B_{d,0}=B_d$, $C_{d,0}=C_d$, and $H_{d,0}=H_d$.
When these fields act on a lowest conformal space, I also use
$e,H,f$ for their weight-preserving zero modes; brackets with a field
mean the corresponding flavor action, and brackets between operators
mean their supercommutator.

Here is an explicit formula fixing the top-field normalization.
For any invariant homogeneous polynomial $P$ of degree $d$, set
$P_a=\partial P/\partial X^a$ and
$P_{ab}=\partial^2P/\partial X^a\partial X^b$. Then
\begin{equation}\label{fnd:top-formula}
 H(P)=\frac1dP_{ab}b^ap^b-\frac{d-1}{d^2}P_a\partial Y^a
                     +\frac1{d^2}P_{ab}Y^a\partial X^b.
\end{equation}
Indeed $\beta P=-P_ab^a$, $\delta P=-P_ap^a$, and
$\beta\delta P=P_{ab}b^ap^b-P_a\partial Y^a$; the last term is
the first Taylor coefficient of the double pole in $b(z)p(w)$.
Equation \eqref{fnd:top-formula} follows from
$FP=P_aY^a$ and \eqref{fnd:normalization}. Euler's formula
$X^aP_{ab}=(d-1)P_b$ proves that its flavor raising mode vanishes.
Closure of $P$ under $Q$ is invariance under the infinitesimal
adjoint action; closure of every descendant follows from the
commuting superderivations. This proves closure on the chain level.

For the superconformal-primary convention, put $m=N^2-1$,
$q_X=\operatorname{tr}X^2$ and $\Delta_X=\sum_a\partial_a^2$.
For $3\le d\le N$ replace $P_d$ by
\begin{equation}\label{fnd:harmonic}
 \widehat P_d=\sum_{k=0}^{\lfloor d/2\rfloor}
 \frac{(-1)^kq_X^k\Delta_X^kP_d}
 {4^kk!\prod_{a=0}^{k-1}(d+m/2-a-2)}.
\end{equation}
The denominators are nonzero. For a homogeneous $u$ of degree $\ell$,
$\Delta_X(q_X^ku)=q_X^k\Delta_Xu+2k(2\ell+m+2k-2)q_X^{k-1}u$;
consecutive terms of \eqref{fnd:harmonic} therefore cancel, giving
$\Delta_X\widehat P_d=0$. The only positive affine mode on a pure
$X$ bottom that can fail to vanish is
$f_{(1)}P=-\Delta_XP/2$. The supercurrent products with $P$
have only simple poles, and
\[
 T(z)P(w)\sim\frac{dP(w)}{2(z-w)^2}
                    +\frac{\partial P(w)}{z-w}.
\]
Thus $\widehat P_d$ is a small $\mathcal N=4$
superconformal-primary bottom. For its top, put $L_n=T_{(n+1)}$.
The $L_1$ contributions from $\beta\delta P/d$ and
$\partial FP/d^2$ are $-FP/d$ and $+FP/d$, while all higher
possible contributions are multiples of $\Delta_XP$. More explicitly,
\[
 G_{Y,(1)}\delta P=-FP,\qquad
 G_{Y,(2)}\delta P=\Delta_XP,\qquad
 G_{Y,(n)}\delta P=0\quad(n\ge3).
\]
These equations and $[L_n,G_{Y,(r)}]=((n+1)/2-r)G_{Y,(n+r)}$
prove the Virasoro primary conditions for the top component.
The fermionic components are also Virasoro primary by their elementary
operator products; they are superconformal descendants of the bottom.
The raw and harmonic
conventions will be shown below to have identical strong spans.

\begin{lemma}[Extremal flavor sector]\label{fnd:extremal}
Every flavor spin $j$ at conformal weight $h$ in $C_N$ satisfies
$j\le h$. The entire subspace with flavor weight $m_f=h$ consists
of invariant polynomials in undifferentiated $X$, has ghost number
zero, and survives unchanged in $\mathcal H_N$. It is generated by
$P_2,\ldots,P_N$ under normal multiplication.
\end{lemma}
\begin{proof}
Each creation letter has $h-m_f\ge0$, with equality only for an
undifferentiated $X$. In fact $h\ge j+|\operatorname{gh}|$.
Consequently no ghost-number $-1$ preimage can have the weight
and flavor charge of an extremal pure-$X$ polynomial. These
polynomials are cycles by invariance and are not boundaries.
Restriction of an invariant polynomial to traceless diagonal matrices
is injective because matrices with distinct eigenvalues are dense.
Its restriction is symmetric in the eigenvalues. Symmetric polynomials
are generated by elementary symmetric polynomials: cancel their
lexicographically largest monomials successively. Averaging a lift over
the symmetric group proves the same assertion on the trace-zero
hyperplane. Newton's identities change the elementary symmetric
generators triangularly to $P_2,\ldots,P_N$. They are independent
because they are the coefficients of arbitrary monic trace-zero
characteristic polynomials. Finally $XX$ has no contraction, so
normal multiplication is exactly polynomial multiplication.
\end{proof}

\subsection{A free cyclic calculation}
\label{fnd:cyclic-section}

Give the relative PBW complex the filtration
\begin{equation}\label{fnd:filtration}
 \deg_FX=\deg_FY=1,\qquad \deg_Fb=2,\qquad
 \deg_Fc=0,\qquad\deg_F\partial=0.
\end{equation}
At weight $h$ it is bounded between $0$ and $2h$. Every Wick
contraction removes degree two, so it is a good vertex filtration.
One contraction of the cubic BRST current preserves its degree;
additional contractions lower it. The principal differential is
therefore the classical relative BRST differential. Constant gauge
invariants commute with passage to the associated graded: every
fixed-weight representation is finite dimensional and the gauge
algebra is reductive.

Use divided jets $X_r=\partial^rX/r!$, and similarly for $Y_r,b_r,c_r$.
Let $\mathcal T$ be the vector space of signed cyclic words of length
at least two in $X_r,Y_r,b_r$ ($r\ge0$) and $c_r$ ($r\ge1$).
No rank-dependent trace identities are imposed. Its differential is
\begin{equation}
\begin{aligned}
qX_r&=-\sum_{i=1}^r[c_i,X_{r-i}],&
 qY_r&=-\sum_{i=1}^r[c_i,Y_{r-i}],\\
 qb_r&=\sum_{i=0}^r[X_i,Y_{r-i}]
          -\sum_{i=1}^r[c_i,b_{r-i}]_{\rm super},&
 qc_r&=-\sum_{i=1}^{r-1}c_ic_{r-i}.
\end{aligned}
\label{fnd:cyclic-differential}
\end{equation}
Cyclic quotients here are vector-space quotients by supercommutators.
They are not commutative algebra quotients. Translation acts by
$DX_r=(r+1)X_{r+1}$, and analogously on the other letters.
Before passing to cyclic words $[q,D]=-\operatorname{ad}(c_1)$;
on cyclic words this is zero.

\begin{lemma}[Stable trace coordinates]\label{fnd:stable-traces}
At weights at most $h$, if $N\ge2h$, evaluation of cyclic words
as matrix traces identifies the classical relative complex with
$(\operatorname{Sym}_{\rm super}\mathcal T,q)$ in that range.
At arbitrary $N$ the same trace map is surjective on chains.
\end{lemma}
\begin{proof}
For $m\le N$, a $GL_N$-equivariant endomorphism of
$(\mathbb C^N)^{\otimes m}$ is a unique linear combination of
factor permutations. A diagonal torus forces its value on a tensor
of $m$ distinct basis vectors to be such a combination. Subtracting
that combination gives a multilinear map vanishing on all independent
$m$-tuples, a dense set, hence everywhere. Evaluation on the original
tensor proves independence. Index contraction identifies the
permutation cycles with products of traces. Polarization, followed by
symmetrization or antisymmetrization of repeated slots, supplies exactly
the graded cyclic and supercommutative identifications.

For spanning at arbitrary $N$, the commutant of factor permutations
in $\operatorname{End}((\mathbb C^N)^{\otimes m})$ is the space of
symmetric tensors in $\operatorname{End}(\mathbb C^N)^{\otimes m}$.
Polarization spans it by $a^{\otimes m}$; density allows $a$ to be
invertible. The image of $\mathbb C[S_m]$ equals its double
commutant: average a Hermitian form, decompose into irreducibles,
and on each isotypic summand the two commutants are the endomorphism
algebras of the representation and multiplicity factors.
The elementary matrices in the first factor are obtained by averaging
matrix coefficients over the finite group. Thus the commutant of
$GL_N$ is the permutation image at every rank, proving trace spanning.

Center each matrix letter by subtracting its scalar trace divided by
$N$. This is an invertible triangular change in trace coordinates;
setting all scalar traces to zero deletes precisely the length-one
traces. Every letter has weight at least $1/2$, so weights at most
$h$ involve at most $2h$ letters. This proves the stable assertion
for the entire truncated complex. Its differential is
\eqref{fnd:cyclic-differential}; terms with undifferentiated $c$ are
constant conjugations and cancel in traces.
\end{proof}

\begin{theorem}[The complete cyclic calculation]\label{fnd:cyclic}
Let $\widetilde{\mathcal T}$ include length-one words and omit the empty word. Its cohomology has a basis consisting
of all ordinary derivatives of the single-trace Casimir short
multiplets of degrees $d\geq1$. The degree-one multiplet consists
of its two even bottom components and its two odd components;
it has no top component. Removing length-one words removes
exactly this degree-one multiplet and all of its derivatives.
Consequently $H(\mathcal T,q)$ consists of the derivative towers
of the Casimir multiplets of degrees $d\geq2$, with no further
classes in any ghost degree or weight.
\end{theorem}

I give the algebraic comparison and the representatives separately.
In particular, I do not identify translation by simply applying
an associative-algebra resolution map: that would lose translation.

\subsubsection{An explicit associative resolution}
Consider the coaugmented coalgebra
\[
 C=\Lambda(u,v)\otimes\Gamma(t),\qquad
 |u|=|v|=-1,\quad |t|=0.
\]
Here $u,v,t$ are primitive, and $\Gamma(t)$ has basis $t^{[r]}$
with coproduct
$\Delta t^{[r]}=\sum_{i=0}^r t^{[i]}\otimes t^{[r-i]}$.
The tensor product of coalgebras uses the graded signs.
Give $u,v,t$ weights $1/2,1/2,1$, respectively.
On $R=T(s\overline C)$, where $s$ raises degree by one, set
\[
 q(sc)=-\sum_{\overline\Delta c}
       (-1)^{|c'|}(sc')(sc'').
\]
Coassociativity makes the two ways of splitting a coalgebra
element cancel in $q^2$, so this is a differential. Under
\[
 s(ut^{[r]})=X_r,\quad s(vt^{[r]})=Y_r,\quad
 s(uvt^{[r]})=b_r,\quad s(t^{[r]})=c_r\ (r\geq1),
\]
its four formulas are exactly the previously displayed formulas
for $q$. For example the terms $u\otimes v-v\otimes u$ give
$qb_0=X_0Y_0-Y_0X_0$, and the terms
$t^{[i]}\otimes ut^{[j]}+ut^{[j]}\otimes t^{[i]}$ give
$-c_iX_j+X_jc_i$.

Let
\[
 A=\C[x,y]\otimes\Lambda(\theta),\qquad
 |x|=|y|=0,\quad |\theta|=1,
\]
with zero differential and weights $1/2,1/2,1$.
There is a morphism of differential graded algebras
\[
 \pi:R\longrightarrow A,\qquad
 X_0\mapsto x,\quad Y_0\mapsto y,\quad c_1\mapsto\theta,
\]
sending all other generators to zero.

\begin{lemma}
The map $\pi$ is a quasi-isomorphism. All constructions and
comparisons in this proof respect weight.
\end{lemma}
\begin{proof}
Here are explicit resolution arguments for this fact.
The twisting map $C\to A$ sends $u,v,t$ to $x,y,\theta$ and
all other basis elements to zero. Its twisted free $A$-module
$C\otimes A$ is the tensor product of the two one-variable
Koszul resolutions for $x,y$ and the resolution
\[
 \cdots\xrightarrow{\ \theta\ }\Lambda(\theta)
 \xrightarrow{\ \theta\ }\Lambda(\theta)
 \xrightarrow{\ \theta\ }\Lambda(\theta)
 \longrightarrow\C\longrightarrow0.
\]
This last resolution is exact since both the image and the
kernel of multiplication by $\theta$ equal $\C\theta$.
The $x$ and $y$ resolutions are exact because multiplication
by either indeterminate is injective. Tensor products over
$\C$ preserve their contractions. Thus $C\otimes A$ resolves
the augmentation module $\C$. For this statement the resolution
index of $u^\epsilon v^\eta t^{[r]}$ is $\epsilon+\eta+r$;
its displayed cohomological degree is the total degree after
retaining the internal odd degree of $\theta$.

For completeness, use the normalized bar coalgebra $BA$ with
basis $[a_1|\cdots|a_m]$, $a_i\in\overline A$. Its twisted
module $BA\otimes A$ resolves $\C$: the usual extra terminal
slot contracts its augmented complex. The coalgebra map
$\kappa:C\to BA$ is the sum over iterated reduced coproducts,
applying the twisting map in each slot. Thus
\[
 \kappa(uv)=[x|y]-[y|x],\qquad
 \kappa(t^{[r]})=[\theta|\cdots|\theta].
\]
The cancellation of adjacent products says directly that
$\kappa$ is a chain map. It extends to a map between the two
free $A$-module resolutions just described and induces the
identity on their augmentation. Such a map is a chain homotopy
equivalence: recursively lift a map in each nonnegative
resolution index using exactness in the next target index,
and lift the difference from the identity in the same way to
construct the homotopies. These lifts can be taken separately
in each internal weight. Tensoring these homotopies with the
augmentation module proves that $\kappa:C\to BA$ is a
quasi-isomorphism of underlying complexes.

The cobar construction preserves this quasi-isomorphism here.
Indeed, use the descending filtration by the number of cobar
letters (at least a specified number). The associated differential is the differential of
the underlying coalgebra; the map on each associated tensor
power is a quasi-isomorphism by the tensor-product contraction
argument over $\C$. At fixed weight this filtration is finite,
so the comparison gives a quasi-isomorphism $\Omega C\to\Omega BA$.

Finally $\Omega BA\to A$ is a quasi-isomorphism by the following
elementary contraction. Filter by the total number of $A$-entries
in all its bar blocks. The bar differential lowers this number;
the associated cobar differential changes an internal bar into
a separator between two cobar letters. For a list of $m$ entries,
the choices of the $m-1$ separators give the tensor product of
$m-1$ two-term complexes with differential the identity.
It is contractible for $m\geq2$. Only the empty list and lists
of one entry remain, giving exactly $A$ under the counit.
All filtrations are finite at a fixed positive weight, since
every entry has weight at least $1/2$. The composite counit
is the displayed map $\pi$.
\end{proof}

\subsubsection{Why the derived cyclic quotient is computed by forms}
I spell out the two cyclic-homology comparisons needed here.
For an augmented graded algebra $S$, write $\overline{CC}_\bullet(S)$
for its cyclic chain complex with the contribution of the scalar algebra
removed, and write $\overline{HC}(S)$ for the homology of its total
complex. A Hochschild slot lowers total degree by one; thus a form has
total degree equal to its internal degree minus its form degree.
When the internal differential is omitted, the subscript in
$\overline{HC}_n(S)$ denotes cyclic homological degree.

\begin{lemma}[Cyclic comparison in this positive-weight setting]
For the differential graded tensor algebra $R$ above,
\[
 H\bigl(R/(\C+[R,R]_{\rm super}),q\bigr)
 \simeq\overline{HC}(A).
\]
The right side means cyclic homology with its total grading.
\end{lemma}
\begin{proof}
First ignore the internal differential of a tensor algebra
$T(V)$. It has the following length-one bimodule resolution:
\[
 0\longrightarrow T(V)\otimes V\otimes T(V)
 \longrightarrow T(V)\otimes T(V)
 \longrightarrow T(V)\longrightarrow0.
\]
The first map sends $a\otimes v\otimes b$ to
$av\otimes b-a\otimes vb$, with the graded tensor convention.
Cutting a word at each of its positions gives a telescoping
proof of exactness. After taking the bimodule trace, the
degree-$m$ part of its differential is $1-\rho$ on $V^{\otimes m}$,
where $\rho$ is signed cyclic rotation. Therefore
\[
 HH_0(T(V))_m=(V^{\otimes m})_{\rho},\qquad
 HH_1(T(V))_m=(V^{\otimes m})^{\rho},\qquad HH_{n\geq2}=0.
\]
The cyclic boundary from $HH_0$ to $HH_1$ is, directly from
its insertion-and-rotation formula, the cyclic norm
$1+\rho+\cdots+\rho^{m-1}$. For $m>0$ this is an isomorphism
from coinvariants to invariants: its inverse is the quotient
map divided by $m$. The exact sequence obtained by deleting
the first column of the cyclic bicomplex now gives
\[
 \overline{HC}_0(T(V))=T(V)/(\C+[T(V),T(V)]),\qquad
 \overline{HC}_{n>0}(T(V))=0.
\]
Equivalently, projection of the reduced cyclic complex to
its cyclic-word quotient is a quasi-isomorphism.

Restore the internal differential $q$. Apply the preceding
calculation first in the double complex, keeping internal
degree as the second grading. It identifies the resulting
complex with $(R/(\C+[R,R]),q)$. There is no convergence
problem: at a fixed weight the internal degrees are bounded;
normalized bar slots have positive weight; and at each fixed
total degree only finitely many cyclic columns can occur.

The algebra quasi-isomorphism $R\to A$ induces a cyclic
quasi-isomorphism. To see this without a separate invariance
assertion, apply the internal differential first in the same
cyclic bicomplex. Each normalized Hochschild chain group is
a finite sum, at a fixed weight, of tensor products of the
underlying augmented complexes. Tensor-product contractions
show that the induced map is a quasi-isomorphism on every
such group. The bounded comparison just described then gives
the assertion on the total cyclic complexes.
\end{proof}

\begin{lemma}[The required supercommutative calculation]
In positive weight,
\[
 \overline{HC}(A)=
 \bigoplus_{k\geq0}\Omega_A^k/d\Omega_A^{k-1},
 \qquad
 \Omega_A=\C[x,y]\otimes\Lambda(\theta,dx,dy)\otimes\C[\tau],
 \quad \tau=d\theta.
\]
The weight of $dx,dy$ is $1/2$, that of $\theta,\tau$ is $1$,
and total degree is internal $\theta$-degree minus form degree.
In particular $dx,dy$ have total degree $-1$, $\theta$ has
degree $1$, and $\tau$ has degree zero.
\end{lemma}
\begin{proof}
The bimodule resolution of $A$ is explicit. For each of
$x_L-x_R,y_L-y_R$ take its two-term Koszul resolution. For
$\theta_L-\theta_R$ take the repeated multiplication resolution
used above; it is exact because, after a linear change of the
two exterior generators, multiplication by their difference
has equal kernel and image. Their tensor product resolves
$A$ over $A\otimes A^{\rm op}$. Tensoring with $A$ kills all
three differences and leaves zero differential. Its terms
are precisely
$A\otimes\Lambda(dx,dy)\otimes\C[d\theta]$.

The map from normalized Hochschild chains to these terms is
\[
 a_0[a_1|\cdots|a_k]\longmapsto
 \frac1{k!}a_0\,da_1\cdots da_k.
\]
Adjacent-product terms cancel, so the Hochschild differential
maps to zero. The signed antisymmetrizations in $x,y$ and
symmetrizations in the suspended odd variable $\theta$
identify the displayed bimodule resolution with the bar
resolution; on its generators this map is the identity.
The same free-resolution comparison argument therefore makes
the displayed Hochschild map a quasi-isomorphism.
Applying the insertion-and-rotation formula for the cyclic
boundary gives exactly $d$; its $k+1$ equal summands account
for the factorial normalization. Thus the full cyclic
calculation reduces to the mixed complex $(\Omega_A,0,d)$.

In this mixed complex, cyclic degree $k$ consists of
$\Omega_A^k\oplus\Omega_A^{k-2}\oplus\cdots$. Its top summand
contributes $\Omega_A^k/d\Omega_A^{k-1}$; the lower summands
contribute de Rham cohomology. Those contributions vanish in
positive weight. Indeed, for the Euler vector field
$E=x\partial_x+y\partial_y+\theta\partial_\theta$,
$d\iota_E+\iota_Ed$ is multiplication by the positive total
polynomial-and-form degree. Division by this integer is an
explicit contraction. Only the scalar constant survives in
degree zero, and it has already been removed. This proves
the formula.
\end{proof}

\subsubsection{The four families and the actual representatives}
Put $\mathscr P=\C[x,y]$ and $\omega=x\,dy-y\,dx$.
The preceding form quotient has the following particularly
useful vector-space description:
\begin{equation}
 \bigoplus_k\Omega_A^k/d\Omega_A^{k-1}
 \simeq
 \C[\tau]\otimes
 \bigl((\mathscr P/\C)\oplus\theta \mathscr P\oplus \mathscr P\omega\oplus\theta \mathscr P\omega\bigr).
 \label{fnd:forms-decomposition}
\end{equation}
Multiplication by $\tau$ is well defined on the quotient,
since $d\tau=0$. Here is a direct proof of its freeness and
of the absence of missing classes. Write $d_0$ for the
ordinary de Rham differential in $x,y$. The relations
\[
 d(\theta\beta)=\tau\beta-\theta d_0\beta
\]
eliminate all $\theta$-free positive powers of $\tau$.
The remaining relations on the $\theta$-free part are exactly
$d_0\Omega_{\mathscr P}$. There are no further relations on the
$\theta$-part: applying $d_0$ to a term already eliminated
produces $d_0^2=0$. Thus the form-degree-$k$ quotient is
\[
 \bigl(\Omega_{\mathscr P}^k/d_0\Omega_{\mathscr P}^{k-1}\bigr)
 \ \oplus\ 
 \theta\bigoplus_{a=0}^{\min(k,2)}\Omega_{\mathscr P}^a\tau^{k-a}.
\]
The first summand is zero for $k\geq2$. Moreover
\[
 \Omega_{\mathscr P}^1=d_0\mathscr P\oplus \mathscr P\omega.
\]
For a homogeneous one-form of positive Euler degree this
follows from $d_0\iota_{E_0}+\iota_{E_0}d_0=\deg$, where
$E_0=x\partial_x+y\partial_y$, and
$\iota_{E_0}\Omega_{\mathscr P}^2=\mathscr P\omega$. The intersection is zero
because $\iota_{E_0}d_0f=(\deg f)f$ whereas
$\iota_{E_0}\omega=0$.
Finally, for $f$ homogeneous of degree $m$,
\[
 \tau f=\theta d_0f,\qquad
 \tau(f\omega)=\theta d_0(f\omega)
     =(m+2)\theta f\,dx\,dy.
\]
These nonzero maps, together with multiplication on the
$\theta$-summands, prove exactly the free decomposition \eqref{fnd:forms-decomposition}.

Let $\mathscr P_m$ denote the homogeneous polynomials of degree $m$.
As a flavor representation it is the spin-$m/2$ representation.
The four free families have the following quantum numbers:
\[
\begin{array}{c|c|c|c|c}
\text{family}&\text{parameter}&h&j&\operatorname{gh}\\\hline
 \mathscr P_d&d\geq1&d/2&d/2&0\\
 \theta \mathscr P_{d-1}&d\geq1&(d+1)/2&(d-1)/2&1\\
 \mathscr P_{d-1}\omega&d\geq1&(d+1)/2&(d-1)/2&-1\\
 \theta \mathscr P_{d-2}\omega&d\geq2&(d+2)/2&(d-2)/2&0
\end{array}
\]
Each power of $\tau$ increases $h$ by one, preserving $j$ and
ghost number. This proves a complete graded dimension formula;
it does not yet identify the actual translation on the cyclic
complex with multiplication by $\tau$.

To identify representatives, take the actual raw single-trace
Casimir fields and their short descendants. Equations \eqref{fnd:normalization}--\eqref{fnd:top-formula} say they are $Q$-closed, and each is homogeneous
of PBW length $d$. A single BRST contraction raises this length by one. Two contractions
lower it by one; three contractions leave no field at the current
insertion and have no residue. Closure therefore holds separately
at the two output lengths, and in particular they are $q$-closed. All their derivatives are $q$-closed on the cyclic
quotient. Their highest flavor components are, up to nonzero
normalizations,
\[
 \operatorname{tr} X^d,\qquad \operatorname{tr} X^{d-1}b,\qquad
 \operatorname{tr} X^{d-1}p,\qquad H_d\ (d\geq2),
\]
with $H_d$ given by \eqref{fnd:top-formula}.

There is a useful independence test requiring no homology
comparison map. Replace every noncommuting jet letter by an
ordinary supercommuting indeterminate with the same name.
This defines a linear map from the cyclic quotient to the
supercommutative jet algebra, and every $q$-image maps to zero:
the commutators vanish, while the terms in
$\sum_i c_i c_{r-i}$ cancel in pairs. This is only a classical
test map, not a claim about a rank-one quantum theory.
The displayed representatives restrict to
\[
 X^d,\qquad X^{d-1}b,\qquad X^{d-1}p,\qquad
 (d-1)X^{d-2}\bigl(bp-d^{-1}X\partial Y+d^{-1}Y\partial X\bigr).
\]
Every ordinary derivative of each expression is nonzero.
For example the coefficient involving $b_0c_{r+1}$ in the
$r$-th derivative of the last expression is a nonzero multiple
of $X_0^{d-2}$.

At fixed $(h,j,\operatorname{gh})$ there is one candidate highest vector for
each of ghost numbers $1$ and $-1$. At ghost number zero there
are at most two: a derivative of the bottom of degree $2j$,
and a derivative of the top of degree $2j+2$. The latter has
nonzero $bp$ terms and the former has none, so these two are
independent even after the test map. Consequently no linear
combination of these highest vectors is exact. Finite-dimensional
flavor representation theory gives the same independence for
all their components. The complete dimension formula above
now proves that these actual derivatives exhaust the cyclic
cohomology. No identification of $D$ with $\tau$ was assumed.

Lastly, the cyclic image of $q$ on a length-one word is zero,
and $q$ increases the length of every other word. Therefore
the length-one space is a direct summand complex with zero
differential. Its basis is $\operatorname{tr} X_r,\operatorname{tr} Y_r,\operatorname{tr} b_r$ and
$\operatorname{tr} c_{r+1}$, exactly the degree-one multiplet and its
derivatives. Deleting it proves the assertion for
$\mathcal T$ and completes the theorem.

\subsection{Universal quantum lifting and finite-rank truncation}

Let $\mathscr R=\mathbb C[z,z^{-1}]$ be the coefficient ring,
distinct from the cobar algebra $R$ used in the cyclic calculation.
Define the free-trace module
$\mathscr C=\mathscr R\otimes\operatorname{Sym}_{\rm super}\mathcal T$.
Wick contraction with matrix propagator
$\delta^i_l\delta^k_j-z^{-1}\delta^i_j\delta^k_l$, assigning $z$
to each closed index loop, gives all its vertex operations.
At any fixed input and mode these are finite Laurent-polynomial
expressions. The explicit general contraction formula is recorded in
\eqref{of:wick-formula}; it applies equally to the formal parameter $z$.
The BRST residue is defined by the same rule. Its principal differential
is $q$ and all its higher-contraction terms lower
\eqref{fnd:filtration}.

These operations satisfy the vertex and differential axioms over $\mathscr R$.
Indeed any fixed coefficient of an axiom is a finite Laurent-polynomial
combination of trace monomials. For every sufficiently large integer
$N$ these monomials are independent by Lemma~\ref{fnd:stable-traces},
and the axiom holds in the ordinary free-field algebra. Each Laurent
coefficient vanishes at infinitely many integers and therefore vanishes
identically. Evaluation at $z=N$ consequently defines a surjective
map of differential vertex superalgebras to $C_N$ at every $N\ge2$.
This argument concerns finite identities whose coefficients are Laurent
polynomials; it imposes no finite-weight bound on the subsequent theorem.

\begin{lemma}[Filtered quantum lifting]\label{fnd:lifting}
The cohomology of $\mathscr C$ is strongly generated over $\mathscr R$ by
the complete raw Casimir multiplets of degrees $d\ge2$.
Every singular product of two such components consequently has a
chain identity
\begin{equation}\label{fnd:universal-chain}
 a_{(n)}b=\mathcal P(\partial^k s_d)+QU,\qquad n\ge0,
\end{equation}
where $\mathcal P$ is a finite parenthesized normal polynomial and
$U$ is a relative trace polynomial, both over $\mathscr R$.
\end{lemma}
\begin{proof}
Split the classical cyclic complex over $\mathbb C$, separately in
weight, filtration and ghost number, into its cohomology representatives
and contractible pairs. The symmetric superalgebra of a contractible
pair has no nonconstant cohomology: the elementary contraction divided
by total polynomial degree contracts each positive-degree component.
Tensoring these contractions proves
\[
 H(\operatorname{gr}_F\mathscr C,q)
 =\mathscr R\otimes\operatorname{Sym}_{\rm super}H(\mathcal T,q).
\]
No inversion of a nonconstant polynomial in $z$ is involved.
Theorem~\ref{fnd:cyclic} identifies its generators, and
\eqref{fnd:normalization} gives actual closed quantum representatives.
Given a closed chain, express its leading class as a differential
polynomial in these representatives. Subtract a normal lift of that
polynomial and a BRST image lifting its exact leading remainder.
The filtration strictly decreases. At fixed conformal weight it is
bounded by \eqref{fnd:filtration}, so this process terminates.
The construction is homogeneous in ghost number: for an input of
ghost number $g$, the normal polynomial has ghost number $g$ and its
BRST primitive has ghost number $g-1$. Applying this construction to
$a_{(n)}b$ proves \eqref{fnd:universal-chain}.
\end{proof}

\begin{lemma}[Reduction of the complete tower]\label{fnd:tower}
For $N\times N$ matrices, every component of every raw multiplet of degree $d>N$
is a normal differential polynomial in components of degrees $2,\ldots,N$.
\end{lemma}
\begin{proof}
Newton's identities and Cayley--Hamilton express
$P_d=\Phi_d(P_2,\ldots,P_N)$ for $d>N$. Because $XX$ is regular,
this is an actual vertex identity with any fixed parentheses.
Apply the flavor and supercurrent zero modes to this identity,
retaining parentheses. To see explicitly that their action stays
inside the span of complete multiplets, temporarily use
$B'=\beta P$, $C'=\delta P$, $K'=\beta\delta P+d^{-1}\partial FP=dH(P)$.
The relations \eqref{fnd:super-derivations} give
\[
\begin{array}{c|cccc}
 &\alpha&\beta&\gamma&\delta\\\hline
 B'&0&0&\partial P&-K'+d^{-1}\partial FP\\
 C'&-\partial P&K'-d^{-1}\partial FP&0&0\\
 K'&\frac{d-1}{d}\partial B'&d^{-1}\partial FB'&
        \frac{d-1}{d}\partial C'&d^{-1}\partial FC'
\end{array}
\]
where an entry is the action of its column on its row.
Commuting each derivation past powers of $F$ gives all components.
The normal differential span is thus stable under precisely the
derivations needed to produce the higher multiplets. Applying them
factor by factor to $\Phi_d$ proves the assertion. This is a vertex
derivation argument; it uses no commutative chain rule for composite
normal products.
\end{proof}

\begin{theorem}[Uniform strong closure]\label{fnd:closure}\label{thm:strong-closure}
For every $N\ge2$, the normal differential span $A_N$ of all components
of the Casimir short multiplets of degrees $2,\ldots,N$ is a conformal
vertex subalgebra of $\mathcal H_N$. The statement holds for either
the raw or the harmonic primary convention.
\end{theorem}
\begin{proof}
Specialize the actual identity \eqref{fnd:universal-chain} at $z=N$.
Only powers of $N$ occur in denominators; hence every positive rank
is allowed. Its primitive remains invariant and contains no
undifferentiated $c$. Lemma~\ref{fnd:tower} then reduces all higher
multiplets on the right to degrees at most $N$. Thus every singular
product of generators belongs to their normal differential span in
cohomology. The iterate identity and the derivation rule for translation
extend closure to arbitrary normal words by induction on total conformal
weight and the number of factors. This also provides any chosen
right-nested spanning convention.

For completeness, the triangular harmonic change preserves the strong
span rather than merely the generated vertex algebra. Filter the raw
strong span by word cost, twice the sum of the undifferentiated
component weights. A singular product lowers this cost, since its
weight is at least one smaller than the sum of its input weights;
the iterate identity extends this estimate to all normal words.
The associated graded multiplication is therefore supercommutative.
Each correction in \eqref{fnd:harmonic} is $q_X$ times an invariant
polynomial of strictly smaller degree; hence it is a polynomial in
lower Casimirs. Applying the displayed derivations makes a triangular
change of complete multiplets whose factors have smaller individual
weights. In the commutative associated graded this change is invertible.
Lifting its inverse and subtracting the leading symbols decreases word
cost. The filtration is bounded at each weight, so induction proves
equality of the two normal differential spans.
\end{proof}

This theorem constructs the algebra whose exhaustion is established
below. In particular, $\mathcal H_N/A_N$ is a graded $A_N$-module.
If it is nonzero, let $r$ be its least weight and $F$ its entire
weight-$r$ space. It is finite dimensional, $r>0$, and
Lemma~\ref{fnd:extremal} gives
\begin{equation}\label{fnd:first-missing}
                       j<r
 \quad\text{for every flavor spin occurring in }F.
\end{equation}
This conclusion retains every ghost degree. It follows from the
extremal sector of the chain complex and does not require a choice
of a preferred cohomological representative.

\section{Odd transport at every degree and the invariant annihilator}
\label{sec:odd}

I now control the integer zero modes on an arbitrary finite lowest
conformal space $F\neq0$ of an untwisted, lower-bounded module for
$A_N$. Its conformal weight is denoted by $r$. Flavor is integrable,
but $F$ may contain several spins, with arbitrary multiplicities.
The proof produces a nonzero subspace preserved by the entire
zero-mode algebra on which every odd zero mode vanishes. This is the
input that will allow the subsequent exclusion argument to retain the
full flavor representation.

Set $p_i=o(W_{4,i})$ and
\[
 O^{B,(n)}_i=o(B_{2n+1,i}),\qquad
 O^{C,(n)}_i=o(C_{2n+1,i}),\qquad n\geq1,
\]
so each family has spin $n$. Write $J=(e,-H/2,-f)$, and use
$[H,e]=2e$, $[e,f]=H$. All brackets in this section are operator
supercommutators; a bracket with an even operator is an ordinary
commutator. Out-of-range tensor components mean zero.
The physical zero mode is used only for fields of integral weight.
The raw trace fields at degrees greater than $N$ are retained: they
are well-defined fields, expressed in the prescribed generators by
the characteristic polynomial and its supersymmetry descendants.

The argument moves from a finite generating set for the lowest-space
algebra to a simultaneous induction for the full odd tower, and then
to a radical argument that produces an invariant common kernel.
Appendix~\ref{app:odd-fields} supplies the field identities at each stage.

\subsection{A finite generating set for the lowest-space algebra}
\label{odd:zero-generators}

Let $\mathcal A(F)\subset\operatorname{End}(F)$ be the associative
algebra of all integral-weight field zero modes. I first establish
\begin{equation}
 \mathcal A(F)=\left\langle
   1,e,H,f,p_0,\ldots,p_4,
   O^{B,(1)}_0,O^{B,(1)}_1,O^{B,(1)}_2,
   O^{C,(1)}_0,O^{C,(1)}_1,O^{C,(1)}_2,L_0
                  \right\rangle .
 \label{odd:fifteen-generators}
\end{equation}
Here $L_0=r1$ on $F$; its inclusion makes the origin of the central
parameter explicit. Dependent generators at small ranks cause no
problem.

For even $a,b\geq2$ and odd $d\geq3$, a single contraction gives
\begin{align}
 [o(P_a),o(W_{b,1})]&=a\,o(P_{a+b-2}),\notag\\
 [o(P_a),o(B_{d,1})]&=a\,o(B_{a+d-2,0}),\notag\\
 [o(P_a),o(C_{d,1})]&=a\,o(C_{a+d-2,0}).
 \label{odd:upward}
\end{align}
There is only one $Y$ in the right field. The scalar-projector
corrections are respectively
$-aP_{a-1}P_{b-1}/N$, $-aP_{a-1}B_{d-1,0}/N$ and
$-aP_{a-1}C_{d-1,0}/N$. The factors have regular mutual OPE and
half-integral weights. In a physical zero-mode product their energy
shifts are nonzero half-integers of opposite signs; the lowering factor
can be commuted to the right and annihilates $F$. Thus all three
corrections vanish. Taking $a=4$ and lowering in flavor proves that
all even bottom zero modes follow from $J,p$, and every odd zero mode
follows from the cubic odd triplets by quartic commutators.

The integer-weight tops are fixed by the arbitrary-length mixed
identity in Appendix~\ref{of:mixed-ope}. For even $m\geq2$, put
\[
 K_{m+1}=\{O^{B,(1)}_0,o(C_{m+1,1})\}
              -\{O^{B,(1)}_1,o(C_{m+1,0})\}.
\]
Its zero-mode consequence is
\begin{equation}
 K_{m+1}=\frac{m+2}{m+1}\left(
 o(H_{m+2})-\frac N4o(P_m)
 -\frac18\sum_{\substack{a+b=m\\a,b\geq2\ {\rm even}}}
                      o(P_a)o(P_b)\right).
 \label{of:mixed-top}
\end{equation}
The pairs in the sum are ordered. To make its specialization
transparent, the zero-mode bracket is $o(O_0)+o(O_1)$ in
\eqref{of:cm1-o0} and \eqref{of:cm1-o1}; the third OPE coefficient has binomial coefficient
zero. Normal splitting gives
\[
 o(:G_XC_{m,0}:)=-o(G_{X(0)}C_{m,0})=\tfrac12o(P_m),
 \qquad o(E_2^{[m]})=-o(E_0^{[m]}),
\]
since $G_{X(0)}C_{m,0}=\partial P_m/m$. Translation gives
$o(\partial P_m)=-mo(P_m)/2$ and
$o(\partial^2P_m)=m(m+2)o(P_m)/4$. The single-trace coefficient is
\[
 -\frac{m+2}{2N}+\frac{N(m+2)^2}{4m(m+1)}
       -\frac{N(m+2)}{2m}+\frac{m+2}{2N}
       =-\frac{N(m+2)}{4(m+1)}.
\]
Odd--odd degree pairs vanish by the regular half-integral argument.
For even $a,b$, the coefficient of $o(P_a)o(P_b)$ is
\[
 \frac{m+2}{m(m+1)}
 \left(\frac{(a+1)(b+2)-ab}{4}-\frac{m+1}{2}\right)
       =-\frac{(m+2)b}{4m(m+1)}.
\]
Commutativity of the pure-$X$ zero modes replaces $b$ by $m/2$ in
the ordered sum and proves \eqref{of:mixed-top}. On any subspace
annihilated by all odd zero modes it reduces to the following identities,
with the degree-two identity obtained directly from $H_2=-T$:
\begin{equation}
\begin{gathered}
 o(H_2)=-r,\\
 o(H_{m+2})=\frac N4o(P_m)
     +\frac18\sum_{\substack{a+b=m\\a,b\geq2\ {\rm even}}}
                               o(P_a)o(P_b),\qquad m\geq2\text{ even}.
\end{gathered}
 \label{of:fullodd-top}
\end{equation}

It remains to pass from strong generators to all fields. Formula
\eqref{of:normal-modes} expresses the zero mode of a normal product
as a product of integral zero modes, when present, and zero modes of
singular coefficients of strictly smaller conformal weight. Derivatives
contribute their scalar translation factors. Strong closure,
Theorem~\ref{thm:strong-closure}, and induction on conformal weight
therefore reduce every strong monomial to integral generator zero
modes. For half-integral factors only the lower-weight singular
coefficients remain. Equations \eqref{odd:upward} and
\eqref{of:mixed-top} now prove \eqref{odd:fifteen-generators}.

\subsection{The field input and an injective transport system}

The following identities hold on the whole lowest space. They are
proved in Appendix~\ref{app:odd-fields} before any odd-annihilator
condition is imposed. Define the spin-one tensor $K$ by
\[
 K_0=\{O^{B,(1)}_0,O^{C,(1)}_1\}
                 -\{O^{B,(1)}_1,O^{C,(1)}_0\},
 \quad K_1=[f,K_0]/2,\quad K_2=[f,K_1].
\]
Then
\begin{align}
 \{O^{B,(1)}_i,O^{B,(1)}_j\}
 &=\{O^{C,(1)}_i,O^{C,(1)}_j\}=0,\notag\\
 \{O^{B,(1)}_i,O^{C,(1)}_j\}
 &=-\{O^{B,(1)}_j,O^{C,(1)}_i\},\notag\\
 [K_i,O^{\alpha,(1)}_j]&=[K_i,K_j]=0.
 \label{odd:cubic}
\end{align}
The commuting $K_i$ are nilpotent. Indeed their finite joint generalized
spectrum is preserved by conjugation by the connected flavor group
$SL_2(\mathbb C)$. A connected group fixes a finite set pointwise,
and spin one has no nonzero invariant covector. Every joint character
is therefore zero. Simultaneous triangularization gives a nilpotent
ideal generated by $K_0,K_1,K_2$ in their commutative algebra.

Write
\begin{equation}
 \Delta=2r+\frac{N^2}{4},\qquad \gamma=8\Delta-8,\qquad
 d_n=\frac{n(n-1)(4\Delta-n^2)}{2n+1}.
 \label{odd:parameters}
\end{equation}
The whole-lowest quartic identities are
\begin{equation}
 [p_1,p_4]-3[p_2,p_3]=\gamma f+2NK_2,
 \qquad [p_a,K_i]=(2i-a)T_{a+i-1},
 \label{odd:quartic}
\end{equation}
where $T$ is a spin-two tensor. For both odd families at every $n\geq1$,
\begin{equation}
 [p_0,O^{\alpha,(n)}_0]=0,
 \quad [p_1,O^{\alpha,(n)}_0]=-2nO^{\alpha,(n+1)}_0,
 \quad [p_2,O^{\alpha,(n)}_0]=-4nO^{\alpha,(n+1)}_1.
 \label{odd:first-three}
\end{equation}
The second identity is the upward contraction; the third is the
arbitrary-index field calculation \eqref{of:uqo3}--\eqref{of:uqo11}.
Their barred versions follow from the actual relative chain derivation,
Subsection~\ref{of:chain-derivation-section}.

\begin{lemma}[Injectivity of transport]
\label{odd:transport-injection}
For $m\geq1$, let $X_{\ell b}$, $0\leq\ell\leq4$,
$0\leq b\leq2m$, and $Y_{it}$, $0\leq i\leq2$,
$0\leq t\leq2m+2$, take values in any complex vector space.
Out-of-range entries are zero. If
\begin{equation}
 (2i-a)X_{a+i-1,b}+(4b-2ma)Y_{i,a+b-1}=0
 \quad(0\leq a\leq4,\ 0\leq i\leq2,\ 0\leq b\leq2m),
 \label{odd:injection-system}
\end{equation}
then $X=Y=0$.
\end{lemma}
\begin{proof}
For $m\geq2$, the pairs $(a,i)=(0,0),(4,2)$ first give
$Y_{0,t}=0$ for $0\leq t\leq2m-1$ and $Y_{2,t}=0$ for
$3\leq t\leq2m+2$. The pairs $(0,1),(4,1)$ give
\[
 X_{0,b}=-2bY_{1,b-1},\qquad
 X_{4,b}=2(b-2m)Y_{1,b+3}.
\]
The pairs $(1,0),(3,2)$ consequently imply
\[
 Y_{1,t}=0\quad\text{for}\quad
 t\in[0,2m-2]\cup[4,2m+2].
\]
If $m\geq3$, these intervals cover the entire range. If $m=2$, put
$z=Y_{1,3}$. The four equations with the indicated entries give
$X_{4,0}=-8z$, $Y_{2,2}=-2z/3$, $X_{1,3}=2z$, and
$X_{1,3}+8z=0$, so $z=0$. Next $(1,1)$ kills $X_1$, while
$(0,2)$ gives $X_{1,b}=-bY_{2,b-1}$; taking $b=1,2,3$
kills the remaining entries of $Y_2$. Similarly $(3,1)$ kills
$X_3$, while $(4,0)$ gives $X_{3,b}=(b-2m)Y_{0,b+3}$;
$b=2m-3,2m-2,2m-1$ kills the remaining entries of $Y_0$.
The displayed equations kill $X_0,X_4$, and $(1,2)$ kills $X_2$.
All the chosen indices are valid also for $m=2$.

At $m=1$, put $Z_{\ell j}=2Y_{j\ell}$. Add and subtract the
system with its $i,b$ interchange. The arrays $X+Z$ and $X-Z$
satisfy, respectively,
\[
 (2i-a)V_{a+i-1,j}\mathbin{\pm}(2j-a)V_{a+j-1,i}=0.
\]
For the plus sign, $i=j$ kills every entry except
$V_{40},V_{21},V_{02}$, which the triples
$(a,i,j)=(3,0,2),(3,0,1),(1,0,2)$ kill in order.
For the minus sign the pairs $(i,j)=(0,1),(0,2),(1,2)$ give
\[
\begin{gathered}
 V_{00}=V_{11}=V_{10}=V_{01}=V_{02}=V_{20}=0,\\
 V_{30}=3V_{21},\qquad V_{12}=-V_{30}=3V_{21},\\
 V_{40}=2V_{31},\quad V_{31}=0,\quad V_{22}=-V_{40}/3,\quad
 V_{32}=V_{41}=V_{42}=0.
\end{gathered}
\]
Thus all entries vanish. Both systems are injective and $X=Y=0$.
\end{proof}

The relevant consequence retains every lower return. If
$[K,O^{\alpha,(n)}]=0$ and
\begin{equation}
 [p_a,O^{\alpha,(n)}_b]
  =(4b-2na)O^{\alpha,(n+1)}_{a+b-1}+L^{\alpha,n}_{ab},
 \qquad [K_i,L^{\alpha,n}_{ab}]=0,
 \label{odd:transfer-hypothesis}
\end{equation}
then Jacobi gives \eqref{odd:injection-system} with
$X_{\ell b}=[T_\ell,O^{\alpha,(n)}_b]$ and
$Y_{it}=[K_i,O^{\alpha,(n+1)}_t]$. Consequently
\begin{equation}
 [T,O^{\alpha,(n)}]=0,\qquad [K,O^{\alpha,(n+1)}]=0.
 \label{odd:transfer}
\end{equation}
This uses commutation of the lower return with $K$, and does not
require that return to vanish.

\subsection{All-degree returns by simultaneous induction}

For one family suppress $\alpha$. At $n\geq2$ define the actual
operators
\begin{align}
 X_n&=[p_3,O^{(n)}_0]+6nO^{(n+1)}_2,
 &E_n&=X_n-d_nO^{(n-1)}_0,\notag\\
 X_{n,1}&=[f,X_n]/(2n-2),
 &E_{n,1}&=[f,E_n]/(2n-2),\notag\\
 Z_n&=[p_4,O^{(n)}_0]+8nO^{(n+1)}_3-4X_{n,1}.
 \label{odd:errors}
\end{align}
The first three columns \eqref{odd:first-three} imply
$[e,X_n]=[e,E_n]=0$ and $[H,E_n]=2(n-1)E_n$;
similarly $[e,Z_n]=0$, $[H,Z_n]=2(n-2)Z_n$.
In particular $[e,E_{n,1}]=E_n$. These are highest vectors
inside $\operatorname{End}(F)$, with no assumption on its
multiplicity spaces.

For clarity, the five first-column brackets determine the full return.
Writing $A_{ij}=[p_i,O^{(n)}_j]$, covariance gives
\begin{equation}
 A_{i,j+1}=
 \frac{[f,A_{ij}]-(4-i)A_{i+1,j}}{2n-j},
 \qquad 0\leq j<2n,\quad A_{5,j}=0.
 \label{odd:all-columns}
\end{equation}
Every denominator in this range is nonzero. The last two columns
are retained in \eqref{odd:errors}. In particular lowering gives
\begin{align}
 A_{21}&=-4(n-1)O^{(n+1)}_2-X_n/n,\notag\\
 A_{12}&=(8-2n)O^{(n+1)}_2+3X_n/[n(2n-1)],\notag\\
 A_{03}&=12O^{(n+1)}_2-6X_n/[n(n-1)(2n-1)].
 \label{odd:lower-once}
\end{align}
A further lowering, with $V=O^{(n+1)}_3$, gives
\begin{align}
 A_{31}&=(4-6n)V+\frac{n-3}{n}X_{n,1}-\frac{Z_n}{2n},\notag\\
 A_{22}&=(8-4n)V-\frac{4(n-2)}{n(2n-1)}X_{n,1}
                      +\frac{Z_n}{n(2n-1)},\notag\\
 A_{13}&=(12-2n)V+\frac{3(3n-5)}{n(n-1)(2n-1)}X_{n,1}
                      -\frac{3Z_n}{2n(n-1)(2n-1)}.
 \label{odd:lower-twice}
\end{align}

Suppose only that $[K_2,O^{(n-1)}_0]=0$ at the preceding index.
For $n\geq3$, use the upward relation and Jacobi with
\eqref{odd:quartic} to obtain
\begin{align}
 [p_4,O^{(n)}_0]={}&\frac{3}{2(n-1)}
 \bigl([p_2,[p_3,O^{(n-1)}_0]]
               -[p_3,[p_2,O^{(n-1)}_0]]\bigr)\notag\\
 &+\gamma O^{(n-1)}_1
              -\frac{[p_1,[p_4,O^{(n-1)}_0]]}{2(n-1)}.
 \label{odd:jacobi-step}
\end{align}
Before the stated commutation is used, the right side also contains
$N[K_2,O^{(n-1)}_0]/(n-1)$. Thus this step uses no centrality at
the degree being produced. Substitution of \eqref{odd:errors} at
$n-1$ gives
\begin{align}
 A_{40}={}&-9A_{22}+6A_{31}+4A_{13}
  +\left(\gamma-\frac{2(n+2)d_{n-1}}{n-1}\right)O^{(n-1)}_1
  +\mathcal R_n,\notag\\
 \mathcal R_n={}&
 \frac{3[p_2,E_{n-1}]-4[p_1,E_{n-1,1}]-[p_1,Z_{n-1}]}{2(n-1)}.
 \label{odd:remainder}
\end{align}
Let
\begin{equation}
 \delta_n=-\frac{2(n-3)(n+1)(2n+1)}{n(n-1)(2n-1)},\qquad
 q_n=\frac{(2n+1)(n+1)}{(2n-1)(n-1)}.
 \label{odd:recursion-coefficients}
\end{equation}
Insert \eqref{odd:lower-twice}; the $V$ coefficient cancels, and
\[
 \delta_nd_n=\gamma-\frac{2(n+2)d_{n-1}}{n-1}
\]
is an identity of rational functions in the unrestricted index $n$
and parameter $\Delta$. I obtain
\begin{equation}
 q_nZ_n+\delta_nE_{n,1}=\mathcal R_n.
 \label{odd:recursion}
\end{equation}
Raising therefore gives, for every $n\geq4$,
\begin{equation}
 E_n=\frac{[e,\mathcal R_n]}{\delta_n},\qquad
 Z_n=\frac{\mathcal R_n-\delta_nE_{n,1}}{q_n}.
 \label{odd:solved-recursion}
\end{equation}
Both denominators are nonzero. The only exceptional index is $n=3$,
where $\delta_3=0$ and $q_3=14/5$.

The finite starting identities are exact field identities. The cubic
return is \eqref{of:cubic-return}; the quintic Jacobi calculation gives
$E_2=Z_2=0$. At $n=3$, \eqref{odd:recursion} gives $Z_3=0$.
The explicit septic identity \eqref{of:septic-return} gives
\begin{equation}
 E^{\alpha}_3=-\frac{30}{7}O^{\alpha,(1)}_0K_0,
 \qquad Z^{\alpha}_3=0.
 \label{odd:septic-error}
\end{equation}
The conversion factor from $X_3$ to the third coupling is $-14/5$,
as is obtained immediately from \eqref{odd:lower-once}. This one
septic identity supplies the value not determined at the resonance.

Let $U_m$ be the unital associative algebra generated by both odd
families through index $m$, and $U$ the algebra generated by all raw
odd tensors. They are subalgebras of the finite-dimensional
$\operatorname{End}(F)$. I prove simultaneously
\begin{equation}
 [K,O^{\alpha,(n)}]=[T,O^{\alpha,(n)}]=0,
 \quad E_n,Z_n\in U_{n-2}\ (n\geq3),
 \quad[p_i,U_m]\subset U_{m+1}.
 \label{odd:bootstrap-result}
\end{equation}
At $n=1$, cubic centrality and the pure upward return meet
\eqref{odd:transfer-hypothesis}; hence $[T,O^{(1)}]=0$ and
$[K,O^{(2)}]=0$. At $n=2$, the only lower return is cubic,
so the same argument gives $[T,O^{(2)}]=0$ and
$[K,O^{(3)}]=0$. At $n=3$, the lower terms are quintic and
flavor descendants of \eqref{odd:septic-error}. They commute with
$K$, by the preceding steps and \eqref{odd:cubic}. Therefore
$[T,O^{(3)}]=0$, $[K,O^{(4)}]=0$. These complete return tables
also prove $[p_i,U_m]\subset U_{m+1}$ for $m=1,2,3$.

For $n\geq4$, assume $K$ commutes with tensors through $n$, $T$
with tensors through $n-1$, $E_q,Z_q\in U_{q-2}$ for $3\leq q<n$,
and $[p_i,U_m]\subset U_{m+1}$ for $m<n$.
The preceding-index commutation justifies \eqref{odd:jacobi-step}.
The errors at $n-1$ lie in $U_{n-3}$; the already proved normalization
at smaller indices puts $\mathcal R_n$ in $U_{n-2}$.
Flavor preserves each $U_m$, so \eqref{odd:solved-recursion} gives
$E_n,Z_n\in U_{n-2}$. The complete return obtained from
\eqref{odd:all-columns} has upward part
$(4b-2na)O^{(n+1)}_{a+b-1}$ and lower part made of
$O^{(n-1)},E_n,Z_n$ and their flavor descendants. That part lies
in $U_{n-1}$ and commutes with every $K_i$. Lemma
\ref{odd:transport-injection} now gives the next centralities.
The same complete return, with the derivation rule, gives
$[p_i,U_n]\subset U_{n+1}$. This closes the induction.
In particular
\begin{equation}
 [p_i,U]\subset U,
 \qquad K_i\in Z(U)\quad\hbox{are nilpotent}.
 \label{odd:normalization}
\end{equation}
Every step is an identity for raw fields at arbitrary degree. If an
input tensor becomes zero on a particular $F$, its return equation
continues to hold; no step divides by a tensor or by its matrix entries.

\subsection{The semisimple quotient of the odd algebra}

Let $\mathfrak r=\operatorname{rad}U$ be the Jacobson radical of the
underlying ordinary unital associative algebra, defined using all its
ordinary simple left modules. The parity automorphism preserves their
annihilators, so $\mathfrak r$ is graded. Three elementary facts suffice:
\begin{equation}
 \mathfrak r^{\dim F}=0,\qquad
 [P,U]\subset U\Longrightarrow[P,\mathfrak r]\subset\mathfrak r,
 \qquad z\in Z(U),\ z^k=0\Longrightarrow z\in\mathfrak r.
 \label{odd:radical-facts}
\end{equation}
For the first, take a composition series of the faithful $U$-module
$F$; each radical element lowers it. For the second,
$\exp(t\operatorname{ad}P)$ preserves $U$ and permutes annihilators
of its simple modules. For the third, a central operator on a finite
simple module is scalar, since any eigenspace is invariant; nilpotence
sets that scalar to zero. Every simple module is finite dimensional
because it is a quotient of the regular module. A nilpotent two-sided
ideal annihilates every simple module. It follows also that
$U/\mathfrak r$ has no nonzero nilpotent two-sided ideal: its preimage
would be an extension of two nilpotent ideals and hence nilpotent.

Only after \eqref{odd:normalization} has been proved do I apply these
facts. They imply $K_i\in\mathfrak r$, and all current and quartic
normalizers preserve $\mathfrak r$. The initial error
\eqref{odd:septic-error} lies in $\mathfrak r$; induction through
\eqref{odd:remainder}--\eqref{odd:solved-recursion} puts every
$E_n,Z_n$ there. On $\overline U=U/\mathfrak r$ the complete first
column is consequently
\begin{align}
 [p_0,O^{\alpha,(n)}_0]&=0,\notag\\
 [p_1,O^{\alpha,(n)}_0]&=-2nO^{\alpha,(n+1)}_0,\notag\\
 [p_2,O^{\alpha,(n)}_0]&=-4nO^{\alpha,(n+1)}_1,\notag\\
 [p_3,O^{\alpha,(n)}_0]&=-6nO^{\alpha,(n+1)}_2
                              +d_nO^{\alpha,(n-1)}_0,\notag\\
 [p_4,O^{\alpha,(n)}_0]&=-8nO^{\alpha,(n+1)}_3
                              +4d_nO^{\alpha,(n-1)}_1.
 \label{odd:linear-table}
\end{align}
Here $O^{\alpha,(0)}=0$. These equations include every zero-input
degree, which will be essential now.

If both cubic images vanish, upward transport kills all odd images.
Otherwise a nonzero family has indices $1,\ldots,M$: a zero index
would force all higher indices to vanish. The maximum exists because
the flavor representation on $\overline U$ is finite. Apply the fourth
line of \eqref{odd:linear-table} at the zero-input degree $M+1$.
It gives $d_{M+1}O^{\alpha,(M)}_0=0$, hence
\begin{equation}
 4\Delta=(M+1)^2=L^2,\qquad L=M+1\geq2.
 \label{odd:cutoff}
\end{equation}
Every other nonzero family has the same $M$, since $\Delta$ is common.

I construct the resulting Lie action directly on the odd quotient.
In the defining $L$-dimensional space take
\[
\begin{gathered}
 e=\sum_{i=1}^{L-1}E_{i,i+1},\qquad
 f=\sum_{i=1}^{L-1}i(L-i)E_{i+1,i},\\
 H=\operatorname{diag}(L-1,L-3,\ldots,1-L),\qquad C=(L^2-1)/4.
\end{gathered}
\]
The spin-$n$ tensors $\mathsf R^{(n)}$ with highest components
$\mathsf R^{(n)}_0=2e^n$, $1\leq n<L$, form a basis of
$\mathfrak{sl}_L$: their distinct flavor types are independent,
traceless, and their dimensions sum to $L^2-1$. Put
\[
 \mathsf p=(2e^2,-e(H+1),H^2/2-2C/3,f(H-1),2f^2).
\]
They satisfy exactly \eqref{odd:linear-table}. For an explicit
verification, use $[f,e^n]=-ne^{n-1}(H+n-1)$ and
$ef=C-H^2/4+H/2$ to obtain
\begin{align}
 \mathsf R^{(n)}_1&=-e^{n-1}(H+n-1),\notag\\
 \mathsf R^{(n+1)}_2&=e^{n-1}\left(
 \frac{H^2}{2}+(n-1)H+\frac{n^2(n-1)-2C}{2n+1}\right),\notag\\
 [\mathsf p_3,2e^n]&=ne^{n-1}
 \{4C-3H^2-6(n-1)H-2(n-1)(2n-1)\}.
 \label{odd:matrix-third}
\end{align}
Subtracting $-6n\mathsf R^{(n+1)}_2$ leaves
$d_n\mathsf R^{(n-1)}_0$, since $4C+1=L^2$.
The highest, first and second columns follow by the same one-step
commutators. For the last column, at $n\geq2$,
\begin{align}
 \mathsf R^{(n+1)}_3&=-\tfrac12e^{n-2}(H+n-2)
 \left(\frac{H^2}{2}+(n-2)H+\frac{n(n-1)^2-6C}{2n+1}\right),\notag\\
 [2f^2,2e^n]&=4ne^{n-2}(H+n-2)
 \left(\frac{H^2}{2}+(n-2)H+(n-1)^2-2C\right).
 \label{odd:matrix-fourth}
\end{align}
Their difference is $4d_n\mathsf R^{(n-1)}_1$ as required.
At $n=1$, compute directly
$[2f^2,2e]=-8f(H-1)=-8\mathsf R^{(2)}_3$.
The upward relation and flavor lowering show that the current and
quartic matrices generate $\mathfrak{sl}_L$; when $L=2$, the
quartic matrices vanish and the currents alone suffice.

Map each of two copies of this matrix basis to its respective odd
family in $\overline U$. Equation \eqref{odd:linear-table} and flavor
lowering intertwine the generating matrices with the actual
normalizer derivations. Formally, take the free Lie algebra on the
current and quartic symbols. Its map to these defining matrices is
surjective. A word in the kernel acts as zero on both model adjoint
modules, and therefore on every actual odd image. Its actual action
is a derivation; because the odd images generate $\overline U$, that
derivation is zero everywhere. Hence the action factors through
\begin{equation}
 \mathfrak{sl}_L\longrightarrow
          \operatorname{Der}_{\bar0}(\overline U),
 \qquad
 \vartheta_B,\vartheta_C:\mathfrak{sl}_L
           \longrightarrow\overline U_{\bar1}
 \quad\hbox{are equivariant}.
 \label{odd:adjoint-action}
\end{equation}
This construction permits overlap of the two images and does not use
an even-action theorem on an already existing odd annihilator.

\subsection{Principal tensors and the common kernel}

\begin{lemma}[Principal tensor generation]
\label{odd:principal-tensor}
Let $\mathfrak s\subset\mathfrak{sl}_L$ be the principal
$\mathfrak{sl}_2$ above. Under the diagonal adjoint action,
\[
 U(\mathfrak{sl}_L)(\mathfrak s\otimes\mathfrak s)
          =\mathfrak{sl}_L\otimes\mathfrak{sl}_L.
\]
\end{lemma}
\begin{proof}
Let $\mathcal V$ denote the submodule on the left and $\mathfrak h$
the diagonal traceless Cartan. Cartan-weight projection of
$H\otimes e$ gives $H\otimes E_{i,i+1}$ individually, since the
simple-root weights are distinct. Such projection is polynomial in
commuting Cartan adjoint operators and preserves $\mathcal V$.
Likewise $E_{i,i+1}\otimes H\in\mathcal V$.
Weyl representatives are products of root exponentials and preserve
$\mathcal V$. For a fixed root $E_{ab}$ I obtain
$H_w\otimes E_{ab}$ and $E_{ab}\otimes H_w$, with the entries of
$H_w$ in slots $a,b$ any consecutive pair $(t,t-2)$ from the
principal diagonal, and all remaining entries arbitrarily permuted.
These diagonals span $\mathfrak h$. For $L\geq3$, their differences
span the traceless subspace of the remaining $L-2$ slots, of dimension
$L-3$. Modulo it, two successive pairs have determinant
\[
 \det\begin{pmatrix}t&t-2\\t-2&t-4\end{pmatrix}=-4.
\]
For $L=2$, the Cartan is already one-dimensional. Thus
$h\otimes E_{ab},E_{ab}\otimes h\in\mathcal V$ for every $h$.

Write $\alpha_{ab}(h)=h_a-h_b$ and $H_{ab}=E_{aa}-E_{bb}$.
Applying $E_{cd}$ to $h\otimes E_{ab}$ gives
\[
 -\alpha_{cd}(h)E_{cd}\otimes E_{ab}
                     +h\otimes[E_{cd},E_{ab}].
\]
For nonopposite roots the second term is zero or a Cartan--root
term already known. Choosing $\alpha_{cd}(h)\neq0$ supplies every
nonopposite root--root tensor. For opposite roots and
$\alpha_{ab}(h)=0$, the same action supplies $h\otimes H_{ab}$.
These tensors, together with $H\otimes H$, span
$\mathfrak h\otimes\mathfrak h$: with the trace form, a linear
functional annihilating all the former tensors has the form
$\beta(h,k)=\operatorname{tr}(hTk)$ with every $H_{ab}$ an eigenvector
of $T$. Simple roots form a basis and each sum of consecutive simple
roots is a root, so connectedness forces $T=cI$. The remaining
functional is the trace form, which does not annihilate $H\otimes H$,
since $\operatorname{tr}H^2=L(L^2-1)/3\neq0$. This also proves the
one-dimensional Cartan case. Finally the root action above, with
opposite roots and $\alpha_{ab}(h)\neq0$, supplies the opposite
root--root tensors. All tensor types are now present.
\end{proof}

For $\alpha,\beta\in\{B,C\}$ the bilinear map
\[
 q_{\alpha\beta}(x\otimes y)=
    \vartheta_\alpha(x)\vartheta_\beta(y)
       +\vartheta_\beta(y)\vartheta_\alpha(x)
\]
is equivariant, since \eqref{odd:adjoint-action} acts by even
derivations. The cubic relations \eqref{odd:cubic}, and
$K_i\in\mathfrak r$, make this map zero on
$\mathfrak s\otimes\mathfrak s$. Lemma
\ref{odd:principal-tensor} makes it zero everywhere.
Thus a finite spanning list of all odd images anticommutes pairwise,
including its squares. The algebra $\overline U$ is a quotient of
the unital exterior algebra on this list. Its augmentation ideal is
nilpotent, hence zero by \eqref{odd:radical-facts}. Therefore every
raw odd operator belongs to $\mathfrak r$.

\begin{theorem}[Invariant full odd annihilator]
\label{odd:invariant-kernel}\label{odd:common-kernel}
On every finite lowest space $F\neq0$ as above, the two-sided ideal
of $\mathcal A(F)$ generated by its odd zero modes is nilpotent.
There exists a nonzero $\mathcal A(F)$-invariant subspace
$W\subset F$ annihilated by every odd zero mode.
\end{theorem}
\begin{proof}
The even generators in \eqref{odd:fifteen-generators} normalize
$\mathfrak r$, and the odd generators lie in $\mathfrak r$.
Commuting these generators past radical factors proves
$\mathcal A(F)\mathfrak r=\mathfrak r\mathcal A(F)$ and
\[
 (\mathcal A(F)\mathfrak r)^{\dim F}
             \subset\mathcal A(F)\mathfrak r^{\dim F}=0.
\]
A last nonzero member of the sequence
$F,\mathfrak r F,\mathfrak r^2F,\ldots$ is killed by
$\mathfrak r$. Hence $W=\ker_F\mathfrak r$ is nonzero. The
normalization makes it invariant under all even generators, and
all odd generators kill it. Equation \eqref{odd:fifteen-generators}
then makes it invariant under the entire zero-mode algebra.
\end{proof}

\section{Exclusion of additional lowest spaces}
\label{sec:descent}

I now turn the relations among the Casimir fields into a statement about
all conformal weights.  The relevant object is the complete lowest space
of a module, including every flavor type and every multiplicity.  This is
essential: an even zero mode can mix different flavor representations,
and an intermediate state at energy one half above the lowest space need
not itself satisfy lowest-space relations.

\begin{theorem}\label{thm:lowest-exclusion}
Let $N\ge5$, and let $M$ be a lower-bounded untwisted module for the
Casimir vertex algebra $A_N$.  Suppose that its lowest conformal weight is
$r$, and that $M_r$ and $M_{r+1/2}$ are finite-dimensional integrable
flavor modules.  There is no nonzero subspace $W\subset M_r$ such that
\begin{enumerate}
\item every physical zero mode of every integer-weight field preserves $W$;
\item every odd physical zero mode annihilates $W$;
\item every flavor spin $j'$ occurring in $W$ satisfies $j'<r$.
\end{enumerate}
\end{theorem}

The theorem is applied after Theorem~\ref{odd:common-kernel}.  Its
proof uses two complementary forms of finiteness.  Matrix
Cayley--Hamilton identities give a finite boundary for raw trace degrees.
Finite flavor support then turns the even zero modes into a finite
matrix Lie algebra.  These facts constrain different operators; neither
is a cutoff on conformal weight.

\subsection{The whole-input descent}

For a field $U$ of half-integral weight, write
$U^+=U_{(\mathrm{wt}(U)-3/2)}|_W$.  This maps $W$ into the actual space
$M_{r+1/2}$.  Set $T_c=P_c^+$ for odd $c\ge3$.  I use trace-degree
indices for
\begin{equation}
\begin{gathered}
 R_s=o(P_s)|_W,\qquad Q_s=\frac1s[f,R_s],\qquad
 \Delta=2r+\frac{N^2}{4},\\
 R_2=2e,\qquad Q_2=-H,\qquad R_4=p_0,\qquad Q_4=p_1.
\end{gathered}
 \label{low:operators}
\end{equation}
Here and below $s$ in an even-trace sum is a positive even integer.
All such sums are finite on $W$.

The parity-even generators of the zero-mode algebra have integral flavor
spin, and the odd zero modes vanish on $W$.  I may therefore restrict to one flavor-parity component
of $W$.  Let $j$ be its largest spin, and let $W_m$ denote its weight-$m$
space, with $H|_{W_m}=2m$.  Put
\begin{equation}
 k=\max\{c\le N:c\text{ odd}\},\qquad
 \operatorname{depth}(T_cx)=\frac{k-c}{2}+j-m
 \quad(x\in W_m).
 \label{low:depth}
\end{equation}
In particular $k\ge5$.  A lower highest-weight summand of spin $j-t$
enters at depth $t$; it is included from the outset.

I first establish the boundary.  On every $v\in W_j$, the ordered
Cayley--Hamilton and top identities in
Appendix~\ref{df:sec-boundary} give, with $d=k+2>N$,
\begin{align}
 T_dv&=0,& T_dfv&=\frac{2jd}{k}T_kv,\notag\\
 H_d^+v&=\frac{(k^2-1)/8-r-j/d}{k}T_kv,&
 B_{k+1}^+v&=C_{k+1}^+v=0.
 \label{low:boundary}
\end{align}
The first polar of $C_{k+1}^+$ is retained.  The cubic middle channel,
evaluated with that polar and evaluated by its field formula, respectively,
yields
\begin{equation}
 -\frac{2j}{k^2}T_kv
 =\frac{d}{k+1}H_d^+v-
       \frac{Nd(k-1)}{4k^2}T_kv.
 \label{low:boundary-middle}
\end{equation}
Consequently
\begin{equation}
 \left(r+\frac{(k^2-1)(2N-k)}{8k}-\frac jk\right)T_kv=0.
 \label{low:boundary-positive}
\end{equation}
The coefficient is positive since $r>j$ and $N\ge k\ge5$.
Thus $T_k$ vanishes on the entire space $W_j$.

Let
$\mathcal Y_t=-[u^{2t}]\exp(-\sum_{s\text{ even}}R_su^s/s)$.
The first-half mode of the determinant identity is the ordered recursion
\begin{equation}
 \frac{T_c}{c}
 =\sum_{t\ge1,\ c-2t\ge3}
         \frac{T_{c-2t}}{c-2t}\mathcal Y_t
 \quad(c>N),\qquad [H,\mathcal Y_t]=2t\mathcal Y_t.
 \label{low:raw-ch}
\end{equation}
Each term has the same depth as its left-hand side and a strictly smaller
raw degree.  Repeated substitution terminates at actual degrees at most
$N$.  At those degrees there are no negative depths, and depth zero
consists only of $T_k|_{W_j}$.  Hence all nonpositive raw depths vanish.

Suppose that every smaller depth has vanished, and put
$b=k-2\delta\ge3$ at the current depth.  Define maps on the whole input by
\begin{equation}
 \Phi_bx=\frac{T_cx}{c},\qquad
 Y_{b,B}x=B_{c+1}^+x,\qquad Y_{b,C}x=C_{c+1}^+x,
 \quad c=b+2(j-m),\quad x\in W_m.
 \label{low:maps}
\end{equation}
Their common output flavor is $j+b/2$.  Introduce the input operator
$a=b+2j-H$, and put
\begin{align}
 A&=N+2f+\sum_s(s+1)R_s,&
 B&=2f+p_2+\sum_sR_s,\notag\\
 T&=p_{2,\mathrm{out}}-\frac{2N}{3},&
 \mathcal R(z)&=-3f-\frac N2+
             \sum_sR_s\left(-\frac12+\frac{s}{z-s}\right),\notag\\
 \mathcal D(z)&=\Delta z^2+
          \frac14[A,B](z+3)+\frac12A\mathcal R(z).
 \label{low:ordered-operators}
\end{align}
Every function of $a$ acts first, on the right of the neighbouring
operator.  The spectra of $a$ and $a-s$ consist of nonzero odd integers.
No commutativity between $a$ and $A$ or $B$ is used.

The complete field calculations in
Appendices~\ref{df:sec-source} and~\ref{df:sec-half-transport} give, for
each of the two current source maps $Y$,
\begin{equation}
 YA=0,\qquad TY=YB,\qquad
 Y\left(\Delta a^2+\frac14[A,B](a+3)\right)=0.
 \label{low:source-rows}
\end{equation}
The assumptions are precisely preceding raw-depth vanishing and the
three hypotheses of the theorem.  In particular the derivation of
$TY=YB$ evaluates normal products on the specified image of $\Phi_b$;
it does not apply a lowest-space relation on $M_{r+1/2}$.
Since $YA=0$ and $YBA=TYA=0$, the commutator term vanishes.
The invertibility of $a^2$ and $\Delta>0$ imply $Y=0$.
Thus both current sources vanish at the new depth.

After this elimination the cubic and quartic equations become
\begin{equation}
 T\Phi_b=\Phi_bB,\qquad
 \Phi_bA=K\Phi_ba,\qquad \Phi_b\mathcal D(a)=0,
 \label{low:bottom-rows}
\end{equation}
where the actual output operator is
\begin{equation}
 K=\frac12\bigl(\{S_0,\bar S_2\}-\{S_2,\bar S_0\}\bigr),
 \qquad S_i=o(B_{3,i}),\quad \bar S_i=o(C_{3,i}).
 \label{low:k-output}
\end{equation}
For $b\ge5$, the independently derived adjacent row in
Appendix~\ref{df:sec-adjacent} yields
\begin{equation}
 K^2\Phi_b=4\Delta\Phi_b,\qquad [K,T]\Phi_b=0.
 \label{low:output-spectrum}
\end{equation}
The same conclusion at $b=3$ follows from the genuine degree-two
fermion identities in Appendix~\ref{df:sec-terminal}.  There the
adjacent row initially contains a return $\mathfrak F:W\to W_j$.
Appendix~\ref{df:sec-terminal-reduction} expresses it, on
$x\in W_{j-(q-2)/2}$ with even $q\ge4$, as
\begin{equation}
 \mathfrak Fx=\frac{q+2}{q+1}
 \left[-\frac{q(q+1)}2\left([Q_q,p_2]-2(q-2)o(W_{q+2,2})\right)
       +q(q-2)\left(\Delta-\frac{q^2}{16}\right)R_{q-2}\right]x.
 \label{low:terminal-return}
\end{equation}
The even third-channel identity, proved at every even degree in
Appendix~\ref{df:sec-even-third}, sets this expression to zero.
The separate highest-coordinate computation gives
$\mathfrak F|_{W_j}=0$.  Hence $\mathfrak F=0$ on all of $W$,
and~\eqref{low:output-spectrum} also holds at the terminal depth.

If $\Phi_b\ne0$, its finite-dimensional image is invariant under $K,T$.
On that image $K^2=4\Delta$ and $[K,T]=0$.  Choose a left eigenvector
of $T$ in one nonzero $K$ eigenspace and pull it back through $\Phi_b$.
This gives $0\ne\ell\in W^*$ and scalars $x,\lambda$ satisfying
\begin{equation}
 \ell B=x\ell,\qquad \ell(A-\lambda a)=0,
                 \qquad\lambda^2=4\Delta=N^2+8r.
 \label{low:necessary-row}
\end{equation}
The pullback is nonzero because $\Phi_b$ is surjective onto its image.
This argument allows arbitrary multiplicities and Jordan blocks.

\subsection{The even action and the Newton obstruction}

I next exclude~\eqref{low:necessary-row}.  Assume first that the flavor
action on $W$ is nontrivial.  Write
$R^{(n)}_i=o(W_{2n,i})|_W$.  The even third channel and its Jacobi
consequence, proved in Appendices~\ref{df:sec-even-third}
and~\ref{df:sec-even-fourth}, give
\begin{align}
 [p_1,R^{(n)}_0]&=-2nR^{(n+1)}_0,\notag\\
 [p_2,R^{(n)}_0]&=-4nR^{(n+1)}_1,\notag\\
 [p_3,R^{(n)}_0]&=-6nR^{(n+1)}_2+
       \frac{n(n-1)(4\Delta-n^2)}{2n+1}R^{(n-1)}_0,\notag\\
 [p_4,R^{(n)}_0]&=-8nR^{(n+1)}_3+
       \frac{4n(n-1)(4\Delta-n^2)}{2n+1}R^{(n-1)}_1.
 \label{low:even-table}
\end{align}
Finite flavor support supplies a last nonzero tensor, say $R^{(L-1)}$.
The third-channel relation at the next index gives $4\Delta=L^2$.
Since $r>0$, this integer satisfies $L>N$.

The full table~\eqref{low:even-table}, together with flavor covariance,
identifies these actual operators with a representation
\begin{equation}
 \rho:\mathfrak{sl}_L\longrightarrow\operatorname{End}(W),\qquad
 R_{2s}=2\rho(e_{\mathrm{def}}^s)\quad(1\le s<L),
 \qquad R_{2s}=0\quad(s\ge L).
 \label{low:sl-action}
\end{equation}
A proof by matrix tensors and generator brackets is given in
Appendix~\ref{df:sec-even-lie}.  In particular this is an actual Lie
homomorphism.  The power on the right is taken in the defining matrix
representation before applying $\rho$.

Let $M_N=\lfloor N/2\rfloor$.  The determinant and its top component give
\begin{align}
 \mathcal E(z)&=\exp\left(-\sum_{s\ge1}
                      \frac{R_{2s}}{2s}z^s\right),&
 \deg\mathcal E&\le M_N,\notag\\
 \left(r+\frac{M_N(N-M_N)}2\right)[z^{M_N}]\mathcal E&=0.
 \label{low:newton-bound}
\end{align}
Appendix~\ref{df:sec-newton} proves the top formula for an arbitrary
number of factors with fixed normal-product parentheses.  Its scalar
coefficient is positive, so $\deg\mathcal E\le M_N-1$.

For a simple composition factor of the representation~\eqref{low:sl-action},
write its dominant highest weight as
$\Lambda=\sum_{u=1}^{L-1}m_u\omega_u$ and put
$q=\sum_um_u u$.  Let $\rho_G$ denote the integrated
$SL_L(\C)$ action on this composition factor.  The identity
\begin{equation}
 \mathcal E(z)=\rho_G(I-ze_{\mathrm{def}})
 \label{low:newton-group}
\end{equation}
and an explicit highest-weight minor give a matrix coefficient
$(-z)^q$; hence $q\le M_N-1$.  The defining matrices
\begin{equation}
 C_\varepsilon=2f_{\mathrm{def}}+
       2\sum_{s=1}^{L-1}(2s+1)e_{\mathrm{def}}^s+
                  \varepsilon L H_{\mathrm{def}},\qquad\varepsilon=\pm1,
 \label{low:spectral-matrix}
\end{equation}
have distinct eigenvalues
\begin{equation}
 \{L^2+L-2\}\ \cup\
 \{-L^2+L-2+2Li:0\le i\le L-2\}.
 \label{low:defining-spectrum}
\end{equation}
All are congruent to $-2$ modulo $L$.  Appendix~\ref{df:sec-spectrum}
proves~\eqref{low:defining-spectrum} for every $L$ by an invariant
polynomial flag; it follows that
\begin{equation}
 \operatorname{Spec}\rho(C_\varepsilon)\subset-2q+L\mathbb Z
 \label{low:spectral-congruence}
\end{equation}
on the simple factor.  On the other hand, the second equation
in~\eqref{low:necessary-row}, with $\lambda=\varepsilon L$, requires
$\lambda(b+2j)-N$ to be an eigenvalue of this operator.  A composition
series makes its characteristic polynomial the product of the
characteristic polynomials of its simple factors.  Thus the required
eigenvalue occurs on one such factor and forces $2q\equiv N\pmod L$.
This is impossible because
\begin{equation}
                0\le2q\le2M_N-2<N<L.
 \label{low:congruence-contradiction}
\end{equation}
Consequently $\Phi_b=0$ when the flavor action is nontrivial.

If the flavor action is trivial, $e=H=f=0$, and covariance kills every
positive-spin even tensor.  Thus $j=0$, $A=N$, and $a=b$.
Equations~\eqref{low:bottom-rows} and~\eqref{low:output-spectrum} give
\begin{equation}
 \left(\frac{N^2}{b^2}-N^2-8r\right)\Phi_b=0.
 \label{low:scalar-branch}
\end{equation}
Its coefficient is negative for $b\ge3$, $r>0$, so $\Phi_b=0$ in
this case as well.  The determinant recursion~\eqref{low:raw-ch}
propagates the new vanishing to every raw degree at the same depth.
Induction down to $b=3$ therefore proves the whole-map equality
$\Phi_3=0$.

\subsection{The terminal supercharge}

It remains to use the actual degree-two fermion.  Put
$a=3+2j-H$, $z=a-2$, and $\mathcal L=Aa^{-1}$.  Define
$Y'_Fx=F_{c-1}^+x$ for $x\in W_m$, $c=a|_{W_m}$ and $F=B,C$.
The terminal field calculation retains its source-square return and
then cancels its transport defect.  With $\mathfrak F=0$ established by
\eqref{low:terminal-return}, it gives
\begin{gather}
 Y'_FA=0,\qquad Y'_F\mathcal L=0,\qquad
 TY'_F=Y'_F(B-2\mathcal L),\notag\\
 Y'_F\left[\mathcal D(z)+\mathcal L J_0+
       \frac12([B,\mathcal L]-2\mathcal L^2)(a+1)\right]=0,\notag\\
 J_0=-\mathcal R(z)+\frac32A.
 \label{low:last-source}
\end{gather}
Every factor keeps its displayed order.  The transport reduces to
$TY'_F=Y'_FB$, and therefore
$Y'_F[B,\mathcal L]=TY'_F\mathcal L-Y'_F\mathcal LB=0$.
The other terms with left factor $Y'_F\mathcal L$ vanish as well.
Thus $Y'_F\mathcal D(z)=0$.  Also
$Y'_F[A,B]=Y'_FAB-TY'_FA=0$, so
\begin{equation}
                  0=\Delta Y'_Fz^2.
 \label{low:last-elimination}
\end{equation}
The spectrum of $z$ is positive and odd.  Hence both maps $Y'_F$ vanish.
For $0\ne v\in W_j$, their genuine endpoint is
$Y'_Cv=\bar G_{X(0)}v$.  Lowestness gives $G_{Y(1)}v=0$, while the
small $\mathcal N=4$ superconformal relation gives
\begin{equation}
 0=\{G_{Y(1)},\bar G_{X(0)}\}v=(r-H/2)v=(r-j)v.
 \label{low:final-contradiction}
\end{equation}
This contradicts $r>j$ and proves Theorem~\ref{thm:lowest-exclusion}.

\section{Exhaustion of the relative cohomology}
\label{sec:exhaustion}

I now combine the preceding results. The argument applies to the full
cohomology as a graded vector space; no Euler characteristic is taken.

\begin{proof}[Proof of Theorem~\ref{thm:main}]
By Theorem~\ref{thm:strong-closure}, the normal differential span $A_N$
is a conformal vertex subalgebra of $\mathcal H_N$. In particular,
$M=\mathcal H_N/A_N$ is an $A_N$-module. Suppose that $M\ne0$.
The relative creation variables have positive half-integral conformal
weights, so each weight space is finite dimensional and the spectrum
is bounded below. The vacuum lies in $A_N$. There is consequently a
least missing weight $r>0$ and a nonzero finite-dimensional bottom
$F=M_r$.

Every mode lowering conformal weight annihilates $F$. The flavor
zero modes act as $\mathfrak{sl}_2$, and the extremal-sector statement
of Lemma~\ref{fnd:extremal} gives $j<r$ for every spin $j$ occurring
in $F$. This uses the entire relative complex: an extremal state is
an invariant polynomial in undifferentiated $X$, and all such
polynomials already belong to $A_N$.

For $N\ge5$, Theorem~\ref{odd:common-kernel} supplies a nonzero
subspace $W\subset F$ annihilated by every odd weight-preserving
mode and stable under the even weight-preserving modes. To apply the
lowest-space relations, form the actual $A_N$-submodule generated
by $W$. Its weight-$r$ space is exactly $W$. Indeed, order any
word of modes by commuting modes of negative energy shift to the
right. They annihilate the original bottom. A surviving word of
total energy shift zero contains only zero-shift modes; each preserves
$W$. The commutator formula replaces an interchange by shorter words
of the same total energy shift, so induction on the number of modes
justifies the ordering: order first by word length and then by the
number of inverted pairs of negative and nonnegative shifts. Words of positive shift cannot contribute
to weight $r$. Thus the generated module has the required bottom,
with the same strict inequality $j<r$. Its weight-$(r+1/2)$ space
is a subspace of $M_{r+1/2}$ and is finite dimensional.

Theorem~\ref{thm:lowest-exclusion} now rules out this nonzero lowest
space. Its hypotheses retain every flavor block, both odd families,
and all possible sources in the descent. This contradiction proves
$M=0$ for $N\ge5$. The explicit endpoint arguments in
appendix~\ref{sec:small-rank} give the same contradiction for
$N=2,3,4$.

It follows that $\mathcal H_N=A_N$ for every $N\ge2$.
All identities are homogeneous in cohomological degree. Modes of the
odd generators may change that degree, and the argument has retained
them throughout. Since $A_N$ is a graded subspace, the vanishing of the
whole quotient proves the equality separately in every cohomological
degree; no concentration in degree zero is assumed.
The triangular change of complete multiplets in
Theorem~\ref{thm:strong-closure} identifies the raw and harmonic strong
spans. Finally, the iterate identity and translation rule put every
normal word into the right-nested form of
\eqref{intro:strong-span}. This proves the stated strong-generation
theorem in all conformal weights.
\end{proof}

The least-weight argument exhausts the unbounded conformal spectrum,
while the arbitrary-degree identities control the entire odd tower
before the invariant kernel is constructed. The three smallest
matrix sizes supply the endpoints of the field presentation. For
$N\ge5$, the proof is uniform with $N$ as a parameter.

\section{Protected operator content and physical consequences}
\label{sec:physics}

\subsection{A complete set of protected local building blocks}

Theorem~\ref{thm:main} identifies the local building blocks of the
Schur algebra at finite rank. For a Casimir degree $d$, the complete
short multiplet has the following components:
\begin{equation}
\begin{array}{c|c|c|c|c}
 &h&SU(2)_F\text{ spin}&\text{parity}&\text{multiplicity}\\\hline
 \text{bottom}&d/2&d/2&\text{even}&d+1\\
 \text{first fermion}&(d+1)/2&(d-1)/2&\text{odd}&d\\
 \text{second fermion}&(d+1)/2&(d-1)/2&\text{odd}&d\\
 \text{top}&(d+2)/2&(d-2)/2&\text{even}&d-1
\end{array}
\label{phys:multiplets}
\end{equation}
The two fermionic families have ghost numbers $-1$ and $+1$.
There are $4d$ displayed components at degree $d$, and hence
$2N(N+1)-4$ in the preferred generating set. Its largest conformal
weight is $(N+2)/2$. The spectrum of the algebra nevertheless extends
to arbitrarily high weight through normal products and derivatives.
This finite bound concerns generators and has no implication that
the number of primary operators is finite.

Completeness says more than the existence of these familiar multiplets.
A class containing derivatives or vector-multiplet letters might, in
principle, define an additional protected generator invisible in the
pure scalar invariant ring. The theorem excludes this possibility in
the full relative quantum cohomology. I include both fermionic
families and every top component in the proof. The scalar bottom
fields alone do not span the full cohomology.
Thus the invariant degrees organize a complete quantum operator
sector after their supersymmetric completion.

This gives a precise relation between half-BPS data and the larger
Schur sector. The Casimir bottoms are the highest flavor components of
protected scalar multiplets. Their short descendants, followed by
normal multiplication and holomorphic differentiation, reach every
Schur cohomology class. The resulting statement concerns equality in
cohomology. Four-dimensional operators that differ by an exact term
or combine into the same twisted class are naturally identified in
this description. Within that sector, no further primitive local
input is required.

\subsection{Finite rank is implemented through relations}

The distinction between stable trace expressions and finite-rank
cohomology is visible directly in the proof. Before imposing matrix
identities there is a Casimir multiplet at every degree $d\ge2$.
For $N\times N$ matrices, Cayley--Hamilton expresses the higher bottom traces in
terms of degrees $2,\ldots,N$. Its short-supermultiplet descendants
give quantum identities for the corresponding higher components.
The exhaustion theorem establishes that this reduction does not
leave room for an additional family of generators elsewhere in the
cohomology.

Consequently, the finite-rank protected algebra is organized by
a fixed list of multiplets and their rank-dependent relations.
The theorem fixes the generating fields throughout the spectrum,
while the trace identities organize the independent composites.
It also explains why the many possible derivative and fermion
structures in the BRST complex can be organized in terms of a much
smaller set of physical data. Their redundancy is a consequence of
quantum cohomology and the field identities, rather than a purely
classical restriction to commuting scalars.

In the holographic setting \cite{Maldacena1998,WittenAdS1998}, this
statement has a direct protected interpretation. Trace multiplets
provide the natural single-particle labels at large rank, and products
provide multiparticle operators. At a finite value of $N$, the trace
cutoff reorganizes these labels through exact identities. For the
Schur local algebra, the theorem shows that finite-rank completion
requires no additional primitive generating multiplets beyond the Casimir list.
This supplies an algebraic description of the finite-rank reorganization
of this protected sector. The appropriate observables are its twisted
local correlators and graded spaces of states.

This interpretation connects the local algebra to the protected bulk
constructions of refs.~\cite{BonettiAdS2018,CostelloTwisted2021}
and to the worldsheet description of Schur operators
\cite{LiWorldsheet2024}. In the terminology of the cohomological
analysis of ref.~\cite{Chang2024}, products of the trace multiplets
organize the Schur graviton sector. The completeness statement fixes
the local cohomological content of this organization at finite rank:
trace identities change the independent combinations of such fields,
but introduce no new primitive family. This is a statement about
protected local operators, whose multiplication and differentiation
remain meaningful even when a large-rank particle interpretation is
no longer the most useful description.

\subsection{Counting states before taking an index}

Strong generation also controls the unsigned dimensions of the
protected spectrum. I define the ghost-refined series
\[
 Z_N(q,y)=\sum_{h,g}\dim\mathcal H_N^{g}[h]\,q^hy^g.
\]
Every coefficient is finite. Strong generation implies the
coefficientwise bound
\begin{equation}
 Z_N(q,y)\ \preccurlyeq\
 \prod_{d=2}^{N}\prod_{n=0}^{\infty}
 \frac{(1+yq^{(d+1)/2+n})^d
       (1+y^{-1}q^{(d+1)/2+n})^d}
 {(1-q^{d/2+n})^{d+1}(1-q^{(d+2)/2+n})^{d-1}}.
 \label{phys:bound}
\end{equation}
The inequality is coefficientwise in $q^h y^g$, for
$h\in\tfrac12\mathbb Z_{\ge0}$ and $g\in\mathbb Z$; the vacuum
has weight zero. To prove the bound,
assign a generating component of weight $h$ the cost $2h$ and give
translation cost zero. A singular product $a_{(t)}b$, $t\ge0$,
has an expression of cost at most $2(h_a+h_b-t-1)$, so costs at
least two less than the two input factors. Sesquilinearity gives
the same decrease for derivatives, and the iterate identity extends
it to normal words. Interchanging two factors changes a normal
word by terms of lower filtration. The associated graded algebra
is therefore a quotient of the differential supercommutative algebra
on the displayed generators. Each even derivative contributes a
geometric series and each odd derivative an exterior factor.
Multiplying their contributions gives \eqref{phys:bound}. Quotienting
by relations can only reduce each multigraded dimension.

The Schur index is instead a signed trace, with fermionic parity
inserted. Its relation to the chiral algebra is an important link to
localization and protected counting
\cite{Gadde2013,PanPeelaers2019,DedushenkoFluder2020}. Equation
\eqref{phys:bound} is an unsigned spanning bound, established before
any cancellation between bosons and fermions. In particular, the
proof of completeness controls classes that a signed trace could
pair in its coefficients. The theorem gives a complete space of
candidate monomials on which character and null-relation calculations
can now be based at any rank.

The exact index formulas of
refs.~\cite{Bourdier2015,Bourdier2016,Hatsuda2023} offer detailed
counting data with which this organization can be compared. The
series $Z_N(q,y)$ retains cohomological degree, whereas the refinements
used in the Schur and Macdonald indices have their specified
four-dimensional charge assignments \cite{Song2017}. Keeping these
gradings explicit gives a concrete order of operations: construct
the differential normal monomials, impose their relations, and then
take the desired graded trace. Equation~\eqref{phys:bound} controls
the first step uniformly, without requiring the cancellations of the
last step to reveal the size of the underlying spaces.

\subsection{Protected products and the bootstrap}

Once the singular operator products among the complete generating
fields and the required relations are specified, the vertex-algebra
identities recursively determine products involving composite
fields. Normal ordering is defined by subtracting singular terms,
and differentiation determines products of descendants. Theorem
\ref{thm:main} ensures that these operations reach every field of the
protected algebra. Thus a computation formulated in this generating
set cannot omit an independent Schur generator at a later conformal
weight.

For four-dimensional correlators, the reconstruction applies to their
meromorphic twisted sector. This is the part that enters the
protected crossing equations studied in the $\mathcal N=4$
superconformal bootstrap \cite{Bootstrap2013,Bootstrap2017}.
Completeness makes the choice of protected generating fields uniform
in rank and allows higher-weight protected contributions to be
organized systematically as composites. It is a structural input
to the bootstrap, alongside the operator-product coefficients and
null relations that determine a particular correlator.

The same perspective sharpens the relation to Higgs-branch geometry.
The bottom generators retain the invariant-polynomial organization,
while the fermions, derivatives, and quantum normal-ordering terms
retain information beyond the commutative scalar ring. This
organization is compatible with the role of free-field realizations
and filtrations in relating vertex algebras to Higgs-branch effective
descriptions \cite{BeemHiggs2019}. The complete generating set is a
natural starting point for studying the associated variety and the
null states that govern differential equations for characters
\cite{BeemRastelli2018}.

In this respect, the Hilbert-scheme construction and free-field
embedding of ref.~\cite{AKM2023}, and the general relation between
quasi-lisse geometry and modular differential equations
\cite{ArakawaKawasetsu2018}, provide complementary geometric
information. The operator-product analyses of
refs.~\cite{GaberdielLi2025,BonettiUniversal2025} address how
associativity constrains the same multiplet pattern. The present
theorem identifies a complete set of fields for such an analysis in
the relative gauge-theory cohomology. It thereby gives a common local
language for comparing geometric constructions, explicit products,
and protected correlation functions.

\subsection{Discussion}

The Schur algebra retains an infinite spectrum of protected local
operators. Its simplest scalar classes come from the Casimir invariants
of one adjoint scalar, with degrees fixed by the gauge group.
I have shown that this finite Casimir
structure extends to a complete set of strong generators for
$\mathcal N=4$ $SU(N)$ super Yang--Mills theory, proving the $SU(N)$
strong-generation assertion of Conjecture~3 in ref.~\cite{Beem2013}. The theorem concerns the full standard relative
quantum BRST cohomology: it holds for every $N\ge2$, every conformal
weight, and every cohomological degree. Both fermionic families and all
flavor representations enter the argument. Thus the connection between
the invariant degrees $2,\ldots,N$ and the protected operator algebra
extends beyond the existence of distinguished scalar classes. Their
complete short multiplets generate the entire Schur cohomology under
normal multiplication and holomorphic differentiation.

This result identifies a finite structure within an unbounded protected
spectrum. Increasing conformal weight permits progressively more
complicated derivatives, products, and primary combinations, while the
list of generating multiplets remains fixed. Completeness separates
these two kinds of complexity. The appearance of a new primary at high
weight does not require a new strong generator: its cohomology class is
already a normal polynomial in the Casimir multiplet components and
their derivatives. The invariant degrees therefore organize the
construction of every protected local field, even when its expression
in the original gauge-theory letters has no simple scalar form. This is
a statement about the structure of the operator algebra throughout its
spectrum, rather than a truncation of that spectrum.

The proof explains why this organization survives quantization. Its
first mechanism constructs the Casimir algebra through cyclic trace
cohomology, quantum lifting, and finite-rank matrix identities. Its
second mechanism tests whether any cohomology could remain outside that
algebra. A hypothetical first missing weight would define a finite
lowest space, on which the arbitrary-degree transport identities and
the lowest-space obstruction apply. The exclusion retains the
fermionic modes, derivative contributions, and non-scalar flavor
representations that are absent from a calculation confined to scalar
invariants. The explicit normal-ordering corrections and BRST
primitives provide the link between the trace identities and the
quantum fields. In this way, the finite-rank identities and the
unbounded conformal spectrum enter one argument: the former construct
the candidate algebra, and the latter is exhausted by ruling out its
first possible omission.

For protected counting, this completeness gives a description before
bosonic and fermionic contributions are combined into a signed trace.
The unsigned multigraded bound in \eqref{phys:bound} expresses the
result directly at the level of spaces of states. Every class is a finite linear combination of differential normal
monomials in the Casimir multiplet components; relations determine
their linear dependences. This supplies a uniform starting point for character
calculations at arbitrary rank and complements the index and
localization methods of
refs.~\cite{Gadde2013,PanPeelaers2019,DedushenkoFluder2020}. In
particular, cohomological completeness controls the individual classes
whose contributions may cancel in an index. It is this operator-level
information that allows protected counting and protected products to be
organized within the same generating framework.

The corresponding consequence for correlators is a complete choice of
local input for the meromorphic sector. Once the generator operator
products and the requisite relations are specified, the vertex-algebra
identities propagate that information to composite fields and their
derivatives. Strong generation guarantees that this reconstruction
reaches every Schur class, so an analysis at a later conformal weight
cannot require an additional independent generating multiplet. This
makes the Casimir organization useful for the protected part of the
$\mathcal N=4$ bootstrap
\cite{Bootstrap2013,Bootstrap2017}, where meromorphic data constrain
four-dimensional correlation functions. It also places the invariant
polynomials and their supersymmetric descendants in a common quantum
framework, compatible with the geometric role of filtrations and
free-field realizations in
refs.~\cite{BeemHiggs2019,BeemRastelli2018}.

The recent geometric and operator-product constructions make this
comparison especially concrete
\cite{AKM2023,GaberdielLi2025,BonettiUniversal2025}. The Casimir
degrees label the fields simultaneously in the invariant-theoretic,
free-field, and supersymmetric descriptions. The theorem establishes that this shared Casimir pattern
is exhaustive within the gauge-theory cohomology, providing a complete
generating set for comparisons of these constructions. The
successive weights enrich the composite operator content while
remaining accessible from the same finite family of multiplets.

At finite $N$, the holographic reading is correspondingly precise.
Trace multiplets give the familiar large-rank single-particle labels,
while their products describe the associated multiparticle
organization. The theorem shows how this organization closes within
the local Schur algebra when finite-rank trace identities become
operative. The protected spectrum is reorganized through relations
among the Casimir multiplets and their composites, with no additional
primitive generating family needed to complete this sector. In the
setting of the $AdS_5/CFT_4$ correspondence
\cite{Maldacena1998,WittenAdS1998}, this provides an exact algebraic
account of that finite-rank reorganization for twisted local
observables. The infinitely many protected fields retain a finite
Casimir structure because quantum cohomology relates their scalar,
fermionic, and derivative descriptions.

The generating theorem opens the way to explicit presentations of the
relations and proofs of their completeness at all $N$. Other simple gauge algebras and defect modules offer further
settings in which to explore the same interplay of supersymmetry,
invariant theory, and quantum cohomology. For the local Schur algebra
considered here, the organization is complete: the Casimir short
multiplets reach every protected class through their normal products
and derivatives, at every conformal weight.

\acknowledgments
\addcontentsline{toc}{section}{Acknowledgments}
I thank Lanzhou University for internal funding, and China Three Gorges University and Westlake University for their hospitality and assistance. I am also grateful to the Yau Mathematical Sciences Center at Tsinghua University and the organizers of “Quantum Information in Quantum Gravity 2026” (QIQG 2026) for their hospitality during the course of this work. OpenAI’s GPT-6 was used as an auxiliary tool for some lengthy algebraic calculations, as well as for occasional suggestions and consistency checks during the development of the mathematical proofs.

\section*{Data Availability}
No datasets were generated or analysed during the current study.

\appendix
\section{Field identities for odd transport}
\label{app:odd-fields}

This appendix supplies the field calculations used in Section~\ref{sec:odd}.
All trace expressions without an intervening colon denote simultaneous
normal ordering of the elementary oscillators. Colons between composite
fields denote the vertex normal product, with the displayed parentheses.
The distinction matters in the finite identities below.
All matrix letters are traceless unless a superscript $\mathrm{full}$ is
present. The ghosts remain traceless also in the auxiliary full-boson
calculation. Ordinary derivatives are denoted by $\partial$.
The normalized flavor components of a highest vector $A_0$ of spin $s$
are $A_i=(\operatorname{ad}f)^iA_0/(2s)^{\underline{i}}$.
Here $x^{\underline{i}}=x(x-1)\cdots(x-i+1)$.
For coupled normal products I write
\begin{equation}
 \mathcal N_k(A,B)=\sum_{i=0}^k(-1)^i\binom ki:A_iB_{k-i}:.
 \label{of:normal-coupling}
\end{equation}
A nested coupling is polarized before taking the next coupling. I use
$W_{d,i}$ for the components of $P_d$, and $B_{d,i},C_{d,i}$ for those of
$B_d,C_d$. In finite tables $B,D,E$ abbreviate the degree $3,5,7$ odd
multiplets, respectively. $T$ denotes the stress tensor and $H_d$ the raw
Casimir top; neither is the Cartan generator $H$.

I begin with the Wick and normal-mode formulas and the relative
chain derivation. The ensuing field identities establish cubic
nilpotence and the quartic transport channels, followed by the
cubic, quintic, and septic initial returns used in the simultaneous
induction. The final subsection proves the mixed cubic--odd product
at arbitrary degree. These finite initial identities and
arbitrary-degree identities have distinct roles in
section~\ref{sec:odd}.

\subsection{A finite Wick formula and physical zero modes}

I first give a common algebraic rule for every finite calculation in this
appendix. It is useful both for the bosonic OPEs and for the relative BRST
primitives. Let $a^{[u]}=\partial^u a/u!$ for
$a\in\{X,Y,b,c\}$. Put
\begin{equation}
 \kappa_{XY}=1,\qquad \kappa_{YX}=-1,\qquad
 \kappa_{bc}=\kappa_{cb}=1,
 \qquad
 \mathsf P^{ik}_{jl}=\delta^i_l\delta^k_j-
                      N^{-1}\delta^i_j\delta^k_l,
 \label{of:propagator}
\end{equation}
with all other $\kappa$ zero. Thus a contraction of
$(a^{[u]})^i{}_j(z)$ with $(d^{[v]})^k{}_l(w)$ is
\[
 \kappa_{ad}(-1)^u\binom{u+v}{u}
        \mathsf P^{ik}_{jl}(z-w)^{-u-v-1}.
\]
For products of matrix traces, expand each trace into matrix entries,
retaining their written order. Let $M$ run over all partial matchings
between left and right letters with nonzero $\kappa$, with no vertex used
twice. Let $d(M)=\sum_{(i,j)\in M}(u_i+v_j+1)$, and let
$\epsilon(M)$ be the Koszul sign of moving the two members of each
contracted odd pair together and deleting them. The sign is independent
of the deletion order, since each contraction has even parity. With
$L_M$ the unmatched left vertices and $R_M$ the unmatched right vertices,
for every integer $k\geq-1$, the $k$th vertex product is
\begin{equation}
\begin{split}
 A_{(k)}B={}&\sum_M\epsilon(M)
 \prod_{(i,j)\in M}\left[
 \kappa_{a_i d_j}(-1)^{u_i}\binom{u_i+v_j}{u_i}
                    \mathsf P_{ij}\right] \\
 &\quad\times
 \sum_{\substack{t_i\geq0\ (i\in L_M)\\
            \sum t_i=d(M)-k-1}}
 \prod_{i\in L_M}\binom{u_i+t_i}{t_i}
 \Bigl[\prod_{i\in L_M}a_i^{[u_i+t_i]}
                   \prod_{j\in R_M}d_j^{[v_j]}\Bigr]_{\rm PBW}.
\end{split}
\label{of:wick-formula}
\end{equation}
The original matrix indices are summed after inserting every projector;
the empty matching contributes only at $k=-1$. A negative required
Taylor degree gives zero. An empty trace is $N$, and a one-letter
traceless trace is zero. Formula~\eqref{of:wick-formula} is a finite
sum with explicitly specified coefficients. For a normal product
containing a composite factor one first expands that factor using the
same formula; hence it also fixes nested normal-order corrections.

To prove the formula, split each elementary field into creation and
annihilation parts and commute the annihilation parts to the right.
At each interchange the elementary supercommutator either leaves the
ordered pair or replaces it by its scalar contraction. Induction on
the number of left letters therefore gives exactly all partial
matchings of matrix entries, each once, with the stated Koszul sign
and contraction product.
Taylor expansion of an unmatched left jet is
$a^{[u]}(w+z)=\sum_{t\geq0}\binom{u+t}{t}z^t a^{[u+t]}(w)$.
The coefficient of $(z-w)^{-k-1}$ imposes the displayed total Taylor
degree. This proves the identity before any matrix-index contraction;
summing indices proves it for traces. In particular the same formula
applied to $A=j_{\rm BRST}=\operatorname{tr}(cXY-cYX-ccb)$ and $k=0$
computes $QB$. There is no matrix-rank inference in this argument.

For homogeneous $A,B$ of weights $a,b\geq1$ with $w=a+b\in\mathbb Z$,
write $\varepsilon=(-1)^{|A||B|}$. On a lowest conformal space,
\begin{equation}
\begin{split}
 o(:AB:)={}&
 \begin{cases}\varepsilon o(B)o(A),&a,b\in\mathbb Z,\\
 0,&a,b\in\mathbb Z+\tfrac12,\end{cases}\\[-1mm]
 &+\varepsilon\sum_{j=0}^{\lceil a\rceil-2}
       \sum_{t=0}^{w-2-j}\binom{w-2-j}{t}\,o(B_{(t)}A).
\end{split}
\label{of:lowest-normal-modes}
\end{equation}
Indeed the normal-mode sum at index $w-1$ has terms
$A_{(j)}B_{(w-2-j)}$ for $j<0$, which kill the bottom, and reversed
terms for $j\geq0$. The latter kill it for $j>a-1$. The endpoint
$j=a-1$ exists precisely when $a$ is integral and gives the first line.
For the remaining indices, commute the lowering mode of $B$ to the
right and use its mode commutator. This gives the second line.
In a coupling, reversing the tensors adds $(-1)^k$.
Also
\begin{equation}
 o(\partial^tA)=(-1)^ta^{\overline{t}} o(A),\qquad
 a^{\overline{t}}=a(a+1)\cdots(a+t-1).
 \label{of:translation}
\end{equation}
These formulas explain why an arbitrary half-integral pair cannot be
replaced by zero: its contraction sum must first be evaluated.

The corresponding first-half formula follows from the same normal-mode
sum. To state both cases at once, for $\sigma\in\{0,\tfrac12\}$ define
$D^{[\sigma]}=D_{(\operatorname{wt}D-1-\sigma)}|_E$ whenever its index
is integral, and write $D^+=D^{[1/2]}$. For fields $A,B$ of weights
$a,b\geq1$, put $w=a+b$ and assume $w-1-\sigma\in\mathbb Z$.
Then, on the original lowest input $E$,
\begin{equation}
\begin{split}
 (:AB:)^{[\sigma]}={}&\varepsilon
  \sum_{\substack{j\in\mathbb Z_{\geq0}\\a-1-\sigma\leq j\leq a-1}}
       B_{(w-2-\sigma-j)}A_{(j)}\big|_E\\
 &+\varepsilon\sum_{j=0}^{\lceil a-\sigma\rceil-2}
       \sum_{t=0}^{w-2-\sigma-j}
       \binom{w-2-\sigma-j}{t}(B_{(t)}A)^{[\sigma]}.
\end{split}
\label{of:normal-modes}
\end{equation}
At $\sigma=0$ this is \eqref{of:lowest-normal-modes}. At $\sigma=1/2$
the first line is $\varepsilon B^+o(A)$ if $a$ is integral, and
$\varepsilon o(B)A^+$ if $a$ is half integral; in the latter case
$o(B)$ acts on the first-half output and is not specialized as an
operator on the bottom. Derivatives satisfy
\begin{equation}
 (\partial^tD)^{[\sigma]}=(-1)^t
                  (\operatorname{wt}D-\sigma)^{\overline{t}}D^{[\sigma]}.
 \label{of:shifted-translation}
\end{equation}
For completeness, the mode of $:AB:$ is the sum of
$A_{(j)}B_{(w-2-\sigma-j)}$ for $j<0$, and reversed terms for $j\geq0$.
The first group kills $E$, since the right mode has shift
$j+1+\sigma-a<0$. In the second group $j>a-1$ also kills $E$.
The surviving indices with $j\geq a-1-\sigma$ give the first line.
At the remaining indices the $B$ mode lowers the original input;
commuting it past $A_{(j)}$ yields the second line by the mode
commutator formula. This derivation uses only the lowest property of
$E$, and makes no such assertion about its first-half image.

\begin{lemma}[Scalar-vacuum projection]
\label{of:scalar-projection}
Adjoin scalar symplectic bosons $x,y$ in their vacuum module, and put
$X^{\rm full}=X+xI$, $Y^{\rm full}=Y+yI$. If $F$ is any integral-weight
field in the tensor product, then on its lowest space
\[
 o(F)|_{E\otimes|0\rangle}
       =o(\pi_{\rm vac}F)|_E\otimes1.
\]
For the raw trace fields and their flavor polars, $\pi_{\rm vac}F$ is
exactly the corresponding traceless field. Their integer zero modes,
and consequently their commutators on this space, agree.
\end{lemma}
\begin{proof}
The scalar vacuum module has nonnegative weights and one-dimensional
weight zero. An integral zero mode preserves total conformal weight.
Positive scalar output weight would require a matter output below its
bottom and is therefore impossible. For a raw normal-ordered trace,
every term containing a scalar oscillator has positive scalar weight,
so its vacuum projection is zero. This remains true after flavor
polarization. Projection is taken after normal ordering and is not
asserted to preserve arbitrary vertex products. The BRST differential
acts only on the traceless tensor factor, so it commutes with projection.
\end{proof}

\subsection{The relative chain derivation}\label{of:chain-derivation-section}
On the actual BRST complex put $p=\partial c$ and
\begin{equation}
 \mathcal D=-\frac12\bigl(\operatorname{tr}:c\partial c:\bigr)_{(0)}.
 \qquad
 \mathcal Db=p,\quad
 \mathcal Dp=\mathcal DX=\mathcal DY=\mathcal Dc=0,
 \quad [\mathcal D,Q]=0.
 \label{of:chain-derivation}
\end{equation}
Here $\mathcal D$ is even, has conformal degree zero and ghost
degree two.  It is a zero-mode derivation of all vertex products.
To check its sign, contract $:c\partial c:$ with $b$.
The $c,b$ contraction passes the remaining odd $\partial c$ and
gives $-p$ in the residue; the $\partial c,b$ double pole has
coefficient $-c$ and its Taylor residue gives another $-p$.
The factor $-1/2$ therefore gives $\mathcal Db=p$.
The scalar parts of the adjoint projector vanish against the
traceless ghost fields.  The other elementary actions are zero.
On the actual BRST current
$j=\operatorname{tr}(cXY-cYX-ccb)$ this yields
\[
 \mathcal Dj=-\operatorname{tr}(ccp)
                 =-\frac13\partial\operatorname{tr}(c^3).
\]
The last equality uses graded cyclicity: rotating one odd letter
past the other two has positive sign.  Since a derivative has
zero zeroth mode, the mode Jacobi identity proves
$[\mathcal D,Q]=0$.

The derivation preserves the relative subcomplex.
For every adjoint component,
$[\mathcal D,b^a_{(0)}]=p^a_{(0)}=(\partial c^a)_{(0)}=0$.
It also commutes with the gauge zero modes, since its defining
trace field is a gauge singlet.  Thus the defining field need
not itself be relative for its zero mode to act on the relative
complex.  Its conformal degree zero preserves the conformal
grading of the chain complex.  The barred identity is obtained
at the field level, so no action of $\mathcal D$ on an arbitrary
lowest-space module is assumed.  It commutes with scalar-vacuum
projection, because it acts only on the ghost factor.

\subsection{Cubic odd nilpotence}
Let $E$ be any finite lowest conformal space of weight $r$.
Use the normalized cubic fields
\[
\begin{gathered}
 B=(\operatorname{tr}X^2b,\tfrac12\operatorname{tr}(XY+YX)b,
                      \operatorname{tr}Y^2b),\qquad C=B|_{b\mapsto p},\\
 J=(e,-H/2,-f),\qquad G=(G_X,G_Y),\qquad
 \bar G=(\bar G_X,\bar G_Y).
\end{gathered}
\]
Write the same letters for their zero modes when acting on $E$.
For weight-two fields set
\begin{equation}
\begin{gathered}
 Z(A,B)=A_{(0)}B-\tfrac12\partial(A_{(1)}B),\\
 K_0=Z(B_0,C_1)-Z(B_1,C_0),\qquad
 K_1=\tfrac12f_{(0)}K_0,\qquad K_2=f_{(0)}K_1.
\end{gathered}
 \label{of:cubic-mixed}
\end{equation}
These $K_i$ are weight-three fields forming a flavor spin-one tensor.
Their zero modes are the antisymmetric mixed cubic anticommutators.
The all-level spin-two mixed identity is given explicitly in
\eqref{of:cubic-mixed-spin-two}.

\begin{proposition}
On $E$,
\begin{equation}
\begin{gathered}
 \{B_i,B_j\}=\{C_i,C_j\}=0,\qquad
 \{B_i,C_j\}=-\{B_j,C_i\},\\
 [K_i,B_j]=[K_i,C_j]=[K_i,K_j]=0.
\end{gathered}
 \label{of:cn2}
\end{equation}
The associative augmentation ideal generated by the six operators
$B_i,C_i$ in their unital algebra is nilpotent.  Consequently
\begin{equation}
 W=\bigcap_i\ker B_i\cap\bigcap_i\ker C_i\ne0
 \label{of:cn3}
\end{equation}
is flavor-stable.  There is a finite flavor-stable filtration of $E$
whose successive quotients have all $B_i,C_i,K_i$ equal to zero.
\end{proposition}

\paragraph{Exact trace identities.}
All the following identities are obtained by \eqref{of:wick-formula},
with the complete relative primitives displayed explicitly. The only
relations used in collecting the resulting matrix entries are graded
cyclicity, supercommutation of distinct trace factors and tracelessness.
Thus they hold in the free relative complex at every $N\geq2$.

The spin-two components vanish:
\[
 Z(B_0,B_0)=Z(C_0,C_0)=Z(B_0,C_0)=0,\qquad
 Z(B_0,C_1)+Z(B_1,C_0)=0.
\]
The scalar components are
\begin{subequations}\label{of:cubic-scalar}
\begin{align}
Z(B_0,B_2)-Z(B_1,B_1)&=-\tfrac6N:G_XG_Y:-\tfrac12Q\operatorname{tr} bbb,
 \label{of:cubic-scalar-b}\\
Z(C_0,C_2)-Z(C_1,C_1)&=-\tfrac6N:\bar G_X\bar G_Y:
                         -\tfrac32Q\operatorname{tr} bpp,
 \label{of:cubic-scalar-c}\\
Z(B_0,C_2)-2Z(B_1,C_1)+Z(B_2,C_0)
 &=-\tfrac6N\bigl(:G_X\bar G_Y:-:G_Y\bar G_X:+\partial T\bigr)
                         -\tfrac32Q\operatorname{tr} bbp.
 \label{of:cubic-scalar-bc}
\end{align}
\end{subequations}
The small-$\mathcal N=4$ OPE gives
$G_{X(0)}G_Y=\bar G_{X(0)}\bar G_Y=0$ and
$G_{X(0)}\bar G_Y-G_{Y(0)}\bar G_X=-2T$.
For a weight-$3/2$ first factor the lowest normal zero mode is
$o(:ab:)=-o(a_{(0)}b)$, so all right-hand sides in \eqref{of:cubic-scalar} have zero
mode zero.  Flavor lowering proves the first three relations of \eqref{of:cn2}.

Here is the additional identity needed for the remaining commutators.
For weight-three $A$ and weight-two $F$ define
\[
 \mathcal C(A,F)=A_{(0)}F-\tfrac23\partial(A_{(1)}F)
                         +\tfrac16\partial^2(A_{(2)}F),
 \qquad
 \mathcal S_F=\mathcal C(K_0,F_2)-2\mathcal C(K_1,F_1)
                                  +\mathcal C(K_2,F_0).
\]
Then $o(\mathcal C(A,F))=[o(A),o(F)]$.  Put
\[
 D=G_{X(0)}C_1-G_{Y(0)}C_0,\qquad D_1=f_{(0)}D,
 \qquad U\mathbin\cdot V=U_0V_2-2U_1V_1+U_2V_0.
\]
The two chain identities are
\begin{subequations}\label{of:cn5}
\begin{align}
\mathcal S_B&=-\tfrac{16}N:(\partial J)\mathbin\cdot B:
             +\tfrac8N:J\mathbin\cdot\partial B:
             +\tfrac{12}N(:G_XD_1:-:G_YD:)+QU_-,
 \label{of:cn5-b}\\
\mathcal S_C&=-\tfrac{16}N:(\partial J)\mathbin\cdot C:
             +\tfrac8N:J\mathbin\cdot\partial C:
             +\tfrac{12}N(:\bar G_XD_1:-:\bar G_YD:)+QU_+.
 \label{of:cn5-c}
\end{align}
\end{subequations}
Explicit primitives are given below so that \eqref{of:cn5} can be verified
without guessing an exact term.  Every juxtaposed word means its trace;
all derivatives in this display are ordinary derivatives:
\begin{subequations}\label{of:cn6}
\begin{align}
U_-={}&\operatorname{tr}\bigl(4bbbp
 +2(\partial X)bYb-X(\partial b)Yb
 +2Xb(\partial Y)b-XbY(\partial b)\notag\\
&+3(\partial X)Ybb-3Xbb(\partial Y)
  -3X(\partial Y)bb+3(\partial X)bbY\notag\\
&-\tfrac12XY(\partial b)b+\tfrac12X(\partial b)bY
  +\tfrac12XYb(\partial b)-\tfrac12Xb(\partial b)Y\bigr),
 \label{of:cn6-minus}\\
U_+={}&\operatorname{tr}\bigl(5bpbp
 +\tfrac52(\partial X)Ybp-\tfrac52Xbp(\partial Y)
 -\tfrac72X(\partial Y)bp+\tfrac72(\partial X)bpY\notag\\
&+\tfrac12XYb(\partial p)-\tfrac12Xb(\partial p)Y
 +\tfrac72(\partial X)Ypb-\tfrac72Xpb(\partial Y)\notag\\
&-\tfrac52X(\partial Y)pb+\tfrac52(\partial X)pbY
 +2(\partial X)bYp+2Xp(\partial Y)b\notag\\
&-X(\partial b)Yp-XpY(\partial b)
 -\tfrac12XY(\partial p)b+\tfrac12X(\partial p)bY\notag\\
&+2Xb(\partial Y)p+2(\partial X)pYb
 -XbY(\partial p)-X(\partial p)Yb\bigr).
 \label{of:cn6-plus}
\end{align}
\end{subequations}
The expressions above use ordinary derivatives. In the normalized-jet
convention of \eqref{of:wick-formula}, $c^{[2]}=\partial p/2$;
this fixes the factors accompanying $\partial p$. The two formulas give
the primitive for each ghost degree separately.

For a weight-one current $J$ and weight-two $F$, splitting the normal
mode on $E$ gives
\begin{equation}
 o(:(\partial J)F:)=-o(F)o(J),\qquad
 o(:J\partial F:)=-2o(F)o(J).
 \label{of:cn7}
\end{equation}
Thus the first two terms of each identity in \eqref{of:cn5} cancel.  Set
\[
 \mathcal L(A,G)=A_{(0)}G-\partial(A_{(1)}G)
                       +\tfrac12\partial^2(A_{(2)}G).
\]
Two further exact, non-cohomological Wick identities are
\begin{equation}
 \mathcal L(D_1,G_X)-\mathcal L(D,G_Y)=0,\qquad
 \mathcal L(D_1,\bar G_X)-\mathcal L(D,\bar G_Y)=0.
 \label{of:cn8}
\end{equation}
Indeed the only surviving normal-mode summand on $E$ in $o(:GD:)$
is $D_{(2)}G_{(0)}$, and commuting the lowering operator $D_{(2)}$
to the right identifies it with $o(\mathcal L(D,G))$.
Equations \eqref{of:cn5}--\eqref{of:cn8} therefore prove
\begin{equation}
 [K_0,B_2]-2[K_1,B_1]+[K_2,B_0]=0
 \quad\hbox{and the same equation with }B\hbox{ replaced by }C.
 \label{of:cn9}
\end{equation}

\paragraph{The algebraic consequence.}
Choose Cartesian spin-one coordinates.  After an immaterial common
normalization of $K$, the relations already proved read
$\{B_i,C_j\}=\epsilon_{ijk}K_k$.
Write $A_{i\ell}=[B_i,K_\ell]$.  Super Jacobi with $B_i,B_j,C_k$
gives
\[
 \epsilon_{jk\ell}A_{i\ell}+\epsilon_{ik\ell}A_{j\ell}=0.
\]
Taking $i=j$ shows $A_{i\ell}=0$ for $i\ne\ell$; taking distinct
$i,j,k$ shows $A_{ii}=A_{jj}$.  Hence $A_{i\ell}=\delta_{i\ell}S$
for an odd scalar operator $S$.  Equation \eqref{of:cn9} is its nonzero scalar
contraction, so $S=0$.  The identical argument works for $C$.
Since each $K$ is a linear combination of $\{B_i,C_j\}$, all $K$ also
commute with one another.  This proves \eqref{of:cn2}.

The three commuting $K$ have a finite joint generalized spectrum in
$\mathbb C^3$.  The finite-dimensional flavor representation integrates
to $SL_2(\mathbb C)$, and conjugation acts on this spectrum through the
spin-one action.  A connected group acts trivially on a finite set;
a spin-one representation has no nonzero fixed covector.  Thus every
joint character is zero.  Simultaneous triangularization now shows
that the commuting ideal $\mathfrak k=(K_0,K_1,K_2)$ is nilpotent:
$\mathfrak k^s=0$ for some $s$ (one can take $s\le\dim E$).

Using \eqref{of:cn2}, reorder any word in the six odd generators to a sum of
words containing each odd generator at most once, multiplied by a
monomial in $K$.  The reordering only replaces two odd factors by one
$K$.  A word of odd-generator length at least $2s+6$ therefore vanishes.
The augmentation ideal $I=(B_i,C_i)$ satisfies $I^{2s+6}=0$.
If $I^mE$ is its last nonzero power, then $I^mE\subset W$, proving
\eqref{of:cn3}.  The powers $I^aE$ are flavor-stable and their successive
quotients have trivial cubic action, as claimed.

\paragraph{The raw quartic top.}
Put $V=\operatorname{tr}(b[X,\partial X])$. With
\[
 Z=B_{0(0)}C_1-B_{1(0)}C_0
       -\tfrac12\partial(B_{0(1)}C_1-B_{1(1)}C_0),
 \qquad L=:H\partial e:-:(\partial H)e:,
\]
one has
\begin{equation}
 Z=\frac43H_4-\frac4N:G_X\bar G_X:
       +\frac2N L+\left(-\frac N3+\frac2N\right)\partial^2e
       +\frac13QV.
 \label{of:uc5}
\end{equation}

Let $H_{4,0}=H_4$ be the raw quartic top, and let
$H_{4,1}=f_{(0)}H_{4,0}/2$, $H_{4,2}=f_{(0)}H_{4,1}$.
Equation \eqref{of:uc5} and direct lowest-mode splitting give
\[
 o(:G_X\bar G_X:)=e,\qquad
 o(:H\partial e:-:(\partial H)e:)=0,\qquad
 o(\partial^2e)=2e.
\]
For the first equality use $G_{X(0)}\bar G_X=\partial e$;
for the second, both summands have zero mode $-eH$.
The $N^{-1}$ terms in \eqref{of:uc5} consequently cancel, leaving
\begin{equation}
 o(H_{4,i})=\tfrac34K_i+\tfrac N2J_i\qquad\hbox{on }E.
 \label{of:cn10}
\end{equation}
Thus $W$ and every $I^aE$ above are also invariant under the raw
quartic top zero modes, and these modes act as $NJ_i/2$ on each
successive quotient.  Equivalently the even operators $J_i,K_i$
form the current $\mathfrak{sl}_2$ semidirect its abelian adjoint
module; their abelian ideal acts nilpotently in every finite-dimensional
representation.  This statement concerns the quartic top and makes
no corresponding assertion for its spin-two bottom.

\paragraph{All-level forms of the two second-channel identities.}
The highest mixed cubic OPE is particularly simple:
\begin{equation}
 \begin{aligned}
 B_{3,0(1)}C_{3,0}&=P_4-\frac1N:P_2P_2:,\\
 B_{3,0(0)}C_{3,0}&=\frac12\partial
                         \left(P_4-\frac1N:P_2P_2:\right).
 \end{aligned}
 \label{of:cubic-mixed-spin-two}
\end{equation}
There are no higher poles. Only $b$ can contract with $p$, producing
the double pole; the first Taylor term of the two surviving source
$X$ letters is one half the derivative of its coefficient. The scalar
projector gives the displayed two-trace term. Flavor lowering gives
all components of this spin-two identity.

The cubic second-channel quartic field is
\begin{equation}
\begin{split}
 &\sum_{i=0}^2(-1)^i\binom2i
 \left(W_{4,i(0)}B_{3,2-i}
          -\frac12\partial(W_{4,i(1)}B_{3,2-i})\right)\\
 &\qquad=-\frac{20}{N}\partial B_{3,0}
       +\frac{10}{N}\mathcal N_1(G,W_3)
       +\frac{10}{3}Q\operatorname{tr}(X^2bb).
\end{split}
\label{of:quartic-cubic-field}
\end{equation}
To verify the equality directly, use
$W_{4,2}=2\operatorname{tr}(X^2Y^2)/3+
\operatorname{tr}(XYXY)/3$ and the normalized first polars.
There are one and two bosonic contractions. Their simple-pole
Taylor terms and the derivative of the double pole, expanded with
\eqref{of:wick-formula}, leave the three displayed terms;
$Q\operatorname{tr}(X^2bb)$ is obtained from the two choices of the
contracted $b$ and the two adjacent $X$ positions. Its primitive is
relative. This is a field identity before lowest-space specialization.
Its barred version follows by the chain derivation, including the
image $\operatorname{tr}(X^2pb+X^2bp)$ of the primitive.

\subsection{The quartic spin-one relation}
Let $E$ be a finite lowest space of conformal weight $r$ for the
Casimir vertex subalgebra; no scalar-flavor assumption is made.
Write $P_i=W_{4,i}$, let $t_i$ be the normalized polarizations of
$(\operatorname{tr} X^2)^2$, and put $R_i=P_i-t_i/4$.
The same letters will denote their zero modes on $E$. Set
$\Omega=ef+fe+H^2/2$ and use the cubic anticommutator vector $K_i$
of \eqref{of:cubic-mixed}.

\begin{proposition}
For every $N\ge2$,
\begin{equation}
 [R_0,R_3]-3[R_1,R_2]
 =-3(R_2e+R_1H-R_0f)+\Omega e+2N K_0
        +\left(\frac{13}{2}-2N^2-16r\right)e.
 \label{of:qr1}
\end{equation}
This is an identity on the full lowest space. It neither assumes that
$K_i=0$ nor that the cubic common kernel is preserved by $R_i$.
\end{proposition}

\paragraph{Field identity.}
For weight-two fields define
$Z(A,B)=A_{(0)}B-\tfrac12\partial(A_{(1)}B)$, so that
$o(Z(A,B))=[o(A),o(B)]$.
Put $w_i=W_{3,i}$ and
\[
\begin{gathered}
 \mathcal J=:eP_2:+:HP_1:-:fP_0:,
 \qquad \mathcal W=:w_0w_2:-:w_1w_1:,\\
 \mathcal C=:e(:ef:):+\tfrac14:e(:HH:):.
\end{gathered}
\]
Let $H_4$ denote the raw quartic top of \eqref{of:uc5}. The exact
field identity is
\begin{align}
 Z(R_0,R_3)-3Z(R_1,R_2)
 ={}&-3\mathcal J-2\mathcal C-\frac{18}{N}\mathcal W
     +\frac{8N}{3}H_4+\frac{54}{N}\partial P_1\notag\\
 &-16:eT:-14:G_X\bar G_X:\notag\\
 &+a_N:H\partial e:+b_N:(\partial H)e:
       +c_N\partial^2e+Q\mathcal U,
 \label{of:qr2}
\end{align}
where
\[
 a_N=\frac{198}{N^2}-\frac{15}{N}-\frac{295}{8}
                +10N-\frac{5N^2}{8},\qquad
 b_N=\frac{18}{N^2}+\frac{15}{N}+\frac{303}{8}
                -10N+\frac{5N^2}{8},
\]
\[
\begin{aligned}
 c_N&=-\frac{288}{N^2}+\frac{223}{4}-\frac{5N^2}{3},\\
 \mathcal U&=8\operatorname{tr}(bYX^3)
       +\frac{16}{3}\operatorname{tr}(bXYX^2)
       -\frac23\operatorname{tr}(bX^2YX)
       +\frac{10N}{3}\operatorname{tr}(bX\partial X).
\end{aligned}
\]
All traced words in $\mathcal U$ and the polarizations $t_i$ are
simultaneous PBW products. Products displayed with colons are true
vertex normal products.

Expanding the two sides by \eqref{of:wick-formula} gives equal
matrix-entry coefficients. The single- and double-contraction terms,
with their surviving left-letter Taylor expansions, also give
\begin{equation}
 w_{0(0)}w_2=6P_1+\frac{12}{N}:eH:+3N\partial e,
 \qquad
 w_{1(0)}w_1=\left(\frac8N-2N\right)\partial e.
 \label{of:qr3}
\end{equation}

\paragraph{Lowest-mode reduction.}
A weight-one first factor in a normal product gives the reversed
product of the two zero modes on $E$. A weight-$3/2$ first factor
gives $o(:AB:)=-o(A_{(0)}B)$. Therefore
\[
 o(\mathcal J)=P_2e+P_1H-P_0f,\qquad
 o(\mathcal C)=(fe+H^2/4)e=(\Omega-H)e/2,
\]
\[
 o(\mathcal W)=-6P_1-\frac{12}{N}He
                        +\left(5N-\frac8N\right)e.
\]
Also $o(\partial P_1)=-2P_1$,
$o(:eT:)=re$, $o(:G_X\bar G_X:)=e$, and
\[
 o(:H\partial e:)=o(:(\partial H)e:)=-eH,
 \qquad o(\partial^2e)=2e,
 \qquad o(H_4)=\frac34K_0+\frac N2e.
\]
Substitution into \eqref{of:qr2} cancels the $P_1/N$ terms and
all negative rank powers. Using $He-eH=2e$, the result is
\begin{equation}
 -3(P_2e+P_1H-P_0f)-\Omega e+2NK_0
                  +\left(\frac{19}{2}-2N^2-16r\right)e.
 \label{of:qr4}
\end{equation}
The polarizations of $(\operatorname{tr}X^2)^2$ are affine-current descendants of
$4:e^2:$, so their zero modes satisfy
\[
 o(t_0/4)=e^2,\qquad
 o(t_1/4)=-(He+eH)/4,\qquad
 o(t_2/4)=H^2/4-\Omega/6.
\]
The $\mathfrak{sl}_2$ commutators then give
\[
 P_2e+P_1H-P_0f
 =R_2e+R_1H-R_0f+e-\frac23\Omega e.
\]
Inserting this in \eqref{of:qr4} proves \eqref{of:qr1}.

The corresponding raw relation is
\begin{equation}
 [p_0,p_3]-3[p_1,p_2]=-(8\Delta-8)e+2NK_0,
 \qquad \Delta=2r+N^2/4.
 \label{of:quartic-raw}
\end{equation}
For an explicit conversion, write
$Q_i=o(t_i/4)$, so $Q_0=e^2$, $Q_1=-(He+eH)/4$,
$Q_2=H^2/4-\Omega/6$, and obtain $Q_3,Q_4$ by normalized lowering.
Substitute $R_i=p_i-Q_i$ in the preceding relation and move every
current through $p_i$ with $[e,p_i]=ip_{i-1}$ and
$[f,p_i]=(4-i)p_{i+1}$. The mixed terms cancel its
$-3(R_2e+R_1H-R_0f)$ term, and the pure-current terms cancel
$\Omega e$ and shift the scalar coefficient by $3/2$.
The result is \eqref{of:quartic-raw}; lowering twice gives
$[p_1,p_4]-3[p_2,p_3]=(8\Delta-8)f+2NK_2$.

On a lowest subspace invariant under all integral zero modes and
annihilated by all odd zero modes, $K_i=0$, since each $K_i$ is an
odd anticommutator. Thus the same actual relation restricts to
\begin{equation}
\begin{gathered}
 [p_0,p_3]-3[p_1,p_2]=-(8\Delta-8)e,\\
 [p_1,p_4]-3[p_2,p_3]=(8\Delta-8)f.
\end{gathered}
 \label{of:quartic-relation}
\end{equation}
Both equations act on that invariant input; no restriction on a
first-half output has been imposed.

\subsection{Second channels at arbitrary trace length}
\begin{equation}
 [p_2,o(\operatorname{tr}X^{2m}b)]
       =-4m\,o(B_{2m+3,1}),\qquad [p,O^{\alpha,(m)}]_2=0.
 \label{of:second-odd}
\end{equation}
Let $n=2m\geq2$ and $O_n=\operatorname{tr}(X^n b)$. Write $B^{[n]}$ for its spin-$m$ tensor.
\paragraph{The auxiliary full-boson calculation.}
Adjoin only the scalar symplectic bosons to the traceless theory;
the ghosts remain traceless.  The full bosons have propagator
$\delta^a_d\delta^c_b/(z-w)$.  Their scalar factors are in the
vacuum module.  By Lemma~\ref{of:scalar-projection}, an integer zero mode on the
tensor-product lowest space equals the zero mode of its actual
scalar-vacuum projection $\pi_{\rm vac}$.  The raw integer fields
$W_{4,a},B^{[n]}_i$ project to their traceless counterparts.
The BRST differential is independent of the scalar bosons, so
$\pi_{\rm vac}Q=Q\pi_{\rm vac}$.  This statement is applied after
all normal contractions; it is not a multiplicativity assertion
for arbitrary fields.

Put $P_k=\operatorname{tr}X^k$, with $P_0=N$, and set
$O_l=\operatorname{tr}(X^l b)$.  After vacuum projection,
$P_1=O_0=0$.  These eliminations will occur only in pure-$X$
products, where there are no internal contractions.  Define
\begin{equation}
\begin{split}
 \alpha_i&=\frac{n-2i}{n+2}-\frac23\delta_{i0},\qquad
 V_n=\sum_{i=0}^n\alpha_i\operatorname{tr}(X^{n-i}bX^ib),\\
 Z_n^{\rm full}&=(W_{4,2}^{\rm full})_{(0)}O_n^{\rm full}
       -\frac2{n+2}\partial((W_{4,2}^{\rm full})_{(1)}O_n^{\rm full}).
\end{split}
 \label{of:uqo2}
\end{equation}
The field $V_n$ is relative, has ghost number $-2$, and has
conformal weight $n/2+2$.  Its $Q$ image therefore has the
same weight and ghost number as $Z_n$.

The exact field identity is
\begin{equation}
\begin{split}
 &\pi_{\rm vac}(Z_n^{\rm full}+2n B^{[n+2],\rm full}_1)-QV_n\\
 &\qquad=\frac{2n}{n+2}\sum_{k=2}^{n-1}
 \left(\frac{n-k+2}{k}:(\partial P_k)O_{n-k}:
                       -:P_k\partial O_{n-k}:\right).
\end{split}
 \label{of:uqo3}
\end{equation}
The empty sum at $n=2$ is zero.  On the right all factors are
pure $X,b$, so their PBW and normal products agree.  All fields
outside the $\pi_{\rm vac}$ expression are traceless fields.

\paragraph{Proof by arbitrary-index contraction sums.}
I give the complete contraction counts, including the position
of every differentiated matrix in the trace containing $b$.
Introduce
\begin{equation}
\begin{gathered}
 A_k=\operatorname{tr}(X^{k-1}\partial X)=\partial P_k/k,
 \qquad J_{l,t}=\operatorname{tr}(X^t(\partial X)X^{l-1-t}b),\\
 D_l=\operatorname{tr}(X^l\partial b),\qquad
 \partial O_l=\sum_{t=0}^{l-1}J_{l,t}+D_l.
\end{gathered}
 \label{of:uqo4}
\end{equation}
Set $g_{l,t}=2-\delta_{t0}-\delta_{t,l-1}$ for $l\ge2$.
The target $O_n$ contains one $b$, which never contracts with
the bosonic quartic.  Thus there are at most two contractions.
The quartic polar is
$W_{4,2}=2\operatorname{tr}(X^2Y^2)/3+
\operatorname{tr}(XYXY)/3$.

For the double pole, choose the two contracted target positions
and let $q$ be the number of $X$ letters between them in the arc
without $b$.  There are $n-1-q$ choices.  The two adjacent-source
assignments and the two separated-source assignments give
\begin{equation}
 \frac23\sum_{q=0}^{n-2}(n-1-q)
   \bigl(P_qO_{n-q}+P_{q+2}O_{n-2-q}+P_{q+1}O_{n-1-q}\bigr).
 \label{of:uqo5}
\end{equation}
After projection this is
$2(n-1)NO_n/3+\sum_{k=2}^{n-1}2(n-k)P_kO_{n-k}$.
For the simple-pole Taylor term the derivative is placed on a
surviving source $X$.  In the adjacent-source case its two
possible positions give weight one at the two ends of a
$b$-containing arc and weight two in its interior.  In the
separated case they give one copy of every $J_{l,t}$, or one
copy of $A_k$.  Hence, writing $l=n-k$, the complete simple-pole
double-contraction coefficients are
\begin{equation}
\begin{array}{c|c}
\text{monomial}&\text{coefficient}\\\hline
 N J_{n,t}&\frac23g_{n,t}\\
 P_kJ_{l,t},\ k\ge2,l\ge2&\frac23(g_{l,t}+1)\\
 P_{n-1}J_{1,0}&\frac23\\
 A_kO_l,\ k\ge2,l\ge1&2l+\frac43\\
 P_kD_l\text{ or }ND_n&0
\end{array}
 \label{of:uqo6}
\end{equation}
For example the $A_kO_l$ entry is the sum
$4(l+1)/3+2l/3$: the first summand comes from two adjacent
source survivors and the second from one separated survivor.

It remains to count $QV_n$ independently.  Write
$V_i=\operatorname{tr}(X^ubX^vb)$, $u=n-i,v=i$.
The current $\operatorname{tr}(cXY-cYX)$ contracts its $c$
with either $b$ and its $Y$ with one target $X$.  Taylor expansion
of its surviving $X$ gives exactly
\begin{equation}
\begin{split}
 QV_i\big|_{\partial X}={}&
 \sum_{a=0}^{u-1}\bigl(P_{u-1-a}J_{v+a+1,a}-A_{u-a}O_{v+a}\bigr)\\
 &+\sum_{a=0}^{v-1}\bigl(A_{a+1}O_{n-1-a}-P_aJ_{n-a,u}\bigr)\\
 &+\sum_{a=0}^{u-1}\bigl(P_aJ_{n-a,v}-A_{a+1}O_{n-1-a}\bigr)\\
 &+\sum_{a=0}^{v-1}\bigl(A_{v-a}O_{u+a}-P_{v-1-a}J_{u+a+1,a}\bigr).
\end{split}
 \label{of:uqo7}
\end{equation}
The first two lines contract the first target $b$, and the last
two contract the second.  In each line the difference comes from
the two orders $cXY,cYX$.  These four choices exhaust the
bosonic-current double contractions.  The ghost-current term
$-\operatorname{tr}(ccb)$ contracts both target $b$ letters;
its Taylor term is
\begin{equation}
                 QV_i\big|_{\partial b}=-P_iD_{n-i}+P_{n-i}D_i.
 \label{of:uqo8}
\end{equation}
Scalar-propagator parts cancel in these BRST counts, since the
current is the adjoint commutator current.  Equivalently one can
first split the bosons into full and scalar parts: $Q$ acts only
on their traceless component.  No triple contraction contributes
to a residue: it would contract all three current letters and
leave no source survivor to Taylor-expand.

For $k+l=n$, summing \eqref{of:uqo7}--\eqref{of:uqo8} with the weights $\alpha_i$
now gives particularly short coefficients:
\begin{equation}
\begin{array}{c|l}
P_kJ_{l,t}&
 \alpha_{l-1-t}-\alpha_{k+1+t}-\alpha_{n-t}+\alpha_t
 =\dfrac{4(k+1)}{n+2}-\dfrac23(\delta_{t0}+\delta_{t,l-1})\\[2mm]
A_kO_l&
 2\left(\displaystyle\sum_{i=k}^n\alpha_i-
              \displaystyle\sum_{i=0}^l\alpha_i\right)
 =\dfrac43-\dfrac{4k(l+1)}{n+2}\\[2mm]
P_kD_l&\alpha_l-\alpha_k=\dfrac{2(k-l)}{n+2}.
\end{array}
 \label{of:uqo9}
\end{equation}
The second and third rows here have $k\ge2,l\ge1$.
The first also applies to $k=0$, with $P_0=N$.
At the remaining endpoint $ND_n$, \eqref{of:uqo8} gives
$-4(n-1)/(3(n+2))$.
All equalities in \eqref{of:uqo9} are finite sums of an affine function;
in particular they are identities in the arbitrary indices,
not estimates obtained at selected degrees.

Finally consider the undifferentiated single-contraction terms.
Put
\[
 t_j=\operatorname{tr}(X^jYX^{n+1-j}b).
\]
The quartic gives
\[
 -\frac23\sum_{a=0}^{n-1}(t_a+t_{a+1}+t_{a+2}),
 \qquad 2nB^{[n+2]}_1=\frac{2n}{n+2}\sum_{j=0}^{n+1}t_j.
\]
The classical part of $QV_i$ is
$t_i+t_{n+1-i}-t_{i+1}-t_{n-i}$.
Its weighted sum agrees with the preceding difference. The
coefficients at $j=0$, at $j=1$, in the range $2\le j\le n-1$,
and at $j=n,n+1$ are, respectively,
\begin{equation}
 \frac{4(n-1)}{3(n+2)},\quad
 \frac{2(n-4)}{3(n+2)},\quad
 -\frac4{n+2},\quad
 \frac{2(n-4)}{3(n+2)},\quad
 \frac{4(n-1)}{3(n+2)}.
 \label{of:uqo10}
\end{equation}
At $n=2$ the middle interval is empty.  Single ghost-current
terms cancel the gauge-variation terms containing an
undifferentiated $c$; thus these are all remaining classical terms.
Subtracting the derivative of \eqref{of:uqo5}, with coefficient
$2/(n+2)$, from \eqref{of:uqo6}, and then subtracting \eqref{of:uqo9}, leaves
exactly the right side of \eqref{of:uqo3}.  The $N$ endpoint cancels.
This proves the field identity by arbitrary-index contraction
counts, with every split trace retained.

\paragraph{Lowest zero modes and flavor lowering.}
If $k$ is even, the pure-$X$ factors on the right of \eqref{of:uqo3}
have integer weights and regular mutual OPE.  Their zero modes
are ordered products and translation gives
\begin{equation}
 o(:(\partial P_k)O_l:)=-\frac k2o(P_k)o(O_l),\qquad
 o(:P_k\partial O_l:)=-\frac{l+2}{2}o(P_k)o(O_l).
 \label{of:uqo11}
\end{equation}
The indicated linear combination is zero.  If $k$ is odd, both
weights are half integral.  Their commuting modes have opposite,
nonzero half-integral energy shifts; move the lowering factor
to the right to kill the lowest space.  Both products again
have zero physical zero mode.  The $Q$ term is zero in cohomology.
Since $o(Z_n^{\rm full})$ is the auxiliary zero-mode commutator,
scalar-vacuum restriction proves the first identity of \eqref{of:second-odd}.

The single-contraction highest transports are
$[p_0,B^{[n]}_1]=4B^{[n+2]}_0$ and
$[p_1,B^{[n]}_0]=-nB^{[n+2]}_0$.  Their original scalar terms
are pure-$X$ half-integral products and have zero lowest zero mode
by the same argument.  Lowering once, and using \eqref{of:second-odd}, gives
\begin{equation}
 [p_1,B^{[n]}_1]=(4-n)B^{[n+2]}_1,
 \qquad [p_0,B^{[n]}_2]=8B^{[n+2]}_1.
 \label{of:uqo12}
\end{equation}
Thus their second coupling is $8-2(4-n)-2n=0$, as asserted.

Apply $\mathcal D$ once to the actual field identity \eqref{of:uqo3}.
It fixes all bosonic fields, sends $O_l$ to
$\bar O_l=\operatorname{tr}(X^lp)$, and sends its primitive to
\begin{equation}
\begin{split}
 \mathcal DV_n
 &=\sum_{i=0}^n\alpha_i\operatorname{tr}
       (X^{n-i}pX^ib+X^{n-i}bX^ip)\\
 &=\sum_{i=0}^n(\alpha_i-\alpha_{n-i})
                       \operatorname{tr}(X^{n-i}bX^ip),\\
 \alpha_i-\alpha_{n-i}
 &=\frac{2(n-2i)}{n+2}-\frac23\delta_{i0}+\frac23\delta_{in}.
\end{split}
 \label{of:uqo15}
\end{equation}
This primitive has ghost number zero, as required for its
$Q$ image to represent the barred commutator.  The minus sign
in the second line is the graded cyclic sign of interchanging
the two odd letters.  Consequently the entire field identity
and all its lowest-mode cancellations apply to $\bar O_l$.
This proves \eqref{of:second-odd} for the barred
family without a chain-level ghost swap or an unproved outer
symmetry assertion.

For even $a=2s\geq2$, set $R_s=o(P_{2s})$ and
$Q_s=o(W_{2s,1})=[f,R_s]/(2s)$. The same projection argument gives
the even second channel used below.
\paragraph{The second-polar commutator has no additional term.}
The required second-polar identity is
\begin{equation}
 [R_s,p_2]=4sQ_{s+1}\qquad(s\ge1).
 \label{of:ung3}
\end{equation}
I prove it from the fields, before imposing a representation
of their zero-mode algebra.  Put $a=2s$ and temporarily compute
with full matrix propagators.  The normalized quartic polar is
$W_{4,2}=\frac23\operatorname{tr}(X^2Y^2)
          +\frac13\operatorname{tr}(XYXY)$, with fully normal
ordering understood.  There are at most two contractions against
$P_a$.  Let $F_i^{\rm full}=(W_{4,2}^{\rm full})_{(i)}P_a^{\rm full}$
and write $F_i=\pi_{\rm vac}F_i^{\rm full}$.  In the next display
the raw traces on the right are traceless fields.  The complete
indexed counts, followed by this vacuum projection, give
\begin{equation}
\begin{split}
 F_1={}&\frac{2aN}{3}P_a+
             a\sum_{\substack{b+c=a\\b,c\ge2}}:P_bP_c:,\\
 F_0={}&-2aW_{a+2,1}+\frac{4N}{3}\partial P_a
             +2a\sum_{\substack{b+c=a\\b,c\ge2}}
                              \frac1c:P_b\partial P_c:.
\end{split}\label{of:ung4}
\end{equation}
For completeness, the two-contraction interval sums before combining
are
\[
 \frac{2a}{3}\sum_{q=0}^{a-2}P_qP_{a-q},\qquad
 \frac a3\sum_{q=0}^{a-2}P_{q+1}P_{a-1-q},
\]
with $P_0=N$ and with $P_1=\operatorname{tr}X$ still present.
The first comes from adjacent source $Y$ letters and the second
from separated source $Y$ letters.  The target starting point has
$a$ choices.  In the Taylor coefficient, a surviving trace of degree
$c$ containing two source letters contributes $2\partial P_c/c$;
one containing one source letter contributes $\partial P_c/c$.
Terms involving the degree-one trace contain either $P_1$ or
$\partial P_1$.  They are pure-$X$ normal products, so they have no
internal contractions and their scalar-vacuum projection is zero.
The remaining pure-$X$ factors project termwise.  This proves both
lines of \eqref{of:ung4} by an arbitrary-length count. Throughout the
calculation the full $U(N)$ field $P_1$ is retained until projection.
The count applies to both parities of $a$.

Since the source has conformal weight two, its auxiliary zero-mode
commutator is $o(F_0^{\rm full})+o(F_1^{\rm full})$.
Lemma~\ref{of:scalar-projection} therefore identifies the actual traceless commutator
$[o(W_{4,2}),o(P_a)]$ with $o(F_0)+o(F_1)$.
The single-trace correction cancels because
$o(\partial P_a)=-(a/2)o(P_a)$.  Every pair in \eqref{of:ung4} also cancels.
If $b,c$ are both even, the pure-$X$ fields have regular mutual OPE,
so on the lowest space
\begin{equation}
 o(:P_bP_c:)=o(P_b)o(P_c),\qquad
 o(:P_b\partial P_c:)=-\frac c2o(P_b)o(P_c).
 \label{of:ung5}
\end{equation}
If $b,c$ are both odd, both displayed normal products have zero
integer zero mode.  To see this directly, a term in the normal-mode sum is a
product of two physical modes with opposite conformal shifts.  Each
shift is a nonzero half-integer.  The mutual regular OPE says their
modes commute, so place the factor with negative shift on the right;
it kills the lowest space.  The same argument applies after a
derivative, because its conformal weight still has the same
half-integer parity.  These exhaust
the parity possibilities because $a$ is even.  Hence
$o(F_0)+o(F_1)=-2a\,o(W_{a+2,1})$ on the auxiliary lowest space.
Reversing the commutator and setting $a=2s$ proves \eqref{of:ung3}
in the actual traceless algebra.
All original scalar-propagator corrections are thereby included;
none survives this integer zero-mode restriction.

The same argument, now with only one source $Y$ and hence a single
pole, proves the upward identity.  Together with the regular OPE
of highest pure traces it gives
\begin{equation}
 [R_s,R_t]=0,\qquad [R_s,Q_t]=2sR_{s+t-1}
                          \qquad(s,t\ge1).
 \label{of:ung6}
\end{equation}
Thus \eqref{of:ung3} is a genuine field consequence additional to the highest
upward relation; it has not been assumed from a small matrix block.

\begin{equation}
 [p,R^{(m)}]_2=0,\qquad R^{(m)}_i=o(W_{2m,i}),\quad m\geq1.
 \label{of:second-even}
\end{equation}
Indeed lower $[p_1,R^{(m)}_0]=-2mR^{(m+1)}_0$ once and use
$[p_2,R^{(m)}_0]=-4mR^{(m+1)}_1$ to obtain
$[p_1,R^{(m)}_1]=2(2-m)R^{(m+1)}_1$.
Lowering $[p_0,R^{(m)}_1]=4R^{(m+1)}_0$ then gives
$[p_0,R^{(m)}_2]=8R^{(m+1)}_1$.
The three coefficients add to $8-4(2-m)-4m=0$.

\begin{equation}
 \begin{gathered}
 [p_0,R^{(m)}_1]=4R^{(m+1)}_0,\qquad
 [p_1,R^{(m)}_0]=-2mR^{(m+1)}_0,\\
 [p_2,R^{(m)}_0]=-4mR^{(m+1)}_1,\qquad
 [p,R^{(m)}]_2=0,\qquad m\geq1.
 \end{gathered}
 \label{of:even-leading-channels}
\end{equation}
The first identity is the single-contraction upward relation, the
second its flavor consequence, the third is \eqref{of:ung3}, and
the last was proved above. All are valid on the entire actual lowest
space, including degrees beyond the rank.

In the degree convention $R_d=o(P_d)$ and
$Q_d=o(W_{d,1})=[f,R_d]/d$ for even $d$, these identities include
\begin{equation}
 \begin{gathered}
 [R_s,p_2]=2sQ_{s+2},\qquad [Q_s,R_t]=-tR_{s+t-2},\\
 [R_s,R_t]=0,\qquad [p,R^{(m)}]_2=0
 \quad(s,t\geq2\text{ even},\ m\geq1).
 \end{gathered}
 \label{of:even-first-second}
\end{equation}
The first equation is \eqref{of:ung3} with its index divided by two.
For the second, the sole $Y$ in $W_{s,1}$ contracts with one of the
$t$ target $X$ letters, giving $-tP_{s+t-2}$ and the projector term
$t:P_{s-1}P_{t-1}:/N$. The latter has regular half-integral factors
and zero lowest zero mode. The third equation follows from the
regular pure-$X$ OPE. Thus both independent degree indices are
unrestricted.

\subsection{The pure quartic--central channel}
\paragraph{The whole-lowest cubic-central highest defect vanishes.}
Write $S_i=R^{(3)}_i$ for the sextic tensor and put
\begin{equation}
 C_0=[p_0,p_3]-3[p_1,p_2]= -\gamma J_0+2NK_0,
 \qquad \gamma=16r+2N^2-8.
\label{of:uke3}
\end{equation}
Use normalized spin-one descendants $C_i$; this $C$ is even and
is unrelated to the barred cubic notation in other sections.
The full quartic bracket has the exact two-channel decomposition
$[p_a,p_b]=4(b-a)S_{a+b-1}+c_{ab}C_{a+b-3}$,
where $c_{03}=2/5,c_{12}=-1/5$ and all other coefficients are fixed by antisymmetry and normalized lowering.  Jacobi on $p_0,p_1,p_2$ therefore gives
\begin{equation}
 [p_0,C_0]=20\bigl([p_0,S_2]-2[p_1,S_1]+[p_2,S_0]\bigr)=0.
\label{of:uke4}
\end{equation}
The last equality is \eqref{of:second-even} at $m=3$, not a sextic-zero assumption.
Since $[p_0,J_0]=0$ and $N\ne0$, this implies
\begin{equation}
 [p_0,K_0]=0,
 \qquad \{B^{(5)}_0,C^{(3)}_0\}=0.
\label{of:uke5}
\end{equation}
\paragraph{The exact remaining fourth-channel relation.}
Free Jacobi and normalized lowering give one further identity:
\begin{equation}
 [p,C]_2=-5[p,S]_4,
 \qquad
 [p,S]_4=-\frac{2N}{5}[p,K]_2.
\label{of:uke6}
\end{equation}
The first identity is checked as an equality of free associative
words in $p_0,\ldots,p_4$: take
$S_0=[p_0,p_1]/4$, lower it as spin three, and take the coupled
brackets.  Explicitly, the descendants needed are
\[
 \begin{aligned}
 S_1&=[p_0,p_2]/8,&
 S_2&=[p_1,p_2]/10+[p_0,p_3]/20,\\
 S_3&=[p_1,p_3]/10+[p_0,p_4]/80,&
 S_4&=[p_2,p_3]/10+[p_1,p_4]/20,\\
 C_1&=[p_0,p_4]/2-[p_1,p_3],&
 C_2&=[p_1,p_4]-3[p_2,p_3].
 \end{aligned}
\]
Writing $\mathcal J(u,v,w)=[u,[v,w]]-[v,[u,w]]+[w,[u,v]]$,
direct substitution gives
$[p,C]_2+5[p,S]_4
 =\tfrac54\mathcal J(p_0,p_1,p_4)
   -\tfrac52\mathcal J(p_0,p_2,p_3)=0$.
The second equality of \eqref{of:uke6} follows from \eqref{of:uke3}, since $[p,J]_2=0$.
\begin{equation}
 [p,S]_4=\sum_{i=0}^4(-1)^i\binom4i[p_i,S_{4-i}]=0,
 \qquad S_i=o(W_{6,i}).\label{of:fourth-sextic}
\end{equation}
\paragraph{The fourth quartic--sextic coupling.}
For normalized tensors define the ordered normal coupling
\begin{equation}
 \mathcal N_k(A,B)=\sum_{i=0}^k(-1)^i\binom ki:A_iB_{k-i}:.
 \label{of:qsf3}
\end{equation}
An inner coupling in a nested expression denotes its entire tensor,
obtained by normalized flavor lowering of its displayed highest
component.  Write $J=(e,-H/2,-f)$, let $\mathcal T$ be the stress
tensor, and set $G=(\operatorname{tr}Xb,\operatorname{tr}Yb)$ and
$\bar G=(\operatorname{tr}X\partial c,\operatorname{tr}Y\partial c)$.
The fields $B^{(4)},C^{(4)}$ have highest components
$\operatorname{tr}X^3b,\operatorname{tr}X^3\partial c$.
For $d=3,4$ the top field used here is precisely
\begin{equation}
 H_{d,0}=-\frac1d(\bar G_Y)_{(0)}(G_Y)_{(0)}W_{d,0}
             +\frac1{d^2}\partial(f_{(0)}W_{d,0}).
 \label{of:qsf4}
\end{equation}
It has conformal weight $d/2+1$ and flavor spin $d/2-1$.
These definitions fix all normal-order and derivative conventions.

The weight-four field representing \eqref{of:fourth-sextic} is
\begin{equation}
 \mathcal F=\sum_{i=0}^4(-1)^i\binom4i
 \left((W_{4,i})_{(0)}W_{6,4-i}
       -\frac13\partial((W_{4,i})_{(1)}W_{6,4-i})\right).
 \label{of:qsf5}
\end{equation}
Indeed $o(\partial A)=-\operatorname{wt}(A)o(A)$ and the
zero-mode commutator formula for a weight-two first field gives
$o(\mathcal F)=[p,S]_4$.  The following table specifies sixteen
strong fields $\mathcal O_j$ and Laurent coefficients $c_j$.
\begin{equation}
\renewcommand{\arraystretch}{1.23}
\begin{array}{c|l|l}
j&\mathcal O_j&c_j\\\hline
1&\partial H_4&448/9\\
2&\mathcal N_1(J,\partial^2J)&560/N\\
3&\partial^3J&-2240/(3N^3)+8680/(27N)-2800N/27\\
4&:J\partial\mathcal T:&560/(3N)+280N/9\\
5&:(\partial J)\mathcal T:&896/(3N)+448N/9\\
6&\mathcal N_2(J,\partial W_4)&168/N^2-392/15\\
7&\mathcal N_2(\partial J,W_4)&-336/N^2+784/15\\
8&\mathcal N_3(W_3,W_5)&-56/(3N)\\
9&\mathcal N_1(W_3,H_3)&-56/3\\
10&\mathcal N_3(W_5,W_3)&28/(3N)\\
11&\mathcal N_1(H_3,W_3)&-35/3\\
12&\mathcal N_0(J,\mathcal N_3(W_3,W_3))&28/3\\
13&\mathcal N_2(J,\mathcal N_0(J,\partial J))&1344/(5N)\\
14&\mathcal N_2(\partial J,\mathcal N_0(J,J))&2016/(5N)\\
15&\mathcal N_1(G,C^{(4)})&-14\\
16&\mathcal N_1(\bar G,B^{(4)})&14
\end{array}
 \label{of:qsf6}
\end{equation}
Every field in this table means its flavor highest component.

Here is an explicit relative primitive for the full field equality,
including all scalar projectors.  A bracket $\langle w_1|\cdots|w_t\rangle$
means the PBW product of the indicated matrix traces; it does not
mean a nested normal product.  On a flavor-weight-one PBW state with
$2M$ bosonic matrix letters put
\begin{equation}
 \Pi_1^{(M)}=\prod_{\ell=2}^{M}
       \left(1-\frac{fe}{(\ell-1)(\ell+2)}\right).
 \label{of:qsf7}
\end{equation}
Here $e,f$ act by flavor raising and lowering on fields.
The seventeen seeds and coefficients below all have ghost number
$-1$ and conformal weight four.  Set $M_j=3$ for $j\le6$ and
$M_j=2$ otherwise.
\begin{equation}
\renewcommand{\arraystretch}{1.17}
\begin{array}{c|l|r}
j&V_j&d_j\\\hline
1&\langle XXXX|YYb\rangle&14N/3\\
2&\langle XX|XX|YYb\rangle&7/3\\
3&\langle XXXXYbY\rangle&-56/3\\
4&\langle XX|XXY|Yb\rangle&-28/3\\
5&\langle XXXXbYY\rangle&-56/3\\
6&\langle XXXYXYb\rangle&-56/3\\
7&\langle XX|XY\partial b\rangle&140/9\\
8&\langle XXX|Y\partial b\rangle&-140/27\\
9&\langle XXX(\partial Y)b\rangle&-56N/3\\
10&\langle XX|X(\partial Y)b\rangle&140/9\\
11&\langle XXX|(\partial Y)b\rangle&-56/27\\
12&\langle XX(\partial X)Yb\rangle&56N/3\\
13&\langle XX|(\partial X)Yb\rangle&224/9\\
14&\langle XX(\partial X)|Yb\rangle&-224/9\\
15&\langle XX|(\partial X)bY\rangle&-28/3\\
16&\langle X(\partial X)|XYb\rangle&280/9\\
17&\langle X(\partial X)Y|Xb\rangle&-56/3
\end{array}
 \label{of:qsf8}
\end{equation}
With precisely these conventions the exact identity is
\begin{equation}
 \mathcal F=\sum_{j=1}^{16}c_j\mathcal O_j
       +Q\left(\sum_{j=1}^{17}d_j\Pi_1^{(M_j)}V_j\right).
 \label{of:qsf9}
\end{equation}
This is an identity in the formal-rank traceless Wick algebra,
using the projector $\delta\delta-N^{-1}\delta\delta$ in every
contraction, before taking a lowest space or imposing a finite-rank
trace relation.  Formula~\eqref{of:wick-formula} expands both sides into $62$ PBW monomials, with identical coefficients in $\mathbb Q[N,N^{-1}]$. In performing this expansion, $Q$ is its $k=0$ instance with left field $j_{\rm BRST}$, and every projected seed is expanded by the finite product \eqref{of:qsf7}. This proves \eqref{of:qsf9} directly in the relative complex.

\paragraph{Actual lowest normal modes, with all odd terms retained.}
Apply \eqref{of:normal-modes}, retaining its contraction sum also
for the half-integral pairs and for the odd pairs. All correction fields
have weight at most three, so the cubic identities above suffice for
this reduction.

Put $k=K_0$, $P=er$, and, with the displayed operator ordering, put
\begin{equation}
 L=p_2e+p_1H-p_0f,\qquad
 U=\frac{4e^2f-2eH+eH^2}{3}-e.
 \label{of:qsf11}
\end{equation}
Use the actual cubic normalization $o(H_{4,0})=3k/4+Ne/2$
from \eqref{of:cn10}.  Applying \eqref{of:normal-modes} and the traceless Wick contractions
to \eqref{of:qsf6} gives the following entire zero-mode table:
\begin{equation}
\renewcommand{\arraystretch}{1.23}
\begin{array}{c|l}
j&o(\mathcal O_j)\\\hline
1&-9k/4-3Ne/2\\
2&-2e\\
3&-6e\\
4&-2P\\
5&-P\\
6&-2L\\
7&-L\\
8&-9Nk/2-(18+3N^2)e+36L/N+36P\\
9&-2k-(8/(3N)+4N/3)e-16P/N\\
10&-9Nk+(-480/N^2+324-66N^2)e+72L/N+72P\\
11&-4k+(-80/(3N)+8N/3)e-32P/N\\
12&12NP-72(U+e)/N\\
13&U\\
14&U\\
15&-k-2Ne/3\\
16&k+2Ne/3
\end{array}
 \label{of:qsf12}
\end{equation}
For clarity, the correction fields in \eqref{of:normal-modes} all have weight at
most three.  Their reduction uses the actual fields
$J,\mathcal T,W_4,H_4$, current products, $\mathcal N_k(W_3,W_3)$,
and $\mathcal N_k(G,\bar G)$, modulo explicit BRST primitives.
Current products use the reversed integer endpoint in \eqref{of:normal-modes};
the half-integral products use its contraction sum.  This gives a
direct recursive computation of every entry of \eqref{of:qsf12}, without
a representation classification or an assumption on the cubic kernel.
In particular the two orders of each half-integral pair have been
evaluated separately; their different entries are essential.

Substitution of \eqref{of:qsf6} into \eqref{of:qsf12} gives
\begin{equation}
             \sum_{j=1}^{16}c_j o(\mathcal O_j)=0
 \label{of:qsf13}
\end{equation}
coefficient by coefficient in $k,e,P,L,U$.  The BRST term in
\eqref{of:qsf9} acts trivially in cohomology.  Equations \eqref{of:qsf5} and
\eqref{of:qsf13} prove \eqref{of:fourth-sextic}, and \eqref{of:uke5}--\eqref{of:uke6} prove \eqref{of:quartic-K}.

Together with \eqref{of:uke5}--\eqref{of:uke6}, the vanishing fourth coupling proves that
$[p,K]$ has only spin two. I normalize its components by
\begin{equation}
 [p_a,K_i]=(2i-a)T_{a+i-1},\qquad 0\leq a\leq4,\quad 0\leq i\leq2,
 \label{of:quartic-K}
\end{equation}
where components outside the range $0,\ldots,4$ mean zero. The highest
spin-three channel is \eqref{of:uke5}, and the spin-one channel is zero by
\eqref{of:uke6}. Raising and lowering then determine every remaining coefficient
from one component, proving the normalization in \eqref{of:quartic-K}.

\subsection{The cubic and quintic initial returns}
\label{of:low-returns}
The regular highest OPE, the upward relation and the second-channel
identity at $n=2$ give the complete cubic table
\begin{equation}
 [p_i,O^{\alpha,(1)}_j]=(4j-2i)O^{\alpha,(2)}_{i+j-1}.
 \label{of:cubic-return}
\end{equation}
Its coefficients obey normalized flavor lowering; these three channels
exhaust the tensor product of spins two and one. Put
$A_0=[p_0,p_3]-3[p_1,p_2]+\gamma e=2NK_0$, with
$\gamma=8\Delta-8$, and let $A_i$ be its normalized spin-one descendants.
For either odd family put $D=O^{\alpha,(2)}$, $B=O^{\alpha,(1)}$, and define
\[
 \mathcal E_2=[p_0,D_2]-2[p_1,D_1]+[p_2,D_0],
\]
\[
 \mathcal E_1=[p_0,D_3]-3[p_1,D_2]+3[p_2,D_1]-[p_3,D_0]+\gamma B_0,
\]
\[
 \mathcal E_0=[p_0,D_4]-4[p_1,D_3]+6[p_2,D_2]-4[p_3,D_1]+[p_4,D_0].
\]
Super Jacobi and \eqref{of:cubic-return} give
\begin{equation}
 [A,B]_0=-6\mathcal E_2,\qquad
 [A,B]_1=2\mathcal E_1,\qquad
 [A,B]_2=6\mathcal E_0.
 \label{of:quintic-jacobi}
\end{equation}
Here $[A,B]_k=\sum_i(-1)^i\binom ki[A_i,B_{k-i}]$.
For example, $[A_0,B_0]$ expands into
$-6[p_0,D_2]+12[p_1,D_1]-6[p_2,D_0]$, since
$[p_i,B_0]=-2iD_{i-1}$ and $[e,B_0]=0$.
For the next two equations substitute
$A_1=[p_0,p_4]/2-[p_1,p_3]-\gamma H/2$ and
$A_2=[p_1,p_4]-3[p_2,p_3]-\gamma f$; apply the same Jacobi
expansion and collect the coefficients $1,-3,3,-1$ and
$1,-4,6,-4,1$. The current terms are respectively
$2\gamma B_0$ and zero, giving the stated expressions.
Cubic centrality makes all three left sides zero. The second-channel
zero and these two lower-channel equations prove the entire quintic
return: it has the upward part and only its spin-one return, with
coefficient $d_2=2(4\Delta-4)/5$. In the notation of
Section~\ref{sec:odd}, this is $E_2=Z_2=0$.
The argument is carried out for both actual field families; their
required field identities follow from the chain derivation of
Subsection~\ref{of:chain-derivation-section}.

\subsection{The septic initial return}
Let $n=2m\geq4$ and $B^{[n]}$ have highest component $\operatorname{tr}(X^n b)$.
Use auxiliary full bosons and traceless ghosts,
and project to the scalar vacuum after all contractions.
For $k=3,4$ put
\begin{equation}
 \mathcal H_{k,n}=\pi_{\rm vac}\sum_{a=0}^{k}(-1)^a\binom ka
 \left((W_{4,a}^{\rm full})_{(0)}B^{[n],\rm full}_{k-a}
 -\frac2{n+2}\partial
             ((W_{4,a}^{\rm full})_{(1)}B^{[n],\rm full}_{k-a})\right).
 \label{of:uor1}
\end{equation}
The actual whole-lowest operator is precisely
$o(\mathcal H_{k,n})=[p,B^{[n]}]_k$.  This follows from the
integer zero-mode scalar-vacuum lemma; no equality between the
unprojected full and traceless OPEs is assumed.

\paragraph{The finite index prescription.}
Use source vertices $L=\{0,1,2,3\}$, target matrix vertices
$R=\{0,\ldots,n-1\}$, and a target ghost vertex $*$.
Start with the two oriented cycles $(0,1,2,3)_L$ and
$(0,1,\ldots,n-1,*)_R$.
Choose $q\in\{1,2,3\}$, subsets $A\subset L,C\subset R$
of size $q$, and a bijection $\rho:A\longrightarrow C$.
These choices are each counted once.  For every matched pair
exchange its two outgoing successors in the two-cycle permutation;
then delete all matched vertices.  Read the resulting cycles as
matrix traces, multiplying them in PBW order.  An empty cycle
contributes $N$.  A cycle of length one contributes zero after
scalar-vacuum projection, including the trace of $b$.
This rule is equivalently composition of the initial permutation
with the disjoint transpositions $(a,\rho(a))$, followed by
deletion of all contracted vertices.

Choose a subset $U\subset L\setminus A$ and a subset
$V\subset R\setminus C$ with
\begin{equation}
             |U|=u,\qquad |V|=k-q-u,
             \qquad 0\le u\le k-q.
 \label{of:uor2}
\end{equation}
Assign $Y$ to these uncontracted vertices, $X$ to every other
uncontracted matrix vertex, and $b$ to $*$.  Impossible subset
sizes contribute no summand.  Let $\Phi$ denote the PBW trace
product produced by the preceding cut rule, with these letters.
Only the surviving \emph{source} letters are allowed to receive
Taylor derivatives.  Define $\mathsf L_j\Phi$ to be the sum over
all assignments of nonnegative jet orders of total degree $j$
to those letters, replacing a letter $Z$ of order $d$ by
$\partial^d Z/d!$.  In particular
\[
 \mathsf L_0\Phi=\Phi,\qquad
 \mathsf L_1\Phi=\sum_{v\in L\setminus A}\Phi[\partial Z_v],
\]
and for $q=3$ only one source survivor exists, so
$\mathsf L_2\Phi=\Phi[\partial^2 Z_v/2]$.
Let $\mathsf L_{-1}\Phi=0$.  Set
\begin{equation}
 \mathsf E_q\Phi=\mathsf L_{q-1}\Phi
                       -\frac2{n+2}\partial(\mathsf L_{q-2}\Phi).
 \label{of:uor3}
\end{equation}
The derivative in the second term acts on the entire PBW product,
including the target letters and the ghost.  This distinction
retains all derivative returns.

\paragraph{Closed coefficients for every trace profile.}
With exactly the conventions above, the full formula is
\begin{equation}
\boxed{\quad
 \mathcal H_{k,n}=
 \sum_{q=1}^3\ \sum_{\substack{A\subset L,\ C\subset R\\|A|=|C|=q}}
 \ \sum_{\rho:A\overset\sim\longrightarrow C}
 \ \sum_{u=0}^{k-q}\omega_{k,q,u}(n)
 \ \sum_{\substack{U\subset L\setminus A,\ |U|=u\\
                    V\subset R\setminus C,\ |V|=k-q-u}}
                      \mathsf E_q\Phi .
 \quad}
 \label{of:uor4}
\end{equation}
There is no implicit Wick coefficient in this expression.  Its
coefficients are the following closed rational functions:
\begin{equation}
\begin{split}
 \omega_{3,q,u}(n)
   &=(-1)^u\frac{(n+1)(n+2)(4-u-q)!(n-3+u)!}
                        {4(n-q+2)!},\\
 \omega_{4,q,u}(n)
   &=(-1)^u\frac{(n+1)(4-u-q)!(n-4+u)!}{(n-q+1)!}.
\end{split}
 \label{of:uor5}
\end{equation}
For the third channel the six entries, in a form without
factorials, are
\begin{equation}
\begin{array}{c|ccc}
q\backslash u&0&1&2\\\hline
1&\dfrac{3(n+2)}{2n(n-1)(n-2)}
 &-\dfrac{n+2}{2n(n-1)}&\dfrac{n+2}{4n}\\[2mm]
2&\dfrac{(n+1)(n+2)}{2n(n-1)(n-2)}
 &-\dfrac{(n+1)(n+2)}{4n(n-1)}&\\[2mm]
3&\dfrac{(n+1)(n+2)}{4(n-1)(n-2)}&&
\end{array}
 \label{of:uor6}
\end{equation}
For $q=1$ the trace product has one face; for $q=2$ it has
two faces; for $q=3$ it has either one or three faces.
Empty faces, and hence every power of $N$, are retained by the
cut rule.  The formula includes products in which the ghost
and a differentiated source survivor belong to different faces.
It also includes every position of the remaining $Y$ letters;
they have not been replaced by a commuting-symbol approximation.

\paragraph{Proof of the formula.}
In a summand of the $k$th coupling there are exactly $k$ letters
$Y$ among the two raw tensors.  A contraction removes one $X$
and one $Y$.  If $\ell$ of the $q$ matched source vertices carried
$Y$, then the source polar index was $a=\ell+u$ and the target
polar index was $k-\ell-u$.  Their normalized polar denominators
are $\binom4{\ell+u}$ and $\binom n{k-\ell-u}$.
There are $\binom{q}{\ell}$ ways to place these source $Y$ letters
among the matching pairs; their target partners are then forced.
The contraction sign is $(-1)^\ell$, so together with the coupling
sign $(-1)^{\ell+u}$ it leaves $(-1)^u$.  Thus the coefficient of
the prescribed cut and survivor assignment is exactly
\begin{equation}
 \omega_{k,q,u}(n)=(-1)^u\sum_{\ell=0}^q
 \frac{\binom{q}{\ell}\binom k{\ell+u}}
      {\binom4{\ell+u}\binom n{k-\ell-u}}.
 \label{of:uor7}
\end{equation}
Every Wick matching gives a unique term of this construction,
and conversely; no division by cyclic face multiplicities or
by $q!$ occurs.

For $k=4$ the binomial ratio in the numerator cancels.
Writing $r=4-u$ and using
$\binom ns^{-1}=(n+1)\int_0^1t^s(1-t)^{n-s}\,dt$, the finite
sum becomes
\[
 (-1)^u(n+1)\int_0^1t^{r-q}(1-t)^{n-r}\,dt,
\]
which is the second formula in \eqref{of:uor5}.  For $k=3$ use
$\binom3a/\binom4a=(4-a)/4$ and put $r=3-u$.
The same binomial sum inside the integral has the extra factor
$(r+1-q+qt)/4$.  Integrating gives the first formula in \eqref{of:uor5}.
All factorial arguments in the stated ranges are nonnegative.

There are $q$ simple contractions.  The zeroth OPE mode therefore
needs Taylor degree $q-1$, and the first mode needs degree $q-2$;
this gives \eqref{of:uor3}.  Although a fourth contraction is possible
when $k=4$, it leaves no source survivor.  It contributes only
to the third OPE mode, and contributes neither to the zeroth
nor to the first mode.  It is therefore absent from \eqref{of:uor4}.
These observations exhaust the contractions and prove \eqref{of:uor4}.
Replacing the single target $b$ by $\partial c$ leaves this
bosonic contraction proof unchanged; equivalently apply the
actual raising derivation \eqref{of:chain-derivation} to the field formula.

\paragraph{The septic resonance identity.}
All letters in a trace denote simultaneous PBW normal ordering.
Let $B_d=\operatorname{tr}(X^{d-1}b)$, $W_d=\operatorname{tr}X^d$,
$J_0=\operatorname{tr}X^2/2$, $G_0=\operatorname{tr}Xb$,
and use the raw top fields
\begin{equation}
 H_d=-\frac1d(\bar G_Y)_{(0)}(G_Y)_{(0)}W_d
             +\frac1{d^2}\partial(f_{(0)}W_d).
 \label{of:sor3}
\end{equation}
In the field table $J$ means the current field, while $T$ means the
stress tensor. Define
\begin{equation}
 \mathcal C_k^{a,b}(A,B)
 =\sum_{i=0}^k(-1)^i\binom ki
       :\partial^a A_i\,\partial^b B_{k-i}:.
 \label{of:sor4}
\end{equation}
For $\mathcal H_{3,6}$ in \eqref{of:uor1}, the complete exact identity is
\begin{equation}
 \mathcal H_{3,6}=Q\mathcal V+\sum_{j=1}^{19}c_j\mathcal F_j.
 \label{of:sor5}
\end{equation}
Table~\ref{of:table-resonance-fields} specifies every $c_j$ and $\mathcal F_j$;
table~\ref{of:table-resonance-modes} gives the corresponding whole-lowest
zero modes. In that table $B_i=o(B_{3,i})$, $D_i=o(B_{5,i})$,
$e=o(J_0)$ and $H=[e,f]$, while $r$ is central.
\small
\begin{longtable}{r l l}
\phantomsection\label{of:table-resonance-fields}$j$&$\mathcal F_j$&$c_j$\\\hline
\endfirsthead
\multicolumn{3}{c}{\tablename~\thetable\ (continued)}\\
$j$&$\mathcal F_j$&$c_j$\\\hline
\endhead
\hline
\multicolumn{3}{r}{\textit{Continued on the next page}}\\
\endfoot
\hline
\caption{Fields and coefficients in the degree-seven resonance identity.}\\
\endlastfoot
1&$\partial^2 D$&$\frac{276}{5}-\frac{1}{5}N^{2}$\\
2&$\mathcal C_{0}^{0,2}(J,B)$&$-48N^{-1}+\frac{1}{2}N$\\
3&$\mathcal C_{0}^{1,1}(J,B)$&$-\frac{612}{5}N^{-1}-\frac{19}{2}N$\\
4&$\mathcal C_{0}^{2,0}(J,B)$&$-\frac{204}{5}N^{-1}+4N$\\
5&$\mathcal C_{1}^{0,1}(J,D)$&$-\frac{114}{5}$\\
6&$\mathcal C_{1}^{1,0}(J,D)$&$\frac{342}{5}$\\
7&$\mathcal C_{0}^{0,0}(T,D)$&$-\frac{96}{5}$\\
8&$\mathcal C_{0}^{0,2}(G,W_3)$&$-20N^{-1}+\frac{2}{5}N$\\
9&$\mathcal C_{0}^{1,1}(G,W_3)$&$-\frac{216}{5}N^{-1}-\frac{37}{15}N$\\
10&$\mathcal C_{0}^{2,0}(G,W_3)$&$-\frac{102}{5}N^{-1}+\frac{2}{5}N$\\
11&$\mathcal C_{1}^{0,1}(G,W_5)$&$-\frac{36}{25}$\\
12&$\mathcal C_{1}^{1,0}(G,W_5)$&$\frac{12}{5}$\\
13&$\mathcal C_{0}^{0,0}(G,H_5)$&$\frac{51}{5}$\\
14&$\mathcal C_{1}^{0,1}(W_3,B_4)$&$-\frac{51}{5}$\\
15&$\mathcal C_{1}^{1,0}(W_3,B_4)$&$17$\\
16&$\mathcal C_{1}^{0,1}(W_4,B)$&$-\frac{36}{5}$\\
17&$\mathcal C_{1}^{1,0}(W_4,B)$&$\frac{36}{5}$\\
18&$\mathcal C_{0}^{0,0}(H_3,B_4)$&$\frac{102}{5}$\\
19&$\mathcal C_{0}^{0,0}(H_4,B)$&$16$\\
\end{longtable}
\begin{longtable}{r l}
\phantomsection\label{of:table-resonance-modes}$j$&$o(\mathcal F_j)$\\\hline
\endfirsthead
\multicolumn{2}{c}{\tablename~\thetable\ (continued)}\\
$j$&$o(\mathcal F_j)$\\\hline
\endhead
\hline
\multicolumn{2}{r}{\textit{Continued on the next page}}\\
\endfoot
\hline
\caption{Lowest-space zero modes of the fields in the resonance identity.}\\
\endlastfoot
1&$(12)D_{0}$\\
2&$(6)B_{0}e$\\
3&$(2)B_{0}e$\\
4&$(2)B_{0}e$\\
5&$(-3)D_{1}e+(-\frac{3}{2})D_{0}H$\\
6&$(-1)D_{1}e+(-\frac{1}{2})D_{0}H$\\
7&$(1)D_{0}r+(3)D_{0}$\\
8&$0$\\
9&$0$\\
10&$0$\\
11&$(18)D_{0}$\\
12&$(6)D_{0}$\\
13&$(-\frac{12}{5})D_{0}$\\
14&$(18)D_{0}+(-36N^{-1})B_{0}e$\\
15&$(6)D_{0}+(-12N^{-1})B_{0}e$\\
16&$(-2)B_{1}p_{0}+(2)B_{0}p_{1}+(18)D_{0}$\\
17&$(-2)B_{1}p_{0}+(2)B_{0}p_{1}+(6)D_{0}$\\
18&$(-8)D_{0}+(8N^{-1})B_{0}e$\\
19&$(\frac{3}{4})B_{0}K_{0}+(18N^{-1}+\frac{1}{2}N)B_{0}e+(-\frac{27}{2})D_{0}$\\
\end{longtable}
\normalsize
Every normal-mode correction is included in table~\ref{of:table-resonance-modes}. For example,
$o(:TD:)=D_0(r+3)$, rather than just $rD_0$; also
\begin{equation}
 o(:H_4B:)=\frac34B_0K_0+
       \left(\frac N2+\frac{18}N\right)B_0e-\frac{27}2D_0.
 \label{of:sor6}
\end{equation}
The last identity uses the actual whole-lowest relation
$o(H_{4,i})=3K_i/4+NJ_i/2$.

Here is a completely specified primitive in \eqref{of:sor5}. For a PBW trace word
$w$ of flavor weight two with $d$ matter letters $X,Y$
(each differentiated matter letter counted once; ghosts not counted), define
\begin{equation}
 \Pi_2w=\prod_{s=3}^{d/2}
       \left(1-\frac{f_{(0)}e_{(0)}}{(s-2)(s+3)}\right)w.
 \label{of:sor7}
\end{equation}
Only $d=4,6$ occur in the primitive seeds: this is the identity for $d=4$ and
$1-f_{(0)}e_{(0)}/6$ for $d=6$. A bracket $[u]$ below is the
PBW trace of the displayed word; juxtaposed brackets are PBW trace
products. The ghost letters are odd, and their displayed order is part
of the definition. Set $\mathcal V=\sum_w v_w\Pi_2w$ with the following
coefficients in table~\ref{of:table-resonance-primitive}. All jets here are first derivatives.
\small
\begin{longtable}{l l}
\phantomsection\label{of:table-resonance-primitive}$w$&$v_w$\\\hline
\endfirsthead
\multicolumn{2}{c}{\tablename~\thetable\ (continued)}\\
$w$&$v_w$\\\hline
\endhead
\hline
\multicolumn{2}{r}{\textit{Continued on the next page}}\\
\endfoot
\hline
\caption{Coefficients of the relative BRST primitive in the resonance identity.}\\
\endlastfoot
$[X X X X X Y b b]$&$-2$\\
$[X X X X X b Y b]$&$\frac{16}{5}$\\
$[X X X X X b][Y b]$&$-\frac{3}{5}N^{-1}+\frac{13}{10}N$\\
$[X X][X X X b][Y b]$&$-\frac{73}{20}+\frac{6}{5}N^{2}$\\
$[X X X][X X b][Y b]$&$-\frac{19}{10}+\frac{1}{10}N^{2}$\\
$[X X X X X b b Y]$&$-\frac{9}{5}$\\
$[X X X X][X b b Y]$&$\frac{3}{10}N$\\
$[X X X X][X Y b b]$&$-\frac{3}{10}N$\\
$[X X X X][X b][Y b]$&$-\frac{49}{40}-\frac{9}{20}N^{2}$\\
$[X X][X X][X b b Y]$&$\frac{67}{80}-\frac{3}{80}N^{2}$\\
$[X X][X X][X Y b b]$&$-\frac{67}{80}+\frac{3}{80}N^{2}$\\
$[X X][X X][X b][Y b]$&$-\frac{3}{20}N^{-1}$\\
$[X X X X Y X b b]$&$-\frac{69}{10}$\\
$[X X X X Y b X b]$&$-\frac{8}{5}$\\
$[X X X X Y b][X b]$&$-\frac{3}{10}N^{-1}-\frac{9}{5}N$\\
$[X X X X b X Y b]$&$\frac{61}{10}$\\
$[X X X X b][X Y b]$&$\frac{3}{10}N^{-1}-\frac{3}{5}N$\\
$[X X][X X b][X Y b]$&$-\frac{3}{5}-\frac{3}{20}N^{2}$\\
$[X X X X b X b Y]$&$\frac{11}{2}$\\
$[X X X][X b X b Y]$&$-\frac{1}{10}N$\\
$[X X X][X X b b Y]$&$-\frac{1}{10}N$\\
$[X X X][X X Y b b]$&$\frac{1}{10}N$\\
$[X X X][X Y b X b]$&$\frac{1}{10}N$\\
$[X X X X b][X b Y]$&$\frac{3}{10}N^{-1}-\frac{3}{5}N$\\
$[X X][X X b][X b Y]$&$-\frac{3}{5}-\frac{3}{20}N^{2}$\\
$[X X X][X b][X b Y]$&$-\frac{4}{5}$\\
$[X X X][X Y b][X b]$&$\frac{4}{5}$\\
$[X X X X b Y X b]$&$\frac{59}{10}$\\
$[X X][X X b Y X b]$&$-\frac{3}{20}N^{-1}+\frac{9}{10}N$\\
$[X X][X X b X Y b]$&$\frac{3}{20}N^{-1}-\frac{9}{10}N$\\
$[X X][X X X Y b b]$&$-\frac{3}{20}N^{-1}+\frac{9}{10}N$\\
$[X X][X X X b b Y]$&$\frac{3}{20}N^{-1}-\frac{9}{10}N$\\
$[X X][X X Y b X b]$&$-\frac{3}{20}N^{-1}+\frac{9}{10}N$\\
$[X X][X X b X b Y]$&$\frac{3}{20}N^{-1}-\frac{9}{10}N$\\
$[X X X X b Y][X b]$&$-\frac{3}{10}N^{-1}-\frac{9}{5}N$\\
$[X X X Y X b X b]$&$-\frac{23}{10}$\\
$[X X X Y b X X b]$&$-\frac{7}{2}$\\
$[X X X Y X X b b]$&$-\frac{23}{10}$\\
$[X X X b X X Y b]$&$\frac{71}{10}$\\
$[X X X b][X X Y b]$&$-\frac{12}{5}N$\\
$[X X X b X X b Y]$&$\frac{59}{10}$\\
$[X X X b][X X b Y]$&$-\frac{12}{5}N$\\
$[X X X b X Y X b]$&$\frac{17}{10}$\\
$[X X X X b \partial b]$&$-\frac{2}{5}N$\\
$[X X X][X b \partial b]$&$-\frac{17}{10}$\\
$[X X X X][b \partial b]$&$\frac{9}{8}+\frac{3}{20}N^{2}$\\
$[X X][X X][b \partial b]$&$\frac{1}{16}N+\frac{3}{80}N^{3}$\\
$[X X][X X \partial b b]$&$\frac{17}{20}+\frac{3}{20}N^{2}$\\
$[X X][X X b \partial b]$&$-\frac{7}{20}-\frac{3}{20}N^{2}$\\
$[X X X b X \partial b]$&$-\frac{5}{2}N$\\
$[X X X \partial X b b]$&$\frac{27}{10}N$\\
$[X X][X \partial X b b]$&$-1$\\
$[X X X][\partial X b b]$&$\frac{7}{10}$\\
$[X X X b \partial X b]$&$\frac{3}{5}N$\\
$[X X][X b \partial X b]$&$-\frac{31}{10}$\\
$[X X \partial X X b b]$&$-\frac{19}{10}N$\\
$[X X \partial X][X b b]$&$\frac{27}{10}$\\
$[X X \partial X b X b]$&$\frac{21}{10}N$\\
$[X X \partial X b][X b]$&$\frac{21}{10}$\\
$[X X b X \partial X b]$&$-\frac{12}{5}N$\\
$[X X b][X \partial X b]$&$\frac{3}{5}$\\
$[X X b X b \partial X]$&$\frac{8}{5}N$\\
$[X \partial X X b X b]$&$\frac{7}{10}N$\\
\end{longtable}
\normalsize
These 63 explicitly specified terms are relative and gauge invariant;
$Q\mathcal V$ therefore vanishes in BRST cohomology. They require neither
a matrix-rank specialization nor a finite-rank trace-basis assumption.

The tables give a direct proof at the unique resonance degree. To
verify the field equality, expand $\mathcal H_{3,6}$ with \eqref{of:uor4}--\eqref{of:uor7},
each normal product with \eqref{of:wick-formula}, and each term of
$Q\mathcal V$ with its $j_{\rm BRST}$ specialization. The projection
$\Pi_2$ is either $1$ or $1-fe/6$, so every term is specified by the
displayed tables. To organize the coefficient comparison, let
$(d,g,\ell)$ record the numbers of matter letters and ghost letters,
and the total jet order, respectively. The projected field
$\mathcal H_{3,6}$ has precisely the profiles
$(8,1,0)$, $(6,1,1)$ and $(4,1,2)$: one, two and three
matter contractions leave eight, six and four matter letters.
The expansion of $Q\mathcal V$ and of the nineteen strong fields can
also contain $(4,3,1)$; these three-ghost terms cancel between them.
Collecting the coefficients of the cyclic trace words within each of
these four profiles, with every differentiated position retained,
gives \eqref{of:sor5}. These profiles exhaust the contractions of
the displayed inputs. In particular, the primitive seeds have either
six matter letters and no jet or four matter letters and one jet;
their BRST images are relative, so every surviving $c$ is differentiated.
The contraction rule then gives exactly the four profiles above,
with at most three contractions and Taylor degree at most two.
The zero-mode entries in table~\ref{of:table-resonance-modes} follow from
\eqref{of:normal-modes}; all singular correction fields have weight at
most four and are reduced by the preceding cubic and quartic identities.
For instance the only correction to the reversed endpoint in $:TD:$
is $D_{(0)}T+D_{(1)}T$, whose zero mode is $3D_0$.
Consequently the coefficient of $D_0$ after summing the nineteen rows is
$-12(N^2+8r-9)/5$, the coefficient of $B_0K_0$ is $12$, and those of
$B_0e,D_1e,D_0H,B_ip_j$ are zero. I obtain the actual operator identity
\begin{equation}
 [p,O^{\alpha,(3)}]_3
       =-\frac{12}{5}(4\Delta-9)O^{\alpha,(2)}_0
                         +12O^{\alpha,(1)}_0K_0.
 \label{of:septic-return}
\end{equation}
The barred identity is obtained by applying the relative chain derivation
to the field equality, including its primitive, before taking zero modes.

\subsection{The mixed cubic--odd OPE at arbitrary degree}
\label{of:mixed-ope}
All formulas below hold in the free relative matrix BRST system at
arbitrary rank $N\ge2$. Let $m\ge2$, set $d=m+1$, and put
\[
\begin{gathered}
 V_i^{[m]}=\operatorname{tr}(bX^i(\partial X)X^{m-1-i}),\\
 E_0^{[m]}=:eW_{m,1}:+\tfrac12:HP_m:,\qquad
 E_2^{[m]}=:(\partial e)W_{m,1}:+\tfrac12:(\partial H)P_m:.
\end{gathered}
\] Define
\[
 U_m=\frac{m+2}{m(m+1)}\sum_{i=0}^{m-1}
                 \left(i-\frac{m-1}{2}\right)V_i^{[m]}.
\]
Empty sums are zero. In sums with $a+b=m$ below, both $a,b\ge2$
and the pairs are ordered. The three nonzero OPE coefficients are
\begin{subequations}\label{of:cm1}
\begin{align}
O_0={}&\frac{m+2}{m+1}H_{m+2}
        -\frac{m+2}{N}:G_XC_{m,0}:
        -\frac2N E_2^{[m]}+QU_m
        +\frac{N(m+2)}{m^2(m+1)}\partial^2P_m\notag\\
&+\frac{m+2}{m(m+1)}\sum_{a+b=m}
    \left(\frac{a+1}{b}:P_a\partial^2P_b:
                        -:(\partial P_a)(\partial P_b):\right),
 \label{of:cm1-o0}\\
O_1={}&-\frac2N E_0^{[m]}
       +\left(\frac{N(m+2)}{m^2}-\frac{m+2}{Nm}\right)\partial P_m
       +\frac{m+2}{m}\sum_{a+b=m}\frac1b:P_a\partial P_b:,
 \label{of:cm1-o1}\\
O_2={}&\left(\frac{N(m+2)}m-\frac{2(m+2)}N\right)P_m
                 +\frac{m+2}m\sum_{a+b=m}:P_aP_b:.
 \label{of:cm1-o2}
\end{align}
\end{subequations}
Here $O_j=(B_{3,0})_{(j)}C_{m+1,1}
             -(B_{3,1})_{(j)}C_{m+1,0}$.

\begin{proof}
I give a coefficient calculation with an arbitrary word length;
no interpolation in $m$ or in matrix rank is used.
Write $D=\partial X$, $X_{(2)}=\partial^2X/2$, and put
\[
 I_j=\operatorname{tr}(bX^jpX^{m-j}),\quad
 J_j=\operatorname{tr}(YX^jDX^{m-j}),\quad 0\le j\le m,
 \qquad K=\operatorname{tr}(X^{m+1}\partial Y).
\]
The single-contraction, leading-length part of $O_0$ is
\begin{equation}
\begin{aligned}
&\frac{m+2}{2m}(I_0+I_m)
      +\frac{m+2}{m}\sum_{j=1}^{m-1}I_j
      -\frac{m+2}{N}(G_XC_{m,0})_{\rm PBW}\\
 &\quad-\frac{m-2}{2m}(J_0+J_m)
       +\frac2m\sum_{j=1}^{m-1}J_j-K
       -\frac2N(E_2^{[m]})_{\rm PBW}.
\end{aligned}
\label{of:cm2}
\end{equation}
For example the $B_{3,0}C_{m+1,1}$ contraction gives the adjacent
pair $I_j+I_{j+1}$ with coefficient $1/m$, while
$-B_{3,1}C_{m+1,0}$ gives it with coefficient $1/2$.
The two endpoint indices occur once, and each interior index twice.
The corresponding $b,p$ contraction leaves the derivative of the
left factor: its endpoint $J$ coefficient is $1/m-1/2$, its
interior coefficient is $2/m$, and its $K$ coefficient is $-1$.
The two traceless-projector contributions are exactly those
shown in \eqref{of:cm2}.

Direct application of $qb=[X,Y]$ and $qD=-[p,X]$ gives
\[
 qV_i^{[m]}=I_i-I_{i+1}+J_i-J_{i+1}.
\]
Moreover the raw top is
\[
 H_{m+2}=\sum_{j=0}^{m}I_j
           +\frac1{m+2}\sum_{j=0}^{m}J_j
           -\frac{m+1}{m+2}K.
\]
The adjacent differences of the affine coefficients in $U_m$
therefore show that \eqref{of:cm2} equals the leading PBW
part of the first four terms in $O_0$ in \eqref{of:cm1-o0}.

It remains to count double contractions. Define, for $b\ge1$,
\[
 A_b=\operatorname{tr}(X^{b-1}D),\quad
 K_b=\operatorname{tr}(X^{b-1}X_{(2)}),\quad
 L_{b,j}=\operatorname{tr}(X^jDX^{b-2-j}D),\quad 0\le j\le b-2,
\]
with $A_0=0$, $P_0=N$, and $P_1=A_1=K_1=0$.
The double contraction in $O_2$ has one uncontracted left $X$.
If the other arc has length $a$, its coefficient is $(m+2)/m$.
The first summand contributes $2/m$ (two left $X$ slots), and
the second contributes $1$ (the left $Y$ slot and its $m$ possible
partners). The two single traceless-projector terms contribute
$-2(m+2)P_m/N$; the double-projector term leaves a one-letter
trace and vanishes. Thus the full double-contraction expression is
\begin{equation}
 \frac{m+2}{m}\sum_{a=0}^{m-1}P_aP_{m-a}
                              -\frac{2(m+2)}NP_m.
 \label{of:cm3}
\end{equation}
This is $O_2$. For its contributions to $O_1$ and $O_0$, replace
its distinguished uncontracted left $X$, respectively, by $D$
and by $X_{(2)}$. In particular the pure derivative correction
in $O_0$ is
\begin{equation}
 \frac{m+2}{m}\sum_{a=0}^{m-1}P_aK_{m-a}
                              -\frac{2(m+2)}NK_m.
 \label{of:cm4}
\end{equation}

For a simultaneous PBW expression $F$, write $sF$ for the contribution
to $QF$ from exactly two contractions with the full BRST current.
Equivalently, in \eqref{of:wick-formula} take $A=j_{\rm BRST}$,
$B=F$, and $k=0$, and retain only matchings with $|M|=2$, including
their Taylor coefficients and both terms of each traceless projector.
I use the same notation in Appendix~\ref{app:descent-fields}.
The weighted BRST double-contraction sum needed here is as follows.
Put $\lambda_i=i-(m-1)/2$ and
$\widetilde U_m=\sum_i\lambda_iV_i^{[m]}$. Then
\begin{equation}
 s\widetilde U_m=
 \sum_{a=0}^{m-1}\left[
 (m-2a-1)P_aK_b-(a+1)P_a\sum_{j=0}^{b-2}L_{b,j}
                                      +abA_aA_b\right],\quad b=m-a.
 \label{of:cm5}
\end{equation}
Here is the full single-word cut formula, which also proves the
weighted count without an endpoint convention. For $0\le i\le m-1$,
\begin{equation}
\begin{aligned}
sV_i^{[m]}={}&P_{m-1-i}K_{i+1}-P_iK_{m-i}\\
 &+\sum_{t=0}^{i-1}
   \left(A_{m-1-t}A_{t+1}-P_tL_{m-t,i-t-1}\right)\\
 &+\sum_{t=i+1}^{m-1}
   \left(P_{m-1-t}L_{t+1,i}-A_tA_{m-t}\right).
\end{aligned}
\label{of:cm5a}
\end{equation}
Indeed the BRST current must contract its $c$ with the unique $b$
and its $Y$ with either $D$ or an undifferentiated $X$. Cutting
$\operatorname{tr}(bRZS)$, where $Z$ is the contracted target letter, gives
\[
 \operatorname{tr}(S)\operatorname{tr}(X^{[\mathrm{Taylor}]}R)
                 -\operatorname{tr}(R)\operatorname{tr}(X^{[\mathrm{Taylor}]}S).
\]
Here the surviving left $X$ has Taylor order two when $Z=D$
and order one when $Z=X$. The $Z=D$ term is the first line of
\eqref{of:cm5a}; contracting an $X$ before or after $D$ gives
respectively its second or third line. A zero-length arc contributes
$P_0=N$. The $-1/N$ projector parts cancel because replacing any
one slot of $\operatorname{tr} c[X,Y]$ by a scalar matrix gives zero.

Multiply \eqref{of:cm5a} by $\lambda_i$ and sum. The coefficient
of $P_aK_b$ is $\lambda_{m-1-a}-\lambda_a=m-2a-1$.
For each $P_aL_{b,j}$ there are exactly two contributions, whose
coefficient is
\[
 \lambda_j-\lambda_{a+j+1}=-(a+1).
\]
Their indices lie between $0$ and $m-1$. Finally the ordered
$A_aA_b$ coefficient is
\[
 \sum_{i=a}^{m-1}\lambda_i-\sum_{i=0}^{a-1}\lambda_i=ab,
 \qquad a+b=m,
\]
since $\sum_i\lambda_i=0$. These three sums prove
\eqref{of:cm5} for arbitrary $m$.

Finally,
\[
 \partial P_b=bA_b,\qquad
 \partial^2P_b=b\left(2K_b+\sum_{j=0}^{b-2}L_{b,j}\right).
\]
Substitute \eqref{of:cm5} into \eqref{of:cm4} and use these
identities. This gives exactly the displayed $\partial^2P_m$
and two-trace derivative terms in $O_0$. The normal-order
corrections in $:G_XC_{m,0}:$ and $E_2^{[m]}$ are respectively
$K_m$ and $(m+2)K_m/2$; their coefficients account for the last
term of \eqref{of:cm4}. Similarly the correction in
$E_0^{[m]}$ is $(m+2)\partial P_m/(2m)$; together with the
$D$ version of \eqref{of:cm3} it gives the displayed $O_1$.
This completes the uniform proof.
\end{proof}

\section{Field identities and the complete lowest-space descent}
\label{app:descent-fields}

This appendix proves the identities used in Section~\ref{sec:descent}.
The propagator, Wick sum, and normal-mode formula are given in
\eqref{of:propagator}, \eqref{of:wick-formula}, and
\eqref{of:normal-modes}, respectively. Every calculation below is a
finite sum of these contractions with its word length left arbitrary.
I write $W_{d,i}$ for normalized flavor polars of $P_d$, so that
$[f,W_{d,i}]=(d-i)W_{d,i+1}$, and use the analogous normalization for
$B_{d,i},C_{d,i}$.  When an integer trace degree is even, $R_d=o(P_d)$
and $Q_d=[f,R_d]/d$.  A field $Q_d$ appearing inside a normal product
instead means $W_{d,1}=\operatorname{tr}(X^{d-1}Y)$; the two conventions
agree after taking its zero mode.  For half-integral weight,
$U^+=U_{(\mathrm{wt}(U)-3/2)}$ denotes the energy-$1/2$ mode.
Every commutator involving such a map retains its input and output
actions.  Throughout the descent $W\subset M_r$ satisfies the hypotheses
of Theorem~\ref{thm:lowest-exclusion}; its ambient first-half space is
$M_{r+1/2}$.

The calculation follows five stages. Sections~\ref{df:sec-kernels}--\ref{df:sec-boundary}
establish the remainder kernels, cubic channels, and finite trace
boundary. Sections~\ref{df:sec-quartic-low}--\ref{df:sec-odd-quartic}
derive the complete quartic field equation at arbitrary odd degree.
Sections~\ref{df:sec-raw-transport}--\ref{df:sec-adjacent}
transport this equation on the whole lowest input and its actual
first-half image. Sections~\ref{df:sec-terminal}--\ref{df:sec-even-fourth}
resolve the degree-two endpoint and close the even channels.
Finally, sections~\ref{df:sec-even-lie}--\ref{df:sec-spectrum}
identify the even Lie action and establish the Newton and spectral
obstruction. The order follows the dependencies of the proof in
section~\ref{sec:descent}; every stage includes both its field identities
and their normal-mode evaluation.

\subsection{Exhaustive remainder kernels and transfer between fermion families}
\label{df:sec-kernels}

\paragraph{The formal domain and independence of spectator profiles.}
Work first in the formal commutative algebra on cyclic matrix words,
with the usual cyclic sign for odd letters. The only words needed in
the kernel calculations below contain at most one odd letter, denoted
$p$, so there is no extra odd permutation ambiguity. The words
$X,X_1,X_2,Y,Y_1,p,p_1$ are treated as independent matrix letters;
$X_i=\partial^iX/i!$ and $Y_1=\partial Y$ when evaluating fields.
The unmarked pure traces are $P_d=\operatorname{tr}X^d$.

In a differential image containing $p$, its trace is distinguished.
A coefficient can be compared separately for each distinguished
cyclic word and each multiset of the remaining unmarked trace factors.
This is a comparison in the formal trace algebra, before imposing
finite-matrix trace identities. Such a formal identity evaluates to an
identity for every matrix size. If independence is instead formulated
through generic matrices, it has an elementary stable-size proof:
assign a distinct matrix-index vertex to every letter position and
use disjoint directed cycles for the prescribed trace factors. The
monomial containing exactly their directed-edge matrix entries occurs
only for that cyclic-word multiset. Repeated cycles contribute a
nonzero integer multiplicity. Matrix size at least the total number
of positions suffices. Thus no finite-rank trace relation is used to
infer the kernel.

\paragraph{The one-jet, one-$Y$ kernel.}
For a trace with $d\ge2$ matter letters, put
\[
 D_* =\operatorname{tr}(X^{d-1}Y_1),\qquad
 D_j=\operatorname{tr}(YX^jX_1X^{d-2-j})\quad(0\le j\le d-2).
\]
The part of the BRST differential containing the first ghost jet is
obtained by replacing a differentiated matter letter by its commutator
with $p$. Define
\[
 T_i=\operatorname{tr}(pX^iYX^{d-1-i})\quad(0\le i\le d-1).
\]
Choosing the common overall sign of this differential so that
$qD_*=T_{d-1}-T_0$, direct cyclic rotation gives
\begin{equation}
 qD_j=T_{d-2-j}-T_{d-1-j}.
 \label{df:eq-fdk1}
\end{equation}
Hence, for $R=\alpha D_*+\sum_{j=0}^{d-2}a_jD_j$, the coefficients
at successive $T_i$ give
\begin{equation}
 a_{d-2}=\alpha,\qquad
 a_{d-2-i}=a_{d-1-i}\ (1\le i\le d-2),\qquad
 a_0=\alpha.
 \label{df:eq-fdk2}
\end{equation}
Thus every $a_j=\alpha$, for every length $d$. Its kernel is exactly
the derivative of $\operatorname{tr}(YX^{d-1})$.
Both endpoints are included, so the recurrence determines the kernel for every length.

With two traces, a jet on a pure-$X$ spectator gives
$\partial P_a\,Q_b$, where $Q_b=\operatorname{tr}(YX^{b-1})$;
it is already closed. A jet on the $Y$ trace is, by \eqref{df:eq-fdk2}, a
multiple of $P_a\partial Q_b$. There can be no cancellation of
the nonclosed parts between these two cases: the first is closed
by itself, while the second has a distinguished output trace
containing both $p$ and $Y$. Different spectator multisets also
cannot cancel.

Flavor raising replaces $Y$ by $X$, up to a common nonzero
normalization. On a single trace it sends $\partial Q_d$ to
$\partial P_d$, so no single-trace one-jet remainder is flavor
highest. On two traces of distinct degrees $a,b$, its two output
profiles are $P_a\partial P_b$ and $P_b\partial P_a$.
Its kernel is spanned by the ordered expressions
\begin{equation}
 D_{ab}=P_a\partial Q_b-Q_a\partial P_b,
 \qquad D_{ba}=P_b\partial Q_a-Q_b\partial P_a.
 \label{df:eq-fdk3}
\end{equation}
For $a=b$, there is one such kernel vector, $D_{aa}$.
Passing from formal PBW products to the displayed vertex normal
order only adds shorter pure-$X$ two-jet terms: each of these
products has one $Y$, so a reordering contraction removes that
$Y$ and one $X$. These shorter terms must be retained, and are
classified next.

\paragraph{The pure-$X$ two-jet kernel, with endpoints exposed.}
For $d\ge2$, write
\[
 K_d=\operatorname{tr}(X^{d-1}X_2),\quad
 L_j=\operatorname{tr}(X^jX_1X^{d-2-j}X_1),\quad
 S_i=\operatorname{tr}(pX^iX_1X^{d-1-i}).
\]
Here $0\le j\le d-2$, $0\le i\le d-1$, and
$L_j=L_{d-2-j}$. The possible $p_1$ term in $qK_d$ is the
trace of $X^{d-1}[X,p_1]/2$ and is zero. Direct cyclic rotation
therefore gives
\begin{equation}
 qK_d=S_{d-1}-S_0,\qquad
 qL_j=S_j-S_{j+1}+S_{d-2-j}-S_{d-1-j}.
 \label{df:eq-fdk4}
\end{equation}
Use a symmetric coefficient list $b_j=b_{d-2-j}$ when writing
$R=\alpha K_d+\sum_{j=0}^{d-2}b_jL_j$. This is a unique
representation: split the coefficient of a two-element orbit
equally between its two indices; retain the full coefficient at
a fixed middle index. Then
\begin{equation}
 qR=(2b_0-\alpha)S_0
   +2\sum_{i=1}^{d-2}(b_i-b_{i-1})S_i
   +(\alpha-2b_{d-2})S_{d-1}.
 \label{df:eq-fdk5}
\end{equation}
Consequently $qR=0$ if and only if $b_i=\alpha/2$ for every
$i$. The kernel is exactly
\begin{equation}
 2K_d+\sum_{j=0}^{d-2}L_j=\frac1d\partial^2P_d.
 \label{df:eq-fdk6}
\end{equation}
The same argument includes $d=2$, where the list has one entry.

A trace with one jet only is, by cyclicity,
$A_d=\operatorname{tr}(X^{d-1}X_1)=\partial P_d/d$ and is
closed. Thus at total jet order two every multitrace monomial
either has both jets in one distinguished trace, to which
\eqref{df:eq-fdk5} applies with unchanged spectator factors, or has one
jet in each of two traces, giving a product $A_aA_b$.
There are no other placements. In the first case a differential
output has both $p$ and $X_1$ in that same distinguished trace.
It cannot cancel a differential from a different trace degree
or different spectator multiset. In the second case the
differential is identically zero. Repeated spectator degrees
merely collect equal monomials; they create no new kernel.
Therefore the complete multitrace kernel is spanned by
\begin{equation}
 (\partial^2P_a)\prod_iP_{d_i},\qquad
 (\partial P_a)(\partial P_b)\prod_iP_{d_i}.
 \label{df:eq-fdk7}
\end{equation}
For pure-$X$ fields the BRST current admits no two target
contractions: it has one $X$ and one $Y$, while the target
contains only $X$ jets. Thus this is also the exact quantum
closure condition, not merely the leading part of an unchecked
spectral sequence.

\paragraph{Why the quartic calculation has no additional trace profiles.}
In the full-propagator quartic third channel, one, two, and three
contractions leave respectively the two-$Y$ zero-jet sector,
the one-$Y$ one-jet sector, and the pure-$X$ two-jet sector.
Their maximum numbers of nonempty traces are respectively one,
two, and three. There is no fourth contraction in this channel:
the two factors together contain only three $Y$ letters.
The zero-jet primitive has one trace; its two-contraction terms
have at most two traces. Its three-contraction term has no
surviving source letter to Taylor-expand and hence no residue.
The first-jet primitives have either one trace or one spectator
trace. Their two-contraction terms have at most three traces.
A contraction joining the spectator to the marked trace lowers,
rather than raises, the number of faces. The canonical $PH$,
$BC$, and $D$ columns admit at most one cross-contraction, which
joins their two traces. These facts bound the complete final
pure-$X$ remainder by three nonempty traces.

The ghost coefficient removal is likewise an arbitrary-length
path elimination. With $I_j=\operatorname{tr}(bX^jpX^{b-2-j})$,
the ghost part of the differential of
$\operatorname{tr}(bX^iX_1X^{b-3-i})$ is, in the convention
of \eqref{df:eq-qd12}, $I_i-I_{i+1}$. Its image is exactly the lists of
coefficients having sum zero. Subtracting the average $h_{ab}$
in \eqref{df:eq-uet3} leaves precisely such a list. The affine coefficients
in \eqref{df:eq-qd12} solve its successive-difference equations, including
the two endpoints; $b=2$ has no difference to remove.
After removing both the same-trace $bp$ and the separate-trace
$BC$ columns, the longest remaining ghost-free sector is the
one-$Y$ one-jet sector, so \eqref{df:eq-fdk1}--\eqref{df:eq-fdk3} apply. Subtracting
those columns leaves exactly \eqref{df:eq-fdk7}. This gives a terminating classification by length.

Degree-one center columns require one additional ordering rule.
Before scalar-vacuum projection a column with one degree-one
factor can have one scalar contraction with the other factor.
Its surviving term is a single pure-$X$ trace with total jet
order two; there are no untouched spectator traces in this
two-column contraction. It is therefore included in the
single-trace coefficient fixed by the Ward identity, rather
than deleted before normal ordering. Pure-$X$ products with an
uncontracted $P_1$ project to zero. Since
$L_1\partial^2P_d=2(d+1)\partial P_d$ and $d\ge2$, the
single-trace Ward equation determines its coefficient uniquely.
The corresponding pair and triple cut coefficients are evaluated in Appendices~\ref{df:sec-two-face} and~\ref{df:sec-three-face}.

\paragraph{Family transfer without an operator on the module.}
Let $\delta$ be the even relative-chain derivation
\begin{equation}
 \delta b=p,\qquad \delta p=\delta X=\delta Y=\delta c=0.
 \label{df:eq-fdk8}
\end{equation}
Its construction, $[\delta,Q]=0$, and preservation of the
relative conditions are given by the explicit current calculation
in \eqref{of:chain-derivation}. It has conformal and flavor degree zero. It fixes
all bottom fields and their flavor polars and sends $B_d$ to
$C_d$, and $G_X,G_Y$ to $\bar G_X,\bar G_Y$.
It also fixes every raw $H_d$: in its ghost term, applying
$\delta$ produces
$\sum_i\operatorname{tr}(pX^ipX^{d-2-i})$, whose terms
cancel under $i\leftrightarrow d-2-i$ by odd cyclicity.
The same assertion holds for all flavor polars. Consequently
$\delta$ fixes $A,B,a,\mathcal L,\mathcal D,\widetilde{\mathcal D}$,
and the input maps $E,M,\mathfrak F$ used in the terminal comparison.

Here is the precise transfer principle used on a chosen module
$M$. A chain or mode identity is first an identity in the
universal mode algebra of the cohomology fields. Its specialization
to a lowest input is a finite algebraic derivation using some
declared vanishing right factors. If those right-factor relations
are stable under $\delta$, applying $\delta$ to that derivation
gives another valid specialized identity. This does not require
an endomorphism $\delta:M\to M$, or preservation of the chosen
input subspace by such an endomorphism.

For the present descent the allowed relations are stable:
all negative energy-shift modes kill a lowest vector, and
$\delta$ preserves energy shift; all odd integer zeros kill
$W$, and $\delta$ is even; zeros preserve $W$; flavor weights
above $j$ vanish, and $\delta$ preserves flavor; earlier bottom
depths consist of pure-$X$ fields fixed by $\delta$; both current
source families vanish, with $\delta Y_B=Y_C$, $\delta Y_C=0$.
Thus every discarded factor in Appendix~\ref{df:sec-source}, Appendix~\ref{df:sec-half-transport}, Appendix~\ref{df:sec-adjacent-source}, and Appendix~\ref{df:sec-terminal}
has a discarded image for an explicitly stated reason.

In particular apply this principle directly to the complete
$B$-family terminal equation, with its returns still present:
\[
 Y'_B\widetilde{\mathcal D}=\frac{G_X}{N}\mathfrak F,
 \qquad TY'_B=Y'_B(B-2\mathcal L).
\]
It gives
\begin{equation}
 Y'_C\widetilde{\mathcal D}=\frac{\bar G_X}{N}\mathfrak F,
 \qquad TY'_C=Y'_C(B-2\mathcal L).
 \label{df:eq-fdk9}
\end{equation}
The same procedure applies directly to the complete adjacent source row
$Y'_B\widetilde{\mathcal D}=0$ and to the linear current-source
transport. These are the actual opposite-family equations needed
in Section~\ref{sec:descent}, including the genuine degree-two endpoint.

\subsection{The cubic downward channel at arbitrary length}
\label{df:sec-cubic-down}

For odd $d\ge3$, set $m=d-2$ and use the cubic fermion polars
$B_{3,i}$, $i=0,1,2$.  The first subscript is the trace degree and
the second is the polar index; $B_a$ with a single subscript retains
its trace-degree meaning.  The normalized bosonic polars are
\[
 W_{d,1}=\operatorname{tr}(X^{d-1}Y),\qquad
 W_{d,2}=\frac1{d-1}\sum_{i=0}^{d-2}
                 \operatorname{tr}(X^iYX^{d-2-i}Y).
\]
I derive the coefficient at arbitrary word length. Put
\[
 O_j=(B_{3,0})_{(j)}W_{d,2}-2(B_{3,1})_{(j)}W_{d,1}+(B_{3,2})_{(j)}W_{d,0},
 \qquad
 L=:P_{d-1}G_Y:-:W_{d-1,1}G_X:,
\]
where $G_X=\tr Xb$ and $G_Y=\tr Yb$. Define the PBW primitive
\begin{equation}
 U=\frac{d+1}{d-1}\sum_{i=0}^{m}
            \left(i-\frac m2\right)\tr(bX^ibX^{m-i}).
 \label{df:eq-bd2}
\end{equation}
I need only the following exact field information:
\begin{subequations}\label{df:eq-bd3}
\begin{align}
O_1&=\frac{d(d+1)}{d-1}\sum_{a=0}^{d-2}
             :P_a B_{d-1-a}:
               -\frac{2d(d+1)}N B_{d-1},
 \label{df:eq-bd3-o1}\\
O_0&=QU+\frac{2(d+1)}N L+A\partial B_{d-1}+R,
 \qquad
 A=2(d+1)\left(\frac N{d-1}-\frac dN\right),
 \label{df:eq-bd3-o0}
\end{align}
\end{subequations}
where $P_0=N$, $P_1=B_1=0$ in the degree notation and
$R$ is a linear combination of
\begin{equation}
 :P_a\partial B_b:,\qquad :(\partial P_a)B_b:,
 \qquad a,b\ge2,\quad a+b=d-1.
 \label{df:eq-bd4}
\end{equation}
These products are regular: every factor contains only $X,b$
and their indicated derivatives. I now prove all the
information in \textnormal{\eqref{df:eq-bd3}--\eqref{df:eq-bd4}} with the complete regular remainder evaluated below.

For the single-contraction part put
$T_j=\tr(bX^jYX^{d-1-j})$, $0\le j\le d-1$.
The endpoint and interior coefficients are respectively
\begin{equation}
 -\frac{(d+1)(d-2)}{d-1},\qquad \frac{2(d+1)}{d-1}.
 \label{df:eq-bd5}
\end{equation}
Indeed $B_{3,0}W_{d,2}$ contributes $2/(d-1)$ at each endpoint
and $4/(d-1)$ at each interior position.
The two terms in $-2B_{3,1}W_{d,1}$ cancel at the endpoints
and contribute $2$ at each interior position.
Finally $B_{3,2}W_{d,0}$ contributes $-d$ at each endpoint.
The single traceless-projector terms from the same three
summands are, in order,
\[
 -\frac4N G_XW_{d-1,1},\qquad
 \frac2N G_YP_{d-1}-\frac{2(d-1)}N G_XW_{d-1,1},
 \qquad \frac{2d}N G_YP_{d-1}.
\]
Their sum is $2(d+1)L_{\rm PBW}/N$.
Directly applying $qb=[X,Y]$ gives
\[
 q\tr(bX^ibX^{m-i})
     =T_{m-i+1}-T_{m-i}-T_{i+1}+T_i.
\]
The affine coefficients in \textnormal{\eqref{df:eq-bd2}} have successive
difference one and change sign under $i\mapsto m-i$.
Their weighted sum gives exactly \textnormal{\eqref{df:eq-bd5}}.
Thus the leading part of $O_0$ is
$qU+2(d+1)L_{\rm PBW}/N$.

The double-contraction term has one uncontracted left $b$.
For each resulting arc length $a$, the full-projector
coefficients from the three summands in $O_j$ are, in order,
\[
 \frac2{d-1},\qquad 2,\qquad d.
\]
For the first summand the two assignments of its two left
$X$ slots to the target $Y$ slots give $2/(d-1)$.
For the second, each of its two left cyclic orderings gives
one contribution at each target arc: the $Y,X$ contraction
minus sign cancels the displayed minus sign. For the third,
the $d$ choices of the first target slot give $d$ at each arc.
Their sum is $d(d+1)/(d-1)$.
Summing over the $d-1$ arcs gives $d(d+1)$. A single
scalar projector collapses those arcs to the same word.
Each of the two choices of a single scalar projector contributes
$-d(d+1)B_{d-1}/N$; the double scalar projector leaves a
one-letter $b$ trace and vanishes.
This proves the displayed $O_1$ identity. The contribution of
these contractions to $O_0$ replaces the surviving left $b$
by $\partial b$. In particular its single-trace coefficient of
$\tr((\partial b)X^m)$ is
\begin{equation}
 F=\frac{d(d+1)N}{d-1}-\frac{2d(d+1)}N.
 \label{df:eq-bd6}
\end{equation}

Here and below, $s$ denotes the two-contraction contribution of the
full BRST current defined before \eqref{of:cm5}, applied to the
displayed simultaneous PBW expressions.
The only $\partial b$ terms in $QU$ come from the ghost part
$-\tr ccb$ of the BRST current, contracted with both target
$b$ letters. For a single word their full expression is
\begin{equation}
 \left(s\tr(bX^ibX^{m-i})\right)_{\partial b}
       =P_{m-i}\tr((\partial b)X^i)
                  -P_i\tr((\partial b)X^{m-i}).
 \label{df:eq-bd7}
\end{equation}
The two arcs carry opposite signs from the order of the two
odd contractions; the surviving left $b$ has Taylor order one.
The scalar-projector pieces cancel because the ghost Lie
bracket has zero contraction with a scalar matrix.
Only $i=0,m$ in \textnormal{\eqref{df:eq-bd7}} have an empty arc and hence
contribute to a single trace. Their weighted sum in
\textnormal{\eqref{df:eq-bd2}} is
$N(d+1)(d-2)/(d-1)$.
The normal-order correction in $L$ contains $b\partial X$
and contains no $\partial b$: its left factors have only $X,Y$,
whereas its right factor contains the uncontracted $b$.
Therefore the coefficient of $\tr((\partial b)X^m)$ in
$O_0-QU-2(d+1)L/N$ equals
\[
 F-\frac{N(d+1)(d-2)}{d-1}=A.
\]

Here is a short kernel argument that identifies all remaining
terms needed in \textnormal{\eqref{df:eq-bd3}}. At length $d-1$ every term
has only $X,b$, exactly one $b$, and exactly one derivative;
there are at most two traces. The expression is $Q$-closed,
since $O_0,L$ are products of closed fields and $Q^2=0$.
Its leading differential is therefore $q$-closed. In a single
trace with $m$ copies of $X$, use the basis
\[
 D_*=\tr((\partial b)X^m),\qquad
 D_j=\tr(bX^j(\partial X)X^{m-1-j}),\quad 0\le j<m,
\]
and put $I_j=\tr(bX^jpX^{m-j})$.
The components of $q$ with two odd letters are
\[
 (qD_*)_{bp}=I_m-I_0,
 \qquad(qD_j)_{bp}=I_j-I_{j+1}.
\]
Hence a $q$-closed linear combination must have the same
coefficient on $D_*$ and all $D_j$. Its sole possibility is
that coefficient times $\partial B_{m+1}$.
For two traces, $q$ acts on each trace separately. The same
argument applies to the trace containing $b$; a derivative
on the pure-$X$ trace gives the other possibility
$(\partial P_a)B_b$. Thus the two-trace kernel is contained
in the span \textnormal{\eqref{df:eq-bd4}}.
This comparison is made in the formal cyclic trace algebra,
before imposing finite-rank trace identities. Independence of
its coefficient words can also be checked by polarizing all
letters and using sufficiently large matrices; the resulting
rational functions of the formal matrix-size parameter then vanish identically.
No independence assumption at the particular matrix size $N$ is used.
I have proved \textnormal{\eqref{df:eq-bd3}--\eqref{df:eq-bd4}} at every matrix size.

\begin{equation}
 \boxed{\quad
 R=\frac{2(d+1)}{d-1}
   \sum_{\substack{a+b=d-1\\a,b\ge2}}
     \bigl((a+1):P_a\partial\beta_b:
                   -b:(\partial P_a)\beta_b:\bigr).
 \quad}
\label{df:eq-mbd1}
\end{equation}
In particular this is a complete field identity before taking
modes, with the same explicit relative primitive $U$ as in \eqref{df:eq-bd3}.

Here $\beta_a=\operatorname{tr}(X^{a-1}b)$.\par
\paragraph{An indexed proof of the remainder.}
The first-jet kernel calculation in \eqref{df:eq-bd3}--\eqref{df:eq-bd4} already shows
that $R$ is in the span of the two displayed kinds of terms.
They contain only $X,b$ and their derivatives and hence have
regular mutual contractions. Two coefficient counts determine
the answer, with no finite-degree extrapolation.

First look at $P_a\tr(X^{b-1}\partial b)$, where
$a+b=m+1=d-1$ and $a,b\ge2$. The double matter contraction
in $O_0$ has coefficient $d(d+1)/(d-1)$. Formula \eqref{df:eq-bd7} shows
that its coefficient in $sU$ is
\[
 \frac{d+1}{d-1}
       \{(m-a)-m/2-(a-m/2)\}
 =\frac{d+1}{d-1}(m-2a).
\]
The normal-order correction in $L$ contains no $\partial b$.
Subtracting therefore gives
$2(d+1)(a+1)/(d-1)$, which is the coefficient of
$:P_a\partial\beta_b:$ in \eqref{df:eq-mbd1}.

For the other coefficient, write $\dot X=\partial X$ and
$U_i=\tr(bX^ibX^{m-i})$. In the matter part of $sU_i$,
contract a chosen $b$ and a chosen $X$. With the chosen $b$
cyclically placed first, the cut of a word $\tr(b R X S)$ is
\begin{equation}
 \tr(S)\tr(\dot X R)-\tr(R)\tr(\dot X S).
\label{df:eq-mbd2}
\end{equation}
The same relative sign follows directly from the commutator
in the BRST current. Scalar-projector terms vanish because
that commutator has zero contraction with the identity matrix.
If the first $b$ in $U_i$ is chosen, a pure differentiated
trace of degree $a$ occurs with coefficient $1/a$ when the
chosen $X$ is in its first block and $i\ge a$, and with
coefficient $-1/a$ when it is in the second block and
$i\le m-a$. The factor $1/a$ uses
$\partial P_a=a\tr(X^{a-1}\dot X)$.

Choosing the second $b$ amounts to replacing $i$ by $m-i$
and inserting the minus sign from moving one odd letter past
the other. Its contribution is the same as the first. Thus the
coefficient of $(\partial P_a)\beta_b$ in $sU_i$ is precisely
\begin{equation}
 \frac2a\bigl(\mathbf1_{i\ge a}-\mathbf1_{i\le m-a}\bigr).
\label{df:eq-mbd3}
\end{equation}
Since $b=m-a+1$,
\[
 \sum_{i=a}^m(i-m/2)=ab/2,
 \qquad
 \sum_{i=0}^{m-a}(i-m/2)=-ab/2.
\]
Multiplying \eqref{df:eq-mbd3} by the coefficients of $U$ gives
$2(d+1)b/(d-1)$ in $sU$.
There is no such two-trace term in the direct double contraction
of $O_0$, whose remaining left letter is $b$; its derivative
falls on that letter. The normal-order correction to $L$ is
single-trace: its scalar-projector term has a one-letter $b$
trace and vanishes. The coefficient in $R$ is consequently
$-2(d+1)b/(d-1)$, as claimed. Together with the first count
and the first-jet kernel description, this proves \eqref{df:eq-mbd1}
for every odd $d$ and every matrix size.

\paragraph{The first-half operator identity.}
For even $a$ write
\[
 P_a^0=o(P_a)|_E,\qquad w_{a,1}=o(\tr(X^{a-1}Y)).
\]
For odd $b$ write
$b_b=o(\beta_b)$. A superscript $+$ means the first-half map
of \eqref{of:normal-modes}, so $P_a^+$ is defined for odd $a$ and $\beta_b^+$
for even $b$. Put
\[
 G_X=\beta_2=\tr(Xb),\qquad G_Y=\tr(Yb).
\]
The commutator below includes both its actions,
on the input lowest space and on the output half-level space:
\[
 \mathcal D_d=
 [o(B_{3,0}),W_{d,2}^+]
 -2[o(B_{3,1}),W_{d,1}^+]
 +[o(B_{3,2}),W_{d,0}^+].
\]
Then on every lowest space $E$,
\begin{equation}
\begin{aligned}
 \mathcal D_d={}&\frac{N(d+1)}{d-1}\beta_{d-1}^+
  +\frac{2(d+1)}N
       \bigl(G_Y^+P_{d-1}^0-G_X^+w_{d-1,1}\bigr)\\
 &+\frac{d+1}{d-1}
   \sum_{\substack{a+b=d-1\\a,b\ge2\ \mathrm{even}}}
                         (a+1)\beta_b^+P_a^0\\
 &-\frac{d+1}{d-1}
   \sum_{\substack{a+b=d-1\\a,b\ge3\ \mathrm{odd}}}
                         bP_a^+b_b .
\end{aligned}
\label{df:eq-mbd4}
\end{equation}
Every product on the right has its zero mode acting on $E$
first. In the commutator definition, $o(B_{3,i})$ on
the output means the same physical zero mode, without an
assumption that the output is a lowest space.

Indeed the mode commutator formula gives
\[
 \mathcal D_d=(O_0)_{((d-1)/2)}|_E
                       +(O_1)_{((d-3)/2)}|_E.
\]
For even $a,b$, the two derivative factors in \eqref{df:eq-mbd1} contribute
$-b/2$ and $-a/2$, respectively. Their combined coefficient is
$-(d+1)b/(d-1)$. Adding the coefficient
$d(d+1)/(d-1)$ from $O_1$ gives
$(d+1)(a+1)/(d-1)$. For odd $a,b$, the derivative factors
are instead $-(b+1)/2$ and $-(a-1)/2$. Adding $O_1$ now gives
$-(d+1)b/(d-1)$. This proves the two sums in \eqref{df:eq-mbd4}, using
regularity to move the zero factor to the right.

The complete first-half mode of the remaining normal product is
\begin{equation}
 L^+=G_Y^+P_{d-1}^0-G_X^+w_{d-1,1}
                     -\frac{d(d-3)}2\beta_{d-1}^+.
\label{df:eq-mbd5}
\end{equation}
Apply \eqref{of:normal-modes} with $a=(d-1)/2$, $b=3/2$. Only the zeroth
OPE coefficient is present, and its binomial sum is $(d-3)/2$.
The two OPEs are $G_{Y,(0)}P_{d-1}=-(d-1)\beta_{d-1}$
and $G_{X,(0)}W_{d-1,1}=\beta_{d-1}$, which give \eqref{df:eq-mbd5}.
The remaining single-trace coefficients combine exactly as in
\eqref{df:eq-bd3}: if
\[
 F=\frac{d(d+1)N}{d-1}-\frac{2d(d+1)}N,
 \qquad A=2(d+1)\left(\frac N{d-1}-\frac dN\right),
\]
their sum is
$F-(d-1)A/2-d(d+1)(d-3)/N=N(d+1)/(d-1)$.
This completes the proof of \eqref{df:eq-mbd4}.
Applying the chain derivation of \eqref{df:eq-fdk8} and the relation-stability
argument following it proves the $C$-family identity on the same module.

\subsection{The cubic middle channel and its complete normal-mode evaluation}
\label{df:sec-cubic-middle}

All formulas below hold in the free relative matrix BRST system at
arbitrary matrix size $N\ge2$. Let $m\ge2$, set $d=m+1$, and put
\[
\begin{gathered}
 V_i^{[m]}=\operatorname{tr}(bX^i\partial X X^{m-1-i}),\qquad
 E_0^{[m]}=:eW_{m,1}:+\tfrac12:HP_m:,\\
 E_2^{[m]}=:(\partial e)W_{m,1}:+\tfrac12:(\partial H)P_m:.
\end{gathered}
\]
Here the field denoted by $H$ in these two expressions is the flavor
Cartan current, whereas $H_d$ is the raw top field. Define
\[
 U_m=\frac{m+2}{m(m+1)}\sum_{i=0}^{m-1}
                 \left(i-\frac{m-1}{2}\right)V_i^{[m]}.
\]
Empty sums are zero. In sums with $a+b=m$ below, both $a,b\ge2$
and the pairs are ordered. The three nonzero OPE coefficients are
\begin{subequations}\label{df:eq-cm1}
\begin{align}
O_0={}&\frac{m+2}{m+1}H_{m+2}
        -\frac{m+2}{N}:G_XC_{m,0}:
        -\frac2N E_2^{[m]}+QU_m
        +\frac{N(m+2)}{m^2(m+1)}\partial^2P_m\notag\\
&+\frac{m+2}{m(m+1)}\sum_{a+b=m}
    \left(\frac{a+1}{b}:P_a\partial^2P_b:
                        -:(\partial P_a)(\partial P_b):\right),
 \label{df:eq-cm1-o0}\\
O_1={}&-\frac2N E_0^{[m]}
       +\left(\frac{N(m+2)}{m^2}-\frac{m+2}{Nm}\right)\partial P_m
       +\frac{m+2}{m}\sum_{a+b=m}\frac1b:P_a\partial P_b:,
 \label{df:eq-cm1-o1}\\
O_2={}&\left(\frac{N(m+2)}m-\frac{2(m+2)}N\right)P_m
                 +\frac{m+2}m\sum_{a+b=m}:P_aP_b:.
 \label{df:eq-cm1-o2}
\end{align}
\end{subequations}
Here $O_j=(B_{3,0})_{(j)}C_{m+1,1}-(B_{3,1})_{(j)}C_{m+1,0}$.

\begin{proof}
I give a coefficient calculation with an arbitrary word length;
no interpolation in $m$ or in matrix size is used.
Write $D=\partial X$, $X_{(2)}=\partial^2X/2$, and put
\[
 I_j=\tr(bX^jpX^{m-j}),\quad
 J_j=\tr(YX^jDX^{m-j}),\quad 0\le j\le m,
 \qquad K=\tr(X^{m+1}\partial Y).
\]
The single-contraction, leading-length part of $O_0$ is
\begin{equation}
\begin{aligned}
&\frac{m+2}{2m}(I_0+I_m)
      +\frac{m+2}{m}\sum_{j=1}^{m-1}I_j
      -\frac{m+2}{N}(G_XC_{m,0})_{\rm PBW}\\
 &\quad-\frac{m-2}{2m}(J_0+J_m)
       +\frac2m\sum_{j=1}^{m-1}J_j-K
       -\frac2N(E_2^{[m]})_{\rm PBW}.
\end{aligned}
\label{df:eq-cm2}
\end{equation}
For example the $B_{3,0}C_{m+1,1}$ contraction gives the adjacent
pair $I_j+I_{j+1}$ with coefficient $1/m$, while
$-B_{3,1}C_{m+1,0}$ gives it with coefficient $1/2$.
The two endpoint indices occur once, and each interior index twice.
The corresponding $b,p$ contraction leaves the derivative of the
left factor: its endpoint $J$ coefficient is $1/m-1/2$, its
interior coefficient is $2/m$, and its $K$ coefficient is $-1$.
The two traceless-projector contributions are exactly those
shown in \textnormal{\eqref{df:eq-cm2}}.

Direct application of $qb=[X,Y]$ and $qD=-[p,X]$ gives
\[
 qV_i^{[m]}=I_i-I_{i+1}+J_i-J_{i+1}.
\]
Moreover the raw top is
\[
 H_{m+2}=\sum_{j=0}^{m}I_j
           +\frac1{m+2}\sum_{j=0}^{m}J_j
           -\frac{m+1}{m+2}K.
\]
The adjacent differences of the affine coefficients in $U_m$
therefore show that \textnormal{\eqref{df:eq-cm2}} equals the leading PBW
part of the first four terms in $O_0$ in \eqref{df:eq-cm1-o0}.

It remains to count double contractions. Define, for $b\ge1$,
\[
 A_b=\tr(X^{b-1}D),\quad
 K_b=\tr(X^{b-1}X_{(2)}),\quad
 L_{b,j}=\tr(X^jDX^{b-2-j}D),\quad 0\le j\le b-2,
\]
with $A_0=0$, $P_0=N$, and $P_1=A_1=K_1=0$.
The double contraction in $O_2$ has one uncontracted left $X$.
If the other arc has length $a$, its coefficient is $(m+2)/m$.
The first summand contributes $2/m$ (two left $X$ slots), and
the second contributes $1$ (the left $Y$ slot and its $m$ possible
partners). The two single traceless-projector terms contribute
$-2(m+2)P_m/N$; the double-projector term leaves a one-letter
trace and vanishes. Thus the full double-contraction expression is
\begin{equation}
 \frac{m+2}{m}\sum_{a=0}^{m-1}P_aP_{m-a}
                              -\frac{2(m+2)}NP_m.
 \label{df:eq-cm3}
\end{equation}
This is $O_2$. For its contributions to $O_1$ and $O_0$, replace
its distinguished uncontracted left $X$, respectively, by $D$
and by $X_{(2)}$. In particular the pure derivative correction
in $O_0$ is
\begin{equation}
 \frac{m+2}{m}\sum_{a=0}^{m-1}P_aK_{m-a}
                              -\frac{2(m+2)}NK_m.
 \label{df:eq-cm4}
\end{equation}

Using the same two-contraction operator $s$, the weighted BRST sum
needed here is as follows. Put $\lambda_i=i-(m-1)/2$ and
$\widetilde U_m=\sum_i\lambda_iV_i^{[m]}$. Then
\begin{equation}
 s\widetilde U_m=
 \sum_{a=0}^{m-1}\left[
 (m-2a-1)P_aK_b-(a+1)P_a\sum_{j=0}^{b-2}L_{b,j}
                                      +abA_aA_b\right],\quad b=m-a.
 \label{df:eq-cm5}
\end{equation}
Here is the full single-word cut formula, which also proves the
weighted count without an endpoint convention. For $0\le i\le m-1$,
\begin{equation}
\begin{aligned}
sV_i^{[m]}={}&P_{m-1-i}K_{i+1}-P_iK_{m-i}\\
 &+\sum_{t=0}^{i-1}
   \left(A_{m-1-t}A_{t+1}-P_tL_{m-t,i-t-1}\right)\\
 &+\sum_{t=i+1}^{m-1}
   \left(P_{m-1-t}L_{t+1,i}-A_tA_{m-t}\right).
\end{aligned}
\label{df:eq-cm5a}
\end{equation}
Indeed the BRST current must contract its $c$ with the unique $b$
and its $Y$ with either $D$ or an undifferentiated $X$. Cutting
$\tr(bRZS)$, where $Z$ is the contracted target letter, gives
\[
 \tr(S)\tr(X^{[\mathrm{Taylor}]}R)
                 -\tr(R)\tr(X^{[\mathrm{Taylor}]}S).
\]
Here the surviving left $X$ has Taylor order two when $Z=D$
and order one when $Z=X$. The $Z=D$ term is the first line of
\textnormal{\eqref{df:eq-cm5a}}; contracting an $X$ before or after $D$ gives
respectively its second or third line. A zero-length arc contributes
$P_0=N$. The $-1/N$ projector parts cancel because replacing any
one slot of $\tr c[X,Y]$ by a scalar matrix gives zero.

Multiply \textnormal{\eqref{df:eq-cm5a}} by $\lambda_i$ and sum. The coefficient
of $P_aK_b$ is $\lambda_{m-1-a}-\lambda_a=m-2a-1$.
For each $P_aL_{b,j}$ there are exactly two contributions, whose
coefficient is
\[
 \lambda_j-\lambda_{a+j+1}=-(a+1).
\]
Their indices lie between $0$ and $m-1$. Finally the ordered
$A_aA_b$ coefficient is
\[
 \sum_{i=a}^{m-1}\lambda_i-\sum_{i=0}^{a-1}\lambda_i=ab,
 \qquad a+b=m,
\]
since $\sum_i\lambda_i=0$. These three sums prove
\textnormal{\eqref{df:eq-cm5}} for arbitrary $m$.

Finally,
\[
 \partial P_b=bA_b,\qquad
 \partial^2P_b=b\left(2K_b+\sum_{j=0}^{b-2}L_{b,j}\right).
\]
Substitute \textnormal{\eqref{df:eq-cm5}} into \textnormal{\eqref{df:eq-cm4}} and use these
identities. This gives exactly the displayed $\partial^2P_m$
and two-trace derivative terms in $O_0$. The normal-order
corrections in $:G_XC_{m,0}:$ and $E_2^{[m]}$ are respectively
$K_m$ and $(m+2)K_m/2$; their coefficients account for the last
term of \textnormal{\eqref{df:eq-cm4}}. Similarly the correction in
$E_0^{[m]}$ is $(m+2)\partial P_m/(2m)$; together with the
$D$ version of \textnormal{\eqref{df:eq-cm3}} it gives the displayed $O_1$.
This completes the uniform proof.
\end{proof}

Let $N\ge2$ and let $E$ be the entire lowest conformal space of a
lower-bounded untwisted $A_N$-module, with weight $r$.  No assumption is made on its
flavor support or its odd zero modes.  If a field $A$ has
half-integral weight, put
\[
 A^+=A_{(h_A-3/2)}|_E:E\longrightarrow M_{r+1/2}.
\]
For an integer-weight field put $A^0=o(A)|_E$.  Products such as
$G_X^+C_m^0$ below are compositions in the indicated order, with
the zero mode acting on $E$ first.  All raw degree subscripts
remain meaningful beyond $N$ by the actual trace identities.

For odd $m\ge3$ write
\[
 W_m=P_m^+,\qquad P_a=\tr X^a,\quad
 C_m=\tr(X^{m-1}p),\quad G_X=\tr(Xb).
\]
Use the normalized fermion polars
\[
 B_{3,0}=\tr(X^2b),\qquad B_{3,1}=\tfrac12\tr((XY+YX)b),
 \qquad C_{m+1,0}=\tr(X^mp),
\]
\[
 C_{m+1,1}=\frac1m\sum_{i=0}^{m-1}
                     \tr(X^iYX^{m-1-i}p).
\]
The raw top convention is
\[
 H_d=-\frac1d\bar G_{Y(0)}G_{Y(0)}P_d
                         +\frac1{d^2}\partial f_{(0)}P_d.
\]
Here $e=\tfrac12\tr X^2$, $H=-\tr XY$, $f=-\tfrac12\tr Y^2$,
$G_Y=\tr(Yb)$, and $\bar G_Y=\tr(Yp)$.
Thus $B_{3,i}$ has weight $2$, $C_{m+1,i}$ has weight $(m+2)/2$,
and $H_{m+2}$ has weight $(m+4)/2$. Define a map out of $E$ by
\begin{equation}
 \mathcal K_m=
 \{(B_{3,0})_{(1)},(C_{m+1,1})_{((m-1)/2)}\}|_E
 -\{(B_{3,1})_{(1)},(C_{m+1,0})_{((m-1)/2)}\}|_E.
 \label{df:eq-mcm1}
\end{equation}
The anticommutators are full module operators before restriction;
in particular their second summands are not discarded.  Then
\begin{equation}
 \boxed{
 \begin{aligned}
 \mathcal K_m={}&\frac{m+2}{m+1}H_{m+2}^+
       -\frac{N(m+2)(m-1)}{4m^2}W_m
       -\frac{m+2}{N}G_X^+C_m^0\\
 &-\frac{m+2}{4m}
    \sum_{\substack{a+b=m\\a\ge2\ \mathrm{even}\\b\ge3\ \mathrm{odd}}}
         \frac{b-1}{b}\,W_bP_a^0 .
 \end{aligned}}
 \label{df:eq-mcm2}
\end{equation}
Thus the scalar middle-channel identity has exactly two additional
types of terms on general flavor support.  The current-first
terms cancel identically, even though neither term need vanish
separately.

\paragraph{The unprojected field identity used in the reduction.}
Set
\[
 E_0^{[m]}=:eW_{m,1}:+\tfrac12:HP_m:,\qquad
 E_2^{[m]}=:(\partial e)W_{m,1}:
                         +\tfrac12:(\partial H)P_m:,
\]
where $W_{m,1}=\tr(X^{m-1}Y)$, and put
\[
 U_m=\frac{m+2}{m(m+1)}\sum_{i=0}^{m-1}
  \left(i-\frac{m-1}{2}\right)
       \tr\bigl(bX^i(\partial X)X^{m-1-i}\bigr).
\]
The exact coefficients of
$O_j=(B_{3,0})_{(j)}C_{m+1,1}-(B_{3,1})_{(j)}C_{m+1,0}$ are
\begin{subequations}\label{df:eq-mcm3}
\begin{align}
O_0={}&\frac{m+2}{m+1}H_{m+2}
   -\frac{m+2}{N}:G_XC_m:-\frac2N E_2^{[m]}+QU_m
   +\frac{N(m+2)}{m^2(m+1)}\partial^2P_m\notag\\
&+\frac{m+2}{m(m+1)}\sum_{\substack{a+b=m\\ a,b\ge2}}
       \left(\frac{a+1}{b}:P_a\partial^2P_b:
                       -:(\partial P_a)(\partial P_b):\right),
 \label{df:eq-mcm3-o0}\\
O_1={}&-\frac2N E_0^{[m]}
     +\left(\frac{N(m+2)}{m^2}-\frac{m+2}{Nm}\right)\partial P_m
     +\frac{m+2}{m}\sum_{\substack{a+b=m\\ a,b\ge2}}
                             \frac1b:P_a\partial P_b:.
 \label{df:eq-mcm3-o1}
\end{align}
\end{subequations}
These are the identities \eqref{df:eq-cm1-o0} and \eqref{df:eq-cm1-o1} just proved.  In
particular their validity has no scalar-lowest-space hypothesis.
The supercommutator formula
\[
 [A_{(u)},B_{(v)}]_{\mathrm s}
       =\sum_{j\ge0}\binom uj(A_{(j)}B)_{(u+v-j)}
\]
with cubic VOA index $u=1$ and $v=(m-1)/2$ gives
\begin{equation}
 \mathcal K_m=(O_0)_{((m+1)/2)}|_E
                         +(O_1)_{((m-1)/2)}|_E.
 \label{df:eq-mcm4}
\end{equation}
There is no contribution from $O_2$, since $\binom12=0$.
The exact term $QU_m$ acts as zero.  I now calculate every
other term of \textnormal{\eqref{df:eq-mcm4}} without discarding zero modes.

\paragraph{Current terms cancel on every lowest space.}
For an even current $J$ of weight one and any half-integral-weight
field $V$, normal-mode splitting on $E$ gives
\begin{equation}
 (:JV:)^+=V^+J^0,\qquad
 (:(\partial J)V:)^+=-V^+J^0.
 \label{df:eq-mcm5}
\end{equation}
For the first formula the normal-ordered sum with current index
$k<0$ has current energy at least one, so the right field has
negative energy and kills $E$.  In the second sum, current
indices $k>0$ kill $E$, leaving exactly $k=0$ on the right.
For the second formula use
$(\partial J)_{(k)}=-kJ_{(k-1)}$: the $k=0$ term is zero,
the $k=1$ term has zero current energy and coefficient $-1$,
and every $k>1$ term has a lowering current on the right.
Consequently
\begin{equation}
 (E_2^{[m]})_{((m+1)/2)}|_E
       =-(E_0^{[m]})_{((m-1)/2)}|_E.
 \label{df:eq-mcm6}
\end{equation}
Their coefficients in \eqref{df:eq-mcm3-o0} and \eqref{df:eq-mcm3-o1} are equal, so
their sum in \textnormal{\eqref{df:eq-mcm4}} vanishes identically.

\paragraph{The odd normal product retains a right zero mode.}
The first-half mode of $:G_XC_m:$ has only one surviving
normal-split term before reordering:
\[
 (:G_XC_m:)^+=-(C_m)_{((m-1)/2)}G_{X(0)}|_E.
\]
Indeed negative VOA indices of $G_X$ give energy at least
$3/2$ and hence force a lowering mode of $C_m$ on the right;
positive indices of $G_X$ give a lowering right mode in the
second sum.  Index zero is its sole remaining index, and the
normal-split sign is negative because both fields are odd.
Their elementary simple-pole product is
\[
 G_{X(0)}C_m=\frac1m\partial P_m.
\]
This follows by Taylor expanding the surviving $X$ after the
double-pole $b,p$ contraction; the scalar-projector contribution
contains $\tr\partial X=0$.  Since the first mode index is
zero, its anticommutator with $o(C_m)$ contains only this
simple-pole product. Explicitly, with $s=(m-1)/2$,
\[
 \{G_{X(0)},(C_m)_{(s)}\}|_E
       =\frac1m(\partial P_m)_{(s)}|_E
       =-\frac{s}{m}W_m.
\]
Moving $(C_m)_{(s)}$ to the right therefore gives
\begin{equation}
 (:G_XC_m:)^+=G_X^+C_m^0+\frac{m-1}{2m}W_m.
 \label{df:eq-mcm7}
\end{equation}
The first term in \textnormal{\eqref{df:eq-mcm7}} is exactly the additional
odd-zero-mode term in \textnormal{\eqref{df:eq-mcm2}}.

\paragraph{Every regular two-trace term is explicit.}
In each pair $a+b=m$, one degree is even and the other odd.
Pure-$X$ factors and all their derivatives have mutually regular
OPEs.  A first-half normal-product mode on $E$ can therefore
survive only with a zero mode of the even-degree factor and
a first-half mode of the odd-degree factor.  To see this,
commute the two factor modes.  A lowering mode can then be
moved to the right and kills $E$; two raising modes have
total energy at least one, exceeding $1/2$.
Consequently
\begin{equation}
 (:P_aP_b:)^+=
 \begin{cases}
 W_bP_a^0,&a\text{ even},\ b\text{ odd},\\
 W_aP_b^0,&a\text{ odd},\ b\text{ even}.
 \end{cases}
 \label{df:eq-mcm8}
\end{equation}
Set $u_a=\lfloor a/2\rfloor$.  At the surviving factor mode,
one derivative multiplies by $-u_a$, and two derivatives by
$u_a(u_a+1)$.  Combining the three ordered-pair terms of
\eqref{df:eq-mcm3-o0} and \eqref{df:eq-mcm3-o1} therefore gives the coefficient
\begin{equation}
 c_{a,b}=\frac{m+2}{m(m+1)}u_b
       \left(\frac{(a+1)(u_b+1)}b-u_a-\frac{m+1}b\right).
 \label{df:eq-mcm9}
\end{equation}
If $a$ is odd and $b$ even, write $a=2A+1,b=2B$.
The bracket is
\[
 \frac{(A+1)(B+1)-(A+B+1)}B-A=0.
\]
If $a$ is even and
$b$ odd, write $a=2A,b=2B+1$.  The bracket is
$-(m+1)/(2b)$, yielding
\begin{equation}
 c_{a,b}=-\frac{(m+2)(b-1)}{4mb}.
 \label{df:eq-mcm10}
\end{equation}
This proves exactly the final sum in \textnormal{\eqref{df:eq-mcm2}};
there is no extra copy from reversing the ordered pair.

Finally put $s=(m-1)/2$.  The single-trace derivative terms
and the scalar term from \textnormal{\eqref{df:eq-mcm7}} give
\begin{align*}
 &\frac{N(m+2)}{m^2(m+1)}s(s+1)
   -\left(\frac{N(m+2)}{m^2}-\frac{m+2}{Nm}\right)s
   -\frac{m+2}{N}\frac{s}{m}\\
 &\hspace{4em}=-\frac{N(m+2)(m-1)}{4m^2}.
\end{align*}
The two $1/N$ terms cancel with no flavor assumption.  Together
with \textnormal{\eqref{df:eq-mcm6}--\eqref{df:eq-mcm10}} this proves
\textnormal{\eqref{df:eq-mcm2}} on the whole space $E$.

\subsection{Elementary half modes and the source-square identity}
\label{df:sec-elementary-half}

The elementary single-contraction and mixed cubic equations are
\begin{equation}
 (B_{3,1})_{(0)}P_d=-dB_{d+1}+\frac dN:G_XP_{d-1}:,
 \qquad (B_{3,1})_{(t)}P_d=0\quad(t\ge1).
 \label{df:eq-fhc8}
\end{equation}
There is one contractible $Y$ in $B_{3,1}$.  Its $d$ target positions give
the first term, and the scalar part of the traceless projector gives the
second.  The mixed cubic highest OPE has the two poles
\begin{equation}
 (B_{3,0})_{(1)}C_d=P_{d+1}-\frac1N:P_2P_{d-1}:,
 \quad (B_{3,0})_{(0)}C_d=\frac2{d+1}\partial P_{d+1}
                           -\frac1N:(\partial P_2)P_{d-1}:.
 \label{df:eq-fhc11}
\end{equation}
These follow from the unique $b,p$ contraction and the Taylor expansion
of the two surviving left $X$ letters.  In the first-half commutator,
the two projector terms cancel, because the derivative on the degree-two
right zero has multiplier $-1$.  Thus, for even $d\ge2$,
\begin{equation}
 \{o(B_{3,0}),C_d^+\}=\frac1{d+1}T_{d+1}.
 \label{df:eq-fhc12}
\end{equation}
On a full odd input, set $S_i=o(B_{3,i})$ and
$A_d=S_1T_d=-dB_{d+1}^++dG_X^+R_{d-1}/N$.
Twice lowering $[S_0,T_d]=0$, and then using \eqref{df:eq-mbd4}, gives
\begin{equation}
 S_2T_d-\frac2d[f,A_d]
 =NB_{d-1}^++\frac{2(d-1)}N
       (G_Y^+R_{d-1}-G_X^+Q_{d-1})
 +\sum_{\substack{a+b=d-1\\ a,b\ge2\ {\rm even}}}(a+1)B_b^+R_a.
 \label{df:eq-sfd5}
\end{equation}
Indeed the normalized second lowering of the cubic coupled expression is
$(d+1)/(d-1)$ times the left-hand side.  The odd--odd right endpoints
in \eqref{df:eq-mbd4} vanish by hypothesis.  This proves the equation on every input
weight, including $d=3$.

Put $Z(A,B)=A_{(0)}B-\tfrac12\partial A_{(1)}B$ for
weight-two fields.  Its zero mode is the full supercommutator
of their zero modes at every level.  The spin-two highest field
identity $Z(B_{3,0},B_{3,0})=0$ and its second flavor lowering give
\[
 \{S_0,S_2\}+2\{S_1,S_1\}=0.
\]
The scalar field identity \eqref{of:cubic-scalar-b}, including its exact term, gives
\[
 \{S_0,S_2\}-\{S_1,S_1\}
                  =-\frac6N o(:G_XG_Y:).
\]
Exact fields act as zero in an $A_N$-module.  Subtracting the two
identities proves the all-level formula
\begin{equation}
 \boxed{\quad S_1^2=\frac1N o(:G_XG_Y:).\quad}
 \label{df:eq-nsr2}
\end{equation}
The right side has not been declared zero on the whole half level.
Apply the even relative-chain derivation
$\delta b=p$, $\delta p=\delta X=\delta Y=0$ twice to the
complete field identity: its two sides acquire the same factor two.
This proves the barred counterpart
$\bar S_1^2=N^{-1}o(:\bar G_X\bar G_Y:)$ without a ghost swap.

\paragraph{The two required lowering supercurrents kill the particular source.}
For a maximal-flavor highest input $v$ and any even $a\ge4$, set
$v_a=C_{a,0((a-2)/2)}v=C_{a,0}^+v$.
The exact two-pole OPEs are
\begin{equation}
\begin{aligned}
 G_{X(1)}C_{a,0}&=P_a,&
 G_{X(0)}C_{a,0}&=a^{-1}\partial P_a,\\
 G_{Y(1)}C_{a,0}&=a^{-1}f_{(0)}P_a,&
 G_{Y(0)}C_{a,0}&=-H_a+a^{-2}\partial f_{(0)}P_a.
\end{aligned}
\label{df:eq-nsr3}
\end{equation}
The first line is the single $b,p$ contraction, including its
Taylor coefficient; its scalar-projector term contains $\tr X=0$.
The second follows by inserting the definition of the raw top $H_a$ and expanding the unique $b,p$ contraction and its first Taylor coefficient.  In the full mode anticommutator the index
of $G$ is one, so poles of order at least three do not contribute.
With $n=(a-2)/2$ and $(\partial F)_{(n+1)}=-(n+1)F_{(n)}$,
these give exactly
\begin{equation}
\begin{aligned}
 \{G_{X(1)},C_{a,0((a-2)/2)}\}&=\tfrac12P_a^0,\\
 \{G_{Y(1)},C_{a,0((a-2)/2)}\}&=-H_a^0
                                      +\frac1{2a}[f,P_a^0].
\end{aligned}
\label{df:eq-nsr4}
\end{equation}
The lowering supercurrents annihilate $v$ because it is a lowest
vector.  Every operator on the right of \eqref{df:eq-nsr4} has strictly
positive flavor weight when $a\ge4$, preserves $W$, and hence
annihilates the chosen highest vector.  Therefore
\begin{equation}
                         G_{X(1)}v_a=G_{Y(1)}v_a=0.
 \label{df:eq-nsr5}
\end{equation}
This proves the two specific mode annihilations; no assertion
about all positive modes on $v_a$ is being inferred from its flavor.

The zero mode of $:G_XG_Y:$ is its VOA mode numbered two.
On any $y\in M_{r+1/2}$, the normal-mode formula reduces to
\begin{equation}
 o(:G_XG_Y:)y
   =-G_{Y(1)}G_{X(0)}y-G_{Y(0)}G_{X(1)}y.
 \label{df:eq-nsr6}
\end{equation}
Indeed in the first normal sum a negative index of $G_X$ puts
a mode of $G_Y$ numbered at least two on the right; it lowers
energy by at least $3/2$ and annihilates this half level.
In the second sum every right $G_X$ index at least two vanishes
for the same reason, leaving precisely indices zero and one,
with the minus sign from two odd fields.  The same-ghost
supercurrents have regular OPE, so
$\{G_{Y(1)},G_{X(0)}\}=0$.
Equations \eqref{df:eq-nsr5}--\eqref{df:eq-nsr6} and then \eqref{df:eq-nsr2} yield
\begin{equation}
 \boxed{\quad S_1^2C_{a,0}^+v=0\quad(a\ge4\text{ even}).\quad}
 \label{df:eq-nsr7}
\end{equation}
For the opposite-family source, the OPEs of $G_X,G_Y$
with $B_{a,0}$ are regular: the $G_Y$ single-contraction terms
cancel in pairs by odd cyclicity, and their scalar projector
contains $\operatorname{tr}b=0$. Applying $\delta$ to these
regular OPEs gives
$\bar G_\alpha{}_{(i)}B_{a,0}=-G_\alpha{}_{(i)}C_{a,0}$.
Thus the same flavor bound kills the two barred lowering
supercurrents on $B_{a,0}^+v$. The barred square just proved
and its full normal split give $\bar S_1^2B_{a,0}^+v=0$.
These are statements about the specified source vectors, not
nilpotence on the entire output.

 \eqref{df:eq-nsr2}, the half-level normal
splitting \eqref{df:eq-nsr6}, and the two exact anticommutators \eqref{df:eq-nsr4}
give the complete map identity
\begin{equation}
 \boxed{\quad
 S_1^2C_{a,0}^+
 =\frac1N G_X^+\left(-H_a^0+\frac1{2a}[f,P_a^0]\right)
                    -\frac1{2N}G_Y^+P_a^0.
 \quad}
 \label{df:eq-csh1}
\end{equation}
Every zero mode on the right acts on the input first.  No odd
annihilator assumption is needed for \eqref{df:eq-csh1}.
Indeed on $y=C_{a,0}^+v$ one has
\[
 o(:G_XG_Y:)y
   =G_{X(0)}G_{Y(1)}y-G_{Y(0)}G_{X(1)}y,
\]
because the same-ghost supercurrents have regular OPE.  Substitute
\eqref{df:eq-nsr4}, whose positive supercurrents annihilate the input lowest
space, to obtain \eqref{df:eq-csh1}.  In particular this proof does not use
the lowest-space cubic exterior relations on the output.

\subsection{The initial boundary from the field determinant}
\label{df:sec-boundary}

Let $N\ge3$ and let $E=M_r$ be a lowest space of a lower-bounded
untwisted $A_N$-module. Suppose $V\subset E$ is a nonzero
submodule for the full zero-mode algebra, with finite-dimensional
flavor support, and satisfies the full odd zero-mode annihilator
condition. In particular every integer-weight raw $B_a,C_a$
zero mode vanishes on $V$. Let $j$ be the largest flavor spin
in $V$ and let $v$ be a highest vector in any one of its spin-$j$
summands, so $Hv=2jv$. I do not require $V$ to be isotypic,
its chosen spin-$j$ summand to be stable under the zero algebra,
or $j$ to be the largest flavor spin in $E$.
All first-half maps in this section are restricted to $V$.
Write $V_j=\{x\in V:Hx=2jx\}$ for its maximal-weight subspace.

Use the conventions
\[
\begin{gathered}
 P_a=\tr X^a,\qquad B_a=-\frac1aG_{Y(0)}P_a,\qquad
 C_a=-\frac1a\bar G_{Y(0)}P_a,\\
 H(P)=-\frac1d\bar G_{Y(0)}G_{Y(0)}P
                         +\frac1{d^2}\partial f_{(0)}P.
\end{gathered}
\]
for a homogeneous invariant polynomial $P$ of degree $d$.
Thus $H_a=H(P_a)$, $H_2=-T$, $P_2=2e$, and
$f_{(0)}P_2=-2H$. Write $W_a=P_a^+$ for odd $a$.

\paragraph{The product formula with its two non-scalar endpoints.}
For even $a\ge2$, odd $b\ge3$, and $d=a+b$, one has
\begin{equation}
 \boxed{\quad
 H(P_aP_b)^+v=
 \left\{\frac{ab(b-1)}{2d(d-2)}
       -\delta_{a,2}\left(\frac{2r}{d}+\frac{2j}{d^2}\right)
 \right\}W_{d-2}v .\quad}
 \label{df:eq-htb1}
\end{equation}
For a product of at least three positive-degree raw traces,
of odd total degree $d$, its top satisfies
\begin{equation}
                  H(P_{a_1}\cdots P_{a_k})^+v=0
                  \qquad(k\ge3,\ a_i\ge2).
 \label{df:eq-htb2}
\end{equation}

Here is a detailed mode calculation of \eqref{df:eq-htb1}. Pure-$X$ factors
and their derivatives have regular mutual OPEs. On a lowest
space their first-half product can have only one factor of
half-integral weight, acting with energy $1/2$, with all other
factors acting by zero modes on the input. Every even raw trace
zero mode has positive flavor weight and preserves $V$, so it
kills $v$. Hence any product of at least two positive-degree
pure-$X$ traces, including derivatives on any factors, has zero
first-half mode on $v$. This assertion is about $v$, not the
whole of $V$.

The exact zero-mode derivation identity is
\begin{equation}
\begin{aligned}
H(P_aP_b)={}&\frac ad:H_aP_b:+\frac bd:P_aH_b:
       +\frac{ab}{d}(:B_aC_b:-:C_aB_b:)\\
 &-\frac1{ad}:\partial f_{(0)}P_a\,P_b:
       -\frac1{bd}:P_a\,\partial f_{(0)}P_b:
       +\frac1{d^2}\partial f_{(0)}(:P_aP_b:).
\end{aligned}
\label{df:eq-htb3}
\end{equation}
In particular its last term must be retained on a non-scalar
highest input.

For completeness, the required OPE coefficients are displayed
below. Put $S=P_{d-2}$, and write
$A_{(0)}B=t\partial S$, $A_{(1)}B=sS$ modulo derivatives
of pure-$X$ products with at least two trace factors. Those
remainder terms have zero first-half mode on $v$, as just
proved. All higher poles vanish. If $u=\operatorname{wt}(A)$,
put $K=u-1$ for integer $u$, and $K=u-1/2$ otherwise.
\begin{equation}
\begin{array}{c|c|c|c}
 (A,B)&t&s&K\\\hline
 (H_a,P_b)&-\dfrac{b(a-1)}{d-2}&-\dfrac{b(a-1)}a&a/2\\[1mm]
 (P_a,H_b)&\dfrac{a(b-1)(2-a)}{b(d-2)}
                  &-\dfrac{a(b-1)}b&a/2-1\\[1mm]
 (B_a,C_b)&\dfrac{a-1}{d-2}&1&a/2\\[1mm]
 (C_a,B_b)&-\dfrac{a-1}{d-2}&-1&a/2\\[1mm]
 (\partial f_{(0)}P_a,P_b)&0&ab&a/2\\[1mm]
 (P_a,\partial f_{(0)}P_b)&ab&ab&a/2-1
\end{array}
 \label{df:eq-htb4}
\end{equation}
They follow from
\[
\begin{gathered}
 H(P)=\frac1d P_{ij}b^ip^j
       -\frac{d-1}{d^2}P_i\partial Y^i
       +\frac1{d^2}P_{ij}Y^i\partial X^j,\\
 (P_a)_i(P_b)_i
       =ab\left(P_{d-2}-\frac1N P_{a-1}P_{b-1}\right),
\end{gathered}
\]
in orthonormal traceless coordinates. Taylor expansion gives the
simple poles, skew symmetry gives the reversed rows, and
translation gives the last two rows. Scalar-projector pieces
are among the pure-$X$ remainders already discarded on $v$.

Normal-mode splitting, including all right zero-mode endpoints,
gives for each row the correction
\begin{equation}
 \left\{K\frac{d-3}{2}t-\binom K2s\right\}W_{d-2}v.
 \label{df:eq-htb5}
\end{equation}
Indeed skew symmetry gives
$(-1)^{|A||B|}B_{(0)}A=(s-t)\partial S$ and
$(-1)^{|A||B|}B_{(1)}A=sS$. Summing the finite normal-mode
commutators yields \eqref{df:eq-htb5}, equivalently the two binomial
differences in \eqref{of:normal-modes}. Inserting all six rows with their
coefficients in \eqref{df:eq-htb3} gives
$ab(b-1)/(2d(d-2))$.

The right zero-mode endpoints require a separate check.
For even $a$, $H_a$ is flavor highest of weight $a/2-1$.
Its zero mode preserves $V$, so it kills $v$ unless $a=2$;
then $H_2^0v=-rv$. This contributes
$-\delta_{a,2}(2r/d)W_bv$.
The endpoint in $:P_aH_b:$ vanishes because $P_a^0v=0$.
The two odd products have right integer-weight fermion zero
modes, which annihilate $V$ by hypothesis.
Finally $f_{(0)}P_a$ has flavor weight $a/2-1$; even though
it need not be a highest field, its zero mode kills $v$ when
$a>2$, by stability of $V$. For $a=2$,
\[
 o(\partial f_{(0)}P_2)v=-o(f_{(0)}P_2)v=4jv.
\]
Thus $-(ad)^{-1}:\partial f_{(0)}P_a\,P_b:$ contributes
$-\delta_{a,2}(2j/d)W_bv$. The other derivative product
has right endpoint $P_a^0v=0$.

It remains to evaluate the total derivative in \eqref{df:eq-htb3}.
Set $Q=:P_aP_b:$. Regularity gives $Q^+v=0$ and
$Q^+fv=W_bP_a^0fv$. If $a>2$, $P_a^0$ raises flavor
weight by $a/2>1$ and hence kills $fv$. If $a=2$,
\[
 Q^+fv=4jW_bv,\qquad
 (f_{(0)}Q)^+v=[f,Q^+]v=-4jW_bv.
\]
Because $Q$ has weight $d/2$, its derivative multiplies
the first-half mode by $-(d-1)/2$. The last term in \eqref{df:eq-htb3}
therefore contributes
$\delta_{a,2}\,2j(d-1)W_bv/d^2$.
Combining the two $j$ terms gives
$-\delta_{a,2}\,2jW_bv/d^2$, proving \eqref{df:eq-htb1}.

\paragraph{Why three or more trace factors give zero.}
Let $Q=P_{a_1}\cdots P_{a_k}$, $k\ge3$, with odd total
degree. Its first-half mode is identically zero on $E$ if
at least three factors have odd degree. Otherwise exactly one
factor has odd degree, and every term of $Q^+$ has that
first-half map followed by the zero modes of the $k-1$ even
factors. Their total flavor weight is at least $k-1\ge2$.
They preserve $V$ and consequently kill both $v$ and $fv$.
In either case
\begin{equation}
 Q^+v=Q^+fv=0,\qquad (\partial f_{(0)}Q)^+v=0.
 \label{df:eq-htb6}
\end{equation}
This separately addresses the derivative-flavor term; its
vanishing does not follow merely from $Q^+v=0$.

Apply the two supercharge derivations to the regular product
$Q$, using the fixed-parenthesis expansion \eqref{df:eq-stz2}. Its normal sums keep each untouched spectator in place, exactly as proved below for \eqref{df:eq-stz5}. A resulting summand has either one $H$ factor, two odd
fermion factors, or one differentiated $f_{(0)}P$ factor,
with all other factors remaining pure-$X$ traces. In its
normal-mode calculation a singular contraction can join at
most two of the original trace factors: an $H$ or $f_{(0)}P$
factor has only one $Y$ letter, while the fermion pair has
only one possible $b,p$ contraction. After this contraction
all surviving factors contain only $X$ and derivatives, so
there can be no further singular contraction. Such corrections
therefore retain at least $k-1\ge2$ pure-$X$ trace factors,
whose first-half maps kill $v$.

Terms with no such contraction are handled by moving their
integer zero modes to the right. Positive-flavor zero modes
kill the highest vector in the stable $V$; the integer odd
zero modes kill all of $V$. The only zero-flavor exceptions
are $H_2=-T$ and $\partial f_{(0)}P_2=2\partial(-H)$.
Each of these acts by a scalar on $v$ and multiplies the
first-half map of at least two remaining pure-$X$ factors,
which is zero. Together with \eqref{df:eq-htb6}, this proves \eqref{df:eq-htb2}.

\paragraph{The highest-input Cayley--Hamilton top identity.}
For every odd $d>N$, the degree-$d$ coefficient of
\[
 \det(1-tX)=\exp\left(-\sum_{a\ge2}\frac{P_a}{a}t^a\right)
\]
gives
\[
 P_d=\frac d2\sum_{\substack{a+b=d\\ a,b\ge2}}\frac{P_aP_b}{ab}
                   +\text{products of at least three traces}.
\]
Apply $H$ linearly, then use \eqref{df:eq-htb1}--\eqref{df:eq-htb2}. Pair the
ordered terms with $a$ even and $b$ odd. Their constant sum is
\[
 \sum_{b=3,5,\ldots,d-2}\frac{b-1}{2(d-2)}
                    =\frac{(d-1)(d-3)}{8(d-2)}.
\]
The $a=2$ endpoint contributes $-r/(d-2)-j/(d(d-2))$.
Consequently
\begin{equation}
 \boxed{\quad
 H_d^+v=\frac{(d-1)(d-3)/8-r-j/d}{d-2}\,W_{d-2}v,
                  \qquad d>N\text{ odd}.\quad}
 \label{df:eq-htb7}
\end{equation}
This includes the scalar formula at $j=0$.

\paragraph{The determinant and the highest fermion boundary.}
Let $\mathcal E(t)=\det(1-tX)$ and
$\mathcal W(t)=\sum_{c\ge3\ {\rm odd}}T_ct^c/c$.
Mutual regularity gives
\begin{equation}
 \mathcal E_{\rm odd}^+(t)
 =-\mathcal W(t)\exp\left(-\sum_{s\ge2\ {\rm even}}\frac{R_s}{s}t^s\right).
 \label{df:eq-fhc2}
\end{equation}
For a product of pure traces, a lowering mode commutes to the right and
annihilates the input.  Total energy $1/2$ therefore permits one odd
first-half factor and only even zero factors.  Expanding the exponential
proves the identity including every multiplicity.  Since $\deg\mathcal E\le N$,
for odd $d>N$ and $v\in V_j$ it gives
\begin{equation}
 T_dv=0,\qquad T_dfv=\frac{2jd}{d-2}T_{d-2}v.
 \label{df:eq-fhc4}
\end{equation}
If $N$ is odd, the adjugate identity
\begin{equation}
 \det(1-tX)\sum_{d\ge1}C_dt^{d-1}
        =\operatorname{tr}\bigl(p\,\operatorname{adj}(1-tX)\bigr)
 \label{df:eq-fhc14}
\end{equation}
has degree at most $N-1$ on the right.  Its coefficient of $t^N$, at
the first-half mode on $v$, gives $C_{N+1}^+v=0$: every other term has
a positive even zero or an odd zero on the right.  Applying the same
argument to $b$ gives the $B$ equation.

If $N$ is even, put $d=N+1$ and $q=d-2$.  Twice lowering $T_dv=0$
and using $S_iV=0$ gives
\[
 (S_0W_{d,2}^+-2S_1W_{d,1}^++S_2T_d)v
      =\frac{2(d+1)}{d(d-1)}S_1T_dfv
      =-\frac{4j(d+1)}{d-1}B_{d-1}^+v.
\]
The field equation \eqref{df:eq-mbd4} gives $N(d+1)B_{d-1}^+v/(d-1)$ on its
right: every additional right factor has positive flavor weight at
$d\ge5$.  Hence $(N+4j)B_N^+v=0$, and likewise $C_N^+v=0$.
Thus for $k=\max\{c\le N:c\text{ odd}\}$, both $B_{k+1}^+v$ and
$C_{k+1}^+v$ vanish.  Their first polar is retained:
$kC_{k+1,1}^+v=-C_{k+1}^+fv$.  Equation \eqref{df:eq-fhc12} then gives
\[
 \mathcal K_kv=-\frac1{k(k+2)}T_{k+2}fv
              =-\frac{2j}{k^2}T_kv.
\]
Combining this with \eqref{df:eq-mcm2} and the determinant-top boundary above gives
\eqref{low:boundary-positive}.

\subsection{The first two quartic channels and their three-lowering identity}
\label{df:sec-quartic-low}

Write $p_i=o(W_{4,i})$ and $T_a=P_a^+$.  The input is an arbitrary lowest conformal space.\par

On an arbitrary lowest conformal space, for every odd $a\ge3$,
\begin{equation}
\boxed{\begin{aligned}
 [p_2,T_a]={}&-2aW_{a+2,1}^++\frac{2N}{3}T_a\\
 &+a\sum_{\substack{b+c=a\\ b\ge2\ \mathrm{even},\ c\ge3\ \mathrm{odd}}}
             \frac1cT_cP_b^0
       +\frac{2a}{N}W_{3,1}^+P_{a-1}^0.
\end{aligned}}\label{df:eq-qbl1}
\end{equation}
The related first-polar identity is
\begin{equation}
 [p_1,T_a]=-aT_{a+2}+\frac aN T_3P_{a-1}^0,
 \qquad [p_0,T_a]=0.
 \label{df:eq-qbl2}
\end{equation}
These are full commutators; their right zero modes have not been
removed.

Here is an indexed field proof of \eqref{df:eq-qbl1}.  Work first with full
matrix contractions and then project the scalar summand to its
Fock vacuum.  Put $F_t=(W_{4,2})_{(t)}P_a$ in that auxiliary
calculation.  The normalized source is
\[
 W_{4,2}=\frac23\tr(X^2Y^2)+\frac13\tr(XYXY).
\]
Single contraction gives $-2aW_{a+2,1}$ in $F_0$.  For the
adjacent $Y$ pair, marking its first target endpoint gives $a$
choices, and the target interval between endpoints has lengths
$s=0,\ldots,a-2$.  The two surviving source $X$ letters lie on
the same resulting trace.  For the separated $Y$ pair the two
source letters lie on different resulting traces.  Consequently
\begin{equation}
 F_1=\frac{2aN}{3}P_a
           +a\sum_{\substack{b+c=a\\ b,c\ge2}}:P_bP_c:.
 \label{df:eq-qbl3}
\end{equation}
For clarity, before combining the two source types their sums
are $(2a/3)\sum_{s=0}^{a-2}P_sP_{a-s}$ and
$(a/3)\sum_{s=0}^{a-2}P_{s+1}P_{a-1-s}$, where $P_0=N$
and $P_1=0$.  This also fixes the endpoint coefficient in \eqref{df:eq-qbl3}.

Taylor order one differentiates only the surviving source letters.
If a trace has degree $c$ and contains two source letters, their
contribution is $2\partial P_c/c$; if it contains one source
letter, it is $\partial P_c/c$.  The same interval sum yields
\begin{equation}
 F_0=-2aW_{a+2,1}+\frac{4N}{3}\partial P_a
         +2a\sum_{\substack{b+c=a\\ b,c\ge2}}\frac1c:P_b\partial P_c:.
 \label{df:eq-qbl4}
\end{equation}
This is an arbitrary-length count.  It involves no inferred
coefficient from finitely many degrees.

The desired mode combination is $F_0^++F_1^+$.  Its single-trace
derivative coefficient is $2N/3$.  In an ordered regular pair
with $b$ even and $c$ odd, the derivative contributes
$-a(c-1)/c$, so together with $F_1$ the coefficient is $a/c$.
For $b$ odd and $c$ even the corresponding coefficient is
$-a+a=0$.  This proves the first three terms of \eqref{df:eq-qbl1}.

The one-scalar contraction in the double-local product is
\[
 -\frac{2a}{N(z-w)}W_{3,1}(z)P_{a-1}(w).
\]
Its normal-product contribution and simple-pole contribution
cancel by the cubic-first form of \eqref{of:normal-modes}, except for the exact
right endpoint $-(2a/N)W_{3,1}^+P_{a-1}^0$ in the difference
between the auxiliary and traceless calculations.  The two-scalar
term is proportional to $P_2(z)P_{a-2}(w)/(z-w)^2$; its two
relevant modes cancel because a derivative on the current $P_2$
contributes $-1$.  Thus the last term of \eqref{df:eq-qbl1} has the displayed
positive sign.  Single contraction proves \eqref{df:eq-qbl2} in the same way.
As a covariance check, commuting \eqref{df:eq-qbl1} with $e$ gives exactly
twice \eqref{df:eq-qbl2}.

Three flavor lowerings of $[p_0,T_c]=0$ also give the identity
\begin{equation}
\begin{gathered}
 \mathcal K_c=\frac{c+2}{c(c-1)(c-2)}
 \{-3(\operatorname{ad}f)^2A_c+3(c+1)(\operatorname{ad}f)B_c-c(c+1)C_c\},\\
 A_c=[p_1,T_c],\qquad B_c=[p_2,T_c],\qquad C_c=[p_3,T_c].
\end{gathered}
 \label{df:eq-glq3}
\end{equation}
It is obtained by expanding each lowering with
$[f,p_i]=(4-i)p_{i+1}$ and $[f,W_{c,i}]=(c-i)W_{c,i+1}$;
the coefficient of each of the four terms in the alternating third
coupling is thereby fixed.  Both input and output actions remain in
$\operatorname{ad}f$.

\subsection{The quartic residue: primitives and all one-jet columns}
\label{df:sec-quartic-primitives}

Fix an integer $k\ge4$, set $m=k-2$, and define
\begin{equation}
 O_j=\sum_{i=0}^3(-1)^i\binom3i(W_{4,i})_{(j)}W_{k,3-i},\qquad
 \mathcal M=\sum_{i=0}^2(-1)^i\binom2i:W_{3,i}W_{k-1,2-i}:.
 \label{df:eq-qd1}
\end{equation}
In the formulas in this subsection $M$ in a normal product denotes
$\mathcal M$.  I first evaluate field coefficients, independently of
the parity of the mode that will subsequently be taken.  The displayed distinct-orbit list is for odd $k$; the labelled sums before pairing hold at every $k$.  At even $k$ the unique middle orbit has half the paired interior coefficient.  Its explicit evaluation is included in \eqref{df:eq-uet4}.

Let $S_{a,i}$ denote the average of all matrix words of length
$a$ with $i$ copies of $Y$. Matrix differentiation gives
\[
 \partial_XW_{a,i}=(a-i)S_{a-1,i},\qquad
 \partial_YW_{a,i}=iS_{a-1,i-1}.
\]
Consequently the single-contraction full-projector part of $O_0$
is
\begin{equation}
 3(k+2)\sum_{i=0}^{2}(-1)^i\binom2i
                \tr(S_{3,i}S_{k-1,2-i}),
 \label{df:eq-qd2}
\end{equation}
and its scalar-projector part is $-3(k+2)M_{\rm PBW}/N$.
For example the coefficient on the $i$th product of matrix
symmetrizations in \textnormal{\eqref{df:eq-qd2}} is
\[
 (-1)^i\binom3i(4-i)(3-i)
 +(-1)^i\binom3{i+1}(i+1)(k-2+i)
 =3(k+2)(-1)^i\binom2i.
\]
Put $T_j=\tr(YX^jYX^{k-j})$, so $T_j=T_{k-j}$.
The coefficients of the distinct $T_j$ in
\textnormal{\eqref{df:eq-qd2}} are
\begin{equation}
\begin{cases}
 \dfrac{2k(k+2)}{k-1},&j=0,\\[2mm]
 \dfrac{k(k+2)(k-5)}{(k-1)(k-2)},&j=1,\\[2mm]
 -\dfrac{6k(k+2)}{(k-1)(k-2)},&2\le j\le(k-1)/2.
\end{cases}
 \label{df:eq-qd3}
\end{equation}
Here is a direct count. With $n=k-1$, the three normalized
matrix products in \textnormal{\eqref{df:eq-qd2}}, before its outer
factor $3(k+2)$, contribute
\[
 \frac{2}{n(n-1)}\sum_{a=0}^{n-2}(n-1-a)T_a,
 \quad -\frac{2}{3n}\sum_{a=0}^{n+1}c_aT_a,
 \quad \frac{2T_0+T_1}{3},
\]
where $c_a$ counts pairs $u+v=a$ with
$0\le u\le2$, $0\le v\le n-1$.
Thus $c_0=c_{n+1}=1$, $c_1=c_n=2$, and all interior $c_a=3$.
Using $T_a=T_{k-a}$ gives \textnormal{\eqref{df:eq-qd3}}.

Set
\[
 \gamma=\frac{k(k+2)}{k-1},\qquad
 \alpha=\frac{3k(k+2)}{2(k-1)(k-2)},\qquad
 U_i=\tr(bX^iYX^{k-1-i}),
\]
\begin{equation}
 \lambda_0=\gamma,\quad \lambda_{k-1}=-\gamma,
 \qquad \lambda_i=\alpha(k-1-2i)\quad(1\le i\le k-2),
 \qquad U=\sum_{i=0}^{k-1}\lambda_iU_i.
 \label{df:eq-qd4}
\end{equation}
Since $qU_i=T_i-T_{i+1}$, the differences of these coefficients
are precisely \textnormal{\eqref{df:eq-qd3}}. Hence
\[
 (O_0)_{\rm leading}=qU-\frac{3(k+2)}N M_{\rm PBW}.
\]
The closed remainder
\begin{equation}
 R=O_0-QU+\frac{3(k+2)}N M
 \label{df:eq-qd5}
\end{equation}
has lengths $k$ and $k-2$ only.

Let $I_j=\tr(bX^jpX^{k-2-j})$.
The two empty-arc $X,Y$ contractions of the BRST current give
\begin{equation}
 (sU_i)_{\rm single\ trace,bp}=N(I_{i-1}-I_i),
 \label{df:eq-qd6}
\end{equation}
with missing endpoint indices zero. The two orientations of the commutator current give the displayed difference; scalar-projector terms vanish
because every matrix slot of $\tr c[X,Y]$ annihilates a
scalar matrix. Thus the coefficient of $I_j$ in $R$ is
$N(\lambda_j-\lambda_{j+1})$.
The ghost-bilinear part of $H_k$ is $\sum_j I_j$.
Differences of the $I_j$ are $q$-boundaries of first-jet
primitives, and therefore the coefficient of the raw
single-trace top is
\begin{equation}
 a_kN=\frac{N(\lambda_0-\lambda_{k-1})}{k-1}
          =\frac{2k(k+2)N}{(k-1)^2}.
 \label{df:eq-qd7}
\end{equation}

For a more detailed check, the complete two-trace $bp$ cut
coefficient in $R$ on
$P_a I_j^{[b]}$, where $a+b=k$ and
$I_j^{[b]}=\tr(bX^jpX^{b-2-j})$, is
\begin{equation}
 \lambda_j-\lambda_{a+j+1}.
 \label{df:eq-qd9}
\end{equation}
Indeed a target $X$ before $Y$ in $U_i$ contributes
$P_{i-1-t}I_t$; one after $Y$ contributes $-P_tI_i$
to $sU_i$. The coefficient of $B_aC_b$ in $R$ is
\begin{equation}
 -\sum_{i=0}^{a-1}\lambda_i-\sum_{i=0}^{b-1}\lambda_i
       =-2\gamma-2\alpha(a-1)(b-1)
          \quad(a,b\ge2).
 \label{df:eq-qd10}
\end{equation}
The average of \textnormal{\eqref{df:eq-qd9}} over $j$ is
\begin{equation}
 h_{a,b}=\frac{2\gamma+2\alpha a(b-2)}{b-1},
 \label{df:eq-qd11}
\end{equation}
including $b=2$. The nonconstant part is removed by
\begin{equation}
 \frac{\beta}{b-1}P_a
   \sum_{i=0}^{b-3}\left(i-\frac{b-3}{2}\right)
                  \tr(bX^i(\partial X)X^{b-3-i}),
 \qquad
 \beta=\frac{k(k+1)(k+2)}{(k-1)(k-2)}.
 \label{df:eq-qd12}
\end{equation}
These formulas specify the two-trace ghost sector uniformly;
its bosonic completion and quantum corrections with lower powers of $N$ are
evaluated below.

Write
\[
 \eta=\frac{k(k+1)(k+2)}{(k-1)^2(k-2)},\qquad
 V_i=\tr(bX^i(\partial X)X^{m-1-i}),\quad 0\le i<m.
\]
The single-trace $bp$ differences after removing
$a_kNH_k$ are eliminated by
\begin{equation}
 N\eta\sum_{i=0}^{m-1}\left(i-\frac{m-1}{2}\right)V_i.
 \label{df:eq-qd13}
\end{equation}
For this primitive, the empty-arc part of the indexed identity
\textnormal{\eqref{df:eq-cm5}} is
\begin{equation}
 s\sum_i\left(i-\frac{m-1}{2}\right)V_i
   =N\left((m-1)K_m-\sum_{j=0}^{m-2}L_{m,j}\right)
           +\text{terms with two nonempty traces},
 \label{df:eq-qd14}
\end{equation}
where $K_m=\tr(X^{m-1}\partial^2X/2)$ and
$L_{m,j}=\tr(X^j(\partial X)X^{m-2-j}(\partial X))$.

The terms proportional to $N^2$ in the original $O_j$ have
three contractions and two empty arcs. The contracted three
vertices form a consecutive block on each trace, matched in
reverse cyclic order. There are $4k$ choices of the starts.
If the source block has $i$ copies of $Y$, its contraction
sign cancels $(-1)^i$ in the Clebsch--Gordan coefficient.
The total coefficient is
\begin{equation}
\begin{aligned}
\beta&=4k\sum_{i=0}^3
       \frac{\binom3i^2}{\binom4i\binom{k}{3-i}}\\
 &=k\left(\frac4{\binom k3}
             +\frac9{\binom k2}+\frac6k+1\right)
   =\frac{k(k+1)(k+2)}{(k-1)(k-2)}
   =\eta(m+1).
\end{aligned}
\label{df:eq-qd15}
\end{equation}
The surviving left $X$ has Taylor order two in $O_0$ and
order one in $O_1$. These contributions are therefore
$\beta N^2 K_m$ and
$\beta N^2\partial P_m/m$, respectively.
Subtract \textnormal{\eqref{df:eq-qd13}}. Equations
\textnormal{\eqref{df:eq-qd14}--\eqref{df:eq-qd15}} show that the part proportional to $N^2$
remaining in $O_0$ is
\[
 N^2\eta\left(2K_m+\sum_jL_{m,j}\right)
                         =\frac{N^2\eta}{m}\partial^2P_m.
\]
For odd $k$, write $W_m=P_m^+$.  Its total first-half contribution,
including $O_1$, is
\begin{equation}
\begin{aligned}
N^2\left(\frac\eta m\frac{m^2-1}{4}
              -\frac\beta m\frac{m-1}{2}\right)W_m
 &=-\frac{N^2\eta(m^2-1)}{4m}W_m\\
 &=-\frac{k(k+1)(k+2)(k-3)N^2}
               {4(k-1)(k-2)^2}W_m.
\end{aligned}
\label{df:eq-qd16}
\end{equation}
The nonempty-face contributions have lower powers of $N$ and are evaluated below.

Write $c_{ab}=-2\gamma-2\alpha(a-1)(b-1)$ for
\textnormal{\eqref{df:eq-qd10}}. The remaining coefficient on $D_{ab}$ is
\begin{equation}
 f_{ab}=-\frac{3(k+1)(k+2)}{k-1}
       +2\lambda_a+h_{ab}\frac{b-1}{b},\qquad a+b=k,
 \quad a,b\ge2.
 \label{df:eq-qd19}
\end{equation}
Here is a direct two-contraction derivation, including the
endpoints $a=2$ or $b=2$. Fix the surviving left letters to be
$X,Y$. The flavor sum for two prescribed matched slots is
\[
 \omega_1=-\sum_{i=1}^3
 \frac{\binom3i\binom2{i-1}}{\binom4i\binom{k}{3-i}}
 =-\frac{(k+1)(k+2)}{4k(k-1)}.
\]
For the ordered target monomial
$P_a\tr(X^{b-1}\partial Y)$, cutting an adjacent source pair
has $8k$ incidences and cutting an opposite pair has $4k$.
One can count them by marking the first target endpoint and
then the source cut: in the adjacent case there are four
source edges and two choices of the surviving $Y$ position;
in the opposite case there are two source diagonals and two
choices. The prescribed target arc lengths fix the second
endpoint and the gluing. Thus the coefficient in the original
$O_0$ is $12k\omega_1=-3(k+1)(k+2)/(k-1)$.
For $sU_i$, the derivative-$Y$ part is
\[
 P_{k-1-i}\tr((\partial Y)X^i)
       -P_i\tr((\partial Y)X^{k-1-i}).
\]
Its coefficient in $sU$ on the chosen monomial is
$\lambda_{b-1}-\lambda_a=-2\lambda_a$.
Finally the derivative-$Y$ coefficient of $P_aH_b$ is
$-(b-1)/b$. Subtraction gives \textnormal{\eqref{df:eq-qd19}}.
The $B_aC_b$ columns and the first-jet primitives have no
such derivative-$Y$ term.

After subtracting these ghost and one-$Y$ columns, the remainder is pure
$X$, has two jets and at most three trace faces.  The recurrences in
Appendix~\ref{df:sec-kernels} prove that the listed derivative profiles
exhaust it, including repeated degrees.  The coefficients for its two-
and three-face parts are evaluated next; the remaining single-trace
coefficient is fixed by $L_1O_0=2O_1$.

\subsection{Two nonempty faces of the quartic remainder}
\label{df:sec-two-face}

Let $s\ge2$ be even, $t\ge3$ odd and $c=s+t+2$. Put
\begin{equation}
 \beta=\frac{c(c+1)(c+2)}{(c-1)(c-2)},\qquad
 \epsilon=\frac{\beta}{c-1},\qquad
 M=st+2s+2t+2.
 \label{df:eq-gep1}
\end{equation}
With $P_q=\operatorname{tr}X^q$, the part with the unordered trace
profile $\{s,t\}$ is
\begin{equation}
\boxed{
 R_{s,t}=\frac{\beta N}{s+t+1}\left\{
 \frac{M}{s(s+1)}P_t\partial^2P_s
 +\frac{M}{t(t+1)}P_s\partial^2P_t
 -2(\partial P_s)(\partial P_t)\right\}.}
 \label{df:eq-gep2}
\end{equation}
There is no extra factor for exchanging the two trace faces.
Their degrees have different parity, and the two second-jet
placements in \eqref{df:eq-gep2} are already listed separately.

\paragraph{An indexed derivation for both independent degrees.}
Use the divided jet $X_2=\partial^2X/2$, and write
\begin{equation}
 K_q=\operatorname{tr}(X^{q-1}X_2),\quad
 L_q=\sum_{i=0}^{q-2}\operatorname{tr}
       (X^i\partial X X^{q-2-i}\partial X),\quad
 A_q=\operatorname{tr}(X^{q-1}\partial X).
 \label{df:eq-gep3}
\end{equation}
The quartic third-channel sum has three contractions and one
source survivor in its pure-$X$ two-jet part. For a prescribed
matching of the other three source slots, the normalized polar
sum has weight
\begin{equation}
 \omega=\sum_{i=0}^{3}
   \frac{\binom3i^2}{\binom4i\binom c{3-i}}
 =\frac6{c(c-1)(c-2)}+\frac9{2c(c-1)}+\frac3{2c}+\frac14
 =\frac{\beta}{4c}.
 \label{df:eq-gep4}
\end{equation}
The contraction sign cancels the sign $(-1)^i$ in the channel
sum. One binomial counts source $Y$ positions and the other
comes from the channel; target $Y$ positions are forced.

To obtain $P_tK_s$ with one empty face, mark the face containing
the source survivor. Its target arc has length $s-1$, while the
other two arcs have lengths $t$ and zero. Choose the survivor
among four source slots, the target cut preceding the marked
arc among $c$ slots, and which remaining arc is empty in two
ways. These choices determine all cuts and their reverse-cyclic
gluing, and every such gluing determines those choices. The
other cyclic orientation has only one face and does not enter.
Thus there are $8c$ incidences, with no division when arc
lengths happen to agree: the marked face remains distinguished.
The survivor's Taylor coefficient is $X_2$ and the empty face
is $N$, giving $2\beta N$ on $P_tK_s$. Interchanging the jet
placement gives the same coefficient on $P_sK_t$.

Here are all primitive subtractions contributing to the same power of $N$
and trace profile. For $q=s+t$ and
$a_i=i-(q-1)/2$, consider the first-jet primitive
$\sum_i a_i\operatorname{tr}(bX^i\partial X X^{q-1-i})$.
In its two-contraction BRST term, contract the current's $c$
with $b$ and its $Y$ with one matrix slot. If that slot is
$\partial X$, the coefficient on $P_uK_{q-u}$ is
$a_{q-1-u}-a_u=q-2u-1$. If it is an undifferentiated $X$,
the coefficient on each prescribed $P_uL_{q-u,i}$ is
$a_i-a_{i+u+1}=-(u+1)$. The split-first-jet coefficient is
\begin{equation}
 \sum_{i=u}^{q-1}a_i-\sum_{i=0}^{q-u-1}a_i=u(q-u).
 \label{df:eq-gep5}
\end{equation}
These are the two orientations of the commutator current.
Its scalar part vanishes in each matrix slot, so the primitive
calculation has no additional scalar-projector term.
Taking $u=t$ and $u=s$, and multiplying by the primitive's
coefficient $N\epsilon$, gives the second line below. The
two spectator primitives have coefficients $\beta/(t+1)$ and
$\beta/(s+1)$. Their empty cuts are the same calculation with
$u=0$, namely $N\{(q-1)K_q-L_q\}$. Consequently the complete
coefficient before jet elimination is
\begin{equation}
\begin{split}
 R_{s,t}={}&2\beta N(P_tK_s+P_sK_t)\\
 &-N\epsilon\{(s-t-1)P_tK_s+(t-s-1)P_sK_t
                 -(t+1)P_tL_s-(s+1)P_sL_t+2stA_sA_t\}\\
 &-\frac{N\beta}{t+1}P_s\{(t-1)K_t-L_t\}
  -\frac{N\beta}{s+1}P_t\{(s-1)K_s-L_s\}.
\end{split}\label{df:eq-gep6}
\end{equation}
A contraction with a spectator merges traces and cannot produce
an empty face and two nonempty faces. The zero-jet primitive's
pure-$X$ triple contraction leaves no current survivor to Taylor
expand, so it has no residue. A contraction between two canonical
trace columns merges them and also has no two-trace term proportional
to $N$. This exhausts the contributions at the indicated power of $N$
and trace profile; it is not a claim about the other powers or profiles.

Since $2K_q+L_q=\partial^2P_q/q$ and
$A_q=\partial P_q/q$, \eqref{df:eq-gep6} is \eqref{df:eq-gep2}. For example the
coefficient of $P_tL_s$, divided by $\beta N$, is
\begin{equation}
 \frac{t+1}{s+t+1}+\frac1{s+1}
      =\frac{M}{(s+1)(s+t+1)};
 \label{df:eq-gep7}
\end{equation}
the coefficient of $P_tK_s$ is twice this. The other placement
is symmetric, and the mixed coefficient is
$-2st/(s+t+1)$ in the $A_sA_t$ basis. These three identities
prove the displayed formula as a rational identity in two
independent degrees, with no interpolation in matrix size or trace degree.

\paragraph{The exact first-half action on an arbitrary lowest input.}
Let $x$ belong to a lowest conformal space, and let
$W=T_tP_s^0x$, with the even zero acting first. The pure-$X$
factors have regular OPE. In a first-half mode of their product,
a lowering factor can be commuted to the right and kills $x$;
two raising factors exceed the available half-unit of energy.
Thus the only contribution is this even zero and odd first-half
endpoint. Write $\mathcal V(F)=F^++(L_1F)^+/2$. Translation
has multipliers $-s/2$ and $(1-t)/2$ at these endpoints, and
\[
 L_1\partial P_q=qP_q,\qquad
 L_1\partial^2P_q=2(q+1)\partial P_q.
\]
It follows, keeping the two factors in their actual order, that
\begin{equation}
 \mathcal V(P_t\partial^2P_s)x=-\frac{s^2}{4}W,\quad
 \mathcal V(P_s\partial^2P_t)x=\frac{1-t^2}{4}W,\quad
 \mathcal V((\partial P_s)(\partial P_t))x=-\frac{st}{4}W.
 \label{df:eq-gep8}
\end{equation}
Substitution gives the closed all-degree coefficient
\begin{equation}
\boxed{
 \mathcal V(R_{s,t})x
   =\frac{\beta N(2s+2-3st-2t)}{4t(s+1)}\,T_tP_s^0x.}
 \label{df:eq-gep9}
\end{equation}
Indeed the numerator before cancelling $s+t+1$ is
$-M\{s/(s+1)+(t-1)/t\}+2st$.
Writing $s/(s+1)+(t-1)/t=2-(s+t+1)/(t(s+1))$
proves \eqref{df:eq-gep9} directly.

\subsection{Three nonempty faces and repeated trace degrees}
\label{df:sec-three-face}

Let $s,t\ge2$ be even and $u\ge3$ odd. Set
\begin{equation}
 c=s+t+u+2,\qquad
 \beta=\frac{c(c+1)(c+2)}{(c-1)(c-2)},\qquad
 \sigma=\begin{cases}1,&s\ne t,\\2,&s=t.\end{cases}
 \label{df:eq-gtr1}
\end{equation}
The following sums use the three \emph{labelled} positions
$(q_1,q_2,q_3)=(s,t,u)$, including when $s=t$.
For the unordered profile $\{s,t,u\}$ the exact field is
\begin{equation}
\boxed{\begin{aligned}
 R_{s,t,u}=\frac\beta\sigma\biggl\{&
 \sum_{i=1}^3\frac{\partial^2P_{q_i}}{q_i}
       \prod_{k\ne i}P_{q_k}
       \sum_{j\ne i}\frac{q_j+1}{q_i+q_j+1}\\
 &-2\sum_{1\le i<j\le3}
       \frac{(\partial P_{q_i})(\partial P_{q_j})}
            {q_i+q_j+1}\prod_{k\ne i,j}P_{q_k}
 \biggr\}.
\end{aligned}}
 \label{df:eq-gtr2}
\end{equation}
In particular the repeated-degree case is specified by an
explicit multiplicity, rather than by continuity from distinct
trace labels.

\paragraph{All three-face contractions and primitive subtractions.}
Use the divided-jet symbols $K_q,L_q,A_q$ of \eqref{df:eq-gep3}.
For distinct face lengths, the coefficient on each marked
placement $P_{q_j}P_{q_k}K_{q_i}$ in the original quartic
channel is $2\beta$. Indeed its prescribed matching has
weight $\beta/(4c)$ by the four-term sum \eqref{df:eq-gep4}.
Choose the source survivor in four ways, the first cut of
its marked target arc in $c$ ways, and the cyclic order of
the other two face lengths in two ways. These determine
the reverse-cyclic three-face matching uniquely. Every
matching with that marked face determines these choices.
All faces are now nonempty, so there is no factor $N$.
Forward-cyclic matchings have one face and are excluded.
Taylor order two on the unique survivor gives $K_{q_i}$.

The only primitive subtractions with this trace profile are
the three spectator primitives. With $p$ the spectator and
$q,r$ the other degrees, the primitive coefficient is
$\beta/(q+r+1)$. Its two-contraction cut is
\begin{equation}
\begin{split}
 C_{q,r}={}&(q-r-1)P_rK_q+(r-q-1)P_qK_r\\
          &-(r+1)P_rL_q-(q+1)P_qL_r+2qrA_qA_r.
\end{split}\label{df:eq-gtr3}
\end{equation}
To see every coefficient directly, put $a_i=i-(q+r-1)/2$
in the affine first-jet primitive. The cut with unmarked
degree $r$ gives $a_{q-1}-a_r=q-r-1$ on $P_rK_q$;
the opposite cut gives $r-q-1$. For each same-trace
first-jet pair the coefficient difference is $-(r+1)$
or $-(q+1)$. The two split-first-jet partial sums are
$qr$ each. This proves \eqref{df:eq-gtr3}, equivalently the independent
cut identities \eqref{df:eq-gep5}, before any endpoint specialization.

For distinct labels the unsimplified remainder is therefore
\begin{equation}
 2\beta\sum_{i=1}^3K_{q_i}\prod_{k\ne i}P_{q_k}
 -\beta\sum_{i=1}^3
       \frac{P_{q_i}C_{q_j,q_k}}{q_j+q_k+1},
 \qquad\{j,k\}=\{1,2,3\}\setminus\{i\}.
 \label{df:eq-gtr4}
\end{equation}
The single-trace first-jet primitive has an explicit $N$
and splits into only two traces. The zero-jet primitive's
triple contraction has no Taylor survivor and no residue.
Contractions with a spectator, or between canonical
columns, merge traces and cannot produce this three-trace
profile. No further primitive or canonical contraction
contributes to \eqref{df:eq-gtr4}.

For example the coefficient of $P_tP_uK_s$, divided by
$\beta$, is
\begin{equation}
 2-\frac{s-u-1}{s+u+1}-\frac{s-t-1}{s+t+1}
 =2\left(\frac{u+1}{s+u+1}+\frac{t+1}{s+t+1}\right).
 \label{df:eq-gtr5}
\end{equation}
The coefficient of $P_tP_uL_s$ is the expression in
parentheses, and that of $P_uA_sA_t$ is
$-2st/(s+t+1)$. Substituting
$2K_s+L_s=\partial^2P_s/s$ and
$A_s=\partial P_s/s$ proves \eqref{df:eq-gtr2} for distinct degrees.

If $s=t$, the labelled expression must be divided by two.
For the marked odd face the two orders of the remaining
equal faces coincide, so its original coefficient is
$\beta$. The two labelled even jet placements coincide
as monomials and give coefficient $2\beta$ after division.
For the spectator primitive with odd spectator, the two
equal cuts in $C_{s,s}$ arise from the same index in the
actual cut sum, requiring the same division by two.
For the even spectator the two labelled choices represent
one primitive; division also counts it exactly once.
These exhaust the equal-degree identifications and prove
the factor $1/\sigma$ in \eqref{df:eq-gtr2}.

\paragraph{A closed all-degree endpoint coefficient.}
Let $x$ lie in a lowest conformal space and put
$W=T_uP_s^0P_t^0x$. The two even zeros commute, because
pure-$X$ fields have regular OPE. Regular normal ordering
places both zeros on the input; all other energy splits
vanish. With $\mathcal V(F)=F^++(L_1F)^+/2$, the Ward
multiplier on a twice-differentiated factor of degree $q_i$
is $\delta_i^2-q_i^2/4$, and the multiplier on two
first derivatives is $\delta_i\delta_j-q_iq_j/4$,
where $(\delta_1,\delta_2,\delta_3)=(0,0,1/2)$.
This follows from translation and
$L_1\partial P_q=qP_q$,
$L_1\partial^2P_q=2(q+1)\partial P_q$, with the
undifferentiated spectators retained.

Group the terms of \eqref{df:eq-gtr2} by their pair of differentiated
degrees, counting a second-derivative term once for each
of its two denominators. The even--even pair contributes
$-\beta(s+t)/(4\sigma(s+t+1))$ times $W$.
Each even--odd pair contributes
$-\beta(u-1)/(4\sigma u)$ times $W$: its numerator is
\begin{equation}
 s(u+1)+\frac{u^2-1}{u}(s+1)-2su
       =\frac{(u-1)(s+u+1)}u.
 \label{df:eq-gtr6}
\end{equation}
There are two such pairs, also when the two even degrees
coincide before division by $\sigma$. Hence
\begin{equation}
\boxed{
 \mathcal V(R_{s,t,u})x
 =\frac\beta{4\sigma}
      \left\{\frac2u-2-\frac{s+t}{s+t+1}\right\}
               T_uP_s^0P_t^0x.}
 \label{df:eq-gtr7}
\end{equation}
Its scalar coefficient is strictly negative for the stated
positive degrees. The composed map $T_uP_s^0P_t^0$ must
therefore be retained unless it is separately proved zero.

Every other three-factor parity profile of odd total degree has three
odd factors.  Its first-half action vanishes by regular mode ordering,
because three nonnegative odd-factor energies sum to at least $3/2$.

\subsection{Scalar-propagator compensation in normal products}
\label{df:sec-inverse-rank}

Let \(s\ge2\) be even and \(t\ge3\) odd.  Suppose the only
surviving pure-\(X\) endpoint for this trace profile is
\begin{equation}
 (P_sP_t)^+x=W.
 \label{df:eq-eoi1}
\end{equation}
There are no singular products between the pure-\(X\) factors.
For the ordered pair of degrees \(s_1,s_2\), equal to either
\((s,t)\) or \((t,s)\), set
\begin{equation}
 u=s_1+1,\qquad v=s_2+1,\qquad c=u+v=s+t+2.
 \label{df:eq-eoi2}
\end{equation}
All subsequent coefficients are divided by \(W/N\).  Thus an
overall coefficient of the even right action, such as \(C\),
is retained through \(W\).

\paragraph{Jet endpoints including the Ward action.}
For a pure trace of degree \(q\), put \(\delta_q=0\) if \(q\)
is even and \(\delta_q=1/2\) if it is odd.  Translation acts
on its relevant endpoint by \(e_q=\delta_q-q/2\).
The Ward-combined map is
\(\mathcal V(A)=A^++(L_1A)^+/2\).
The primary identities
\begin{equation}
 L_1\partial P_q=qP_q,\qquad
 L_1\partial^2P_q=2(q+1)\partial P_q
 \label{df:eq-eoi3}
\end{equation}
therefore give
\begin{equation}
 \frac{\mathcal V((\partial^2P_{s_1})P_{s_2})x}{W}
       =d_1:=\delta_{s_1}^2-\frac{s_1^2}{4},
 \qquad
 \frac{\mathcal V((\partial P_{s_1})(\partial P_{s_2}))x}{W}
       =m:=-\frac{st}{4}.
 \label{df:eq-eoi4}
\end{equation}
For example the first equality is
\(e_{s_1}(e_{s_1}-1)+(s_1+1)e_{s_1}
=\delta_{s_1}^2-s_1^2/4\).
The second follows by applying \(L_1\) to either differentiated
factor.  These identities apply in both orders.

\paragraph{The projected-normal-product correction.}
Let \(\Delta\) mean projected full-matrix normal product minus
traceless normal product.  A single scalar propagator gives
\begin{equation}
\begin{aligned}
 \Delta(P_uH_v)&=\frac{u}{vN}
 \left[-\frac{v-1}{2}(\partial^2P_{s_1})P_{s_2}
                  +(\partial P_{s_1})(\partial P_{s_2})\right],\\
 \Delta(B_uC_v)&=\frac1{2N}(\partial^2P_{s_1})P_{s_2},\\
 \Delta(D_{uv})&=\frac cN
 \left[\frac12(\partial^2P_{s_1})P_{s_2}
                  +(\partial P_{s_1})(\partial P_{s_2})\right].
\end{aligned}
\label{df:eq-eoi5}
\end{equation}
For the first line the scalar contraction before expansion is
\[
 \frac{u}{vN}\left\{
 -\frac{(v-1)P_{s_1}(z)P_{s_2}(w)}{(z-w)^2}
 +\frac{P_{s_1}(z)\partial P_{s_2}(w)}{z-w}\right\}.
\]
Taking its regular coefficient proves the formula.
The ghost contraction for the second line is a positive
double pole.  Differentiating the two scalar \(X,Y\)
contractions in
\(D_{uv}=:P_u\partial Q_v:-:Q_u\partial P_v:\)
proves the third.  No column admits two scalar contractions.
Consequently the three Ward-combined responses of \eqref{df:eq-eoi5} are
\begin{equation}
 \Delta_{PH}=\frac uv\left[-\frac{v-1}{2}d_1+m\right],
 \qquad \Delta_{BC}=\frac{d_1}{2},\qquad
 \Delta_D=c\left[\frac{d_1}{2}+m\right].
 \label{df:eq-eoi6}
\end{equation}

\paragraph{An independent computation of the normal scalar response.}
For clarity, this step computes the scalar part of the normal-mode
contractions, rather than assuming that projection preserves normal
products.  In reverse order the scalar poles of \(H_v(z)P_u(w)\)
are
\begin{equation}
 -\frac{u(v-1)}{vN}\frac{P_{s_2}P_{s_1}}{(z-w)^2},
 \qquad
 -\frac{u}{N}\frac{(\partial P_{s_2})P_{s_1}}{z-w}.
 \label{df:eq-eoi7}
\end{equation}
For the reversed ghost column the poles are
\(-P_{s_2}P_{s_1}/N\) at order two and
\(-(\partial P_{s_2})P_{s_1}/N\) at order one; its
normal-mode parity is negative.
The difference of the reversed differentiated \(P,Q\)
contractions has second pole \(cP_{s_1}P_{s_2}/N\)
and zero first pole, while the corresponding undifferentiated
difference is \(-cP_{s_1}P_{s_2}/N\).
All these statements follow by one scalar contraction,
with differentiation before the regular coefficient is taken.

The normal-mode coefficient for a reversed pole of order \(q+1\)
in a product \(A,B\) is
\begin{equation}
 (-1)^{|A||B|}
 \left\{\binom{h_A+h_B-3/2}{q+1}
                  -\binom{h_B-\epsilon_A}{q+1}\right\},
 \quad
 \epsilon_A=\begin{cases}
 1/2,&h_A\in\mathbb Z,\\
 1,&h_A\in\mathbb Z+1/2.
 \end{cases}
 \label{df:eq-eoi8}
\end{equation}
The Ward term is
\((2h_A-1)A_{(0)}B/2\).
These coefficients follow by splitting the normal product
into modes with right index below and above zero, commuting
the latter to the right, and summing the finite binomial
coefficients.  Only \(q=0,1\) occur in \eqref{df:eq-eoi7}.

Here are the resulting finite formulas, so that all signs
and parity choices in that substitution can be checked.
Let
\begin{equation}
\begin{gathered}
 H_*=(c-1)/2,\quad e_1=\delta_{s_1}-s_1/2,\quad
 e_2=\delta_{s_2}-s_2/2,\\
 k_0=H_*-(v/2+1-\epsilon_P),\qquad
 k_1=\binom{H_*}{2}-\binom{v/2+1-\epsilon_P}{2},\\
 \ell_0=H_*-((v+1)/2-\epsilon_B),\qquad
 \ell_1=\binom{H_*}{2}-\binom{(v+1)/2-\epsilon_B}{2},\\
 \kappa_0=(c-3)/2-(v/2-\epsilon_P).
\end{gathered}
\label{df:eq-eoi9}
\end{equation}
For order \((s,t)\), \((\epsilon_P,\epsilon_B)=(1,1/2)\);
for order \((t,s)\) they are \((1/2,1)\).
The positive scalar-propagator normal responses are
\begin{equation}
\begin{aligned}
 R_{PH}
   &=-ue_2k_0-\frac{u(v-1)}v k_1
        +\frac{(u-1)u}{2v}\{-(v-1)e_1+e_2\},\\
 R_{BC}
   &=\ell_0e_2+\ell_1+\frac u2e_1,\\
 R_D
   &=c\left\{k_1-\frac v2\kappa_0
                        +\frac{u-1}{2}(e_1+e_2)\right\}.
\end{aligned}
\label{df:eq-eoi10}
\end{equation}
Inserting \(\delta_s=0,\delta_t=1/2\) in either order gives
\begin{equation}
 R_{PH}=\Delta_{PH},\qquad
 R_{BC}=\Delta_{BC},\qquad
 R_D=\Delta_D.
 \label{df:eq-eoi11}
\end{equation}
These are six polynomial identities in the two independent
degrees after the first equality is multiplied by \(v\). The traceless propagator is the full propagator minus the
positive scalar propagator, so its $1/N$ normal
correction is \(-R\).  Thus \eqref{df:eq-eoi11} proves its termwise
cancellation with \eqref{df:eq-eoi5}, without treating scalar-Fock
projection as an algebra homomorphism.

\subsection{The single-face quartic coefficient}
\label{df:sec-one-face}

Let $k\geq5$ be odd, $m=k-2$, and put
\[
 \gamma=\frac{k(k+2)}{k-1},\quad
 \alpha=\frac{3k(k+2)}{2(k-1)m},\quad
 A_1=-\frac{(k+1)(k+2)}{k-1},\quad
 g=\frac{k(k+1)(k+2)}2.
\]
For every ordered pair $a,b\geq2$ with $a+b=k$, set
\begin{align}
 h_{ab}&=\frac{2\gamma+2\alpha a(b-2)}{b-1},\notag\\
 c_{ab}&=-2\gamma-2\alpha(a-1)(b-1),\notag\\
 f_{ab}&=3A_1+2\alpha(k-1-2a)+h_{ab}\frac{b-1}{b}.
 \label{df:eq-qz1}
\end{align}
They are respectively the coefficients of $:P_aH_b:$,
$:B_aC_b:$, and
$D_{ab}=:P_a\partial Q_b:-:Q_a\partial P_b:$.
Write $s=\lfloor(a-1)/2\rfloor$, $t=\lfloor a/2\rfloor$.
Suppress the common first-half map $W_m=P_m^+$ in the scalar
coefficient calculation, and
write $p_{ab},c'_{ab},d_{ab}$ for the contraction constants obtained from the normal-mode formula:
\begin{align}
 p_{ab}&=\frac{a(b-1)s}{2}
                  \left(\frac{m-1}{m}-\frac{m-s}{b}\right),\notag\\
 c'_{ab}&=\frac t2\left(m-t-\frac{(b-1)(m-1)}m\right),\qquad
 d_{ab}=\frac{k s(m-s)}2.
 \label{df:eq-qz2}
\end{align}
The first formula includes $b=2$ as a contraction constant; the
additional $-r$ from $H_2=-T$ is the already separated energy term.

\paragraph{The Ward data.}
The single-trace $\partial P_m$ coefficients in the $L_1$ images
of the three displayed fields, in simultaneous PBW form, are
\begin{equation}
 \ell^P_{ab}=\frac{a(a-1)(b-1)(2-a)}{mb},\quad
 \ell^C_{ab}=\frac{a(a-1)}m,\quad
 \ell^D_{ab}=\frac{k(a-1)(b+m)}m.
 \label{df:eq-qz3}
\end{equation}
The first two follow by taking $(a-1)(P_a)_{(0)}H_b$ and
$a(B_a)_{(0)}C_b$ from the reversed OPE of the two displayed fields.  Their other
terms are regular pure-$X$ multi-traces.  For the third,
\[
 L_1D_{ab}=b\bigl(:P_aQ_b:-:Q_aP_b:\bigr)
                              +k(a-1)\partial P_m.
\]
The normal-order correction in the parenthesis is
$k(a-1)\partial P_m/m$, giving the last coefficient in \eqref{df:eq-qz3}.
That same parenthesis has first-half action $-ksW_m$.

All the chosen affine-weighted BRST primitives are annihilated by
$L_1$: in the first-jet primitives, $L_1$ replaces $\partial X$
by $X$ and the affine weights sum to zero; zero-jet primitives
have zero $L_1$ action.  Consequently $L_1O_0=2O_1$ fixes the
$N^0$ part $c_{\partial^2}$ of the coefficient multiplying
$\partial^2P_m$ to be
\begin{equation}
 c_{\partial^2}
 =\frac{2g/m-\sum_{a+b=k}
       (h_{ab}\ell^P_{ab}+c_{ab}\ell^C_{ab}
                                +f_{ab}\ell^D_{ab})}{2(m+1)}.
 \label{df:eq-qz4}
\end{equation}
Here $g/m$ is the coefficient of $\partial P_m$ in the $N^0$ part
of the original full-propagator $O_1$.
Its ribbon count is particularly short.  For any three chosen
vertices on each trace, the six contraction bijections have equal
total weight after the flavor sum.  Three give one face and three
give three faces.  Thus the one-face coefficient is half the total
matching count
\[
 24+18(k-2)+6(k-1)(k-2)+k(k-1)(k-2)=k(k+1)(k+2),
\]
which proves the stated value of $g$.  The uncontracted left $X$
has Taylor order one in $O_1$, yielding $\partial P_m/m$.

\paragraph{Eliminating $O_1$ before summing.}
Because $O_1=L_1O_0/2$, a field $F$ in $O_0$ contributes
$F_{\rm half}+(L_1F)_{\rm half}/2$ to the required mode
combination.  In particular $\partial^2P_m$ contributes
$-(m^2-1)W_m/4$, since
$L_1\partial^2P_m=2(m+1)\partial P_m$.
After substituting \eqref{df:eq-qz4}, the remaining scalar coefficient of $N^0$ is
\begin{equation}
 \sum_{a+b=k}R_{ab}-\frac{g(m-1)}{4m},
 \label{df:eq-qz5}
\end{equation}
where
\begin{align}
 R_{ab}={}&h_{ab}\left(p_{ab}-\frac{m-1}{8}\ell^P_{ab}\right)
 +c_{ab}\left(c'_{ab}-\frac{m-1}{8}\ell^C_{ab}\right)\notag\\
 &+f_{ab}\left(d_{ab}-\frac{kbs}{2}
       -\frac{k(a-1)(m-1)}4+\frac{m-1}{8}\ell^D_{ab}\right).
 \label{df:eq-qz6}
\end{align}

\paragraph{An explicit parity-paired sum.}
Each unordered pair has exactly one even member.  Put $a=2j$,
$b=k-2j$, $1\leq j\leq(k-3)/2$.  Direct common-denominator
expansion of \eqref{df:eq-qz1}--\eqref{df:eq-qz3}, \eqref{df:eq-qz6} gives
\begin{equation}
 R_{ab}+R_{ba}=\frac{\gamma}{8m^2}
 \left[-12a^3+36ka^2+(2-4k-30k^2)a
                         +8k^3+4k^2-8k-4\right].
 \label{df:eq-qz7}
\end{equation}
For a short way to check this expansion, before multiplication by
$h_{ab},c_{ab},f_{ab}$ the three reduced brackets in \eqref{df:eq-qz6} have, respectively,
numerators
\[
\begin{array}{c|ccc}
 
   & :P_aH_b: & :B_aC_b: & D_{ab}\\ \hline
 a\text{ even}&-(a-2)(b+1)&a(b-1)&-(a-2)(b-1)\\
 a\text{ odd}&-b(a-1)&b(a-1)&-b(a-1)
\end{array}
\]
with prefactors $a(b-1)/(8mb)$, $1/(8m)$, and $k/(8m)$
in the first, second, and third columns.  Substituting \eqref{df:eq-qz1}
then gives \eqref{df:eq-qz7}.

Finally put $q=(k-3)/2$.  The bracket in \eqref{df:eq-qz7}, after $a=2j$,
is
\[
 -96j^3+144kj^2+(4-8k-60k^2)j+8k^3+4k^2-8k-4.
\]
Using the elementary sums of $j,j^2,j^3$ from $1$ to $q$
gives exactly $(k^2-1)(k-3)(k-2)$.
Consequently
\[
 \sum_{a+b=k}R_{ab}
 =\frac{\gamma(k^2-1)(k-3)(k-2)}{8m^2}
 =\frac{k(k+1)(k+2)(k-3)}{8m}
 =\frac{g(m-1)}{4m}.
\]
This cancels \eqref{df:eq-qz5} identically for every odd $k\geq5$.

\subsection{The complete quartic field equation at every odd degree}
\label{df:sec-odd-quartic}

Let $V$ be a finite, zero-algebra-stable full odd annihilator
in a lowest conformal space of an untwisted module. No
restriction on its flavor support is imposed in this section.
Every operator below is an actual map on this space; integer
zeros act on the input before the displayed first-half map.
Write $P_a=\operatorname{tr}X^a$, $Q_a=W_{a,1}$,
$T_a=P_a^+$ and use the raw tops $H_a$, with $H_2=-T$.
For every odd $c\ge5$, define the full commutator
\begin{equation}
 \mathcal K_c=\sum_{i=0}^3(-1)^i\binom3i
  [(W_{4,i})_{(1)},(W_{c,3-i})_{((c-3)/2)}]\big|_V.
 \label{df:eq-gqc1}
\end{equation}
It is formed before restricting to $V$, so the quartic
action on the input is retained. Set
\begin{equation}
\begin{gathered}
 m=c-2,\qquad g=\frac{c(c+2)}{c-1},\\
 \tau=\frac{3c(c+2)}{2(c-1)(c-2)},\qquad
 \beta=\frac{c(c+1)(c+2)}{(c-1)(c-2)},\\
 h_{ab}=\frac{2g+2\tau a(b-2)}{b-1},\qquad a+b=c,\\
 f_{ab}=-\frac{3(c+1)(c+2)}{c-1}
          +2\tau(c-1-2a)+\frac{b-1}{b}h_{ab}.
\end{gathered}
 \label{df:eq-gqc2}
\end{equation}
The entire first-half field side is the following closed
finite sum; its coefficients contain no unevaluated Wick
coefficient or unknown remainder:
\begin{equation}
\boxed{\begin{aligned}
 \mathcal K_c={}&\frac{2c(c+2)N}{(c-1)^2}H_c^+
 -\frac{g(c+1)(c-3)N^2}{4(c-2)^2}T_{c-2}\\
 &+\sum_{\substack{a+b=c\\a\ge2\ {\rm even},\ b\ge3\ {\rm odd}}}
       \left\{h_{ab}H_b^+P_a^0
           +\frac{f_{ab}}2(Q_b^+P_a^0-T_bQ_a^0)\right\}\\
 &+\sum_{\substack{a+b=c\\a\ge3\ {\rm odd},\ b\ge2\ {\rm even}}}
                  h_{ab}T_aH_b^0\\
 &+\sum_{\substack{s+t=c-2\\s\ge2\ {\rm even},\ t\ge3\ {\rm odd}}}
       \frac{\beta N(2s+2-3st-2t)}{4t(s+1)}T_tP_s^0\\
 &+\sum_{\substack{s+t+u=c-2\\2\le s\le t\ {\rm even}\\u\ge3\ {\rm odd}}}
       \frac\beta{4\sigma_{s,t}}
       \left\{\frac2u-2-\frac{s+t}{s+t+1}\right\}
                                 T_uP_s^0P_t^0\\
 &-\frac{3(c+2)}N\sum_{i=0}^2(-1)^i\binom2i
                           W_{3,i}^+W_{c-1,2-i}^0,
 \qquad \sigma_{s,t}=1+\mathbf1_{s=t}.
\end{aligned}}
 \label{df:eq-gqc3}
\end{equation}
Empty sums vanish. All raw trace degrees have their genuine
finite-$N$ meaning; the identity does not replace $N$ by a
larger matrix size or require independent trace monomials at that size.

\paragraph{The field decomposition and ordered endpoints.}
Here is a derivation assembling all contraction types.
In the projected full-matrix computation, take the quartic
third-channel residue and subtract its BRST primitives.
The zero-jet primitive has coefficients
$\ell_0=g$, $\ell_{c-1}=-g$ and
$\ell_i=\tau(c-1-2i)$ for $1\le i\le c-2$ on
$\operatorname{tr}(bX^iYX^{c-1-i})$.
The single-trace first-jet primitive has coefficient
$N\beta/(c-1)$ times
$\sum_{i=0}^{m-1}(i-(m-1)/2)
  \operatorname{tr}(bX^i\partial X X^{m-1-i})$.
For each ordered $a+b=c$, the spectator primitive is
$\beta P_a/(b-1)$ times the same affine sum of total
degree $b-2$. All are killed by $L_1$, and their BRST
differentials act trivially in cohomology.

Single contractions of the quartic product have successive
cyclic coefficients $\ell_i-\ell_{i+1}$ and are removed
by the first primitive. The split ghost coefficient is
$-\sum_{i<a}\ell_i-\sum_{i<b}\ell_i
 =-2g-2\tau(a-1)(b-1)$.
The coefficients of the remaining top and derivative
columns are $h_{ab},f_{ab}$ in \eqref{df:eq-gqc2}: the derivative
coefficient is the sum of the prescribed two-contraction
coefficient $-3(c+1)(c+2)/(c-1)$, the primitive correction
$2\ell_a$, and the top correction $(b-1)h_{ab}/b$.
The single-trace top coefficient is $2c(c+2)N/(c-1)^2$.
After these subtractions the remainder is pure $X$, has
degree $m$, two jets, and at most three trace factors.
The indexed two-jet kernel and cut calculation preceding
\eqref{df:eq-gep1} prove these assertions at arbitrary degree.

Apply $F\mapsto F^++(L_1F)^+/2$, because the quartic
mode commutator equals the residue plus half its $L_1$
image. Splitting each normal product into its reversed
OPE contractions and its ordered endpoint gives
$H_b^+P_a^0$ for even $a$, and $T_aH_b^0$ for odd $a$.
The fermion-column endpoint has an odd integer zero on
the right and kills $V$. For
$D_{ab}=:P_a\partial Q_b:-:Q_a\partial P_b:$, the Ward
identity retains $b(:P_aQ_b:-:Q_aP_b:)/2$.
Translation then gives endpoint
$(Q_b^+P_a^0-T_bQ_a^0)/2$ for even $a$ and zero for odd
$a$. This is the second and third line of \eqref{df:eq-gqc3}, without
requiring any of those even zeros to vanish.

\paragraph{Every trace profile in the remainder.}
The terms proportional to $N^2$ have one trace. The triple-contraction
matching with two empty faces has $4c$ incidences, hence
coefficient $\beta N^2K_m$. The empty cut of the single-trace
first-jet primitive subtracts
$N^2\beta\{(m-1)K_m-L_m\}/(c-1)$.
Their difference is
$N^2\beta(2K_m+L_m)/(c-1)
 =N^2\beta\partial^2P_m/(m(c-1))$.
Its Ward multiplier is $-(m^2-1)/4$, giving exactly the
single-trace coefficient in the first line of \eqref{df:eq-gqc3}.

The one-trace part proportional to $N^0$ cancels the single-trace
normal contractions. For completeness its elementary
parity sum is as follows. With $q=(c-3)/2$, the combined
contribution per even degree $a=2i$, including its reversed
order, is $g/(8m^2)$ times
\begin{equation}
 -96i^3+144ci^2+(4-8c-60c^2)i+8c^3+4c^2-8c-4.
 \label{df:eq-gqc4}
\end{equation}
Substituting the finite sums of $i,i^2,i^3$ for $1\le i\le q$
gives $(c^2-1)(c-3)(c-2)$. Thus their sum is
$c(c+1)(c+2)(c-3)/(8m)$, exactly the negative of the
Ward-adjusted one-trace remainder. The normal-mode and
Ward coefficients giving \eqref{df:eq-gqc4} are the explicitly derived
ones in \eqref{df:eq-qz1}--\eqref{df:eq-qz6}; they depend only on the pure-$X$
mode $T_m$, not on its input. The cancellation therefore
holds on all of $V$.

The complete two-trace contribution proportional to $N$ is \eqref{df:eq-gep9},
which gives the fourth line of \eqref{df:eq-gqc3}. The three-trace contribution proportional to $N^0$
is \eqref{df:eq-gtr7}, giving the fifth line
with unordered even degrees and their proved multiplicity.
Three odd factors have zero first-half action. These
exhaust the face profiles of a three-contraction quartic
product and of its listed primitives.

\paragraph{Traceless propagators and the original endpoint.}
Canonical projected-minus-traceless corrections and the
$1/N$ normal responses cancel separately in both
orders of each $PH,BC,D$ column, as explicitly calculated
with two independent degrees in \eqref{df:eq-eoi5}--\eqref{df:eq-eoi11}.
This is an equality of ordered first-half maps and applies
to all of $V$; projection is not used as a homomorphism.

In the original quartic product, one scalar contraction
has coefficient $3(c+2)/N$ and leaves the degree-$3$ and
degree-$(c-1)$ second flavor channel. Normal splitting
and skew symmetry cancel its singular terms, leaving
the last line of \eqref{df:eq-gqc3}, with a minus sign because the
traceless propagator subtracts the scalar propagator.
Two scalar contractions leave a current product. Its
two terms cancel by
$(:JV:)^+=V^+J^0$ and
$(:(\partial J)V:)^+=-V^+J^0$ on the whole lowest space.
Three scalar contractions leave a one-letter source trace
and vanish. This proves the formula including all original
projectors and canonical normal-ordering corrections.

\subsection{The raw-degree recursion and complete-input transport}
\label{df:sec-raw-transport}

Let $W$ be an actual finite zero-algebra-stable full odd annihilator
in a lowest space of weight $r$, and let $F$ be its ambient first
half level. Separate the integer and half-integer flavor types if
both occur. Each part is stable: the currents and quartic have
integer spin and the odd zero modes are zero on $W$; these generate
the zero algebra. Work on one nonzero part, and let $j$ be its
largest spin. Write $W_m$ for the flavor weight-$m$ space, so
$j-m$ is a nonnegative integer whenever $W_m\ne0$.

Put $k=\max\{a\le N:a\text{ odd}\}=2M+1$. For odd $a\ge3$
write $T_a=P_a^+$. Assign to $T_ax$, $x\in W_m$, the depth
\begin{equation}
                  D=\frac{k-a}{2}+j-m.
 \label{df:eq-umd1}
\end{equation}
Thus a lower highest vector of spin $j-\ell$ enters at depth
$\ell$. In particular a gap-two lower highest input enters at
depth two, whether or not any first-neighbor map has vanished.

\paragraph{Propagation to raw degrees at fixed matrix size.}
The exact ordered determinant identity is
\begin{equation}
 \frac{T_a}{a}
 =\sum_{\substack{s\ge1\\ a-2s\ge3}}
       \frac{T_{a-2s}}{a-2s}\mathcal Y_s
 \qquad(a>N),
 \qquad [H,\mathcal Y_s]=2s\mathcal Y_s.
 \label{df:eq-umd2}
\end{equation}
Here $H$ is the standard flavor Cartan operator. With the
same convention as in \eqref{low:raw-ch}, define
\[
 \mathcal Y_s
 =-[u^{2s}]
   \exp\left(
     -\sum_{\substack{d\ge2\\ d\ {\rm even}}}
       \frac{R_du^d}{d}
   \right),
 \qquad R_d=o(P_d)|_W.
\]
Thus $\mathcal Y_s$ is the negative of the degree-$2s$
coefficient of the even zero-mode determinant series in
\eqref{df:eq-fhc2}. Its constant coefficient is one, so
extracting the coefficient of odd degree $a>N$ in
\eqref{df:eq-fhc2} gives precisely the positive sign on
the right of \eqref{df:eq-umd2}, with $\mathcal Y_s$
acting on the input first.

Every monomial defining $\mathcal Y_s$ has flavor weight $s$;
therefore $\mathcal Y_s$ maps $W_m$ into $W_{m+s}$,
even on mixed support.
Each term on the right of \eqref{df:eq-umd2} has the same
depth \eqref{df:eq-umd1} and has strictly smaller raw degree.
Iteration consequently expresses every map of raw degree
greater than $N$ in terms of maps of degrees at most $N$
at the same depth. This proves that vanishing at a given
depth at the real degrees propagates to all raw degrees.
At real degrees every depth is nonnegative, so negative
raw depths vanish as well. The degree-one trace is absent
in \eqref{df:eq-umd2}, and is never inserted as a boundary value.

\paragraph{A map on the entire finite input.}
Fix $D$ with $b=k-2D\ge3$ and suppose all smaller depths vanish.
For $x\in W_m$ set $a_m=b+2(j-m)$ and define linear maps,
extended over the weight decomposition,
\begin{equation}
 \Phi_b(x)=\frac{T_{a_m}x}{a_m},\qquad
 \Psi_b(x)=F_{a_m+1,0}^+x.
 \label{df:eq-umd3}
\end{equation}
The second definition applies separately to either fermion family.
It uses no rotation operator on the chosen module. All values of
$\Phi_b$ and $\Psi_b$ have output flavor weight $j+b/2$.
Put $R_s=P_s^0|_W$ for positive even trace degrees $s$.
These are the actual raw highest traces and have flavor weight $s/2$.
They vanish identically for $s>4j$ by finite flavor support.
Define two finite input operators
\begin{equation}
 B=2f+p_2+\sum_{\substack{2\le s\le4j\\s\ {\rm even}}}R_s,\qquad
 A=N\mathbf1+2f+\sum_{\substack{2\le s\le4j\\s\ {\rm even}}}(s+1)R_s.
 \label{df:eq-umd4}
\end{equation}
All sums involving $R_s$ in this subsection use positive even trace degrees; empty
sums are zero. No stable-flavor replacement for $p_2$
or $R_s$ is made. The exact identities at the new depth are
\begin{equation}
 \boxed{\quad p_{2,F}\Phi_b=\frac{2N}{3}\Phi_b+\Phi_b B,
                  \qquad \Psi_b A=0.\quad}
 \label{df:eq-umd5}
\end{equation}
Also $e_F\Phi_b=p_{0,F}\Phi_b=p_{1,F}\Phi_b=0$.

Here is a direct derivation including the endpoint checks.
For $x\in W_m$, preceding-depth vanishing gives
$T_{a_m+2}x=0$. Flavor lowering therefore gives
\begin{equation}
 W_{a_m+2,1}^+x=-\frac{T_{a_m+2}fx}{a_m+2}.
 \label{df:eq-umd6}
\end{equation}
Insert this in the full identity \eqref{df:eq-qbl1}. The scalar projector
contains the right factor $P_{a_m-1}^0x$, of flavor weight
$j+(b-1)/2>j$, and is zero. For a regular term with $R_sx\ne0$
one has $m+s/2\le j$, hence $a_m-s\ge b\ge3$.
Dividing the identity by $a_m$ now gives
\begin{equation}
 p_{2,F}\Phi_b(x)=\Phi_b(p_2x)+2\Phi_b(fx)
                  +\frac{2N}{3}\Phi_b(x)
                  +\sum_s\Phi_b(R_sx),
 \label{df:eq-umd7}
\end{equation}
which is its claimed operator form. The same preceding-depth
argument with $[e,T_a]=[p_0,T_a]=0$ gives the first two
highest conditions. For $p_1$ use \eqref{df:eq-qbl2}: its raw $T_{a+2}$
term is a preceding depth and its projector has the same
impossible right weight. Its right input $p_1x$ also belongs
to a preceding depth.

For the source identity use \eqref{df:eq-fhc8}--\eqref{df:eq-sfd5} at degree
$c=a_m+2$ on $x$. Their left side, because $T_cx=0$, is
$2A_cfx/c$, where $A_c=S_1T_c$ is the map occurring there,
not the finite input operator $A$ in \eqref{df:eq-umd4}.
Its projector $P_{c-1}^0fx$ has flavor weight
$j+(b-1)/2>j$. The two projectors on the direct right side
contain $P_{c-1}^0x$ and $W_{c-1,1}^0x$, whose weights are
$j+(b+1)/2$ and $j+(b-1)/2$, respectively. All vanish on
the actual input. The remaining equality is exactly
\begin{equation}
 2F_{a_m+3,0}^+fx+NF_{a_m+1,0}^+x
       +\sum_{s\ge2\ {\rm even}}(s+1)F_{a_m+1-s,0}^+R_sx=0.
 \label{df:eq-umd8}
\end{equation}
This is $\Psi_b A=0$. Its smallest nonzero raw fermion
degree is at least $b+1\ge4$. Thus the argument includes
the last inner layer $b=3$ without a fictitious trace or
a missing projector. The proof applies to both ghost families
as an identity of fields evaluated on the same input.

\subsection{The full quartic source equation}
\label{df:sec-source}

Let $W$ be an actual finite zero-algebra-stable full odd annihilator
inside a lowest space of weight $r$, and let $j$ be its largest flavor
spin. Work in one flavor-parity component.  No restriction is placed
on the number of flavor types or their multiplicities. Fix a depth
of \eqref{df:eq-umd1}, put $b=k-2D\ge3$, and assume all earlier raw depths vanish.
Here $b$ is odd, $k$ is the largest odd integer at most the actual
matrix size $N\ge5$, and $D\ge1$.  All operators below act on the complete
input $W$.

As in \eqref{df:eq-umd4}, subscripts on $R_s,Q_s$ are
\emph{trace degrees}, so they range over even $s\ge2$.  Set
\begin{equation}
\begin{gathered}
 R_s=o(P_s),\qquad Q_s=\frac{[f,R_s]}s,\qquad
 H=[e,f],\qquad \Delta=2r+N^2/4,\\
 R_2=2e,\qquad R_4=p_0,\qquad Q_2=-H,\qquad Q_4=p_1.
\end{gathered}
 \label{df:eq-ugs1}
\end{equation}
The sums are finite: $R_s=Q_s=0$ for $s>4j$.
For a weight-$m$ input $x$, put
\begin{equation}
 a=b+2(j-m),\quad c=a+2;
 \qquad\text{as an input operator, }a=b+2j-H.
\label{df:eq-ugs2}
\end{equation}
Every function of $a$ written below acts on the right, before its
neighboring noncommuting input operator.  Thus $R_s u(a)$ denotes
$R_s\circ u(a)$ and does not mean $u(a)R_s$.
The spectrum of $a$ is odd and positive.  In particular $a-s$ is
invertible for every even $s$; no divisibility assumption is needed.
The whole source map $Y=\Psi_b$ for either ghost family satisfies
\begin{equation}
 YA=0,\qquad A=NI+2f+\sum_{s\ge2\ {\rm even}}(s+1)R_s
\label{df:eq-ugs3}
\end{equation}
by the actual transport \eqref{df:eq-umd5}.

\paragraph{The new whole-input source row.}
The complete quartic field identities imply the additional equation
\begin{equation}
 \boxed{\quad YZ=0,\qquad
 Z=\Delta a^2+
       \left\{p_3+\sum_{s\ge2\ {\rm even}}(1-3s/2)Q_s\right\}(a+3).\quad}
\label{df:eq-ugs4}
\end{equation}
Thus the operator between braces is
$p_3+2H-5p_1-8Q_6-11Q_8-\cdots$.
This equation holds separately for both ghost families, with every higher trace retained.

I prove more: the complete source-free comparison of the actual
quartic commutator and field side has a finite input operator
$\mathcal D$ satisfying the exact factorization
\begin{equation}
 \boxed{\quad
 \mathcal D=Z+\frac12A\mathcal R,\qquad
 \mathcal R=-3f-\frac N2I+
         \sum_{s\ge2\ {\rm even}}R_s
                            \left(-\frac12+\frac{s}{a-s}\right).
 \quad}\label{df:eq-ugs5}
\end{equation}
Before source removal the field comparison gives $Y\mathcal D=0$;
after both source maps vanish it gives $\Phi_b\mathcal D=0$.
The term $A\mathcal R/2$ remains in the bottom equation.

\paragraph{Complete coefficients before quotienting by the source equation.}
For $2\le s\le t$ even put $\sigma_{s,t}=1+\mathbf1_{s=t}$.
The exact operator specified in \eqref{df:eq-ugs5} is
\begin{equation}
\begin{split}
 \mathcal D={}&-3f^2-2Nf+2ra^2+\frac{N^2}{4}(a^2-1)
                                      +p_3(a+3)\\
 &+\sum_{s\ge2\ {\rm even}} fR_s
                   \left(-\frac{3s+4}{2}+\frac{s}{a-s}\right)\\
 &+\sum_{s\ge2\ {\rm even}} Q_s
                   \left(a+3-\frac{3s(a+2-s)}2\right)\\
 &+N\sum_{s\ge2\ {\rm even}}R_s
                   \left(-\frac{s+2}{4}+\frac{s}{2(a-s)}\right)\\
 &+\sum_{2\le s\le t\ {\rm even}}R_sR_t
 \left[-\frac{s+t+2}{4\sigma_{s,t}}+
       \frac1{2\sigma_{s,t}}
          \left\{\frac{(s+1)t}{a-t}+\frac{(t+1)s}{a-s}\right\}\right].
\end{split}\label{df:eq-ugs6}
\end{equation}
In particular all pure two-trace returns remain in this operator;
their cancellation in the source equation is a consequence of
\eqref{df:eq-ugs3}, not deletion of the original quartic field columns.

Here is a coefficient derivation retaining independent even degrees.
Use $g,\nu,\beta,h_{u,v},f_{u,v}$ of \eqref{df:eq-gqc2}, with its parameter
$c=a+2$, and set
\[
 \kappa=\frac{(c-1)(c-2)}{c(c+2)}.
\]
(The symbol $\nu$ here is the coefficient called $\tau$ in \eqref{df:eq-gqc2}.)
Direct rational simplification gives
\begin{equation}
\begin{split}
 \widehat h(s,c-s):=\kappa h_{s,c-s}
   &=\frac{(3s+2)(c-s)-4(s+1)}{c-s-1},\\
 \widehat f(s,c-s):=\kappa f_{s,c-s}
   &=2-3s+\frac6c-\frac{4(s+1)}{c-s},\qquad
 \kappa\beta=c+1 .
\end{split}\label{df:eq-ugs7}
\end{equation}
All functions in this calculation are evaluated on $x$ first.

The arbitrary-input commutator \eqref{df:eq-glq3}, with the full \eqref{df:eq-qbl1}--\eqref{df:eq-qbl2},
gives, after normalization and before the original projector,
\begin{equation}
 \mathcal C_c=-3f^2-\frac{2N(c+1)}c f
        -\frac{3(c+1)}c\sum_{s\ge2\ {\rm even}}R_sf+(c+1)p_3 .
\label{df:eq-ugs8}
\end{equation}
The preceding-depth assumptions remove $A_cx,A_cfx,B_cx,T_cx$
in that identity.  In particular the regular terms of $B_cx$
have the preceding depth even when the input has several flavor
types.  All terms of $B_cfx$ in \eqref{df:eq-ugs8}, including its even-trace
returns, are retained.

For the field side use the complete middle identity \eqref{df:eq-unq9}, whose
derivation on an arbitrary input uses only the preceding raw bottom
and the full odd condition.  
Use also the exact integer top identities \eqref{of:fullodd-top}:
\begin{equation}
 H_2=-r,\qquad
 H_d=\frac N4R_{d-2}
       +\frac18\sum_{\substack{u+v=d-2\\ u,v\ge2\ {\rm even}}}R_uR_v
                                  \quad(d\ge4\text{ even}).
\label{df:eq-ugs9}
\end{equation}
Substitute these in every quartic field column.  In the leading middle column
the regular factor is precisely
\begin{equation}
 \kappa\frac{2c(c+2)N}{(c-1)^2}\frac{c-1}{4(c-2)}=\frac N2.
\label{df:eq-ugs10}
\end{equation}
This accounts for the leading term $N\sum_s(c-3-s)R_s/2$ on the
field side, before subtraction.

For a fully explicit ledger, write $d=c-s$ and
$\widehat h(u,v)=\kappa h_{u,v}$ also for other pairs $u+v=c$.
After replacing $R_sf$ by $fR_s-sQ_s$, the coefficient of
$fR_s$ in $\mathcal C_c-\mathcal F_c$ is
\[
 -\frac{3(c+1)}c+
       \frac{\widehat h(s,d)(d-1)}{d(d-2)}
                   +\frac{\widehat f(s,d)}2,
\]
and the coefficient of $Q_s$ is
$3s(c+1)/c+d\widehat f(s,d)/2$.
The coefficient of $NR_s$ is the sum of four complete column
contributions:
\begin{equation}
\begin{aligned}
 &-\frac{c-3-s}{2}
 -\frac{\widehat h(s,d)(d-1)(d-3)}{4(d-2)}\\
 &\quad-\frac{\widehat h(c-s-2,s+2)(c-s-2)}4
 -\frac{(c+1)\{2s+2-(3s+2)(c-2-s)\}}{4(s+1)}.
\end{aligned}
 \label{df:eq-ugs11}
\end{equation}
For an ordered symmetric $R_sR_t$ sum the coefficient is
\begin{equation}
\begin{aligned}
 &-\frac{c-3-s-t}{8}
 \left\{\frac{\widehat h(s,c-s)(c-s-1)}{c-s-2}
             +\frac{\widehat h(t,c-t)(c-t-1)}{c-t-2}\right\}\\
 &-\frac{\widehat h(c-s-t-2,s+t+2)(c-s-t-2)}8\\
 &-\frac{c+1}{8}
       \left\{2-\left(2+\frac{s+t}{s+t+1}\right)(c-2-s-t)\right\}.
\end{aligned}\label{df:eq-ugs12}
\end{equation}
To obtain the unordered coefficient, multiply this by
$2/\sigma_{s,t}$.  The three lines come respectively from the
middle substitution, the even top convolution, and the explicit
three-trace quartic field remainder.  Thus the diagonal $s=t$ has its actual
multiplicity, with no extrapolation from distinct degrees.
Equations \eqref{df:eq-ugs7}, \eqref{df:eq-ugs11}, \eqref{df:eq-ugs12} simplify to the coefficients in
\eqref{df:eq-ugs6}.  The remaining scalar coefficients are
$-3$, $-2N$, $2a^2r$, $N^2(a^2-1)/4$ and $a+3$, respectively.
These simplifications are identities of rational functions in the three independent variables $a,s,t$.

The rational simplifications are performed before specializing
small raw degrees.  If a right $R_sx$ is nonzero, its weight gives
$s\le a-b$, hence $a-s\ge b\ge3$.  A nonzero $Q_sx$ or $R_sfx$
gives $s\le a-b+2$, hence $a-s\ge b-2\ge1$.
Likewise a nonzero $R_sR_tx$ implies $a-s-t\ge b$.
Consequently every unavailable term in the original finite quartic field
sums acts as zero.  In particular the middle formula is only used
on nonzero right inputs when its odd degree is at least five.
Apparent factors $d-1$ are cancelled in the rational ledger before
extending the already simplified expression by zero actions.
This justifies \eqref{df:eq-ugs6} at every actual odd $a$, including small
endpoints and raw degrees greater than $N$.

\paragraph{The original last-layer projector cancels on the full input.}
For $b\ge5$ its right weights exceed $j$.  At $b=3$, the original
quartic field projector can retain only its $i=0$ term, namely
\[
 -\frac{3(c+2)}N T_3 W_{c-1,2}^0x.
\]
Indeed its $i$th right factor has weight $j+(b-3)/2+i$.
In \eqref{df:eq-glq3}, the sole possible surviving projector is that of
$A_cf^2x=[p_1,T_c]f^2x$.  It is
\begin{equation}
 -\frac{3(c+2)}{N(c-1)(c-2)}T_3P_{c-1}^0f^2x.
\label{df:eq-ugs13}
\end{equation}
The projectors of $B_cfx$ have right weight
$j+(b-1)/2>j$, and the other endpoints vanish as well.
Moreover $P_{c-1}^0x=P_{c-1}^0fx=0$: their respective right weights
are $j+(b+1)/2$ and $j+(b-1)/2$.
Full flavor covariance therefore gives
$W_{c-1,2}^0x=P_{c-1}^0f^2x/((c-1)(c-2))$.
The two displayed original projectors agree exactly and cancel
before applying a source operation.  This works on every input
weight, not merely the maximal highest vector, and needs no
assumption that a higher trace is zero.

\paragraph{From the actual fields to the source row.}
The comparison of \eqref{df:eq-glq3} and \eqref{df:eq-gqc3}, after the
projectors in \eqref{df:eq-ugs13} cancel, is
$\Phi_b\mathcal D(a)+\mathcal S_{a+2}=0$.  Here on $x\in W_m$,
with $c=a|_{W_m}+2$, the entire source term is
\begin{equation}
\begin{split}
 \mathcal S_cx={}&2NS_1C_{c-1}^+x\\
 &+\sum_{s\ge2\ {\rm even}}
 \frac{\widehat h(s,c-s)(c-s-1)}{c-s-2}
                         S_1C_{c-s-1}^+R_sx.
\end{split}
\label{df:source-sum}
\end{equation}
Only nonzero right inputs are included, so their actual degrees satisfy
$c-s\ge5$.
Apply the first-half operation $S_1$ to this field identity.
On every raw coordinate \eqref{df:eq-fhc8} gives $S_1\Phi_b=-Y$, since its
right projector has weight $j+(b-1)/2>j$.
For a source supplied by a middle substitution on $z=R_sx$,
the two right lowering-supercurrent terms in \eqref{df:eq-nsr4} have minimum
weight
\begin{equation}
          m+s/2+(c-s-3)/2=j+(b-1)/2>j .
\label{df:eq-ugs14}
\end{equation}
The leading middle source has the same formula with $s=0$.
Thus \eqref{df:eq-nsr2}, \eqref{df:eq-nsr6} make its complete $S_1^2$ source term zero.
This is a weight argument for these specified source terms, not a
square-zero assumption on the whole first-half output.  Applying the relative-chain derivation $\delta$ to the complete
$B$-family equation proves the $C$-family equation: $\mathcal D$
is fixed, $\delta Y_B=Y_C$, and every discarded right factor has
a discarded image for the input reasons in \eqref{df:eq-fdk8}--\eqref{df:eq-fdk9}.  Hence $Y\mathcal D=0$
for each family.  After the sources themselves vanish, the same
field comparison gives $\Phi_b\mathcal D=0$.

Finally $[R_s,f]=-sQ_s$ and $[R_s,R_t]=0$ expand
$A\mathcal R/2$ directly.  Its $fR_s$ coefficient is
$-1/2+s/(a-s)-3(s+1)/2$; its $Q_s$ coefficient is
$3s(s+1)/2$.  Its $NR_s$ and unordered $R_sR_t$ coefficients
are exactly those in \eqref{df:eq-ugs6}, and its scalar term is $-N^2/4$.
Subtracting leaves precisely $Z$ in \eqref{df:eq-ugs4}, proving \eqref{df:eq-ugs5}.
Because $YA=0$ on the whole input, right composition by
$\mathcal R$ is legitimate and yields $YZ=0$.
No coefficient function of $a$ is commuted through a trace.

\paragraph{A universal commutator form.}
Write $B=2f+p_2+\sum_sR_s$, as in \eqref{df:eq-umd4}.
The actual identities \eqref{of:even-first-second} give
$[R_s,p_2]=2sQ_{s+2}$ and $[f,R_s]=sQ_s$ in the degree convention
of this section.  Therefore
\begin{equation}
\begin{split}
 [A,B]&=4p_3-2\sum_{s\ge2\ {\rm even}}s^2Q_s
               +2\sum_{s\ge2\ {\rm even}}s(s+1)Q_{s+2}\\
 &=4p_3+\sum_{s\ge2\ {\rm even}}(4-6s)Q_s,\\
 Z&=\Delta a^2+\tfrac14[A,B](a+3).
\end{split}\label{df:eq-ugs15}
\end{equation}
The last equality includes $s=2$: its coefficient is $-8$, exactly
$4-6s$, without an artificial degree-zero trace.
All sums are finite; reindexing uses the zero actual higher traces.
This is an operator commutator on the original input.

\subsection{Transport on the actual first-half image}
\label{df:sec-half-transport}

Use the actual finite full odd input $W$, the current inner degree
$b\ge3$, and preceding raw-depth vanishing of \eqref{df:eq-ugs1}--\eqref{df:eq-ugs3}.
Its maximal flavor spin is $j$ and $\Delta=2r+N^2/4$.
I prove that both complete current source maps vanish if
$\Delta\ne0$.  In particular this holds in the actual problem,
where $r\ge0$ and $N\ge5$.
There is no flavor-count, multiplicity, semisimplicity, or
sextic-zero restriction.

Let $F=M_{r+1/2}$ be the actual first-half level and write
$\Phi=\Phi_b:W\to F$.  The source map for the $B$ family is
$Y=Y_B$, and $S_i=o(B_{3,i})=B_{3,i(1)}$ on the whole module.
The elementary cubic first-half formula \eqref{df:eq-fhc8}, whose right
projector is above maximal flavor weight at this depth, gives
\begin{equation}
                   S_1\Phi=-Y.
\label{df:eq-uhs1}
\end{equation}
Here $S_1$ is an actual cubic \emph{zero mode} on $F$; the
source formula is not a statement that its output is lowest.
The barred identity holds for the other ghost family.
The whole bottom transport is
\begin{equation}
 p_{2,F}\Phi=\frac{2N}{3}\Phi+\Phi B,
 \qquad B=2f+p_2+\sum_{s\ge2\ {\rm even}}R_s.
\label{df:eq-uhs2}
\end{equation}

\paragraph{Four specific lowering half modes annihilate the bottom image.}
Put $G_X=\operatorname{tr}(Xb)$ and $G_Y=\operatorname{tr}(Yb)$.
All four fields
$G_X,G_Y,W_{3,1},W_{3,2}$ have conformal weight $3/2$;
their flavor weights are respectively $1/2,-1/2,1/2,-1/2$.
For any one of them, call it $U$ and write its flavor weight as $q$.
For a weight-$m$ input $x$, set $a=b+2(j-m)$.
Then $\Phi x=T_a x/a$, with
$T_a=P_{a((a-3)/2)}$, has flavor weight $j+b/2$.

The mode $U_{(1)}$ lowers energy by $1/2$, so it kills $x$.
Its commutator with $T_a$ has total energy zero.  More precisely,
\begin{equation}
 [U_{(1)},T_a]_{\rm s}
 =\sum_{t=0}^1\binom1t
       (U_{(t)}P_a)_{((a-1)/2-t)}.
\label{df:eq-uhs3}
\end{equation}
Each coefficient field has integer conformal weight
$(a+1)/2-t$, and the displayed mode is its physical zero mode.
It therefore preserves $W$, by full zero-algebra invariance.
Consequently $U_{(1)}\Phi x$ belongs to $W$ and has flavor weight
$j+b/2+q>j$.  This proves, on the entire image of $\Phi$,
\begin{equation}
 G_{X(1)}\Phi=G_{Y(1)}\Phi
       =W_{3,1(1)}\Phi=W_{3,2(1)}\Phi=0.
\label{df:eq-uhs4}
\end{equation}
This proof uses particular actual zero-mode commutators and the
input weight bound.  It does not assert that arbitrary lowering
modes kill the whole first-half level.

\paragraph{The all-level quartic--cubic defect.}
The exact field identity \eqref{of:quartic-cubic-field}, after taking its physical zero
mode, gives the highest spin-one combination
\[
 C_0:=[p_0,S_2]-2[p_1,S_1]+[p_2,S_0]
       =\frac{40}{N}S_0+\frac{10}{N}o(\mathcal N_0),
\]
\begin{equation}
 \mathcal N_0=:G_XW_{3,1}:-:G_YW_{3,0}:,
 \qquad
 \mathcal N_1=:G_XW_{3,2}:-:G_YW_{3,1}: .
\label{df:eq-uhs5}
\end{equation}
These are all-level identities: the derivative in \eqref{of:quartic-cubic-field} has
zero-mode multiplier $-2$, and its exact field acts trivially
in the actual cohomology module.  No lowest-space normal-product
formula has been applied on $F$.

The raw OPE of $p_0$ with $S_0$ is regular, so
$[p_0,S_0]=0$ on every level.  Its third flavor lowering gives
$[p_3,S_0]+3[p_2,S_1]+[p_1,S_2]=0$.
On the other hand
$C_1=[f,C_0]/2=[p_1,S_2]-2[p_2,S_1]+[p_3,S_0]$.
Since $[f,\mathcal N_0]/2=\mathcal N_1$, these two facts give
\begin{equation}
       [p_2,S_1]=-\frac8N S_1-\frac2N o(\mathcal N_1)
                         \quad\text{on every module level}.
\label{df:eq-uhs6}
\end{equation}
Its right side has not been replaced by zero on the whole $F$.

\paragraph{Evaluation of the normal products on the specific image.}
For an odd field $G$ and even field $V$, both of weight $3/2$,
the normal-mode formula on any $y\in F$ reduces to
\begin{equation}
                    o(:GV:)y=V_{(1)}G_{(0)}y+V_{(0)}G_{(1)}y.
\label{df:eq-uhs7}
\end{equation}
Indeed its VOA mode is numbered two.  In the first normal sum,
$G_{(k)}V_{(1-k)}$ with $k<0$ has a right mode lowering by at
least $3/2$.  In the second sum the right $G_{(k)}$ with $k\ge2$
does the same.  Both kill $F$ by the genuine lower bound $r$;
only $k=0,1$ remain.  Their sign is positive because $V$ is even.

Now take $y=\Phi x$.  By \eqref{df:eq-uhs4}, both $V_{(1)}y$ and $G_{(1)}y$
are zero for the pairs in $\mathcal N_1$.  Thus \eqref{df:eq-uhs7} is simply
$[V_{(1)},G_{(0)}]y$.
The relevant raw OPEs have only one contraction and only a first
pole.  Their exact zero-mode consequences are
\begin{equation}
 [W_{3,2(1)},G_{X(0)}]=-2S_1,
 \qquad [W_{3,1(1)},G_{Y(0)}]=2S_1 .
\label{df:eq-uhs8}
\end{equation}
For an explicit word check, $W_{3,2}=\operatorname{tr}(XY^2)$
and $W_{3,1}=\operatorname{tr}(X^2Y)$.
Contracting either $Y$ in the first trace with the sole $X$ of
$G_X$ gives $-\operatorname{tr}(XYb)-\operatorname{tr}(YXb)$.
Contracting either $X$ in the second trace with the sole $Y$ of
$G_Y$ gives the same two traces with a positive sign.
Their normalized sum is $2B_{3,1}$.
The scalar-projector contractions contain $\operatorname{tr}b=0$,
so there is no omitted $1/N$ term and no higher pole.
It follows that
\begin{equation}
             o(\mathcal N_1)\Phi=-4S_1\Phi,
                    \qquad [p_2,S_1]\Phi=0.
\label{df:eq-uhs9}
\end{equation}
This cancellation is established only on the image where \eqref{df:eq-uhs4}
was proved.  It is not inferred from the lowest-space relation
$[p_2,S_1]=0$.

\paragraph{Closed source transport and complete source elimination.}
Apply $S_1$ to \eqref{df:eq-uhs2} and use \eqref{df:eq-uhs1}, \eqref{df:eq-uhs9}.  The result is
\begin{equation}
 \boxed{\quad p_{2,F}Y=\frac{2N}{3}Y+YB.\quad}
\label{df:eq-uhs10}
\end{equation}
Apply the relative-chain derivation $\delta$ to this complete
$B$-family equation. It fixes $p_2,B$ and sends $Y_B$ to $Y_C$;
the lowestness, full odd and flavor-bound relations used in its
derivation are stable under $\delta$, as proved in \eqref{df:eq-fdk8}--\eqref{df:eq-fdk9}.
This proves the $C$-family equation on the same input without an
operator implementing $\delta$ on the module.
Since $YA=0$ on the entire input, \eqref{df:eq-uhs10} implies
$YAB=0$ and
$YBA=(p_{2,F}Y-2NY/3)A=0$.
Hence $Y[A,B]=0$.
The actual universal source row \eqref{df:eq-ugs4}, in its commutator form
\eqref{df:eq-ugs15}, now reduces to
$0=YZ=\Delta Ya^2$.
Both $\Delta$ and $a^2$ are invertible under the stated hypotheses,
so $Y=0$.  This proves the vanishing of both complete source maps
at every inner depth with preceding raw-depth vanishing, for
arbitrary actual finite flavor support and multiplicities.

\subsection{The adjacent source with the current bottom retained}
\label{df:sec-adjacent-source}

Let $W$ be an actual finite zero-algebra-stable full odd input of
lowest conformal weight $r$, maximal flavor $j$, and one flavor parity.
Fix an odd current degree $b\ge5$. Assume every raw depth strictly
before the current depth vanishes, and both current source maps vanish.
The current bottom is retained. Put
\begin{equation}
 a=b+2j-H,\quad \Phi=\Phi_b,\quad
 \Phi'=\Phi_{b-2},\quad Y'=\Psi_{b-2},\quad
 \Delta=2r+N^2/4,\quad T=p_{2,\mathrm{out}}-2N/3.
 \label{df:eq-unq1}
\end{equation}
On input of weight $m$, write $c=a|_{W_m}$. Thus
$\Phi x=T_cx/c$, whereas $\Phi'x=T_{c-2}x/(c-2)$ and
$Y'x=B_{c-1,0}^+x$ for the $B$ ghost family. All functions
of $a$ are ordered on the right. Use trace-degree subscripts
\begin{equation}
 R_s=P_s^0,\quad Q_s=[f,R_s]/s,\quad
 A=N+2f+\sum_s(s+1)R_s,\quad B=2f+p_2+\sum_sR_s,
 \label{df:eq-unq2}
\end{equation}
where $s\ge2$ is even. All sums act finitely. Denote by
$\mathcal D(z),\mathcal R(z)$ the complete operators of \eqref{df:eq-ugs5}--\eqref{df:eq-ugs6},
with their right-ordered degree variable replaced by $z$. In particular
\begin{equation}
 \mathcal D(z)=\Delta z^2+\tfrac14[A,B](z+3)
                    +\tfrac12A\mathcal R(z),\qquad
 \mathcal R(z)=-3f-N/2+
        \sum_sR_s\left(-\tfrac12+\frac{s}{z-s}\right).
 \label{df:eq-unq3}
\end{equation}
No inverse of an odd source or of $A$ is used.

The exact adjacent source equation is
\begin{equation}
 \boxed{\quad Y'\widetilde{\mathcal D}=0,\qquad
 \widetilde{\mathcal D}=\mathcal D(a-2)+\mathcal L J+
               \tfrac12\bigl([B,\mathcal L]-2\mathcal L^2\bigr)(a+1),\quad}
 \label{df:eq-unq4}
\end{equation}
where
\begin{equation}
 \mathcal L=Aa^{-1},\qquad
 J=6f+2N+\sum_sR_s\left(\frac{3s+4}{2}
                                      -\frac{s}{a-2-s}\right)
       =-\mathcal R(a-2)+\tfrac32A.
 \label{df:eq-unq5}
\end{equation}
This statement holds for either ghost family. It neither assumes nor
concludes $Y'=0$. The two correction terms in \eqref{df:eq-unq4} are essential.

\paragraph{Adjacent transport and the exact denominator.}
For $S_i=o(B_{3,i})$, the actual raw cubic relations imply
\begin{equation}
 S_1\Phi=0,\qquad S_1\Phi'=-Y',\qquad
 T\Phi=\Phi B,\qquad T\Phi'=\Phi'B-2f_{\mathrm{out}}\Phi.
 \label{df:eq-unq6}
\end{equation}
The last equation follows directly from \eqref{df:eq-qbl1}. Indeed its first
polar is
$W_{c,1}^+x=f_{\mathrm{out}}\Phi x-\Phi'fx$.
Its regular sum gives $\Phi'\sum_sR_s$, and its original projector
has right weight $j+(b-3)/2>j$.

The complete cubic boundary identity \eqref{df:eq-fhc8}--\eqref{df:eq-sfd5}, now applied to
$T_c$, gives
\begin{equation}
 Y'A=S_2\Phi\,a,\qquad
 TY'=Y'(B-2\mathcal L),\qquad S_2\Phi=Y'\mathcal L.
 \label{df:eq-unq7}
\end{equation}
For the first equality set $\mathscr A_c=S_1T_c$.
The current source vanishing says $\mathscr A_cx=0$, but
$\mathscr A_cfx=-cY'fx$ must be retained. Hence
$[f,\mathscr A_c]x=cY'fx$. Substitution into
$S_2T_c-2[f,\mathscr A_c]/c$ gives the first equation.
All original right projectors have weight at least
$j+(b-3)/2>j$. Thus the denominator in $\mathcal L$ is the
\emph{current} $a$, and is not $a-2$.
Apply $S_1$ to the last equation of \eqref{df:eq-unq6}, use
$[S_1,f]=-S_2$, and use the actual all-level identity
$[p_2,S_1]\Phi'=0$ to obtain the second equation of \eqref{df:eq-unq7}.

Here that all-level identity is legitimate without any supposition
that $\Phi'$ is a lowest vector. Each of the four specific lowering
half modes $G_{X(1)},G_{Y(1)},W_{3,1(1)},W_{3,2(1)}$ kills
$\Phi'$. Its commutator with the raw first-half field returns to an
integer physical zero mode preserving $W$, and its output flavor
is at least $j+(b-3)/2>j$. The exact normal-mode reduction of \eqref{df:eq-uhs5}--\eqref{df:eq-uhs7}
then gives $[p_2,S_1]\Phi'=0$. The same four lowering modes also
kill $f_{\mathrm{out}}\Phi$: commute through $f$ and $T_c$, or
use their resulting integer zero-mode commutators. The final flavor
has the same lower bound. Consequently
\begin{equation}
 [p_2,S_1]f_{\mathrm{out}}\Phi=0.
 \label{df:eq-unq8}
\end{equation}
In particular no commutator with $S_1$ has simply been discarded.

\paragraph{The full middle formula before deleting its first endpoint.}
For every odd $d\ge5$ and right input $z\in W$, \eqref{df:eq-mcm2} and
\eqref{df:eq-fhc12} give
\begin{equation}
\begin{split}
 H_d^+z={}&\frac{d-1}{d^2(d-2)}
             (f_{\mathrm{out}}T_d-T_df)z
 +\frac{N(d-1)(d-3)}{4(d-2)^2}T_{d-2}z\\
 &+\frac{d-1}{4(d-2)}
       \sum_{s+t=d-2}\frac{t-1}{t}T_tR_sz
 -\frac{d-1}{d-2}S_1C_{d-1,0}^+z.
\end{split}\label{df:eq-unq9}
\end{equation}
To check the first term without a highest-input assumption, write
$C=C_{d-1,0}^+$ and
$C_{d-1,1}^+=(fC-Cf)/(d-2)$. Because the cubic zero modes
annihilate the input, the left side of \eqref{df:eq-mcm2} is
\[
 S_0 C_{d-1,1}^+-S_1C
 =\frac{[f,T_d]}{d(d-2)}-\frac d{d-2}S_1C.
\]
Here $S_0C=T_d/d$ and $[S_0,f]=-2S_1$. Solving \eqref{df:eq-mcm2}
proves \eqref{df:eq-unq9}. In particular, the first term retains the
output action $(d-1)f_{\mathrm{out}}T_d/(d^2(d-2))$.

\paragraph{All current-bottom terms in the quartic comparison.}
Use \eqref{df:eq-glq3} in raw channel $c$, with
$A_c=[p_1,T_c]$, $B_c=[p_2,T_c]$, $C_c=[p_3,T_c]$.
Apart from the original projector treated below, the complete endpoints are
\begin{equation}
\begin{aligned}
 A_cx&=0,\qquad A_cfx=-c(c+2)\Phi fx,\\
 A_cf^2x&=-c(c+2)\Phi'f^2x,\\
 B_cx&=c\Phi\left(2f+2N/3+\sum_sR_s\right)x,\\
 B_cfx&=-2c f_{\mathrm{out}}\Phi fx+2c\Phi'f^2x\\
      &\quad +(2Nc/3)\Phi'fx+c\Phi'\sum_sR_sfx,\\
 C_cx&=c p_{3,\mathrm{out}}\Phi x-c\Phi'p_3x.
\end{aligned}
 \label{df:eq-unq10}
\end{equation}
They follow term by term from \eqref{df:eq-qbl1}--\eqref{df:eq-qbl2} and flavor covariance.
In particular neither $A_cfx$ nor $B_cx$ nor $T_cx$ is zero.
Multiply \eqref{df:eq-glq3} by $\kappa=(c-1)(c-2)/(c(c+2))$.
Its factor before the braces becomes $1/c^2$.
Relative to its usual $\Phi'$ operator, the additional current-bottom
contribution is
\begin{equation}
 f_{\mathrm{out}}\Phi\left[
       6f+\frac{2N(c+1)}c+
               \frac{3(c+1)}c\sum_sR_s\right]x
       -(c+1)p_{3,\mathrm{out}}\Phi x.
 \label{df:eq-unq11}
\end{equation}

On the complete field side \eqref{df:eq-gqc3}, the additional leading-middle
term is $(2N/c)f_{\mathrm{out}}\Phi x$. For $d=c-s$,
the combined additional middle and differentiated-polar term is
\begin{equation}
 f_{\mathrm{out}}\Phi R_s
   \left[\frac{\widehat h(s,d)(d-1)}{d(d-2)}
                         +\frac{\widehat f(s,d)}2\right]x,
 \quad \widehat h=\kappa h,\quad \widehat f=\kappa f.
 \label{df:eq-unq12}
\end{equation}
Indeed $Q_d^+R_sx=f_{\mathrm{out}}\Phi R_sx-\Phi'fR_sx$.
Every other complete quartic field column has the same $\Phi'$ expression
as before: the even top, the two-trace remainder proportional to $N$, and the
three-trace remainder contain no omitted output lowering operator.
The independent rational identity
\begin{equation}
 \frac{3(c+1)}c-
 \frac{\widehat h(s,d)(d-1)}{d(d-2)}
                     -\frac{\widehat f(s,d)}2
       =\frac{3s+4}{2}-\frac{s}{c-2-s}
 \label{df:eq-unq13}
\end{equation}
therefore proves the full field comparison
\begin{equation}
 \Phi'\mathcal D(a-2)+f_{\mathrm{out}}\Phi J
        -p_{3,\mathrm{out}}\Phi(a+1)+\mathcal S_a=0.
 \label{df:eq-unq14}
\end{equation}
All coefficients retain the right ordering. Nonzero right traces
force $c-s\ge b\ge5$, so \eqref{df:eq-unq9} is never used below its degree
range. Unavailable terms in the original finite sums have zero
right action by weight, just as in \eqref{df:eq-ugs11}--\eqref{df:eq-ugs12}.

For $b\ge7$ all original projectors vanish by weight. At $b=5$
the sole possible field projector is
$-3(c+2)T_3W_{c-1,2}^0x/N$.
The sole possible commutator projector is
$-3(c+2)T_3R_{c-1}f^2x/[N(c-1)(c-2)]$.
Since $R_{c-1}x=R_{c-1}fx=0$, full flavor covariance makes these
two original terms exactly equal. They cancel before any cubic
operation. This argument retains every higher-trace return.

\paragraph{Applying the cubic operation to the retained defects.}
Here $\mathcal S_a$ is the sum \eqref{df:source-sum} with $c=a|_{W_m}$. All its terms are specified $S_1C_{d-1,0}^+z$.
Their two lowering-supercurrent right actions have weight at least
$j+(b-3)/2>j$. The full square identity \eqref{df:eq-nsr2}, with its actual
normal split \eqref{df:eq-nsr6}, thus kills the result of applying another
$S_1$ to each of them. This is a statement about these particular
sources, not a square-zero assertion on the entire half level.

The first defect in \eqref{df:eq-unq14} gives
$S_1f_{\mathrm{out}}\Phi J=-Y'\mathcal L J$.
For the second defect use $2p_3=[f,p_2]$ and \eqref{df:eq-unq8}:
\begin{equation}
 S_1p_{3,\mathrm{out}}\Phi
   =\tfrac12\{T S_2\Phi-S_2\Phi B\}
   =\tfrac12Y'\bigl([B,\mathcal L]-2\mathcal L^2\bigr).
 \label{df:eq-unq15}
\end{equation}
Substituting these exact expressions into \eqref{df:eq-unq14} gives \eqref{df:eq-unq4}.
Apply the relative-chain derivation $\delta$ to the complete
equation $Y'_B\widetilde{\mathcal D}=0$. It fixes
$\widetilde{\mathcal D}$ and sends $Y'_B$ to $Y'_C$. The full
odd relations, prior pure-$X$ depth vanishings, both current-source
vanishings and the flavor bounds are stable under $\delta$.
The transfer principle \eqref{df:eq-fdk8}--\eqref{df:eq-fdk9} therefore proves the
other-family equation without a derivation or rotation operator
on the chosen module.

\subsection{Right-ordered subtraction of the two adjacent rows}
\label{df:sec-adjacent}

Assume precisely the actual hypotheses of \eqref{df:eq-unq1}: $b\ge5$ is odd,
every earlier raw depth is zero, and both current source maps are
zero. Retain the current bottom $\Phi=\Phi_b$. All input operators
and right orderings are those of \eqref{df:eq-unq1}--\eqref{df:eq-unq5}. Write
\begin{equation}
 S_i=o(B_{3,i}),\quad \bar S_i=o(C_{3,i}),\quad
 K=\tfrac12\bigl(\{S_0,\bar S_2\}-\{S_2,\bar S_0\}\bigr),
 \qquad T=p_{2,\mathrm{out}}-2N/3.
 \label{df:eq-uac1}
\end{equation}
Thus $K$ is the actual middle component $K_{1,\mathrm{out}}$ of
\eqref{of:cubic-mixed}, with its original normalization. The claim is
\begin{equation}
 \boxed{\quad K^2\Phi=4\Delta\Phi,\qquad [K,p_{2,\mathrm{out}}]\Phi=0.\quad}
 \label{df:eq-uac2}
\end{equation}
No lowest-space reduction has been applied to the image of $\Phi$.
The two equations will instead be used to obtain a joint spectral row on its image.

\paragraph{Exact cubic closure of the input $A$ action.}
The same-ghost regular highest OPE and current source vanishing give
\begin{equation}
 S_0\Phi=\bar S_0\Phi=S_1\Phi=\bar S_1\Phi=0.
 \label{df:eq-uac3}
\end{equation}
For example $S_0\Phi=0$ follows by commuting $S_0$ through the raw
pure-$X$ field $T_a$ and using its zero action on the input. The
spin-two highest mixed cubic field identity and its second flavor
lowering give, on every module level,
\begin{equation}
 \{S_0,\bar S_2\}+4\{S_1,\bar S_1\}
                           +\{S_2,\bar S_0\}=0.
 \label{df:eq-uac4}
\end{equation}
Using \eqref{df:eq-uac3} in \eqref{df:eq-uac4}, the normalization \eqref{df:eq-uac1} gives
$S_0\bar S_2\Phi=K\Phi$. This uses neither the scalar mixed cubic
identity nor a lowest-level normal-product prescription.

Choose the adjacent source of the $C$ ghost family,
$Y'_C x=C_{a-1,0}^+x$. The exact first-half anticommutator \eqref{df:eq-fhc12}
and the full odd input imply
$S_0Y'_C=\Phi$. The barred version of \eqref{df:eq-unq7} is
$Y'_CA=\bar S_2\Phi a$. Applying $S_0$ therefore proves
\begin{equation}
                 \Phi A=K\Phi a,\qquad \Phi \mathcal L=K\Phi,
                 \qquad \mathcal L=Aa^{-1}.
 \label{df:eq-uac5}
\end{equation}
The right factor is the current $a$. The complete transport remains
$\Phi B=T\Phi$.

\paragraph{A degree-shift identity with no commuting approximation.}
Put $E=[B,a]=-4f+\sum_s sR_s$. For $z=a$ or $a-2$ one has
\begin{equation}
             2z\mathcal R(z)-E(z+3)=-Az.
 \label{df:eq-uac6}
\end{equation}
Indeed $zf=f(z+2)$ and $zR_s=R_s(z-s)$. The $f$ coefficient on
the left is $-6(z+2)+4(z+3)=-2z$; the scalar coefficient is $-Nz$;
and the $R_s$ coefficient is
$-(z-s)+2s-s(z+3)=-(s+1)z$. These are exactly the three
coefficients on the right. Denominators $z-s$ have been cancelled
using the actual ordered shift, not by commuting $R_s$ with $z$.
All such operators are defined because $z$ has positive odd
spectrum and $s$ is even.

Set $\mathcal C=[K,T]$ on the output. Equations \eqref{df:eq-uac5} and the
bottom transport give the exact map identities
\begin{equation}
 \Phi[A,B]=\mathcal C\Phi a-K\Phi E,
 \quad \Phi[B,\mathcal L]=-\mathcal C\Phi,
 \quad \Phi \mathcal L^2=K^2\Phi.
 \label{df:eq-uac7}
\end{equation}
For example
$\Phi AB=K\Phi aB=KT\Phi a-K\Phi E$, whereas
$\Phi BA=TK\Phi a$; their difference is the first identity.
No invariant flavor block or diagonalization has been selected.

\paragraph{The current and adjacent rows are independent after projection.}
The complete current row is $\Phi\mathcal D(a)=0$. Substituting
\eqref{df:eq-unq3}, \eqref{df:eq-uac5}--\eqref{df:eq-uac7}, and then \eqref{df:eq-uac6}, gives
\begin{equation}
 4\Phi\mathcal D(a)
       =(4\Delta-K^2)\Phi a^2+\mathcal C\Phi a(a+3)=0.
 \label{df:eq-uac8}
\end{equation}
Right composition with $a^{-1}$ is legitimate, so
\begin{equation}
        (4\Delta-K^2)\Phi a+\mathcal C\Phi(a+3)=0.
 \label{df:eq-uac9}
\end{equation}

The actual adjacent source row \eqref{df:eq-unq4} holds for $Y'_C$.
Apply $S_0$ on its left; its input operator stays on the right,
so $S_0Y'_C=\Phi$ gives $\Phi\widetilde{\mathcal D}=0$.
This is not obtained by falsely setting the current bottom to zero.
Let $z=a-2$. Using the full expression \eqref{df:eq-unq4}, one obtains
\begin{equation}
\begin{split}
4\Phi\widetilde{\mathcal D}
={}&4\Delta\Phi z^2+\mathcal C\Phi z(z+3)\\
 &+K\Phi\{2z\mathcal R(z)-E(z+3)\}+2K^2\Phi z\\
={}&(4\Delta-K^2)\Phi z^2+\mathcal C\Phi z(z+3).
\end{split}\label{df:eq-uac10}
\end{equation}
In the second line use \eqref{df:eq-uac6} and
$K\Phi Az=K^2\Phi az=K^2\Phi(z+2)z$.
Right divide by the invertible $z$ to get
\begin{equation}
        (4\Delta-K^2)\Phi(a-2)+\mathcal C\Phi(a+1)=0.
 \label{df:eq-uac11}
\end{equation}
Subtracting \eqref{df:eq-uac11} from \eqref{df:eq-uac9} gives
$\mathcal C\Phi=-(4\Delta-K^2)\Phi$.
Insert this into \eqref{df:eq-uac9}, with the same right ordering. The result
is $-3(4\Delta-K^2)\Phi=0$, proving both equations in \eqref{df:eq-uac2}.
The nonzero constant $3$ is the difference between the two affine
right coefficients; no spectral or parameter division occurs.

\subsection{The actual degree-two endpoint and its retained returns}
\label{df:sec-terminal}

Assume the actual full odd input $W$, all earlier raw-depth vanishing,
and both current source vanishings, now at current $b=3$.
Put $a=3+2j-H$, $\Phi=\Phi_3$, $T=p_{2,F}-2N/3$.
Use $R_s=P_s^0$, $Q_s=[f,R_s]/s$ for positive even $s$,
$\Delta=2r+N^2/4$, and the right-ordered operators
\[
\begin{gathered}
 A=N+2f+\sum_s(s+1)R_s,\qquad B=2f+p_2+\sum_sR_s,\\
 \mathcal R(z)=-3f-N/2+\sum_sR_s
                    \left(-\tfrac12+\frac{s}{z-s}\right).
\end{gathered}
\]
\[
 \mathcal D(z)=\Delta z^2+\tfrac14[A,B](z+3)
                                  +\tfrac12A\mathcal R(z).
\]
All denominators $z-s$ used here are nonzero odd integers.
For $x\in W_m$ let $c=a|_{W_m}$ and define
\begin{equation}
 Y'_Bx=B_{c-1,0}^+x,\qquad Y'_Cx=C_{c-1,0}^+x.
 \label{df:eq-ute1}
\end{equation}
Their degree-two values on $W_j$ are the genuine small
$\mathcal N=4$ supercurrents.  Define the auxiliary bottom map
$\widehat\Phi x=T_{c-2}x/(c-2)$ for $m<j$, and
$\widehat\Phi|_{W_j}=0$.  The latter is a definition of a zero
boundary coordinate, not insertion of a trace-one field into a
field identity.  Define a map $E:W\to W_j$ by
\begin{equation}
 Ex=R_{c-3}x\quad(c\ge5),\qquad E|_{W_j}=N I.
 \label{df:eq-ute2}
\end{equation}
All degree-dependent coefficients below act on the input first.
Write $G=G_X^+|_{W_j}$ and $W^+=W_{3,1}^+|_{W_j}$ for the
chosen $B$ family; barred formulas hold separately.

\paragraph{The neighboring cubic equation retains its current denominator.}
The exact identity still is
\begin{equation}
       Y'A=S_2\Phi a,\qquad \mathcal L=Aa^{-1},\qquad S_2\Phi=Y'\mathcal L.
 \label{df:eq-ute3}
\end{equation}
Indeed $R_{c-1}x=0$, but $R_{c-1}fx$ may survive.
In \eqref{df:eq-fhc8} the left side gains
$2G_X^+R_{c-1}fx/N$.
Since $R_{c-1}fx=-(c-1)Q_{c-1}x$, this is exactly the
surviving term $-2(c-1)G_X^+Q_{c-1}x/N$ on the right
of \eqref{df:eq-sfd5}.  These two original projectors cancel at every input
weight.  The regular sum has minimum fermion degree two and is
well defined.  Thus \eqref{df:eq-ute3} has no correction.

Also \eqref{df:eq-fhc12} applies at its genuine $d=2$ endpoint and gives
$S_0Y'_C=\Phi$.  The first pole there is
$2\partial P_3/3$, the second pole is $P_3$, so the full
mode coefficient is $1-2/3=1/3$.
The mixed spin-two all-level relation used in \eqref{df:eq-uac4} consequently
proves, also at this terminal layer,
\begin{equation}
                         \Phi A=K\Phi a,\qquad \Phi \mathcal L=K\Phi.
 \label{df:eq-ute4}
\end{equation}
The shifted source is not $-S_1\widehat\Phi$: the correct equation is
\begin{equation}
                       S_1\widehat\Phi=-Y'+GE/N.
 \label{df:eq-ute5}
\end{equation}
For $c\ge5$ this is the original \eqref{df:eq-fhc8} projector.
For $c=3$ it is $0=-G+G$, directly from \eqref{df:eq-ute1}--\eqref{df:eq-ute2}.

\paragraph{Two small raw field identities, proved at their actual endpoints.}
The quartic third channel at $c=3$ has only the following
field residue and second pole.  Let
\[
 U=\tfrac{15}{2}\operatorname{tr}(bYXX-bXXY),\qquad
 M=\sum_{i=0}^2(-1)^i\binom2i:W_{3,i}W_{2,2-i}:.
\]
With $O_t=\sum_{i=0}^3(-1)^i\binom3i
(W_{4,i})_{(t)}W_{3,3-i}$, the complete identities with formal matrix-size parameter $N$ are
\begin{equation}
             O_1=O_2=0,\qquad O_0=QU-15M/N+(15N/2)H_3.
 \label{df:eq-ute6}
\end{equation}
For the one-contraction residue, the cyclic differences of the primitive
are the specialization of \eqref{df:eq-qd4} at $k=3$:
$\lambda_0=15/2$, $\lambda_1=0$, $\lambda_2=-15/2$.
The one-scalar contraction is $-15M/N$.  In the remaining ghost sector,
the two empty-arc coefficients average to $15N/2$ times the raw top
$H_3$, as in \eqref{df:eq-qd6}--\eqref{df:eq-qd7}.
The remaining two-jet pure trace has degree one and is zero.
The second-pole residue has degree three and flavor spin $1/2$;
its cyclic words vanish, since $\operatorname{tr}(X[X,Y])=0$.
The third pole is proportional to $P_1=0$.  This proves all three
equations at the actual endpoint.
The first-half normal correction of $M$ is
$(-6+2+4)W_{3,1}^+=0$, so its first-half mode is exactly
$\sum_i(-1)^i\binom2iW_{3,i}^+W_{2,2-i}^0$.
This proves the actual missing quartic field endpoint without a trace-one term.

Similarly the true degree-two fermions give
\begin{equation}
 B_{3,0(0)}C_{2,1}-B_{3,1(0)}C_{2,0}=\tfrac32H_3,
 \qquad B_{3,0(1)}C_{2,1}-B_{3,1(1)}C_{2,0}=0.
 \label{df:eq-ute7}
\end{equation}
Thus the full middle identity \eqref{df:eq-unq9} has the valid $d=3$ endpoint
\[
                    H_3^+=\tfrac29[f,T_3]-2S_1C_2^+.
\]
The absent term proportional to $N$ has coefficient $d-3=0$, and no regular
sum is present.  For \eqref{df:eq-ute7}, the only contracted odd pair is $b,p$; the remaining single matter contractions give the two terms of the raw $H_3$ with coefficient $3/2$, and their second poles cancel.

\paragraph{The original quartic projectors still cancel, including two survivors.}
Put $R=R_{c-1}$ and keep $Rfx$, which has weight $j$.
The full lowered quartic commutator projector is
\[
 \frac{c+2}{N(c-1)(c-2)}
 \left(6f_FT_3Rf-3T_3Rf^2-6(c+1)W_{3,1}^+Rf\right)x.
\]
The original quartic-field terms at $i=0,1$, using only $Rx=0$, are
\[
 \frac{c+2}{N(c-1)(c-2)}
 \left(6T_3fRf-3T_3Rf^2-6(c-2)W_{3,1}^+Rf\right)x.
\]
Their difference is proportional to
$6([f,T_3]-3W_{3,1}^+)Rfx=0$.
The $i=2$ factor is $Rx=0$.
This argument is valid even at $c=3$, now justified by \eqref{df:eq-ute6}.
It retains the input $f$ in $T_3fRf$ and the output $f_F$
in $f_FT_3Rf$ separately until covariance is used.

Consequently the full adjacent quartic comparison before the
cubic operation remains
\begin{equation}
 \widehat\Phi\mathcal D(a-2)+f_F\Phi J-p_{3,F}\Phi(a+1)
                +\mathcal S=0,\qquad
 J=-\mathcal R(a-2)+3A/2,
 \label{df:eq-ute8}
\end{equation}
where the specified middle sources in $\mathcal S$ have positive
sign: on $x\in W_m$ they are
$2N S_1C_{c-1}^+x+\sum_s\widehat h(s,c-s)
(c-s-1)S_1C_{c-s-1}^+R_sx/(c-s-2)$.
The sum only uses nonzero actual right inputs and has $c-s\ge3$.
On the top coordinate $c=3$, \eqref{df:eq-ute6}--\eqref{df:eq-ute7} verify \eqref{df:eq-ute8}
directly with $\widehat\Phi x=0$; no nonexistent raw field is used.
For an explicit check, its entire top-coordinate value is
\[
 \left\{-\tfrac35T_5f^2+\tfrac65f_FT_5f
       +\tfrac{2N}{3}[f,T_3]
       +\tfrac43(T_3p_3-p_{3,F}T_3)
       +2NS_1C_2^+\right\}x=0.
\]
This is also the direct $c=3$ lowered quartic commutator expression minus \eqref{df:eq-ute6},
with the projectors canceled as above. Here
$\widehat\Phi\mathcal D(1)x$ is retained: although
$\widehat\Phi x=0$, its right lowerings can be nonzero.

\paragraph{The terminal half-level defects and the source-square return.}
Both lowering supercurrents kill $\widehat\Phi$ and $f_F\Phi$.
For $G_Y$ this follows from
$[G_{Y(1)},T_d]=-dB_d^0$ on the input, which is zero by the
full odd hypothesis, even when its flavor is exactly $j$.
For $G_X$ the OPE with $P_d$ is regular.
The barred statements use the corresponding odd integer zeros.
The mode $W_{3,1(1)}$ kills these two maps by their weights.
The two possible even returns are the actual maps
\begin{equation}
 U_*=W_{3,2(1)}\widehat\Phi:W\to W_j,
       \qquad V_*=W_{3,3(1)}\Phi:W\to W_j.
 \label{df:eq-ute9}
\end{equation}
Since $[W_{3,2(1)},f_F]=-W_{3,3(1)}$ and
$W_{3,2(1)}\Phi=0$, the complete normal sum in \eqref{df:eq-uhs7} gives
\begin{equation}
 [p_2,S_1]\widehat\Phi=-2GU_*/N,
       \qquad [p_2,S_1]f_F\Phi=2GV_*/N.
 \label{df:eq-ute10}
\end{equation}
The surviving normal term is $G_X^+W_{3,2(1)}$, after the
usual $-4S_1$ contraction cancels the direct term in \eqref{df:eq-uhs6}.

For $d=c-s$ put
\[
 M_sx=\left(-H_{d-1}^0+\tfrac12Q_{d-1}\right)R_sx,
 \qquad M_0x=\left(-H_{c-1}^0+\tfrac12Q_{c-1}\right)x,
\]
where the $s=0$ notation uses the identity input, not $R_0$.
Every $M_s$ in the actual sum maps to $W_j$.
The complete square \eqref{df:eq-csh1} is
$S_1^2C_{d-1}^+R_sx=GM_sx/N$; its other term has right
weight $j+1$ and vanishes.  Define the explicit map
\begin{equation}
 Z_*x=2NM_0x+
   \sum_s\frac{\widehat h(s,c-s)(c-s-1)}{c-s-2}M_sx.
 \label{df:eq-ute11}
\end{equation}
Explicitly the coefficient of $M_s$ is
$3s+2(c-2)/(c-s-2)$, by the primitive quartic field coefficient;
this sum has $s\ge2$ even and $c-s\ge3$.
Then $S_1\mathcal S=GZ_*/N$.  In particular
\begin{equation}
             M_0|_{W_j}=(r-j)I,
                \qquad Z_*|_{W_j}=2N(r-j)I.
 \label{df:eq-ute12}
\end{equation}
This is the genuine degree-two supercharge return; it cannot be
replaced by the inner-layer square-zero rule.

\paragraph{An exact terminal replacement for the homogeneous adjacent row.}
The bottom transport, including its projector, is
\[
 T\widehat\Phi=\widehat\Phi B-2f_F\Phi+2W^+E/N.
\]
For $c\ge5$ this is \eqref{df:eq-qbl1} with its right $R_{c-3}$ kept.
On $W_j$ it reduces directly to $[f,T_3]=3W_{3,1}^+$.
Applying $S_1$ and using \eqref{df:eq-ute5}, \eqref{df:eq-ute10} yields
\begin{equation}
 TY'=Y'(B-2\mathcal L)+\Xi,
 \quad
 \Xi=\frac1N\left(TGE-GEB-2S_1W^+E+2GU_*\right).
 \label{df:eq-ute13}
\end{equation}
All maps on the right are actual maps with the displayed orders.
The retained sign in $2p_3=[f,p_2]$ now gives
\[
 S_1p_{3,F}\Phi
   =\tfrac12Y'([B,\mathcal L]-2\mathcal L^2)+\tfrac12\Xi \mathcal L+GV_*/N.
\]
Thus the corrected adjacent row, with the same formal operator
$\widetilde{\mathcal D}
=\mathcal D(a-2)+\mathcal L J+
\tfrac12([B,\mathcal L]-2\mathcal L^2)(a+1)$ as in \eqref{df:eq-unq4}, is exactly
\begin{equation}
 \boxed{\quad
 Y'\widetilde{\mathcal D}
   =\frac GN\left(E\mathcal D(a-2)-V_*(a+1)+Z_*\right)
                        -\tfrac12\Xi \mathcal L(a+1).
 \quad}
 \label{df:eq-ute14}
\end{equation}
Its derivation is application of $S_1$ to the actual \eqref{df:eq-ute8},
not continuation of a vanished source-square expression.
Apply $\delta$ to this complete $B$-family identity, with all
returns retained. It fixes $E,\mathcal D,V_*,Z_*,\mathcal L,a$ and sends
$Y'_B,G_X,\Xi_B$ to $Y'_C,\bar G_X,\Xi_C$, respectively.
The transfer principle \eqref{df:eq-fdk8}--\eqref{df:eq-fdk9} proves the $C$-family
identity on the same input; no ghost swap is used.
Equations \eqref{df:eq-ute3}--\eqref{df:eq-ute4} and the original current bottom row
still hold, but \eqref{df:eq-ute14} is inhomogeneous.  The retained terminal maps are evaluated below before the adjacent rows are subtracted.

\paragraph{The transport defect cancels by an additional actual field identity.}
The retained $\Xi$ can in fact be evaluated without any output
lowest-space assumption.  Write $\mathsf J=W_{2,1}$, so that
$o(\mathsf J)=-H$, and put
\begin{equation}
 \mathscr H_cx=\sum_{\substack{2\le s\le c-3\\s\ {\rm even}}}
                    R_{c-3-s}R_sx,\qquad R_0=N I.
 \label{df:eq-ute15}
\end{equation}
The symbol $R_0$ here is a scalar endpoint convention only.
The sum is empty at $c=3$.  The exact two identities are
\begin{equation}
 E Bx=p_2Ex-4Q_{c-1}x+\mathscr H_cx,
 \qquad
 U_*x=-2Q_{c-1}x+\tfrac12\mathscr H_cx-\frac{4j}{N}Ex.
 \label{df:eq-ute16}
\end{equation}
For the first, retain $2R_{c-1}fx$ in $EBx$.  Since
$R_{c-1}x=0$, it equals $-2(c-1)Q_{c-1}x$.
Commuting $R_{c-3}$ past $p_2$ adds $2(c-3)Q_{c-1}x$;
the other right traces give exactly $\mathscr H_cx$.
The same computation at $c=3$ uses $R_2fx=-2Q_2x$ and
$R_0=N I$, with no trace-one operator.

Here is an arbitrary-degree contraction proof of the second identity.
For odd $d\ge3$ let $F_i=(W_{3,2})_{(i)}P_d$ and set
$P_0=N$, $P_1=0$.  The entire two relevant poles are
\begin{equation}
\begin{split}
 F_0={}&-2dQ_{d+1}
    +d\sum_{\substack{s+t=d-1\\ 0\le s\le d-2}}
                           \frac1t:P_s\partial P_t:
    +\frac{2d}{N}:\mathsf J P_{d-1}:,\\
 F_1={}&d\sum_{\substack{s+t=d-1\\ 0\le s\le d-2}}:P_sP_t:
                      -\frac{2d(d-1)}N P_{d-1}.
\end{split}\label{df:eq-ute17}
\end{equation}
Indeed $W_{3,2}=\operatorname{tr}(XY^2)$ has only two
contractible source letters.  A single contraction gives
$-2dQ_{d+1}$.  For a double contraction, the first target
endpoint has $d$ choices and the intermediate target arc
has length $s=0,\ldots,d-2$.  The sole surviving source
$X$ lies in the other trace, of degree $t=d-1-s$;
its Taylor term is $\partial P_t/t$.  One scalar propagator
gives the displayed current product and its double pole;
two scalar propagators leave $\operatorname{tr}X=0$.
These cases exhaust the contractions and give \eqref{df:eq-ute17}.

The physical commutator is $o(F_0)+o(F_1)$.
Pure-$X$ products with two odd degrees have zero integer zero
mode on the lowest space by regular mode ordering.  The even
degrees give $o(:P_s\partial P_t:)=-tR_sR_t/2$.
Moreover $o(:\mathsf J P_t:)=R_t o(\mathsf J)$, so its
second-pole correction converts the projector to
$2d\,o(\mathsf J)R_{d-1}/N$ using
$[o(\mathsf J),R_{d-1}]=-(d-1)R_{d-1}$.
Set $d=c-2$, divide by $d$, and use $o(\mathsf J)E=-2jE$
to obtain \eqref{df:eq-ute16}.  At $c=3$ the proposed expression is
$-2Q_2-4jI=0$ on $W_j$, which proves its compatibility
with the defined zero coordinate without using $P_1$ as a field.

For the other factor in $\Xi$, exact single contraction gives
$[p_2,G_X^+]=-2B_{4,1}^+$.  A separate short field identity is
\begin{equation}
\begin{split}
 (B_{3,1})_{(0)}W_{3,1}
   ={}&-B_{4,1}+\tfrac13 Q\operatorname{tr}(Xbb)
      +\tfrac2N:\mathsf J G_X:-\tfrac1N:P_2G_Y:\\
     &+\left(\tfrac4N-\tfrac{2N}{3}\right)\partial G_X,\\
 (B_{3,1})_{(1)}W_{3,1}
   ={}&\left(\tfrac4N-N\right)G_X.
\end{split}\label{df:eq-ute18}
\end{equation}
These are exact field identities with formal matrix-size parameter $N$, including the
BRST primitive.  The first-half derivative multiplier is $-1$;
current-first normal ordering leaves $G_X^+o(\mathsf J)$
and $G_Y^+R_2$.  Since $R_2W_j=0$ and the odd integer
mode kills $W$, \eqref{df:eq-ute18} gives
\begin{equation}
 S_1W^+=-B_{4,1}^+|_{W_j}-NG/3-4jG/N,
 \quad
 (TG-Gp_2-2S_1W^+)|_{W_j}=8jG/N.
 \label{df:eq-ute19}
\end{equation}
Substituting \eqref{df:eq-ute16} and \eqref{df:eq-ute19} in \eqref{df:eq-ute13} now cancels
all four terms, with their displayed orders.  Consequently
\begin{equation}
 \boxed{\quad\Xi=0,\qquad TY'=Y'(B-2\mathcal L),\qquad
 Y'\widetilde{\mathcal D}
  =\frac GN\left(E\mathcal D(a-2)-V_*(a+1)+Z_*\right).\quad}
 \label{df:eq-ute20}
\end{equation}
The remaining $V_*$ and source-square returns have not vanished.
Apply $\delta$ to the complete equation \eqref{df:eq-ute20}. The return
in parentheses is fixed, and $Y'_B,G_X$ become $Y'_C,\bar G_X$;
\eqref{df:eq-fdk8}--\eqref{df:eq-fdk9} justify this specialization on the same input.
Now apply $S_0$ to that $C$-family equation and use its genuine
degree-two endpoint $S_0\bar G_X^+=T_3/3$ on $W_j$.
Since every right map in parentheses lands in $W_j$, one obtains
the exact terminal bottom equation
\begin{equation}
 \boxed{\quad
 \Phi\left[\widetilde{\mathcal D}
   -\frac1N\{E\mathcal D(a-2)-V_*(a+1)+Z_*\}\right]=0.
 \quad}\label{df:eq-ute21}
\end{equation}
Here maps into $W_j$ are included into $W$ before $\Phi$ acts.

\subsection{Expression of the terminal return through input zero modes}
\label{df:sec-terminal-input}

Keep precisely the assumptions and right-order conventions of
\eqref{df:eq-ute1}--\eqref{df:eq-ute21}.  In particular the current degree is $b=3$,
$c=a|_{W_m}=3+2(j-m)$, and both current sources vanish.
Write
\begin{equation}
 Mx=\left(-H_{c-1}^0+\tfrac12Q_{c-1}\right)x,
 \qquad z=a-2,\qquad R=\mathcal R(a-2),\qquad \mathcal L=Aa^{-1}.
 \label{df:eq-utr1}
\end{equation}
The map $M:W\to W_j$ is the actual $M_0$ of \eqref{df:eq-ute11}.
Let $\Pi$ be the input weight projection onto $W_j$, included
back into $W$. It is not assumed to commute with the zero algebra.

\paragraph{The remaining cubic lowering mode needs only two small poles.}
Let $\bar G_{Y(1)}$ be the lowering half mode of the opposite
supercurrent.  Direct contraction of $\operatorname{tr}(Y^2b)$
with $\operatorname{tr}(Yp)$ has first and second poles
$2\partial\operatorname{tr}(Y^3)/3$ and
$\operatorname{tr}(Y^3)$; the scalar projector contains
$\operatorname{tr}Y=0$.  Since both mode indices are one,
\begin{equation}
 \{S_2,\bar G_{Y(1)}\}=-\tfrac13W_{3,3(1)}.
 \label{df:eq-utr2}
\end{equation}
The coefficient is $-4/3+1$, not the raising-half coefficient
$1/3$ in \eqref{df:eq-fhc12}.  The lowering supercurrent kills $\Phi$:
its action returns to an actual odd integer zero on $W$
(and its output flavor is $j+1>j$).
The opposite-family form of \eqref{df:eq-nsr4} gives on the whole lowest input
\begin{equation}
 \bar G_{Y(1)}Y'_B=-M.
 \label{df:eq-utr3}
\end{equation}
For completeness the sign does not require an abstract exchange
of the two ghost systems.  The true relative-chain derivation
$\delta b=p$, $\delta p=\delta X=\delta Y=0$ commutes with
$Q$.  The OPE of $G_Y$ with $B_d$ is regular: its single
contraction sum pairs the two odd $b$ letters in opposite cyclic
orders and cancels. Applying $\delta$ to this zero OPE gives
$\bar G_Y{}_{(i)}B_d=-G_Y{}_{(i)}C_d$.
The two poles of \eqref{df:eq-nsr3} now prove \eqref{df:eq-utr3}, including its sign.
Combining \eqref{df:eq-utr2}--\eqref{df:eq-utr3} with
$S_2\Phi=Y'_B\mathcal L$, with no action moved through $\mathcal L$, proves
\begin{equation}
                  \boxed{\quad V_*=3M\mathcal L.\quad}
 \label{df:eq-utr4}
\end{equation}
This is an identity for the particular current bottom.  It
uses neither a lowest-space relation on its image nor vanishing
of $S_2\Phi$.

\paragraph{Every even top return has an explicit input expression.}
The full odd hypothesis and \eqref{of:fullodd-top} give, for $c\ge5$,
\begin{equation}
 Mx=\left\{\tfrac12Q_{c-1}-\tfrac N4R_{c-3}
     -\tfrac18\sum_{\substack{u+v=c-3\\ u,v\ge2\ {\rm even}}}
                                  R_uR_v\right\}x.
 \label{df:eq-utr5}
\end{equation}
On $W_j$ the genuine endpoint is $Mx=(r-j)x$, because
$H_2^0=-r$ and $Q_2=-H$.  The sum in \eqref{df:eq-utr5} is ordered.
Every remaining $Q_s$ is $[f,R_s]/s$, and the actual arbitrary
trace tower is recursively determined by
\begin{equation}
 R_2=2e,\quad R_4=p_0,\qquad
 R_{s+2}=[R_4,Q_s]/4\quad(s\ge4\ {\rm even}).
 \label{df:eq-utr6}
\end{equation}
Thus \eqref{df:eq-utr4}--\eqref{df:eq-utr6} express all $V_*$ and top returns through
actual input zeros at any degree; they are not finite-degree fits.

In \eqref{df:eq-ute11} the map $M_s$ is exactly $MR_s$, because its raw
degree is adjusted after $R_s$ acts.  Its coefficient is
$3s+2(a-2)/(a-2-s)$.  Consequently
\begin{equation}
 Z_*=M\left[2N+\sum_sR_s
       \left(3s+\frac{2(a-2)}{a-2-s}\right)\right]
                         =M(3A+2R).
 \label{df:eq-utr7}
\end{equation}
Every rational function remains on the right of the indicated
$R_s$.  Substituting \eqref{df:eq-utr4} and \eqref{df:eq-utr7} gives the complete
simplification of the terminal right return:
\begin{equation}
 \boxed{\quad
 \mathfrak F:=E\mathcal D(a-2)-V_*(a+1)+Z_*
       =E\mathcal D(a-2)+2MR-3M\mathcal L.
 \quad}\label{df:eq-utr8}
\end{equation}
Indeed $3A-3\mathcal L(a+1)=-3\mathcal L$, since $A=\mathcal L a$.
The terminal bottom row is therefore the exact equation
\begin{equation}
                   \Phi(\widetilde{\mathcal D}-\mathfrak F/N)=0.
 \label{df:eq-utr9}
\end{equation}

\paragraph{Two further exact identities, with the endpoint projector retained.}
The trace diagonal satisfies $E^2=NE$, since $E|_{W_j}=NI$.
Moreover
\begin{equation}
 \boxed{\begin{aligned}
 E A&=-4M(a-1)+4\Delta\Pi,\\
 E B&=(p_2-N)E-8M+4\Delta\Pi,
             \qquad \Delta=2r+N^2/4.
 \end{aligned}}\label{df:eq-utr10}
\end{equation}
For the first, on input of raw degree $c$ the lowering term is
$2R_{c-1}fx=-2(c-1)Q_{c-1}x$.  The other terms in $EAx$ are
$NE x+\sum_s(s+1)R_{c-3-s}R_sx$.
Separate the endpoint $s=c-3$ and pair $s$ with $c-3-s$
in the internal commuting sum.  The result for $c\ge5$ is
$(c-1)(NE-2Q_{c-1}+\tfrac12\sum_{u+v=c-3}R_uR_v)x$,
which is $-4M(c-1)x$ by \eqref{df:eq-utr5}.
At $c=3$, $EA=N^2+8j$, while $-4M(a-1)=-8(r-j)$.
Their difference is $N^2+8r=4\Delta$, proving the first
identity with its actual projector.

For the second use \eqref{df:eq-ute16},
$EB=p_2E-4Q_{c-1}+\mathscr H_c$.
At $c\ge5$, $\mathscr H_c=NE+\sum_{u+v=c-3}R_uR_v$;
then \eqref{df:eq-utr5} gives the asserted expression.  At $c=3$ its
two sides differ by precisely the same $4\Delta\Pi$.
No stability of $W_j$ under $A$ or $B$ has been assumed.

\paragraph{The entire return vanishes on the actual highest input.}
The restriction of \eqref{df:eq-utr8} to $W_j$ is:
\begin{equation}
\begin{aligned}
 \mathcal D(1)x&=(2r-2j)x-2Nfx-3f^2x+4p_3x,\\
 E\mathcal D(1)x&=2N(r-5j)x-12p_2x,\\
 V_*x&=N(r-3j)x-3p_2x,\qquad Z_*x=2N(r-j)x.
\end{aligned}\label{df:eq-utr11}
\end{equation}
Here $p_1x=p_0x=0$ by their positive weights.  The exact
covariance identities are $R_2fx=4jx$, $R_4f^2x=12p_2x$,
$R_2p_3x=6p_2x$, and $Q_4fx=-3p_2x$.
For $V_*$ use \eqref{df:eq-utr4} and
$Mfx=-Njx-3p_2x/2$; this retains every quartic return.
Substitution in \eqref{df:eq-utr8} gives
\begin{equation}
                         \mathfrak F\Pi=0.
 \label{df:eq-utr12}
\end{equation}

\subsection{Reduction of the terminal return to the even third channel}
\label{df:sec-terminal-reduction}

Keep the actual full odd input and terminal notation of
\eqref{df:eq-utr1}--\eqref{df:eq-utr10}. The following reduction uses only input zero-algebra identities.
For even trace degree $s$ put $R_s=o(P_s)$,
$Q_s=[f,R_s]/s$, and write $H_s=o(H_s^{\rm raw})$.
Thus $H_2=-rI$. Let $T_s^{(2)}=o(W_{s,2})$ and define
\begin{equation}
 G_k=[Q_k,p_2]-2(k-2)T_{k+2}^{(2)},\qquad k\ge4\text{ even}.
 \label{df:eq-ufr1}
\end{equation}
All rational functions of $a=3+2j-H$ act on the right. The exact
return in \eqref{df:eq-utr8} is
$\mathfrak F=E\mathcal D(a-2)+2M\mathcal R(a-2)-3MAa^{-1}$.
I prove that on $x\in W_{j-(k-2)/2}$,
\begin{equation}
 \boxed{\quad
 \mathfrak F x=\frac{k+2}{k+1}
 \left\{-\frac{k(k+1)}2G_k+
       k(k-2)\left(\Delta-\frac{k^2}{16}\right)R_{k-2}\right\}x.
 \quad}\label{df:eq-ufr2}
\end{equation}
The genuine top input remains $\mathfrak F\Pi=0$ by \eqref{df:eq-utr12}. There is no
formal trace of degree one in this statement.

\paragraph{Precisely the input relations used.}
The relations are flavor covariance, the raw even commutators
\eqref{of:even-first-second}, the full odd top identity \eqref{of:fullodd-top}, and the actual
projector identities \eqref{df:eq-utr10}:
\begin{equation}
\begin{gathered}
 [f,R_s]=sQ_s,\quad [R_s,R_t]=0,\quad
 [Q_s,R_t]=-tR_{s+t-2},\quad [R_s,p_2]=2sQ_{s+2},\\
 H_s=\frac N4R_{s-2}+\frac18
       \sum_{\substack{u+v=s-2\\ u,v\ge2\ {\mathrm{even}}}}R_uR_v\quad(s\ge4),\\
 [A,B]=4p_3+\sum_s(4-6s)Q_s,\qquad
 EA=-4M(a-1)+4\Delta\Pi.
 \end{gathered}\label{df:eq-ufr3}
\end{equation}
The sums are ordered. Both $W_j$ and its inclusion into $W$ are
retained; neither is assumed invariant under the even algebra.
On an input of raw degree $c$, $M=-H_{c-1}+Q_{c-1}/2$,
$E=R_{c-3}$ for $c\ge5$, and $E=N$, $M=r-j$ for $c=3$.
Weight bounds kill every actual factor whose right output has
weight greater than $j$.

\paragraph{Removing the projector and moving the right functions.}
Set $z=a-2$, $R=\mathcal R(z)$, and $\mathcal L=Aa^{-1}$.
The first identity in \eqref{df:eq-utr10} gives, without commuting any
right function through $R$,
\begin{equation}
 \mathfrak F=\Delta Ez^2+\tfrac14E[A,B](a+1)
                   -2MzR+2\Delta\Pi R-3M\mathcal L.
 \label{df:eq-ufr4}
\end{equation}
The actual highest-output part of $R$ is
$\Pi R=E(a-7/2)$. For $c>3$ it is the single trace
$s=c-3$; at $c=3$ it is the scalar $-N/2$.
Hence the two $\Delta$ terms in \eqref{df:eq-ufr4} combine to
$\Delta E(a-3)(a+1)$, including the endpoint.
Since $zf=f(z+2)$ and $zR_s=R_s(z-s)$, the remaining
$M$ terms on raw degree $c$ are exactly
\begin{equation}
 \frac{c+1}{c}\left\{
  6(c-1)M_{c+2}f+N(c-3)M_c
       +\sum_s(c-3-3s)M_{c-s}R_s\right\}.
 \label{df:eq-ufr5}
\end{equation}
The subscripts designate the adjusted raw degree after the
right action, not a separate family of independent maps.

Put $c=k+1$. Multiplying \eqref{df:eq-ufr4} by $(k+1)/(k+2)$ and
using \eqref{df:eq-ufr5} gives the following full expression on $x$:
\begin{equation}
\begin{split}
\frac{k+1}{k+2}\mathfrak F x={}&\biggl[
 (k+1)(k-2)\Delta R_{k-2}+(k+1)R_kp_3\\
 &+(k+1)\sum_{s=2,4,\ldots,k}(1-3s/2)R_{k-s}Q_s\\
 &+6k(-H_{k+2}+Q_{k+2}/2)f
       +N(k-2)(-H_k+Q_k/2)\\
 &+\sum_{s=2,4,\ldots,k-2}(k-2-3s)
                (-H_{k-s}+Q_{k-s}/2)R_s\biggr]x.
\end{split}\label{df:eq-ufr6}
\end{equation}
Here and only for the endpoint in the displayed sum, $R_0=NI$.
Every factor in \eqref{df:eq-ufr6} is an actual zero mode or its composition.

\paragraph{Cancelling the noncommuting returns.}
The weights give $R_kx=Q_{k+2}x=H_{k+2}x=0$.
Lowering $[R_k,p_2]=2kQ_{k+2}$ once yields
\[
 [R_k,p_3]=k(k+1)T_{k+2}^{(2)}-\tfrac k2[Q_k,p_2].
\]
Also $Q_{k+2}fx=-(k+1)T_{k+2}^{(2)}x$.
Thus the $R_kp_3$ and $Q_{k+2}f$ terms give
$-k(k+1)G_kx/2$.
For the other lowering term, differentiate \eqref{of:fullodd-top}, retaining
both orders before using $[Q_u,R_v]=-vR_{k-2}$:
\begin{equation}
 -6kH_{k+2}fx=\left\{
 \frac{3Nk^2}{2}Q_k+\frac{3k}{2}
       \sum_{u+v=k}uR_vQ_u
       -\frac{3k}{4}\sum_{u+v=k}uvR_{k-2}\right\}x.
 \label{df:eq-ufr7}
\end{equation}
All indices in these sums are positive even degrees. For $u+v=k$,
the total coefficient of $R_vQ_u$ in \eqref{df:eq-ufr6} is
\[
 (k+1)(1-3u/2)+3ku/2+(k-2-3v)/2=0.
\]
The endpoint coefficient of $NQ_k$ is also zero.
The scalar from the two reorderings is
\begin{equation}
 -\frac{3k}{4}\sum_{u+v=k}uv
       -\frac12\sum_{v=2,4,\ldots,k-2}v(k-2-3v)
                 =-\frac{k^3(k-2)}{16}.
 \label{df:eq-ufr8}
\end{equation}
This follows from
$\sum v=k(k-2)/4$, $\sum v^2=k(k-2)(k-1)/6$, and
$\sum uv=k(k^2-4)/12$.

The true $H_2=-r$ endpoint contributes $-2(k-2)rR_{k-2}$;
the linear part of $-N(k-2)H_k$ contributes
$-N^2(k-2)R_{k-2}/4$. Together they subtract
$(k-2)\Delta R_{k-2}$ from the first term of \eqref{df:eq-ufr6}.
For $m=k-2$, the remaining ordered pair contribution proportional to $N$ is
\begin{equation}
 -\frac{Nm}{8}\sum_{u+v=m}R_uR_v
       -\frac N4\sum_{u+v=m}(m-3v)R_uR_v=0,
 \label{df:eq-ufr9}
\end{equation}
because the summation is symmetric and the average of $v$ is $m/2$.
The triple coefficient is
$-\frac18\sum_{u+v+w=m}(m-3w)R_uR_vR_w=0$,
by permutation symmetry of the commuting factors. Empty sums
vanish, so these arguments include $k=4$.
This proves \eqref{df:eq-ufr2}.

\paragraph{The exact missing even field identity.}
Put
$\mathcal K_k=\sum_{i=0}^3(-1)^i\binom3i
 [p_i,o(W_{k,3-i})]$.
Three flavor lowerings of $[p_0,R_k]=0$, or directly \eqref{df:eq-glq3},
give
\begin{equation}
 \mathcal K_k=-\frac{k(k+1)(k+2)}{2(k-1)(k-2)}G_k.
 \label{df:eq-ufr10}
\end{equation}
Consequently $\mathfrak F=0$ on every input weight follows if the actual
integer-zero third channel is
\begin{equation}
 \mathcal K_k=\frac{k(k+2)}{k-1}
                     \left(\frac{k^2}{16}-\Delta\right)R_{k-2}
 \qquad(k\ge4\text{ even}).
 \label{df:eq-ufr11}
\end{equation}

\subsection{The even third channel at every degree}
\label{df:sec-even-third}

Let $W$ be an actual zero-algebra-stable full odd annihilator in a
lowest conformal space of weight $r$. No assumption on flavor
support, multiplicity, or semisimplicity is made. All subscripts
below are trace degrees. Put $R_s=o(P_s)$, $Q_s=o(W_{s,1})$,
$p_i=o(W_{4,i})$, $\Delta=2r+N^2/4$. For every even $k\ge4$,
\begin{equation}
 \boxed{\quad
 \sum_{i=0}^3(-1)^i\binom3i[p_i,o(W_{k,3-i})]
 =\frac{k(k+2)}{k-1}
             \left(\frac{k^2}{16}-\Delta\right)R_{k-2}.
 \quad}\label{df:eq-uet1}
\end{equation}
This is an integer-zero identity on $W$. It is not a statement
about the action of the same expression on a first-half image.

\paragraph{Field reduction before specializing modes.}
Write $m=k-2$ and
\begin{equation}
\begin{gathered}
 g=\frac{k(k+2)}{k-1},\qquad
 t=\frac{3k(k+2)}{2(k-1)m},\qquad
 \beta=\frac{k(k+1)(k+2)}{(k-1)m},\\
 a_k=\frac{2k(k+2)}{(k-1)^2},\qquad
 g_0=\frac{k(k+1)(k+2)}2.
\end{gathered}
 \label{df:eq-uet2}
\end{equation}
Let $O_j=\sum_{i=0}^3(-1)^i\binom3i
 (W_{4,i})_{(j)}W_{k,3-i}$. Its physical zero-mode commutator
is $o(O_0)+o(O_1)$; the source weight is two. Temporarily
contract with the full matrix propagator and project onto the
scalar Fock vacuum only after completing Wick contraction.
This gives the same integer commutator as the actual traceless
fields: the raw integer zero modes preserve the tensor-product
lowest space and restrict to the actual raw zeros, as follows.  In a pure-boson full-matrix trace, expand each letter into
its traceless and scalar summands.  At an integer zero mode on
$W\otimes|0\rangle$, a term with a nonvacuum scalar excitation has
strictly positive scalar energy.  All scalar lowering modes annihilate
the vacuum, and regularity moves them to the right.  Thus the traceless
factor would have negative energy and annihilate $W$.  The sole
surviving term is the traceless trace times the scalar vacuum.  Hence
both individual zero modes and their compositions restrict as stated. This uses
only the auxiliary scalar bosons for the original bottom fields.
It does not treat projection as a product homomorphism.

The canonical field reduction uses the following ordered columns
for $a,b\ge2$, $a+b=k$:
\[
 \pi(:P_a^{\rm gl}H_b^{\rm gl}:),\quad
 \pi(:B_a^{\rm gl}C_b^{\rm gl}:),\quad
 \pi(D_{ab}^{\rm gl}),\quad
 D_{ab}=:P_a\partial Q_b:-:Q_a\partial P_b:,
\]
with coefficients
\begin{equation}
\begin{aligned}
 h_{ab}&=\frac{2g+2ta(b-2)}{b-1},\qquad
 c_{ab}=-2g-2t(a-1)(b-1),\\
 f_{ab}&=-\frac{3(k+1)(k+2)}{k-1}
            +2t(k-1-2a)+\frac{b-1}{b}h_{ab}.
\end{aligned}
 \label{df:eq-uet3}
\end{equation}
There is also the single top $a_kNH_k$ and a pure-$X$ two-jet
remainder with at most three nonempty traces.

Here are the parity-independent counts underlying this reduction.
The zero-jet primitive is
$\sum_{i=0}^{k-1}\lambda_i\operatorname{tr}(bX^iYX^{k-1-i})$,
where $\lambda_0=g$, $\lambda_{k-1}=-g$ and
$\lambda_i=t(k-1-2i)$ otherwise. Its classical boundary is the
successive difference of the two-$Y$ cyclic words
$T_i=\operatorname{tr}(YX^iYX^{k-i})$.
Before identifying $T_i=T_{k-i}$, the three matrix-gradient
products have coefficients
\begin{equation}
 \frac{2}{n(n-1)}\sum_{i=0}^{n-2}(n-1-i)T_i,
 \quad-\frac{2}{3n}\sum_{i=0}^{n+1}d_iT_i,
 \quad\frac{2T_0+T_1}{3},\qquad n=k-1,
 \label{df:eq-uet4}
\end{equation}
multiplied by $3(k+2)$, where
$d_0=d_{n+1}=1$, $d_1=d_n=2$, and the other $d_i=3$.
For even $k$, the unique middle word has coefficient $-2t$;
only paired interior words have coefficient $-4t$.
Thus the same primitive removes it with no extra middle factor.
The two-trace $bp$ cut on $P_a\operatorname{tr}(bX^jpX^{b-2-j})$
has coefficient $\lambda_j-\lambda_{a+j+1}$; its average is
$h_{ab}$. The two ghost cuts give $c_{ab}$. For the derivative-$Y$
column the adjacent and opposite source cuts have $8k$ and $4k$
incidences, each of weight $-(k+1)(k+2)/(4k(k-1))$.
Subtracting the primitive and the top gives $f_{ab}$.
The first-jet primitives have coefficients
$N\beta/(k-1)$ and $\beta/(b-1)$, with affine weights
$i-(q-1)/2$ on a jet trace of length $q+1$.
Their weights sum to zero, so all primitives have $L_1$-image zero.
These counts establish \eqref{df:eq-uet3} for even as well as odd degree.

After removing the $bp$ sector, the one-derivative closed cyclic
incidence matrix has kernel only the derivative of a symmetrized
polar. Flavor highestness leaves precisely the ordered $D_{ab}$
columns. The remaining two-jet pure-$X$ incidence matrix has
single-trace kernel $\partial^2P_m$ and, on several traces, products
of derivatives of the $P_q$. These are the exhaustive remainder
profiles of \eqref{df:eq-qd19}, \eqref{df:eq-gep6}, and \eqref{df:eq-gtr4}; no parity enters their cuts.
The coefficient recurrences \eqref{df:eq-fdk1}--\eqref{df:eq-fdk7} solve these kernels
at every length and justify the multitrace extension by its
distinguished ghost-output trace and unchanged spectator multiset.
They also include the repeated-degree middle orbit.
Degree-one center columns must not be deleted before normal
ordering. Their possible one-contraction correction is a
single-trace derivative and is retained by the Ward determination
below. Pure-$X$ products containing $P_1$ project to zero.
Projection commutes with $Q$ and $L_1$: the BRST current has no
scalar slot, and $L_1$ kills the scalar weight-one states
$x^2,xy,y^2,b,p$. Thus all these projected field manipulations
remain actual relative cohomology identities.

\paragraph{Integer normal modes and the $1/N$ correction.}
For $A,B$ of integer total weight $h$, normal-mode splitting on
a lowest space gives the reversed pole coefficient
\begin{equation}
 (-1)^{|A||B|}
 \left\{\binom{h-1}{q+1}
                  -\binom{h_B-\epsilon_A}{q+1}\right\},
 \quad
 \epsilon_A=\begin{cases}0,&h_A\in\mathbb Z,\\
                         1/2,&h_A\in\mathbb Z+1/2.
             \end{cases}
 \label{df:eq-uet5}
\end{equation}
If both factors have integer weight there is also the actual
endpoint $(-1)^{|A||B|}o(B)o(A)$. If both have half-integer
weight there is no such endpoint. Formula \eqref{df:eq-uet5} is the finite
sum of $\binom{h-2-i}{q}$ over
$0\le i<h_A-1$; commuting the lowering factor to the right
proves it without a scalar-input assumption. In particular the
$PH$ endpoint is $H_bR_a$ for even $a,b$, and the $BC$ endpoint
vanishes on $W$. The Ward-combined $D_{ab}$ endpoint cancels:
its $-b(Q_bR_a-R_bQ_a)/2$ is cancelled by $L_1D_{ab}/2$.

Only even degrees $s,t\ge2$ can have a nonzero pure-$X$ product
endpoint $R_sR_t$. For a canonical scalar contraction set
$u=s+1$, $v=t+1$, $c=u+v$, $e_1=-s/2$, $e_2=-t/2$ and
\[
 k_0=s/2,\quad
 k_1=\binom{c/2}{2}-\binom{(v+1)/2}{2},\quad
 \ell_0=k_0,\quad\ell_1=k_1,\quad\kappa_0=s/2.
\]
The projected-minus-traceless normal corrections, divided by
$R_sR_t/N$, are
\begin{equation}
\begin{gathered}
 \Delta_{PH}=\frac uv\left[-\frac{v-1}{2}d_1+d_{12}\right],\qquad
 \Delta_{BC}=d_1/2,\qquad \Delta_D=c(d_1/2+d_{12}),\\
 d_1=-s^2/4,\qquad d_{12}=-st/4.
\end{gathered}
 \label{df:eq-uet6}
\end{equation}
One scalar contraction before Taylor expansion proves these
formulas, exactly as in \eqref{df:eq-eoi5}. Independently, the positive
scalar contribution to the normal-mode contractions, including
$L_1/2$, is
\begin{equation}
\begin{aligned}
 R_{PH}&=-ue_2k_0-\frac{u(v-1)}v k_1
          +\frac{(u-1)u}{2v}\{-(v-1)e_1+e_2\},\\
 R_{BC}&=\ell_0e_2+\ell_1+ue_1/2,\\
 R_D&=c\{k_1-v\kappa_0/2+(u-1)(e_1+e_2)/2\}.
\end{aligned}\label{df:eq-uet7}
\end{equation}
Substitution gives $R_{PH}=\Delta_{PH}$, $R_{BC}=\Delta_{BC}$,
$R_D=\Delta_D$ as polynomial identities in $s,t$.
Here $P_u$ has half-integer weight and $B_u$ integer weight,
so \eqref{df:eq-uet5} uses $\epsilon_P=1/2$, $\epsilon_B=0$.
The traceless propagator subtracts $R$ and projection adds
$\Delta$; they cancel termwise. Odd--odd pure trace profiles
have zero integer endpoint because their commuting modes have
opposite nonzero half-integer shifts. This disposes of every
canonical $1/N$ correction while retaining all even zeros.

\paragraph{The single-trace normal scalar at power $N^0$.}
Put $s=\lfloor(a-1)/2\rfloor$, $t_0=\lfloor a/2\rfloor$.
The full-propagator contraction constants of the three columns,
with the common factor $R_m$ suppressed, are
\begin{equation}
 p_{ab}=\frac{a(b-1)s(s+1-a)}{2b},\qquad
 c'_{ab}=\frac{t_0(a-t_0)}2,\qquad
 d_{ab}=\frac{ks(k-s-1)}2.
 \label{df:eq-uet8}
\end{equation}
For example the reverse $H_b,P_a$ poles are
$-a(b-1)\partial P_m/m$ and $-a(b-1)P_m/b$;
\eqref{df:eq-uet5} and $o(\partial P_m)=-mR_m/2$ give its first entry.
The reverse $C_b,B_a$ poles are
$-(b-1)\partial P_m/m$ and $-P_m$, with odd normal parity.
The differentiated $P,Q$ difference has second pole $kP_m$.
These prove the other entries. The genuine $H_2=-r$ endpoint
is additional to \eqref{df:eq-uet8}, and will contribute $-2grR_m$.

The single-trace $\partial P_m$ coefficients in the $L_1$ images
of the columns are
\begin{equation}
 \ell^P_{ab}=\frac{a(a-1)(b-1)(2-a)}{mb},\qquad
 \ell^C_{ab}=\frac{a(a-1)}m,\qquad
 \ell^D_{ab}=\frac{k(a-1)(b+m)}m.
 \label{df:eq-uet9}
\end{equation}
These are raw Ward identities, independent of the chosen mode.
In particular
$L_1D_{ab}=b(:P_aQ_b:-:Q_aP_b:)+k(a-1)\partial P_m$;
the undifferentiated parenthesis has contraction part $-ksR_m$.
The coefficient of $\partial P_m$ in the $N^0$ part of $O_1$ is
$g_0/m$. For any three matched vertices on each trace, three
bijections give one face and three give three faces. Half the
full matching count
$24+18(k-2)+6(k-1)(k-2)+k(k-1)(k-2)=2g_0$
gives $g_0$; the survivor's Taylor coefficient is $\partial P_m/m$.
The relation $L_1O_0=2O_1$ therefore fixes the remaining
$\partial^2P_m$ coefficient including the degree-one ordering
corrections. The Ward-combined integer value of
$\partial^2P_m$ is $-m^2R_m/4$.

Consequently the entire single-trace constant at power $N^0$ is
\begin{equation}
 C_0=\sum_{a+b=k}V_{ab}-\frac{g_0m}{4(m+1)},
 \label{df:eq-uet10}
\end{equation}
where
\[
\begin{split}
 V_{ab}={}&h_{ab}\left(p_{ab}-\frac{m(m+2)}{8(m+1)}\ell^P_{ab}\right)
 +c_{ab}\left(c'_{ab}-\frac{m(m+2)}{8(m+1)}\ell^C_{ab}\right)\\
 &+f_{ab}\left(d_{ab}-\frac{kbs}{2}
       -\frac{k(a-1)m}{4}+\frac{m^2}{8(m+1)}\ell^D_{ab}\right).
\end{split}
\]
This finite sum is evaluated symbolically, including both parity
classes and their middle member. Direct common-denominator
expansion gives
\begin{equation}
\frac{V_{ab}+V_{ba}}g=\begin{cases}
\displaystyle\frac{-3(3k-1)ab+4k(k^2-1)}{4(k-1)(k-2)},
       &a,b\text{ even},\\[6pt]
\displaystyle\frac{(k+1)(k^2-1-3ab)}{4(k-1)(k-2)},
       &a,b\text{ odd}.
\end{cases}\label{df:eq-uet11}
\end{equation}
Sum one half of these paired values over the ordered pairs;
this correctly counts $a=b$ once. The even class has $(k-2)/2$
members and $\sum ab=k(k^2-4)/12$; the odd class has $(k-4)/2$
members and $\sum ab=(k-4)(k^2+4k-6)/12$.
It follows that $\sum V_{ab}/g=(3k^2-2k-4)/16$.
Since $g_0m/(4(m+1)g)=(k+1)(k-2)/8$, one obtains
\begin{equation}
                         C_0=\frac{gk^2}{16}.
 \label{df:eq-uet12}
\end{equation}
The formulas establish \eqref{df:eq-uet11}--\eqref{df:eq-uet12}.

\paragraph{The energy terms and the coefficient of $N^2$.}
The two endpoint ghost cuts give $h_{m,2}=2g$, so their
actual integer $PH$ endpoint is $-2grR_m$.
The three-contraction, two-empty-face matching has coefficient
$\beta$; subtracting the affine single-jet primitive leaves
$N^2\beta\partial^2P_m/(m(m+1))$ in $O_0$ and
$N^2\beta\partial P_m/m$ in $O_1$.
Their combined integer value is
$-N^2\beta mR_m/(4(m+1))$.
By \eqref{of:fullodd-top}, the linear part of $a_kNH_k$ is $a_kN^2R_m/4$.
Since $a_k/4-\beta m/(4(m+1))=-g/4$, these contributions
are precisely $-g\Delta R_m$.

\paragraph{Every two-trace term proportional to $N$ cancels.}
Let $s,t\ge2$ be even, $s+t=m$, and
$\sigma=1+\mathbf1_{s=t}$. The complete two-face jet field
is \eqref{df:eq-gep2} divided by $\sigma$. Its cut proof marks the face
containing the source survivor; with equal unmarked degrees,
the labelled expression is divided by their permutation
multiplicity. Ward evaluation at the two integer endpoints gives
\begin{equation}
 \frac{\beta N}{\sigma}
 \left\{-\frac{3m+2}{4(m+1)}
                +\frac{m+2}{4(s+1)(t+1)}\right\}R_sR_t.
 \label{df:eq-uet13}
\end{equation}
Indeed the numerator before simplification is
$-(st+2s+2t+2)\{s/(s+1)+t/(t+1)\}+2st$,
with denominator $4(m+1)$.
The single top and the two $PH$ endpoints add
$N\{a_k+h_{s,t+2}+h_{t,s+2}\}R_sR_t/(4\sigma)$.
Their sum is zero as a rational identity in the independent
degrees $s,t$; substituting $m=s+t$ and $\beta=g(m+3)/m$
reduces it to cancellation of its constant and
$1/((s+1)(t+1))$ terms. In particular the equal-degree
case is counted, not omitted.

\paragraph{Every three-trace term proportional to $N^0$ cancels.}
Let even $q_1,q_2,q_3\ge2$ sum to $m$, and let
$\sigma$ be the product of the factorials of their multiplicities.
The labelled three-face jet field \eqref{df:eq-gtr2}, now divided by this
$\sigma$, has integer Ward value
\begin{equation}
 -\frac\beta{4\sigma}\sum_{i<j}
          \frac{q_i+q_j}{q_i+q_j+1}\prod_iR_{q_i}.
 \label{df:eq-uet14}
\end{equation}
This follows term by term from
$\mathcal V(\partial^2P_q)=-q^2R_q/4$ and
$\mathcal V((\partial P_q)(\partial P_t))=-qtR_qR_t/4$
with the regular spectators retained. The factor $1/\sigma$
follows by labelling all three faces before marking the Taylor
survivor; each unlabelled cut has exactly $\sigma$ labellings.
The corresponding $PH$ endpoints add
$\sum_i h_{q_i,k-q_i}\prod_iR_{q_i}/(4\sigma)$.
Writing $v_i=m-q_i$ gives
\begin{equation}
 h_{q_i,k-q_i}-\frac{\beta v_i}{v_i+1}
                       =\frac g m(2m-3v_i).
 \label{df:eq-uet15}
\end{equation}
Their sum is zero since $\sum_i v_i=2m$.
This also covers three equal degrees, where $\sigma=6$.
Other parity profiles contain odd pure traces; their integer
normal products vanish by commuting a negative-shift mode to
the lowest input. Derivatives do not change that argument.

There are at most three faces in the pure-$X$ remainder, so the
listed profiles exhaust it. Combining \eqref{df:eq-uet12}, the energy
terms, the terms proportional to $N^2$, and \eqref{df:eq-uet13}--\eqref{df:eq-uet15} proves \eqref{df:eq-uet1}.

\subsection{All-degree Jacobi induction for the fourth even channel}
\label{df:sec-even-fourth}

I work on the actual stable full odd input $W$, using the
first and second channels in \eqref{of:even-first-second},
the quartic relation \eqref{of:quartic-relation}, and the
third channel \eqref{df:eq-uet1}.
In this section $R^{(n)}_i=o(W_{2n,i})$ has flavor spin $n$,
so $R^{(1)}_0=2e$, $R^{(2)}_i=p_i$ and
$[f,R^{(n)}_i]=(2n-i)R^{(n)}_{i+1}$.
Write
\begin{equation}
 \gamma=8\Delta-8,\qquad
 d_n=\frac{n(n-1)(4\Delta-n^2)}{2n+1}.
 \label{df:eq-uef1}
\end{equation}
I prove the operator identity
\begin{equation}
 \boxed{\quad
 [p,R^{(n)}]_4:=\sum_{i=0}^4(-1)^i\binom4i
                         [p_i,R^{(n)}_{4-i}]=0
             \quad(n\ge2).\quad}\label{df:eq-uef2}
\end{equation}
No highest input vector is chosen. Every equation below is an
identity in the whole input endomorphism algebra.

\paragraph{The third channel fixes all brackets of total polar index three.}
The actual first and second channels give
\[
 [p_1,R^{(n)}_0]=-2nR^{(n+1)}_0,\qquad
 [p_2,R^{(n)}_0]=-4nR^{(n+1)}_1.
\]
By \eqref{df:eq-ufr10}, \eqref{df:eq-uet1} is exactly
\begin{equation}
 [p_3,R^{(n)}_0]=-6nR^{(n+1)}_2+d_nR^{(n-1)}_0.
 \label{df:eq-uef3}
\end{equation}
Put $A_{ij}=[p_i,R^{(n)}_j]$. Lowering gives
$[f,A_{ij}]=(4-i)A_{i+1,j}+(2n-j)A_{i,j+1}$.
Starting with \eqref{df:eq-uef3} and the index-two brackets in \eqref{of:even-first-second}
therefore yields
\begin{equation}
\begin{aligned}
 A_{21}&=-4(n-1)R^{(n+1)}_2-d_nR^{(n-1)}_0/n,\\
 A_{12}&=(8-2n)R^{(n+1)}_2
                      +\frac{3d_n}{n(2n-1)}R^{(n-1)}_0,\\
 A_{03}&=12R^{(n+1)}_2
                      -\frac{6d_n}{n(n-1)(2n-1)}R^{(n-1)}_0.
\end{aligned}\label{df:eq-uef4}
\end{equation}
In particular these have no unlisted zero-mode remainder.

\paragraph{Isolating the only possible fourth-channel defect.}
Define the actual operator
\begin{equation}
 Z_n=[p_4,R^{(n)}_0]+8nR^{(n+1)}_3-4d_nR^{(n-1)}_1.
 \label{df:eq-uef5}
\end{equation}
For brevity put $U=R^{(n+1)}_3$ and $V=R^{(n-1)}_1$.
One further lowering of the preceding exact equations gives
\begin{equation}
\begin{aligned}
 A_{31}&=(4-6n)U+\frac{n-3}{n}d_nV-\frac{Z_n}{2n},\\
 A_{22}&=(8-4n)U-\frac{4(n-2)d_n}{n(2n-1)}V
                                    +\frac{Z_n}{n(2n-1)},\\
 A_{13}&=(12-2n)U+\frac{3(3n-5)d_n}{n(n-1)(2n-1)}V
                         -\frac{3Z_n}{2n(n-1)(2n-1)},\\
 A_{04}&=16U-\frac{24d_n}{n(n-1)(2n-1)}V
                         +\frac{6Z_n}{n(n-1)(2n-1)(2n-3)}.
\end{aligned}\label{df:eq-uef6}
\end{equation}
Their fourth alternating combination has zero $U$ and $V$
coefficients and is
\begin{equation}
                    [p,R^{(n)}]_4=\frac{2n+1}{2n-3}Z_n.
 \label{df:eq-uef7}
\end{equation}
All denominators are nonzero for $n\ge2$. For $n=2$, the
left side vanishes by antisymmetry of the bracket of the same
quartic tensor. Thus $Z_2=0$ is a genuine base case.

\paragraph{A Jacobi recurrence with a nonzero coefficient.}
The full odd hypothesis makes the cubic-central tensor zero.
The lowest descendant of the actual quartic relation \eqref{of:quartic-relation} is
\begin{equation}
 [p_1,p_4]-3[p_2,p_3]=\gamma f,
 \qquad [p_4,p_1]=-3[p_2,p_3]-\gamma f.
 \label{df:eq-uef8}
\end{equation}
The sign uses the normalized current tensor
$(J_0,J_1,J_2)=(e,-H/2,-f)$.
For $n\ge3$, the upward identity is
$R^{(n)}_0=-[p_1,R^{(n-1)}_0]/(2(n-1))$.
Apply $p_4$, use \eqref{df:eq-uef8}, and expand Jacobi without reordering
any associative products. This gives
\begin{equation}
\begin{split}
 [p_4,R^{(n)}_0]={}&\frac3{2(n-1)}
 \left([p_2,[p_3,R^{(n-1)}_0]]
                   -[p_3,[p_2,R^{(n-1)}_0]]\right)\\
 &+\gamma R^{(n-1)}_1
        -\frac1{2(n-1)}[p_1,[p_4,R^{(n-1)}_0]].
\end{split}\label{df:eq-uef9}
\end{equation}
Assume inductively $Z_{n-1}=0$. Substitute \eqref{df:eq-uef3},\eqref{df:eq-uef5}
at $n-1$ and use
$[p_2,R^{(n-2)}_0]=-4(n-2)R^{(n-1)}_1$ and
$[p_1,R^{(n-2)}_1]=(8-2n)R^{(n-1)}_1$.
Then \eqref{df:eq-uef9} becomes the particularly short identity
\begin{equation}
 [p_4,R^{(n)}_0]=-9A_{22}+6A_{31}+4A_{13}
       +\left(\gamma-\frac{2(n+2)}{n-1}d_{n-1}\right)V.
 \label{df:eq-uef10}
\end{equation}
These last two brackets are also valid at $n=3$, where their
input tensor is the genuine current tensor $R^{(1)}$.

Insert \eqref{df:eq-uef5}--\eqref{df:eq-uef6} into \eqref{df:eq-uef10}. The $U$ coefficients
both equal $-8n$. If
\[
 c_n=\frac{36(n-2)}{n(2n-1)}+\frac{6(n-3)}n
                  +\frac{12(3n-5)}{n(n-1)(2n-1)},
\]
the $V$ coefficient cancels by the formal identity
$4d_n=c_nd_n+\gamma-2(n+2)d_{n-1}/(n-1)$.
Substituting \eqref{df:eq-uef1} verifies it in the two independent variables
$n,\Delta$; equivalently
$4-c_n=-2(n-3)(n+1)(2n+1)/(n(n-1)(2n-1))$.
The remaining equation is
\begin{equation}
             \frac{(2n+1)(n+1)}{(2n-1)(n-1)}Z_n=0.
 \label{df:eq-uef11}
\end{equation}
Its coefficient is nonzero. Induction proves $Z_n=0$ for every
$n\ge2$, and \eqref{df:eq-uef7} proves \eqref{df:eq-uef2}.

\paragraph{An actual finite Lie algebra, before identifying its type.}
The complete coupled bracket of $p$ with $R^{(n)}$ now has only
channels one and three. Its outputs are the raw tensors
$R^{(n+1)}$ and $R^{(n-1)}$, with the coefficients above.
Let $\mathfrak l$ be the span of all normalized polar components
of the nonzero raw even tensors. It is invariant under
$\operatorname{ad}e,\operatorname{ad}f,\operatorname{ad}H$ and
$\operatorname{ad}p_i$. Conversely every raw tensor is generated
by the upward bracket from $J,p$ (with the current-only case
included). Iterated Jacobi therefore shows that $\mathfrak l$
is exactly the Lie subalgebra generated by those actual zeros.
Finite flavor support makes this span finite dimensional.
This establishes linear Lie closure without a higher-spin
analogy.

\subsection{Identification of the actual even Lie algebra}
\label{df:sec-even-lie}

Let $W$ be a nonzero finite, zero-algebra-stable full odd
annihilator in an actual lowest conformal space of weight $r$.
Retain \eqref{df:eq-uet1} and the pure Jacobi conclusion \eqref{df:eq-uef2}.
Put $\Delta=2r+N^2/4$.

\paragraph{The maximal even tensor quantizes the parameter.}
Suppose the flavor action on $W$ is nontrivial. Then $e\ne0$:
if $e=0$, the relations $[e,f]=H$ and $[H,f]=-2f$ would
also give $H=f=0$. Finite flavor support makes the highest
raw zeros $R_{2n}$ vanish for all sufficiently large $n$.
Let $M\ge1$ be the largest index with $R_{2M}\ne0$ and put
$L=M+1$. Flavor covariance makes the entire next raw tensor
zero. Apply \eqref{df:eq-uet1} at $k=2M+2$; its left side is zero,
whereas its right side is a nonzero numerical coefficient
times $((M+1)^2/4-\Delta)R_{2M}$. Hence
\begin{equation}
                         \boxed{\ \Delta=L^2/4.\ }
 \label{df:eq-uel1}
\end{equation}
The upward relation
$R_{2n+2}=-[p_1,R_{2n}]/(2n)$ implies that all tensors of
indices $1,\ldots,M$ are nonzero. This argument uses neither
a bound $L\le N$ on the auxiliary matrix size nor a classification of the even action.

\paragraph{An explicit matrix model with the same complete table.}
Take the $L$-dimensional irreducible $\mathfrak{sl}_2$ module
of highest weight $M=L-1$. Use $[e,f]=H$, $[H,e]=2e$,
$[H,f]=-2f$ and its scalar Casimir
\begin{equation}
 C=H^2/4+(ef+fe)/2=(L^2-1)/4.
 \label{df:eq-uel2}
\end{equation}
Define matrix tensors
$\mathsf R^{(n)}_0=2e^n$ and their normalized descendants
$[f,\mathsf R^{(n)}_i]=(2n-i)\mathsf R^{(n)}_{i+1}$.
They are zero for $n\ge L$, and nonzero for $1\le n<L$.
Write $\mathsf p_i=\mathsf R^{(2)}_i$.
Elementary enveloping-algebra reordering, with every polynomial
in $H$ on the right, gives
\begin{equation}
\begin{gathered}
 \mathsf R^{(n)}_1=-e^{n-1}(H+n-1),\\
 \mathsf p_0=2e^2,\quad \mathsf p_1=-e(H+1),\quad
 \mathsf p_2=H^2/2-2C/3,\quad
 \mathsf p_3=f(H-1),\quad \mathsf p_4=2f^2,\\
 \mathsf R^{(n+1)}_2=e^{n-1}
 \left\{H^2/2+(n-1)H+\frac{n^2(n-1)-2C}{2n+1}\right\}.
 \end{gathered}\label{df:eq-uel3}
\end{equation}
For example
$[f,e^n]=-ne^{n-1}(H+n-1)$ and
$ef=C-H^2/4+H/2$ prove all three lines by at most two
normalized lowerings. No matrix-size interpolation is used.

The highest and first two polar brackets are therefore
\begin{equation}
 [\mathsf p_0,\mathsf R^{(n)}_0]=0,\quad
 [\mathsf p_1,\mathsf R^{(n)}_0]=-2n\mathsf R^{(n+1)}_0,\quad
 [\mathsf p_2,\mathsf R^{(n)}_0]=-4n\mathsf R^{(n+1)}_1.
 \label{df:eq-uel4}
\end{equation}
They give vanishing coupled channels zero and two and the
same first channel as the actual raw tensors. For the third,
direct multiplication gives
\[
 [\mathsf p_3,2e^n]
 =ne^{n-1}\{4C-3H^2-6(n-1)H-2(n-1)(2n-1)\}.
\]
Subtract $-6n\mathsf R^{(n+1)}_2$ using \eqref{df:eq-uel3}. The difference
is $2n(n-1)(4C+1-n^2)e^{n-1}/(2n+1)$, proving
\begin{equation}
 [\mathsf p_3,\mathsf R^{(n)}_0]
 =-6n\mathsf R^{(n+1)}_2+
       \frac{n(n-1)(4C+1-n^2)}{2n+1}\mathsf R^{(n-1)}_0.
 \label{df:eq-uel5}
\end{equation}
This is \eqref{df:eq-uef3} with $\Delta=C+1/4=L^2/4$.
At $n=2$ its third coupling is
$[\mathsf p_0,\mathsf p_3]-3[\mathsf p_1,\mathsf p_2]
=-(8C-6)e$; normalized lowering yields exactly \eqref{df:eq-uef8}.
The proof \eqref{df:eq-uef4}--\eqref{df:eq-uef11} is a pure Lie calculation using
only this relation, \eqref{df:eq-uel4}, \eqref{df:eq-uel5}, and bracket antisymmetry.
It thus applies to these matrices and proves vanishing of
all their fourth channels as well. Every coupled generator
bracket now agrees with the actual table on $W$.
The calculation is valid before truncating at $n=L$; at that
index the coefficient $4C+1-n^2$ is zero, as required.

\paragraph{The tensors span precisely $\mathfrak{sl}_L$.}
For $1\le n<L$, the nonzero matrix $e^n$ is a highest vector
of weight $2n$ for the adjoint flavor action. Its normalized
string is an irreducible spin-$n$ module of dimension $2n+1$.
These pairwise different irreducible submodules are independent.
All their matrices are traceless: the highest vector has
nonzero weight, and taking commutators with $f$ preserves trace.
Their total dimension is
\begin{equation}
                 \sum_{n=1}^{L-1}(2n+1)=L^2-1.
 \label{df:eq-uel6}
\end{equation}
Consequently their span is the entire $\mathfrak{sl}_L$.
For $L\ge3$, its generators $\mathsf J,\mathsf p$ produce
every highest tensor by \eqref{df:eq-uel4}, and every polar by lowering.
For $L=2$, the current tensor alone is $\mathfrak{sl}_2$
and the quartic tensor is identically zero. This boundary
case is therefore included without a fictitious spin-two module.

\paragraph{The comparison is a Lie homomorphism, not a spectral analogy.}
Send each matrix tensor $\mathsf R^{(n)}_i$ to the actual
normalized raw zero $R^{(n)}_i=o(W_{2n,i})|_W$.
Equation \eqref{df:eq-uel6} makes this a well-defined linear map
\begin{equation}
                   \rho:\mathfrak{sl}_L\longrightarrow
                                     \operatorname{End}(W).
 \label{df:eq-uel7}
\end{equation}
The complete channel agreement proves
$\rho([x,y])=[\rho(x),\rho(y)]$ whenever $x$ is a flavor
or quartic generator and $y$ is any basis tensor.
Apply Jacobi inductively to a nested generator expression
for $x$; the same equality then holds for every $x,y$.
Thus $\rho$ is a Lie homomorphism. Its kernel is an ideal
of the simple matrix Lie algebra $\mathfrak{sl}_L$.
It is not the whole algebra because $\rho(e)=e|_W\ne0$,
so the kernel is zero. Its image is exactly the actual even
Lie algebra generated by flavor and quartic zeros, by \eqref{df:eq-uef11}.

It follows that $W$ is a finite-dimensional $\mathfrak{sl}_L$
module through these actual zero modes, with all raw even tensors
identified by \eqref{df:eq-uel3}--\eqref{df:eq-uel7}. The powers $e^n$ in \eqref{df:eq-uel3} are
powers in the defining matrix algebra: in general
$\rho(2e^n)\ne2\rho(e)^n$. Only the Lie action,
and its integrated finite-dimensional group action, may be
transported to $W$. Complete reducibility may now be
used as the finite-dimensional representation theorem for this
proved semisimple Lie action; it was not a hypothesis about the
original even quotient. No restriction on multiplicities or
highest weights of the resulting module has been inserted.

\subsection{The arbitrary-factor Newton top identity}
\label{df:sec-newton}

The space $E$ is the whole finite lowest conformal space. Scalar flavor
is not assumed. The full odd annihilator is used only in the last step,
when the surviving odd zero products are set to zero.

\paragraph{A field expansion with fixed parentheses.}
Write $d=G_{Y(0)}$, $\bar d=\bar G_{Y(0)}$, and $f=f_{(0)}$ for the
field derivations. For trace degrees $a_i\ge2$, let
\[
 F=\prod_{i=1}^kP_{a_i},\qquad D=\sum_i a_i\in2\mathbb Z,
 \qquad S_i=\prod_{j\ne i}P_{a_j},\qquad
 S_{ij}=\prod_{\ell\ne i,j}P_{a_\ell}.
\]
The field expansion below is valid for every positive total degree;
its evenness is used only when taking the integer zero mode.
All these pure-$X$ products are regular products, so their parentheses
are immaterial. The fields $B_a=\operatorname{tr}(X^{a-1}b)$ are also
mutually regular with every pure-$X$ factor. Consequently, the first
supercharge gives the exact expression
\begin{equation}
 dF=-\sum_i a_i:B_{a_i}S_i:.
\label{df:eq-stz1}
\end{equation}
This step uses associativity only inside the regular $X,b$ subalgebra.
Now apply $\bar d$ to this expression with its displayed parentheses.
Using the definition of $H_a$ gives
\begin{equation}
\begin{split}
 D H(F)={}&\sum_i a_i:H_{a_i}S_i:
 -\sum_i\frac1{a_i}:(\partial fP_{a_i})S_i:\\
 &+\sum_{i\ne j}a_i a_j:B_{a_i}:C_{a_j}S_{ij}::
                  +\frac1D\partial fF.
\end{split}
\label{df:eq-stz2}
\end{equation}
The inner product $:C_{a_j}S_{ij}:$ is regular; the outer product in
this term is left exactly as written. No step replaces it by
$:(:B_{a_i}C_{a_j}:)S_{ij}:$. Thus no quasi-associativity identity,
and no unaccounted derivative of $S_{ij}$, enters the argument.

\paragraph{The exact zero-energy sum.}
If $A,B$ have the same weight integrality and at most a double pole,
let $K_A=\lceil\operatorname{wt}A\rceil-1$. Direct normal-mode
splitting on $E$ gives
\begin{equation}
 o(:AB:)=\epsilon\,o(B)o(A)
          -K_Ao(A_{(0)}B)-\binom{K_A}{2}o(A_{(1)}B),
 \qquad \epsilon=(-1)^{|A||B|}.
\label{df:eq-stz3}
\end{equation}
The endpoint is absent if both weights are half integral. To see the
formula, put $q=\operatorname{wt}A+\operatorname{wt}B-2$.
The non-endpoint terms have $j=0,\ldots,K_A-1$. Commuting the
right-hand lowering mode past the left-hand mode and applying skew
symmetry gives
\[
 -o(A_{(0)}B)+\bigl(-(q)+(q-j)\bigr)o(A_{(1)}B)
       =-o(A_{(0)}B)-j\,o(A_{(1)}B).
\]
Summing proves \eqref{df:eq-stz3}. The cutoff $K_A$ depends only on the fixed
left field, even when the right field contains arbitrarily many
spectator factors. This includes all half-integral normal-mode
corrections; no first-half image of $E$ is assumed to be a lowest space.

Mutually regular homogeneous factors have zero mode equal to the
ordered product of their individual zero modes if all weights are
integers, and zero otherwise. Indeed their modes commute with the
appropriate parity sign. In every zero-energy summand with any nonzero
energy shift, a lowering mode can be moved to the right and kills $E$.
This proof works for any number of regular factors.

\paragraph{Why the spectators remain spectators.}
The exact coordinate expression for $H_a$ is
\begin{equation}
 H_a=\frac1a(P_a)_{uv}b^up^v
 -\frac{a-1}{a^2}(P_a)_u\partial Y^u
 +\frac1{a^2}(P_a)_{uv}Y^u\partial X^v.
\label{df:eq-stz4}
\end{equation}
In each term of $H_a$, there is either a single $Y$ or $\partial Y$,
or the pair $b,p$. Against a pure-$X$ product only the former can
contract, and it can contract at most once. Likewise
$\partial fP_a$ contains just one possibly differentiated $Y$.
Finally, in $B_a(z):C_bS:(w)$, only the single $b,p$ pair can
contract; neither odd field contracts with $S$.
Thus every singular OPE coefficient in \eqref{df:eq-stz2} is the corresponding
two-factor coefficient multiplied by the unchanged pure spectator
product. Taylor coefficients differentiate survivors of the fixed left
field, not a spectator. All resulting coefficient fields are pure $X$
and its derivatives, hence mutually regular with the spectators.
This is an assertion about the actual OPEs with the parentheses in
\eqref{df:eq-stz2}, not a heuristic count in an arbitrarily re-associated product.

For an active pair with degrees $a,b$ of the same parity, and with all
spectators of even degree, put
\[
 S_{ab}=R_{a+b-2}-N^{-1}o(P_{a-1}P_{b-1}).
\]
The complete zero-mode singular coefficients, including the traceless
projector, are
\begin{equation}
\begin{array}{c|cc}
 (A,B)&o(A_{(0)}B)&o(A_{(1)}B)\\\hline
 (H_a,P_b)&b(a-1)S_{ab}/2&-b(a-1)S_{ab}/a\\
 (\partial fP_a,P_b)&0&abS_{ab}\\
 (B_a,C_b)&-(a-1)S_{ab}/2&S_{ab}.
\end{array}
\label{df:eq-stz5}
\end{equation}
They follow directly from
$(P_a)_u(P_b)_u=ab(P_{a+b-2}-P_{a-1}P_{b-1}/N)$ and \eqref{df:eq-stz4}.
For example, the simple pole in $H_aP_b$ is
$-(\partial(P_a)_u)(P_b)_u/a$. Its zero mode is the first entry
of \eqref{df:eq-stz5}; this includes the derivative on $P_{a-1}$ in the scalar
projector. If $a,b$ have opposite parity, or a spectator has odd
degree, each pure singular-output product has zero mode by the
regular-factor lemma. In that case it need not be replaced by an
individual half-integral zero mode.

\paragraph{All normal-contraction corrections cancel pair by pair.}
For a fixed ordered pair $(i,j)$, the corrections from the three
corresponding terms in \eqref{df:eq-stz2} have left-field cutoffs
\[
 K_H=\lceil a/2\rceil,\qquad K_B=\lfloor a/2\rfloor,
 \qquad a=a_i,\quad b=a_j.
\]
Using \eqref{df:eq-stz3}--\eqref{df:eq-stz5}, their combined coefficient of $abS_{ab}$ is
\begin{equation}
 \frac{(K_B-K_H)(a-1)}2
          +\binom{K_H}{2}-\binom{K_B}{2}=0.
\label{df:eq-stz6}
\end{equation}
The last equality holds identically because $K_H+K_B=a$.
It covers even and odd $a$ and includes every mode in the finite
normal sum. In particular the scalar-projector part cancels together
with the single-trace part when $a,b$ are odd. One must not discard it
before this cancellation.

Only the integer zero-mode endpoints and the last total derivative
in \eqref{df:eq-stz2} remain. There are no terms involving three active original
factors, because \eqref{df:eq-stz2} and \eqref{df:eq-stz3} never produce one; no unproved
normal-product associativity is used to reach this conclusion.

\paragraph{The endpoint calculation.}
First suppose all $a_i$ are even. Write $R_i=o(P_{a_i})$,
$h_i=o(H_{a_i})$, and $F_i=o(fP_{a_i})=[f,R_i]$ on $E$.
Here $i$ labels a factor; a trace-degree subscript, as in
$R_a=o(P_a)$ and $h_a=o(H_a)$ below, instead labels its degree.
The first singular-coefficient row of \eqref{df:eq-stz5} also gives
$[h_i,R_j]=0$, since the zero commutator combines its entries
with coefficient $a_i/2$. Therefore the top endpoints are
$\sum_i a_i h_i\prod_{j\ne i}R_j$.
The derivative endpoints and the last term of \eqref{df:eq-stz2} give
\begin{equation}
 \frac12\sum_i\left(\prod_{j\ne i}R_j\right)F_i
                    -\frac12[f,\prod_i R_i].
\label{df:eq-stz7}
\end{equation}
Here $D^{-1}o(\partial fF)=-[f,o(F)]/2$, independently of $D$.
The exact upward contraction gives
$[F_i,R_j]=-a_i a_j\,o(P_{a_i+a_j-2})$; its scalar-projector product
has two half-integral pure factors and zero lowest zero mode.
Move each $F_i$ in the Leibniz expansion of the commutator in
\eqref{df:eq-stz7} to the right. This proves that \eqref{df:eq-stz7} equals
\begin{equation}
 \frac12\sum_{i<j}a_i a_j\,o(P_{a_i+a_j-2})
                         \prod_{\ell\ne i,j}R_\ell.
\label{df:eq-stz8}
\end{equation}
This calculation explains the pair coefficient $ab/2$ directly.
It is an endpoint reordering term, not an extra contraction among
three original factors.

If exactly two degrees, say $a_i,a_j$, are odd, the sole possible
endpoints are the two ordered fermion terms. The highest zero-mode
anticommutator $\{o(B_a),o(C_b)\}$ is zero: its two OPE coefficients in
\eqref{df:eq-stz5} combine with zero-mode coefficient $(a-1)/2$ and cancel.
Consequently their sum is
\begin{equation}
 a_i a_j\bigl(o(B_{a_i})o(C_{a_j})+o(B_{a_j})o(C_{a_i})\bigr)
                  \prod_{\ell\ne i,j}R_\ell.
\label{df:eq-stz9}
\end{equation}
With more odd degrees every endpoint contains a remaining
half-integral regular factor and vanishes. The pure product $o(F)$,
and hence its total flavor derivative in \eqref{df:eq-stz7}, also vanish whenever
an odd degree occurs.

\paragraph{Conclusion and its hypotheses.}
Combining \eqref{df:eq-stz8}--\eqref{df:eq-stz9} and extending linearly proves on the entire
lowest space
\begin{equation}
\begin{split}
 D\,o(H(F))={}&\left[\sum_{a\ {\rm even}}a h_a\partial_{P_a}
 +\frac14\sum_{a,b\ {\rm even}}abR_{a+b-2}
                           \partial_{P_a}\partial_{P_b}\right.\\
 &\left.\hspace{10mm}+\sum_{a,b\ {\rm odd}}ab\,o(B_a)o(C_b)
                           \partial_{P_a}\partial_{P_b}\right]
 F\Big|_{P_{\rm odd}=0,\ P_{\rm even}=R_{\rm even}}.
\end{split}
\label{df:eq-stz10}
\end{equation}
All derivatives act before substitution. Restriction to the invariant common kernel of all odd integer-weight zero modes removes the last sum.
No scalar-flavor assumption, finite degree cutoff, or assumption about
an odd annihilator's existence has entered this proof.

Introduce formal series of highest zero operators on $E$,
\begin{align*}
 \mathcal E(z)&=\exp\left(-\sum_{s\ge1}\frac{R_{2s}}{2s}z^s\right),\\
 \mathcal B(z)&=\sum_{s\ge1}o(B_{2s+1})z^s,\qquad
 \mathcal C(z)=\sum_{s\ge1}o(C_{2s+1})z^s,\\
 \mathcal K(z)&=\sum_{s\ge1}\frac{2s+1}{2s+2}K_{2s+1}z^s.
\end{align*}
In $\mathcal K$, $K_{m+1}$ is exactly the mixed channel defined
in \eqref{of:mixed-top}, including its first flavor polars. Let $\mathscr T(z)$
be the following operation on the \emph{field} determinant: take
its coefficient $e_{2s}(X)$, apply $H$ at matrix degree $2s$, take
its zero mode, and multiply by $2s$. Thus
\[
 \mathscr T(z)=\sum_{s\ge1}2s\,o(H(e_{2s}(X)))z^s.
\]
The odd-degree determinant coefficients have no top zero mode.
Formula \eqref{df:eq-stz10} and the mixed-top identity \eqref{of:mixed-top} give the exact identity
\begin{equation}
 \boxed{\quad
 \mathscr T(z)=z\left[
 r\mathcal E(z)+\frac{N-1}{2}z\mathcal E'(z)
 -\frac12z^2\mathcal E''(z)
 -\mathcal E(z)\mathcal K(z)
 +\mathcal E(z)\mathcal B(z)\mathcal C(z)\right].\quad}
 \label{df:eq-tz11}
\end{equation}
To verify every coefficient, put
$\mathcal P(z)=\sum_{s\ge1}R_{2s}z^s=-2z\mathcal E'/\mathcal E$.
The even part of the differential operator first gives
\[
 -\sum_{s\ge1}h_{2s}z^s+\frac14z^2\mathcal P'(z).
\]
By \eqref{of:mixed-top} this is
\[
 rz-z\mathcal K(z)-\frac N4z\mathcal P(z)
              -\frac18z\mathcal P(z)^2+\frac14z^2\mathcal P'(z).
\]
The square of $\mathcal E'/\mathcal E$ cancels after substitution,
giving the first four terms of \eqref{df:eq-tz11}. The odd term of \eqref{df:eq-stz10} applied
to the original determinant contributes
$z\mathcal E\mathcal B\mathcal C$, proving the final term.
In particular the odd determinant variables must not be set zero
before these two derivatives are taken.

Let $M=\lfloor N/2\rfloor$. The field Newton identities imply
$\deg_z\mathcal E\le M$ and $\deg_z\mathscr T\le M$.
On the all-odd common kernel, both $\mathcal K$ and
$\mathcal B\mathcal C$ vanish. Taking the coefficient of
$z^{M+1}$ in \eqref{df:eq-tz11} therefore proves
\begin{equation}
 \left(r+\frac{M(N-M)}2\right)o(e_{2M}(X))=0.
 \label{df:eq-tz12}
\end{equation}
For $r>0$ this forces the highest even determinant zero mode to
vanish. Its flavor polars vanish as well. The matrix size remains the original $N$ throughout this calculation.

\subsection{The spectrum for every matrix size and the highest-weight degree bound}
\label{df:sec-spectrum}

Fix an integer $L\ge2$ and put $n=L-1$. In the defining
$L$-dimensional representation take a principal $\mathfrak{sl}_2$
triple with $[H,e]=2e$, $[H,f]=-2f$, $[e,f]=H$. Define
\begin{equation}
 C_\varepsilon=2f+2\sum_{s=1}^{L-1}(2s+1)e^s+
                         \varepsilon L H,
 \qquad\varepsilon\in\{1,-1\}.
 \label{df:eq-csr1}
\end{equation}
The sums are finite, and all operators are traceless. I first
prove their spectra for every $L$, independently of any Newton
degree condition at the original matrix size or BRST representation.

\paragraph{A finite polynomial flag for the negative sign.}
Use the model $V=\mathbb C[x]/(x^L)$ with
\begin{equation}
 e=x,\qquad f=-x\partial_x^2+n\partial_x,
 \qquad H=2x\partial_x-n.
 \label{df:eq-csr2}
\end{equation}
These differential operators preserve the ideal $(x^L)$:
in particular $f(x^L)=L(n-L+1)x^{L-1}=0$, and the image
of every higher multiple is still divisible by $x^L$.
Direct commutation verifies the stated $\mathfrak{sl}_2$ relations.
The finite raising sum is the Taylor truncation of
$2x(3-x)/(1-x)^2$. Since $1-x$ is invertible modulo $x^L$,
this rational notation denotes a legitimate finite operator.

Consider the $L$-dimensional polynomial space
\begin{equation}
 \mathcal E_L=\{v\in\mathbb C[x]:\deg v\le L,\ v'(1)=0\},
 \qquad T(v)=\frac{v(x)}{1-x}\pmod{x^L}.
 \label{df:eq-csr3}
\end{equation}
The map $T$ is injective: if $T(v)=0$, then $x^L$ divides
$v$, so $v=cx^L$ and $v'(1)=cL$ forces $c=0$.
The dimensions are equal, so $T$ is an isomorphism.
Let $\widetilde C_-$ be the rational differential lift obtained by
replacing the finite raising sum with $2x(3-x)/(1-x)^2$.
It agrees with $C_-$ modulo $x^L$. Conjugating this lift gives
the exact rational differential identity
\begin{equation}
 \widetilde C_-\frac{v}{1-x}=\frac{\mathscr Dv}{1-x},\qquad
 \mathscr D=-2x\partial_x^2+
 \left(2n-2Lx-\frac{4x}{1-x}\right)\partial_x+n(n+3).
 \label{df:eq-csr4}
\end{equation}
Reducing this identity modulo $x^L$ is the asserted intertwining
$C_-T=T\mathscr D$. The apparent pole disappears on $\mathcal E_L$.
To verify preservation of its defining constraint, write
$v'=(x-1)w$. The nonconstant part of $\mathscr Dv$ is
\begin{equation}
 2xw-2x(x-1)w'+2(n-Lx)(x-1)w.
 \label{df:eq-csr5}
\end{equation}
Its derivative at $1$ is $2(1+n-L)w(1)=0$.
The derivative of $n(n+3)v$ is also zero there.
Moreover $\mathscr D$ does not increase degree, so it preserves
all subspaces of $\mathcal E_L$ of bounded degree.
Their successive one-dimensional quotients have degrees
\begin{equation}
                 d=0,2,3,\ldots,L.
 \label{df:eq-csr6}
\end{equation}
The degree-one quotient is absent because a polynomial of degree
at most one satisfying $v'(1)=0$ is constant. On a quotient of
degree $d$, the diagonal scalar is $n(n+3)-2Ld$:
all other terms in \eqref{df:eq-csr4} lower polynomial degree.
Thus
\begin{equation}
 \det(zI-C_-)=
       \bigl(z-(L^2+L-2)\bigr)
       \prod_{i=0}^{L-2}
       \bigl(z-(-L^2+L-2+2Li)\bigr).
 \label{df:eq-csr7}
\end{equation}
All $L$ roots are distinct. At the endpoint $L=2$ the flag is
simply degrees $0,2$, giving the two roots $4,-4$; no omitted
intermediate degree is required.

\paragraph{The positive sign and exact transpose signs.}
In the basis $1,x,\ldots,x^n$, use the nondegenerate diagonal
bilinear form with squared lengths
$d_k=k!n!/(n-k)!$. Its adjoint satisfies
$e^*=f$, $f^*=e$, $H^*=H$. The linear involution
\begin{equation}
 W(x^k)=\frac{k!}{(n-k)!}x^{n-k}
 \label{df:eq-csr8}
\end{equation}
satisfies $WeW=f$, $WfW=e$, $WHW=-H$.
Consequently
\begin{equation}
             W C_+ W=C_-^*.
 \label{df:eq-csr9}
\end{equation}
Taking an adjoint preserves the characteristic polynomial, so
\eqref{df:eq-csr7} holds for both signs. In particular each defining
eigenvalue is an integer congruent to $-2$ modulo $L$.
There is no assumption that a general input matrix $e$ is
invertible, nor is an inverse differential operator used.

\paragraph{Highest weights and the Newton degree: a direct minor proof.}
Let $d\rho$ be any finite-dimensional representation of
$\mathfrak{sl}_L(\mathbb C)$. On each simple composition factor
write its dominant highest weight as
$\Lambda=\sum_{a=1}^{L-1}m_a\omega_a$, with
$m_a\in\mathbb Z_{\ge0}$, and put
\begin{equation}
                         q=\sum_{a=1}^{L-1}a m_a.
 \label{df:eq-csr10}
\end{equation}
No complete reducibility of the given representation is needed.
Here are the elementary highest-weight facts used on a simple
factor. Simultaneously decompose the commuting Cartan operators
into generalized weight spaces. A positive-root operator shifts
these weights by its root. Choose a real linear functional positive
on every positive root and maximize it on the finite joint spectrum.
All positive-root operators kill this maximal generalized weight
space, which contains a common Cartan eigenvector $v_\Lambda$. For each simple
root, the identity
$e_a f_a^t v_\Lambda=t(\Lambda(h_a)-t+1)f_a^{t-1}v_\Lambda$
and the last nonzero vector in its finite string give
$\Lambda(h_a)=m_a\in\mathbb Z_{\ge0}$. Reordering root words
by their commutators shows that negative-root monomials on
$v_\Lambda$ span the simple factor. Its highest weight space is
one-dimensional, and every weight differs from $\Lambda$ by an
integral sum of roots. Finite exponentials of root operators,
with these integral Cartan weights, integrate the action to
$SL_L(\mathbb C)$; equivalently one integrates the matrix Lie
algebra action to the simply connected group. Denote this group
action by $\rho$.

Normalize the defining principal $e$ to have entries one just
above the diagonal; this is a diagonal conjugation. Define
\begin{equation}
 \mathcal G_\rho(z)=
 \exp\left(-\sum_{s=1}^{L-1}\frac{z^s}{s}\,d\rho(e^s)\right)
                 =\rho(I-ze).
 \label{df:eq-csr11}
\end{equation}
The equality follows from the finite logarithm
$\log(I-ze)=-\sum e^s z^s/s$ and integration of the Lie algebra
action. It is an identity on $SL_L$ since $\det(I-ze)=1$.
The exponential is finite: each $d\rho(e^s)$ has strictly
positive principal flavor weight, and these commuting operators
are simultaneously nilpotent on the finite representation.

Choose a determinant-one monomial matrix $w$ whose first $L-1$
columns are the basis vectors $v_2,\ldots,v_L$. Its last column
is $(-1)^{L-1}v_1$. The leading $a$ by $a$ minor of
$(I-ze)w$ is exactly
\begin{equation}
                         \Delta_a(z)=(-z)^a
                   \quad(1\le a\le L-1).
 \label{df:eq-csr12}
\end{equation}
Indeed that submatrix is lower bidiagonal with diagonal $-z$.
For $z\ne0$ all these minors are nonzero, so elementary Gaussian
elimination gives a lower--diagonal--upper factorization. Let
$v_\Lambda$ be a highest vector and let $\xi_\Lambda$ extract
its coefficient, normalized to $\xi_\Lambda(v_\Lambda)=1$.
The upper factor fixes the vector, and the lower factor changes
it only by lower weights. Its diagonal factor therefore gives
\begin{equation}
 \xi_\Lambda\bigl(\mathcal G_\rho(z)\rho(w)v_\Lambda\bigr)
       =\prod_{a=1}^{L-1}\Delta_a(z)^{m_a}=(-z)^q.
 \label{df:eq-csr13}
\end{equation}
The diagonal character is the displayed product because the diagonal
entries of a Gauss factorization are $\Delta_a/\Delta_{a-1}$ and
$\Lambda=\sum m_a\omega_a$. This is a polynomial identity, first
established for $z\ne0$.
Hence every polynomial degree bound $\deg\mathcal G_\rho\le d$
forces $q\le d$. If $q=d$, its degree-$d$ coefficient is
nonzero. This proves exactly the needed degree bound, with no
Schur-functor classification, determinant twist, or assertion
that a nilpotent matrix is invertible.

\paragraph{The spectral congruence in every simple factor.}
By \eqref{df:eq-csr7}--\eqref{df:eq-csr9}, $C_\varepsilon$ is conjugate in $SL_L$ to a
traceless diagonal matrix all of whose entries are congruent to
$-2$ modulo $L$. Its simple-root evaluations are divisible by
$L$, and $\omega_a(C_\varepsilon)\equiv-2a\pmod L$.
The weight description above consequently proves
\begin{equation}
       \operatorname{Spec}d\rho(C_\varepsilon)
                          \subset -2q+L\mathbb Z
 \label{df:eq-csr14}
\end{equation}
on that simple factor. The conjugation does not alter the spectrum.
This argument does not assert that arbitrary simultaneous
operators on the module are diagonalizable.

\section{The three small-rank endpoints}
\label{app:small-rank}\label{sec:small-rank}

The uniform argument in the main text uses a Casimir of odd degree
at least five. I give separate all-weight arguments for $N=2,3,4$.
Throughout this appendix the starting algebra is the strongly closed
$A_N$ of Theorem~\ref{thm:strong-closure}. A nonzero quotient
$\mathcal H_N/A_N$ would have a finite-dimensional lowest space $E$
at some $r>0$. The identities below exclude such a space, with no
restriction on its ghost grading or on $r$.

For $N=2$, the Sugawara identity directly excludes a positive
lowest space. For $N=3$ and $N=4$, I first derive the required
relative field identities and then use their complete lowest-space
consequences. The final subsection gives the algebraic verification
of the rank-four quantum corrections used above.

\subsection{The case \texorpdfstring{$N=2$}{N=2}: the Sugawara identity and its module consequence}
\label{sr:rank-two}

Put $A=\operatorname{tr}X^2$, $B=\operatorname{tr}XY$,
$C=\operatorname{tr}Y^2$, $M=[X,Y]$, and
\[
 W=AC-B^2,\qquad
 P=\operatorname{tr}(X\partial Y-(\partial X)Y),\qquad
 Z=\operatorname{tr}(bp),\qquad U=\operatorname{tr}(bM).
\]
These products are simultaneous elementary normal products.
For traceless two-by-two matrices,
$X^2=AI/2$ and $XY+YX=BI$, whence
$\operatorname{tr}M^2=-2W$. Single contractions in $QU$ give
$\operatorname{tr}M^2$; constant gauge transformations cancel. In each ordered pair, the first entry belongs to
$j_{\rm BRST}(z)$ and the second to $U(w)$.
The complete double-contraction residues for matrix size $N$ are
\[
\begin{array}{c|c}
 \text{contracted pairs in the matter part of }j_{\rm BRST}
       &\text{residue}\\\hline
 (c,b),(X,Y)&2N\operatorname{tr}(X\partial Y)\\
 (c,b),(Y,X)&-2N\operatorname{tr}((\partial X)Y)\\
 (X,Y),(Y,X)&2N\operatorname{tr}(bp)
\end{array}
\]
where the two occurrences of each matter field in $[X,Y]$ are included.
The ghost part has no double contraction with this primitive.
Triple contractions have no residue. Thus, at $N=2$,
\[
                         QU=-2W+4P+4Z.
\]
The current normal products give
\[
 :ef:=-\tfrac14AC-\operatorname{tr}((\partial X)Y),\quad
 :fe:=-\tfrac14AC+\operatorname{tr}(X\partial Y),\quad
 :HH:=B^2+P.
\]
Consequently, with $S=:ef:+:fe:+\tfrac12:HH:$,
\begin{equation}\label{sr:sugawara}
                         QU=-4(T-S).
\end{equation}
The primitive is manifestly relative. The flavor level is $-3/2$,
so $S$ is the affine Sugawara tensor.

On an irreducible spin-$j$ summand of the lowest space $E$,
\eqref{sr:sugawara} gives
\begin{equation}\label{sr:sugawara-spectrum}
                         r=2j(j+1).
\end{equation}
For any of the four supercurrents $F$, all $F_{(n)}$ with $n\ge1$
kill $E$, and $F_{(0)}E$ is affine primary: for a current $J$ and
$m\ge1$,
$J_{(m)}F_{(0)}v=(J_{(0)}F)_{(m)}v=0$.
Its possible spins are $j\pm1/2$ and its weight is $r+1/2$.
Applying \eqref{sr:sugawara-spectrum} to either possibility would give
$1/2=2j+3/2$ or $1/2=-2j-1/2$, respectively. Neither is possible.
Thus $F_{(0)}E=0$ for all four supercurrents. The exact identities
\[
 \{G_{X,(0)},\bar G_{X,(1)}\}=-e_{(0)},\qquad
 \{G_{Y,(0)},\bar G_{Y,(1)}\}=f_{(0)}
\]
then make $E$ flavor scalar, and \eqref{sr:sugawara-spectrum} gives
$r=0$, a contradiction. This proves the endpoint for every weight.

\subsection{The case \texorpdfstring{$N=3$}{N=3}: the two required relative identities}
\label{sr:rank-three-fields}

All matrix fields in this subsection are traceless three-by-three
matrices. The propagators, BRST current and completeness identities
are \eqref{fnd:free-ope}--\eqref{fnd:completeness}.
\subsubsection{The weight-$5/2$ relation}
Set
\[
 A=\operatorname{tr}X^2,\quad B=\operatorname{tr}XY,\quad
 C=\operatorname{tr}Y^2,\quad D=\operatorname{tr}X^3,
\]
\[
 E=\operatorname{tr}X^2Y,\quad F=\operatorname{tr}XY^2,\qquad M=[X,Y].
\]
Define
\[
 S_1=\operatorname{tr}(X^2\partial Y),\qquad
 S_2=\operatorname{tr}(XY\partial X+X\partial X Y),\qquad
 D_m=S_2-2S_1,
\]
\[
 F_g=3\operatorname{tr}(Xbp-Xpb),\qquad \mathscr K_3=F_g+D_m=3H_3,
 \qquad U_X=\operatorname{tr}\bigl(b[X^2,Y]\bigr).
\]
Then the following equality holds in the relative complex itself:
\begin{equation}\label{sr:n3-R5}
 :AF:-2:BE:+:CD:+QU_X=\mathscr K_3.
\end{equation}

Here is a direct derivation, including the quantum corrections.
The Cayley--Hamilton identities for a traceless three-by-three matrix are
\[
 X^3-\tfrac12 AX-\tfrac13DI=0,
 \qquad
 X^2Y+XYX+YX^2-BX-\tfrac12 AY-EI=0.
\]
Write $u=\operatorname{tr}(X^2YXY)$ and
$v=\operatorname{tr}(X^3Y^2)$. Taking traces after multiplication by
$XY$ and $Y^2$, respectively, gives
\[
 2u+v=2BE+\tfrac12AF,\qquad v=\tfrac12AF+\tfrac13DC.
\]
Since $\operatorname{tr}(XM^2)=u-v$, it follows that
\begin{equation}\label{sr:n3-classical-R5}
 AF-2BE+CD=-2\operatorname{tr}(XM^2)
\end{equation}
for simultaneous PBW polynomials.

For clarity, the contraction calculation of $QU_X$ can be organized as
follows. The single-contraction terms containing $c$ cancel by gauge
covariance; the contraction of $Qb$ with the coefficient of $b$ gives
$2\operatorname{tr}(XM^2)$. Double contractions give the remaining terms
in the following table; higher contractions have no residue:
\[
\begin{array}{c|l}
\text{remaining elementary fields}&\text{residue for }\mathfrak{sl}_N\\ \hline
 b,p,X&N\operatorname{tr}(Xbp-Xpb)\\
 X,X,\partial Y&2N\operatorname{tr}(X^2\partial Y)\\
 X,Y,\partial X&-N\operatorname{tr}(XY\partial X+X\partial X Y).
\end{array}
\]
For example, the first entry follows by applying the first completeness
identity twice to the exact matrix sum
\[
 \sum_a[\{X,t_a\},[t_a,Z]]
 =N\{X,Z\}-2\operatorname{tr}(XZ)I,
 \qquad \operatorname{tr}X=\operatorname{tr}Z=0,
\]
and pairing the result with $b$ and $Z=p$. The last two entries are the
Taylor coefficients of the leftover $X(z)$ or $Y(z)$ after contracting
$c(z)$ with $b(w)$ and one matter field with $[X^2,Y](w)$.
Equivalently they are obtained by substituting
$A_a=\operatorname{tr}(t_a[X^2,Y])$ in the explicit residue formula
\[
 Q(b^aA_a)=\mu^a A_a+
 f_{abc}\left(A_{a,Y^b}\partial Y^c-A_{a,X^c}\partial X^b\right)
 +f_{abc}b^d p^a A_{d,Y^bX^c},
\]
\[
 \mu=[X,Y],\qquad f_{abc}=\operatorname{tr}(t_a[t_b,t_c]).
\]
This formula is just the three possible kinds of nonvanishing residue:
one contraction, $c$--$b$ plus one matter contraction, and two matter
contractions followed by the first Taylor coefficient of $c(z)$.
Consequently
\begin{equation}\label{sr:n3-QUX}
 QU_X=2\operatorname{tr}(XM^2)+\frac N3F_g-ND_m.
\end{equation}
Finally the regular coefficients in the three two-field OPEs are
\[
 :AF:=AF+2S_2,\qquad :BE:=BE+S_1-S_2,\qquad :CD:=CD-6S_1.
\]
For instance these terms come from one $X(z)$--$Y(w)$ contraction and
the linear Taylor term in the uncontracted factor; double contractions
in these three cases leave no nonconstant field at $z$ and contribute
no regular term. Thus their combination is
$AF-2BE+CD+4D_m$. Combining this observation with
\eqref{sr:n3-classical-R5} and \eqref{sr:n3-QUX} at $N=3$ proves
\eqref{sr:n3-R5}.

\subsubsection{The spin-one cubic-fermion anticommutator}
For the following contraction calculation use the explicitly indicated full-lowering fields
\[
 B_0=\operatorname{tr}(X^2b),\quad \widetilde B_1=\operatorname{tr}(\{X,Y\}b),
 \qquad C_0=\operatorname{tr}(X^2p),\quad
 \widetilde C_1=\operatorname{tr}(\{X,Y\}p).
\]
In particular $\widetilde B_1=f_{(0)}B_0$ and $\widetilde C_1=f_{(0)}C_0$.
Set $\widetilde B_0=B_0$ and $\widetilde C_0=C_0$. For $i,j\in\{0,1\}$ set
\[
 Z_{ij}=(\widetilde B_i)_{(0)}\widetilde C_j-\tfrac12\partial\bigl((\widetilde B_i)_{(1)}\widetilde C_j\bigr),
 \qquad V=\operatorname{tr}(b[X,\partial X]).
\]
The following identity, with an explicit relative primitive, holds:
\begin{equation}\label{sr:n3-BC-spin1}
 Z_{01}-Z_{10}
 =-\frac83:eT:+\frac23\bigl(:H\partial e:-:(\partial H)e:\bigr)
       -\frac43\partial^2e+\frac23QV.
\end{equation}
I give the complete Wick computation.
Put
\[
 d_{abc}=\tfrac12\operatorname{tr}(\{t_a,t_b\}t_c),\qquad
 \kappa=\frac{N^2-4}{2N}.
\]
Completeness gives $\sum_{a,c}d_{abc}d_{dac}=\kappa\delta_{bd}$.
Introduce the simultaneous polynomials
\[
 K=d_{abq}d_{deq}X^aX^bX^dY^e,
 \quad L=d_{abq}d_{deq}(\partial X^a)X^bX^dY^e,
\]
\[
 R=d_{abq}d_{deq}X^aX^b\bigl((\partial X^d)Y^e+X^d\partial Y^e\bigr),
 \quad W=4d_{abq}d_{dak}X^bX^d b^q p^k.
\]
The ghost contraction, the single matter contraction, and their
simultaneous contraction give, respectively,
\begin{align*}
 (B_0)_{(1)}\widetilde C_1&=2K+4\kappa\operatorname{tr}(X\partial X),\\
 (B_0)_{(0)}\widetilde C_1&=4L+W+2\kappa\operatorname{tr}(X\partial^2X),\\
 (\widetilde B_1)_{(1)}C_0&=2K-4\kappa\operatorname{tr}(X\partial X),\\
 (\widetilde B_1)_{(0)}C_0&=2R-W-2\kappa\operatorname{tr}(X\partial^2X).
\end{align*}
These exhaust all contractions, since each OPE has only one possible
$Y$ contraction and one possible $b$--$p$ contraction.
As $\partial K=2L+R$, their prescribed combination is
\begin{equation}\label{sr:n3-BC-PBW}
 Z_{01}-Z_{10}=4L-2R+2W-4\kappa\operatorname{tr}((\partial X)^2).
\end{equation}
In fact the same calculation shows $Z_{10}=-Z_{01}$.

I now reduce the two quartic terms by elementary matrix identities.
Let
\[
 u=\operatorname{tr}(X\partial X),\quad
 v=\operatorname{tr}(Y\partial X),\quad
 w=\operatorname{tr}(X\partial Y),
\]
\[
 t=\operatorname{tr}([X,Y][X,\partial X]),
 \quad J=\operatorname{tr}(b[X,[X,p]]).
\]
For $N=3$, completeness and the derivative of Cayley--Hamilton give
\begin{equation}\label{sr:n3-BC-CH}
 2W=\frac43A\operatorname{tr}(bp)-\frac23J,
 \qquad
 4L-2R=\frac23t+Av-\frac13Aw-\frac23uB.
\end{equation}
Here are intermediate formulas verifying both reductions. Completeness
first gives
\[
 2W=2\operatorname{tr}(b\{X,\{X,p\}\})
          -\frac83\operatorname{tr}(Xb)\operatorname{tr}(Xp).
\]
Use
\[
 X^2p+XpX+pX^2=\tfrac12Ap+\operatorname{tr}(Xp)X
                         +\operatorname{tr}(X^2p)I
\]
and eliminate $XpX$ with $[X,[X,p]]$ to obtain the first formula in
\eqref{sr:n3-BC-CH}. For the second, write
$a=\operatorname{tr}(X^2Y\partial X)$,
$b'=\operatorname{tr}(X^2\partial X Y)$, and
$c'=\operatorname{tr}(XYX\partial X)$. Then
\[
 L=\tfrac14(a+b'+2c')-\tfrac13uB,\quad
 R=\tfrac12(a+b')+\tfrac12Aw-\tfrac13A(v+w),
\]
\[
 a+b'+c'=\tfrac12Av+uB,\qquad t=2c'-a-b'.
\]
Substitution proves the stated reduction.

The remaining normal-ordering corrections are short enough to list in
full. Direct contraction with the BRST current gives, for all $N$,
\begin{equation}\label{sr:n3-QV}
 QV=t-J+N\operatorname{tr}\bigl(X\partial^2X-2(\partial X)^2\bigr).
\end{equation}
The term $t$ comes from $Qb=[X,Y]+\cdots$, and $-J$ from the
$- [p,X]$ term in $Q\partial X$. The only derivative corrections
come from the double contractions of $c(z)$ with $b(w)$ and $Y(z)$
with $X(w)$ or $\partial X(w)$: their unexpanded coefficients are
$-2N\operatorname{tr}(X(z)\partial X(w))/(z-w)^2$ and
$2N\operatorname{tr}(X(z)X(w))/(z-w)^3$, respectively. Their residues
are precisely the last term of \eqref{sr:n3-QV}.
The other required regular coefficients are
\begin{align*}
 :eT:&=\tfrac14 A(w-v)-\tfrac12A\operatorname{tr}(bp)
              +\tfrac14\operatorname{tr}(X\partial^2X)
              -\tfrac12\operatorname{tr}((\partial X)^2),\\
 :H\partial e:&=-Bu+\operatorname{tr}((\partial X)^2)
                        +\tfrac12\operatorname{tr}(X\partial^2X),\\
 :(\partial H)e:&=-\tfrac12A(v+w)
                        +\tfrac12\operatorname{tr}(X\partial^2X),\\
 \partial^2 e&=\operatorname{tr}((\partial X)^2+X\partial^2X).
\end{align*}
Putting these four formulas and \eqref{sr:n3-QV} into the right side
of \eqref{sr:n3-BC-spin1} gives exactly
\eqref{sr:n3-BC-PBW}, using $4\kappa=10/3$.
This proves the field identity without relying on a finite-mode test.

\subsection{The case \texorpdfstring{$N=3$}{N=3}: exclusion of every positive lowest space}\label{sr:rank-three-modules}

The cubic fields used here have the following normalization:
\begin{align*}
w_0&=\operatorname{tr}X^3,&
w_1&=\operatorname{tr}X^2Y,&
w_2&=\operatorname{tr}XY^2,&
w_3&=\operatorname{tr}Y^3,\\
B_0&=\operatorname{tr}X^2b,&
B_1&=\tfrac12\operatorname{tr}(XY+YX)b,&
B_2&=\operatorname{tr}Y^2b,\\
C_0&=\operatorname{tr}X^2p,&
C_1&=\tfrac12\operatorname{tr}(XY+YX)p,&
C_2&=\operatorname{tr}Y^2p.
\end{align*}
Thus $f_{(0)}B_0=2B_1$, and likewise for $C$; this factor of two
is part of my convention. The $w_i$ are even of weight $3/2$,
whereas $B_i,C_i$ are odd of weight two. Set
\[
\mathscr D=(G_X)_{(0)}C_1-(G_Y)_{(0)}C_0,
\qquad
U=\operatorname{tr}\bigl(b\{X,[X,Y]\}\bigr).
\]
The elementary descendant calculation gives $\mathscr D=\mathscr K_3/2=3H_3/2$. The rank-three quintic identity \eqref{sr:n3-R5}, divided by two, is
\begin{equation}\label{sr:n3-R5-input}
R\mathrel{:=}:e w_2:+:H w_1:-:f w_0:-\mathscr D=-\tfrac12QU.
\end{equation}
It implies $R=0$ in relative BRST cohomology. This is the only null
relation needed for the six-zero-mode argument that follows.

Let $A$ denote any conformal vertex subalgebra of that cohomology
containing the displayed fields. The argument applies to any
lower-bounded $A$-module whose lowest conformal-weight space $E$
is finite dimensional. Its conclusion is
\begin{equation}\label{sr:n3-six-zero-modes}
(B_i)_{(1)}E=(C_i)_{(1)}E=0\qquad (i=0,1,2).
\end{equation}
In particular it applies to a nonzero lowest space of $H/A$,
if that quotient is nonzero. The proof does not assume that the
lowest weight is positive, or that $T$ is an affine Sugawara vector.

\subsubsection{Transformation and descent identities}

Write $F_i=B_i$ or $C_i$, and correspondingly put
$(S_X,S_Y)=(G_X,G_Y)$ or $(\bar G_X,\bar G_Y)$.
Direct elementary contraction gives
\begin{align}
(S_X)_{(0)}w_i&=iF_{i-1},&
(S_Y)_{(0)}w_i&=-(3-i)F_i,\label{sr:n3-S-w}\\
e_{(0)}F_i&=iF_{i-1},&
f_{(0)}F_i&=(2-i)F_{i+1},&
H_{(0)}F_i&=(2-2i)F_i.\label{sr:n3-current-F}
\end{align}
Terms outside the indicated index ranges are zero. In the same
notation,
\[
(S_X)_{(0)}e=0,\quad (S_X)_{(0)}H=-S_X,\quad
(S_X)_{(0)}f=-S_Y,
\]
\[
(S_Y)_{(0)}e=-S_X,\quad (S_Y)_{(0)}H=S_Y,\quad
(S_Y)_{(0)}f=0.
\]
There is no parity sign when a supercurrent zero mode passes through
a current or a $w_i$, since those fields are even.

For completeness, the descent of $\mathscr D$ is
\begin{equation}\label{sr:n3-S-D}
(S_X)_{(0)}\mathscr D=-\partial F_0,\qquad
(S_Y)_{(0)}\mathscr D=-\partial F_1.
\end{equation}
Here is a verification fixing all its signs. Denote the translation
operator on states by $\partial$. The small $\mathcal N=4$ OPEs imply
the following identities between odd derivations:
\[
\{(G_X)_{(0)},(\bar G_X)_{(0)}\}
=\{(G_Y)_{(0)},(\bar G_Y)_{(0)}\}=0,
\]
\[
\{(G_X)_{(0)},(\bar G_Y)_{(0)}\}=-\partial,
\qquad
\{(G_Y)_{(0)},(\bar G_X)_{(0)}\}=\partial.
\]
Two unbarred derivations anticommute, as do two barred ones.
Also $(G_X)_{(0)}B_i=(G_Y)_{(0)}B_i=0$ and
$(\bar G_X)_{(0)}C_i=(\bar G_Y)_{(0)}C_i=0$.
These last assertions follow directly because the cubic invariant
tensor is symmetric, whereas the two $b$'s, or two $p$'s, are odd.
Finally, contraction of $b$ with $p$ gives
\[
(G_X)_{(0)}C_0=\tfrac13\partial w_0,
\qquad C_1=-\tfrac12(\bar G_Y)_{(0)}w_1.
\]
Consequently
\begin{align*}
(G_X)_{(0)}\mathscr D
&=(G_Y)_{(0)}(G_X)_{(0)}C_0
=\tfrac13\partial(G_Y)_{(0)}w_0=-\partial B_0,\\
(G_Y)_{(0)}C_1&=-(\bar G_Y)_{(0)}B_1,\\
(G_Y)_{(0)}\mathscr D
&=-(G_X)_{(0)}(G_Y)_{(0)}C_1
=(G_X)_{(0)}(\bar G_Y)_{(0)}B_1=-\partial B_1.
\end{align*}
Applying the barred derivations directly to the definition of $\mathscr D$
gives $-\partial C_0$ and $-\partial C_1$ respectively, using the
same four anticommutators. This proves \eqref{sr:n3-S-D}.

Apply $(S_X)_{(0)}$ and $(S_Y)_{(0)}$ to $R=0$. Zero modes are
superderivations of normal products, so the preceding formulas give
\begin{align}
0={}&2:eF_1:+:HF_0:-:S_Xw_1:+:S_Yw_0:+\partial F_0,
\label{sr:n3-high-desc}\\
0={}&-:S_Xw_2:-:eF_2:+:S_Yw_1:-2:HF_1:
       +3:fF_0:+\partial F_1.
\label{sr:n3-low-desc}
\end{align}
Both sides have weight three and odd parity.

\subsubsection{Direct normal-order reduction on the lowest space}

Every operator of strictly negative conformal degree annihilates $E$.
I record the exact reductions needed here; none sets a normal
product of half-integral-weight fields to zero.

If $a,b$ both have weight $3/2$, the normal-order mode formula is
\[
(:ab:)_{(2)}=
\sum_{i<0}a_{(i)}b_{(1-i)}
+(-1)^{p(a)p(b)}\sum_{i\geq0}b_{(1-i)}a_{(i)}.
\]
On $E$, the first sum and the terms $i\geq1$ in the second sum
vanish. Commuting the surviving $b_{(1)}$ past $a_{(0)}$, and using
$b_{(1)}E=0$, gives
\begin{equation}\label{sr:n3-half-normal}
o(:ab:)|_E=-o(a_{(0)}b)|_E.
\end{equation}
Indeed, $[a_{(0)},b_{(1)}]=(a_{(0)}b)_{(1)}$; the two parity
signs cancel. Thus \eqref{sr:n3-half-normal} holds for either parity.

For a current $J$ and a weight-two field $F$, the same mode split gives
\begin{equation}\label{sr:n3-integer-normal}
o(:JF:)|_E=o(F)J_{(0)}
=J_{(0)}o(F)-o(J_{(0)}F),\qquad
o(\partial F)|_E=-2o(F).
\end{equation}
Only $i=0$ survives in the current-first normal-order formula.
The current is even, so no parity sign occurs in this formula.

On $E$ abbreviate $e_{(0)},H_{(0)},f_{(0)}$ to $e,H,f$, and
$o(F_i)=(F_i)_{(1)}$ to $F_i$. These are operators on $E$;
the order in their products is significant. Applying
\eqref{sr:n3-half-normal}--\eqref{sr:n3-integer-normal} to
\eqref{sr:n3-high-desc} gives
\[
2(eF_1-F_0)+(HF_0-2F_0)+F_0+3F_0-2F_0=0,
\]
or
\begin{equation}\label{sr:n3-K1}
K_1\mathrel{:=}2eF_1+(H-2)F_0=0.
\end{equation}
For the second descendant the two half-integral normal products
contribute $2F_1+2F_1$. Together with the current reorderings, they give
\begin{equation}\label{sr:n3-K-low}
L\mathrel{:=}-eF_2-2HF_1+3fF_0-2F_1=0.
\end{equation}
Using the current action in \eqref{sr:n3-current-F}, direct commutation
in the algebra of operators on $E$ gives
\[
[f,K_1]=2eF_2+2fF_0-4F_1,
\]
and therefore $2L-[f,K_1]=0$ is exactly
\begin{equation}\label{sr:n3-K0}
K_0\mathrel{:=} eF_2+HF_1-fF_0=0.
\end{equation}
The second lowering of \eqref{sr:n3-K1} is
\begin{equation}\label{sr:n3-K1-lowest}
[f,[f,K_1]]=-2HF_2+4fF_1-4F_2=0.
\end{equation}
These computations independently fix the scalar term $-2F_0$ in
\eqref{sr:n3-K1} and the term $-4F_2$ in its second lowering.

\subsubsection{All finite-dimensional spin transitions vanish}

The zero modes $e,H,f$ act on the finite-dimensional vector space $E$
as $\mathfrak{sl}_2$, so $E$ is a direct sum of irreducible spins
$j\in\tfrac12\mathbb Z_{\geq0}$. The three operators $F_i$ transform
as spin one by \eqref{sr:n3-current-F}. Fix an irreducible input
summand $V_j$ and project their outputs onto an irreducible target
summand $V_{j'}$. The resulting map $V_1\otimes V_j\to V_{j'}$
is equivariant. Its only possible targets are $j+1,j,j-1$, omitting
the negative spins and the diagonal target when $j=0$.

Let $v$ be a highest vector in $V_j$, so $ev=0$ and $Hv=2jv$.
If $j'=j+1$, a nonzero equivariant map has $F_0v\ne0$.
Equation \eqref{sr:n3-K1}, together with $eF_1v=F_0v$, gives
\[
0=2(j+1)F_0v,
\]
which is impossible for $j\geq0$.

If $j'=j$, weight considerations give $F_0v=0$, and a nonzero map
has $F_1v\ne0$. Since $eF_2v=2F_1v$, equation \eqref{sr:n3-K0}
gives
\[
0=2(j+1)F_1v,
\]
again impossible.

Finally, if $j'=j-1$, then $j\geq1$ and
$F_0v=F_1v=0$, while a nonzero map has $F_2v\ne0$.
Equation \eqref{sr:n3-K1-lowest} gives
\[
0=-4jF_2v,
\]
which is impossible. The stated nonvanishing properties of a nonzero
map also follow without choosing Clebsch--Gordan coefficients:
successively apply $eF_1v=F_0v$ and $eF_2v=2F_1v$, using the
one-dimensional highest-weight space of the target. If all three
images of $v$ vanish, equivariance under $f$ makes the entire map zero.

Thus every projected map is zero, and $F_0=F_1=F_2=0$ on $E$.
The proof applies separately to $F=B$ and $F=C$, establishing
\eqref{sr:n3-six-zero-modes}. In particular it treats the odd
integer-weight zero modes as genuine operators until the displayed
relations force their vanishing.

\subsubsection{An additional mixed-fermion identity forces a singlet bottom}

I now record the further finite Wick identities needed to exclude
\emph{positive} lowest conformal weights. They are written in the
same normalization $B_1=\frac12\operatorname{tr}(XY+YX)b$ throughout.
Put
\[
V=\operatorname{tr}\bigl(b[X,\partial X]\bigr),\qquad
\mathcal Z=B_{0,(0)}C_1-B_{1,(0)}C_0
-\tfrac12\partial\bigl(B_{0,(1)}C_1-B_{1,(1)}C_0\bigr).
\]
The exact mixed-fermion identity is
\begin{equation}\label{sr:n3-Z-exact}
\mathcal Z=-\tfrac43:eT:
+\tfrac13\bigl(:H\partial e:-:(\partial H)e:\bigr)
-\tfrac23\partial^2e+\tfrac13QV.
\end{equation}
Here $T$ is the actual free-field stress tensor, whose central
charge is $-24$, and it remains the conformal vector in cohomology.
Equation \eqref{sr:n3-Z-exact} is one half of
\eqref{sr:n3-BC-spin1}, because $B_1,C_1$ now have the normalized
flavor convention.
Its exact term has a relative primitive, since $V$ is invariant
and contains no undifferentiated $c$.

For any module, the mode commutator formula and
$o(\partial a)=-\operatorname{wt}(a)o(a)$ give
\begin{equation}\label{sr:n3-Z-zero}
o(\mathcal Z)=
\{B_{0,(1)},C_{1,(1)}\}-\{B_{1,(1)},C_{0,(1)}\}.
\end{equation}
For example,
$\{B_{i,(1)},C_{j,(1)}\}=(B_{i,(0)}C_j)_{(2)}
+(B_{i,(1)}C_j)_{(1)}$; the derivative in the definition of
$\mathcal Z$ supplies the second term with exactly this sign.

I need its action not only on a lowest space, but on any vector
$u$ satisfying
\[
J_{(n)}u=0\ (n\geq1,\ J=e,H,f),\qquad
L_nu=0\ (n\geq1),\qquad L_0u=t u.
\]
The normal-order mode sums give
\[
(:eT:)_{(2)}u=t e_{(0)}u.
\]
For the derivative-current terms, direct splitting gives
\[
(:H\partial e:)_{(2)}u=-e_{(0)}H_{(0)}u,
\qquad
(:(\partial H)e:)_{(2)}u=-e_{(0)}H_{(0)}u.
\]
In the second formula the unique surviving summand has index $i=1$:
$e_{(0)}(\partial H)_{(1)}u=-e_{(0)}H_{(0)}u$.
The $i=0$ summand vanishes because $(\partial H)_{(0)}=0$.
Finally $(\partial^2e)_{(2)}=2e_{(0)}$. Hence, in cohomology,
\begin{equation}\label{sr:n3-Z-primary}
o(\mathcal Z)u=-\tfrac43(t+1)e_{(0)}u.
\end{equation}

Suppose now that the lowest weight $r$ of $E$ is positive.
The six-zero-mode result and \eqref{sr:n3-Z-zero} imply
$o(\mathcal Z)E=0$. Apply \eqref{sr:n3-Z-primary} at $t=r$.
Since $r+1\ne0$, I obtain $e_{(0)}E=0$. The
$\mathfrak{sl}_2$ commutators then imply
\begin{equation}\label{sr:n3-bottom-singlet}
e_{(0)}E=H_{(0)}E=f_{(0)}E=0.
\end{equation}

\subsubsection{Four finite OPE coefficients needed above a singlet bottom}

Set $P=\operatorname{tr}(bbX)$, a relative field of ghost number
$-2$ and weight $5/2$. Elementary Wick contraction gives
\begin{align}
B_{0,(0)}w_2
&=\tfrac43:eG_Y:+\tfrac23\partial G_X-\tfrac13QP,
&B_{0,(1)}w_2&=\tfrac53G_X,\label{sr:n3-B0w2}\\
B_{1,(0)}w_1
&=-:eG_Y:-\tfrac13:HG_X:-\tfrac23\partial G_X+\tfrac13QP,
&B_{1,(1)}w_1&=-\tfrac53G_X.\label{sr:n3-B1w1}
\end{align}
Some explicit intermediate expressions make the quantum terms and
normal-order conventions in these identities checkable. With all
products on the right understood as elementary Wick products,
\begin{align*}
QP&=\operatorname{tr}\bigl(b[X,[X,Y]]\bigr)
+6\operatorname{tr}((\partial X)b)-3\operatorname{tr}(X\partial b),\\
:eG_Y:&=\tfrac12\operatorname{tr}X^2\operatorname{tr}(Yb)
+\operatorname{tr}((\partial X)b),\\
:HG_X:&=-\operatorname{tr}(XY)\operatorname{tr}(Xb)
+\operatorname{tr}((\partial X)b),\\
B_{0,(0)}w_2&=\tfrac23\operatorname{tr}X^2\operatorname{tr}(Yb)
-\tfrac13\operatorname{tr}\bigl(b[X,[X,Y]]\bigr)
+\tfrac53\operatorname{tr}(X\partial b),\\
B_{1,(0)}w_1&=-\tfrac12\operatorname{tr}X^2\operatorname{tr}(Yb)
+\tfrac13\operatorname{tr}(XY)\operatorname{tr}(Xb)
+\tfrac13\operatorname{tr}\bigl(b[X,[X,Y]]\bigr)
-\tfrac53\operatorname{tr}(X\partial b).
\end{align*}
The single contractions give the displayed quartic trace terms;
their reduction uses the polarized traceless $3\times3$
Cayley--Hamilton identity. The double contractions give the
$\pm5/3$ coefficients, using
$\sum_a T^aMT_a=\operatorname{tr}(M)I-\frac13M$ for trace-dual
traceless matrix bases. Taylor expansion of the surviving $b(z)$
gives the corresponding $\partial b$ terms. Substituting the first
three formulas into the last two proves
\eqref{sr:n3-B0w2}--\eqref{sr:n3-B1w1} with their exact primitives.

For $v\in E$, any current $J$ and any weight-$3/2$ field $a$ obey
\begin{equation}\label{sr:n3-Ja-half-raising}
(:Ja:)_{(1)}v=a_{(0)}J_{(0)}v=0.
\end{equation}
Indeed the first normal-order sum contains only $a_{(n)}$ with
$n\geq1$, and in the second sum all $J_{(n)}$ with $n\geq1$
kill $v$. The remaining term vanishes by
\eqref{sr:n3-bottom-singlet}. Also
$(\partial a)_{(1)}=-a_{(0)}$.

Using $B_{i,(1)}v=0$, the commutator formula gives
\[
B_{i,(1)}w_{j,(0)}v
=\bigl((B_{i,(0)}w_j)_{(1)}+(B_{i,(1)}w_j)_{(0)}\bigr)v.
\]
Thus \eqref{sr:n3-B0w2}--\eqref{sr:n3-Ja-half-raising} imply the
two crucial constants
\begin{equation}\label{sr:n3-Bw-constants}
B_{0,(1)}w_{2,(0)}v=(G_X)_{(0)}v,
\qquad
B_{1,(1)}w_{1,(0)}v=-(G_X)_{(0)}v.
\end{equation}

\subsubsection{Excluding a positive singlet bottom}

Put $\mathscr D'=f_{(0)}\mathscr D$. The relation $R=0$ and its flavor lowering give
\[
\mathscr D=:ew_2:+:Hw_1:-:fw_0:,
\qquad
\mathscr D'=:ew_3:+:Hw_2:-:fw_1:.
\]
Consequently $\mathscr D_{(1)}E=\mathscr D'_{(1)}E=0$ by
\eqref{sr:n3-Ja-half-raising}.

The definition of $\mathscr D$, together with the flavor lowering of
$(G_X)_{(0)}C_0=\frac13\partial w_0$, yields
\[
(G_Y)_{(0)}C_0=\tfrac13\partial w_1-\tfrac23\mathscr D,
\qquad
(G_Y)_{(0)}C_1=\tfrac13\partial w_2-\tfrac13\mathscr D'.
\]
For instance the first identity follows by solving
$(G_Y)_{(0)}C_0+2(G_X)_{(0)}C_1=\partial w_1$ together with
the definition of $\mathscr D$; one more flavor lowering gives the second.
Since $C_{i,(1)}E=0$, their mode commutators imply
\begin{equation}\label{sr:n3-CG-constants}
C_{0,(1)}(G_Y)_{(0)}v=-\tfrac13w_{1,(0)}v,
\qquad
C_{1,(1)}(G_Y)_{(0)}v=-\tfrac13w_{2,(0)}v.
\end{equation}
On the other hand $(G_Y)_{(0)}B_i=0$, so
$B_{i,(1)}(G_Y)_{(0)}v=0$.

Apply \eqref{sr:n3-Z-zero} to $u=(G_Y)_{(0)}v$. The last
observations and \eqref{sr:n3-Bw-constants} give
\begin{align}
o(\mathcal Z)(G_Y)_{(0)}v
&=-\tfrac13B_{0,(1)}w_{2,(0)}v
+\tfrac13B_{1,(1)}w_{1,(0)}v\notag\\
&=-\tfrac23(G_X)_{(0)}v.\label{sr:n3-Z-GY-first}
\end{align}
The vector $u$ has conformal weight $r+1/2$. It is annihilated by
all positive current and Virasoro modes: their commutators with
$(G_Y)_{(0)}$ are multiples of $(G_X)_{(n)}$ or $(G_Y)_{(n)}$
with $n\geq1$, which kill $E$. This assertion does \emph{not}
claim that $u$ is a lowest vector for the whole algebra.
Hence \eqref{sr:n3-Z-primary} applies and gives
\begin{equation}\label{sr:n3-Z-GY-second}
o(\mathcal Z)(G_Y)_{(0)}v
=-\tfrac43(r+\tfrac32)(G_X)_{(0)}v,
\end{equation}
where $e_{(0)}(G_Y)_{(0)}v=(G_X)_{(0)}v$ follows from
\eqref{sr:n3-bottom-singlet}. Comparing
\eqref{sr:n3-Z-GY-first} and \eqref{sr:n3-Z-GY-second} yields
\[
(r+1)(G_X)_{(0)}v=0.
\]
Since $r>0$, $(G_X)_{(0)}E=0$.

Finally, the small $\mathcal N=4$ OPE
\[
[G_X{}_{\lambda}\bar G_Y]
=4\lambda^2-\lambda H-T-\tfrac12\partial H
\]
implies
\[
\{(G_X)_{(0)},(\bar G_Y)_{(1)}\}
=-L_0+\tfrac12H_{(0)}.
\]
Both factors on the left annihilate $E$, the second by its
negative conformal degree. On $E$ the right side equals $-r$.
This contradicts $r>0$ and $E\ne0$.

The quotient $\mathcal H_3/A_3$ therefore vanishes.
Theorem~\ref{thm:strong-closure} identifies $A_3$ with the normal
differential span of its degree-two and degree-three multiplets,
proving the full strong-generation assertion for $N=3$.

\subsection{The case \texorpdfstring{$N=4$}{N=4}: exact trace identities and their quantum lifts}\label{sr:rank-four-fields}

The following abbreviations denote simultaneous PBW products, rather than
iterated vertex normal products:
\[
 A=\operatorname{tr}X^2,\ B=\operatorname{tr}XY,\ C=\operatorname{tr}Y^2,
 \qquad D=\operatorname{tr}X^3,\ E=\operatorname{tr}X^2Y,
 \ F=\operatorname{tr}XY^2,
\]
\[
 d_{40}=\operatorname{tr}X^4,\quad d_{31}=\operatorname{tr}X^3Y,
 \quad d_{22}=\tfrac23\operatorname{tr}X^2Y^2+
                     \tfrac13\operatorname{tr}XYXY,\qquad M=[X,Y].
\]
For arbitrary traceless $4\times4$ matrices over a commutative algebra,
\begin{align}
 R_6&=DF-E^2+A d_{22}-2B d_{31}+C d_{40}
                    -\tfrac12A(AC-B^2),\label{sr:n4-R6-def}\\
 R_6&=\tfrac5{12}A\operatorname{tr}M^2
       -2\operatorname{tr}XMXM-3\operatorname{tr}X^2M^2.
       \label{sr:n4-R6-matrix}
\end{align}
Here is a direct algebraic proof.  Cayley--Hamilton for a traceless
$4\times4$ matrix reads
\[
 X^4-\tfrac12 A X^2-\tfrac13D X+
       (\tfrac18 A^2-\tfrac14d_{40})I=0.
\]
Put $a=\operatorname{tr}X^2Y^2$, $b_*=\operatorname{tr}XYXY$, and
\[
 t_0=\operatorname{tr}X^4Y^2,\quad
 t_1=\operatorname{tr}X^3YXY,\quad
 t_2=\operatorname{tr}X^2YX^2Y.
\]
Multiplying Cayley--Hamilton by $Y^2$ and taking the trace gives
\[
 K_0=t_0-\tfrac12 Aa-\tfrac13DF+
                (\tfrac18A^2-\tfrac14d_{40})C=0.
\]
Taking the coefficient of $t$ in Cayley--Hamilton for $X+tY$,
multiplying by $XY$, and taking the trace gives
\[
 K_1=t_0+2t_1+t_2-2Bd_{31}-\tfrac12A(a+b_*)
                  -E^2-\tfrac13DF+\tfrac12AB^2=0.
\]
Expanding $M$ shows
\[
 \operatorname{tr}M^2=2(b_*-a),\quad
 \operatorname{tr}XMXM=2t_2-2t_1,\quad
 \operatorname{tr}X^2M^2=2t_1-t_0-t_2.
\]
The difference between the two sides of
\eqref{sr:n4-R6-matrix} is exactly $K_1-4K_0$.
This proves the identity without an assumption that $X,Y$ commute.

\subsubsection{The complete degree-six quantum correction}
Specify all nesting in the vertex expression
\begin{align*}
 R_6^{\rm n}={}&:DF:-:EE:+:A d_{22}:-2:B d_{31}:+:C d_{40}:\\
 &-\tfrac12:A(:AC:):+\tfrac12:A(:BB:):,
\\
 U_6={}&\tfrac5{12}A\operatorname{tr}(bM)
             -2\operatorname{tr}(bXMX)-3\operatorname{tr}(bX^2M).
\end{align*}
Products inside $U_6$ are simultaneous PBW products.  Define
\[
 \widehat P_4=d_{40}-\tfrac{29}{68}A^2,\qquad
 B_4=-\tfrac14(G_Y)_{(0)}\widehat P_4,\quad
 L_4=(\bar G_Y)_{(0)}B_4,\quad
 \Delta_4=\partial(f_{(0)}\widehat P_4).
\]
Define $\widehat H_4=L_4+\Delta_4/16$; the $T_{(2)}$ anomaly cancels with precisely
this coefficient.  With these conventions the following exact chain identity
holds:
\begin{align}
 R_6^{\rm n}-QU_6={}&4L_4+\Delta_4
 -\tfrac{140}{51}:eT:-\tfrac{10}{17}:G_X\bar G_X:
 -\tfrac{92}{17}:H\partial e:+\tfrac{90}{17}:(\partial H)e:
 +\tfrac{97}{17}\partial^2e\notag\\
 &-\tfrac{67}{6}Q\operatorname{tr}(bX\partial X)
  -\tfrac{77}{6}Q\operatorname{tr}(b(\partial X)X).
 \label{sr:n4-R6-quantum}
\end{align}
Thus the quartic top is explicitly a composite in BRST cohomology.

\subsubsection{A second, spin-$3/2$, degree-seven relation}
There is also the exact classical matrix identity
\begin{align}
 R_7={}&D d_{22}-2E d_{31}+F d_{40}
 -\tfrac14(A^2F-2ABE+B^2D)+\tfrac1{12}(AC-B^2)D,
 \label{sr:n4-R7-def}\\
 R_7={}&-\tfrac32 A\operatorname{tr}(XM^2)
          -\tfrac23 D\operatorname{tr}M^2
          +2\operatorname{tr}(X^3M^2).
 \label{sr:n4-R7-matrix}
\end{align}
A polynomial certificate follows from the first three coefficients
$\mathsf C_0,\mathsf C_1,\mathsf C_2$ of the Cayley--Hamilton
identity for $X+tY$, displayed below. The difference between the two
sides of \eqref{sr:n4-R7-matrix} equals
\begin{align*}
 &\tfrac94\operatorname{tr}(\mathsf C_0Y^2X)
 -\tfrac{17}4\operatorname{tr}(\mathsf C_0YXY)
 +\tfrac54\operatorname{tr}(\mathsf C_1YX^2)
 -\tfrac12\operatorname{tr}(\mathsf C_2X^3)\\
 &\qquad-\tfrac14C\operatorname{tr}(\mathsf C_0X)
 +\tfrac14B\operatorname{tr}(\mathsf C_0Y)=0.
\end{align*}
Expanding $M=[X,Y]$ and cyclically rotating traces verifies this
identity term by term. It proves the relation for noncommuting matrices.
The corresponding primitive and the specified quantum lift are
\begin{align*}
 U_7={}&-\tfrac32 A\operatorname{tr}(bXM)
       -\tfrac23 D\operatorname{tr}(bM)+2\operatorname{tr}(bX^3M),\\
 R_7^{\rm n}={}&:D d_{22}:-2:E d_{31}:+:F d_{40}:
 -\tfrac14:A(:AF:):+\tfrac12:A(:BE:):\\
 &-\tfrac13:B(:BD:):+\tfrac1{12}:A(:CD:):.
\end{align*}
Let
\[
 B_0=\operatorname{tr}X^2b,\quad C_0=\operatorname{tr}X^2p,\quad
 C_1=\tfrac12\operatorname{tr}(XY+YX)p,\quad
 D_3=(G_X)_{(0)}C_1-(G_Y)_{(0)}C_0.
\]
The second exact chain identity is
\begin{align}
 R_7^{\rm n}-QU_7={}&-\tfrac83:TD:
 -\tfrac{16}{3}:e\partial E:-\tfrac83:H\partial D:
 -\tfrac23:(\partial H)D:\notag\\
 &-2:G_XC_0:+2:\bar G_XB_0:
 +\tfrac{19}{3}\partial^2D+\tfrac83:eD_3:\notag\\
 &+9Q\operatorname{tr}(bX^2\partial X)
   +7Q\operatorname{tr}(bX(\partial X)X)\notag\\
 &+4Q\operatorname{tr}(b(\partial X)X^2)
   -7Q\bigl(A\operatorname{tr}(b\partial X)\bigr).
 \label{sr:n4-R7-quantum}
\end{align}
The last primitive again uses a simultaneous PBW product.
\subsubsection{An independent degree-eight spin-two relation}
In the following formulas write $w_0=D,w_1=E,w_2=F$ and
$d_0=d_{40},d_1=d_{31},d_2=d_{22}$.

With the preceding trace normalizations, put
\begin{align*}
 R_8={}&12(d_0d_2-d_1^2)-3(A^2d_2-2ABd_1+B^2d_0)\\
       &+4A(w_0w_2-w_1^2)+(Cw_0^2-2Bw_0w_1+Aw_1^2).
\end{align*}
This has flavor spin two.  For arbitrary traceless four-by-four matrices,
not necessarily commuting, one has
\begin{align}
 R_8={}&-6\operatorname{tr}(X^2MX^2M)
         -3A\operatorname{tr}(X^2M^2)
         -6A\operatorname{tr}(XMXM)\notag\\
       &-6w_0\operatorname{tr}(XM^2)
         +\tfrac12(d_0-A^2)\operatorname{tr}(M^2),
 \qquad M=[X,Y].
 \label{sr:n4-R8-classical}
\end{align}
Here $d_2=(2\operatorname{tr}(X^2Y^2)+\operatorname{tr}(XYXY))/3$.
Thus $R_8$ vanishes on the commuting locus.  It will control the transitions between spins differing by two.

A self-contained Cayley--Hamilton certificate is as follows.  Write
\begin{align*}
 \mathsf C_0={}&X^4-\tfrac A2X^2-\tfrac{w_0}3X
                 +(A^2/8-d_0/4)I,\\
 \mathsf C_1={}&X^3Y+X^2YX+XYX^2+YX^3-BX^2\\
        &-\tfrac A2(XY+YX)-w_1X-\tfrac{w_0}3Y
                 +(AB/2-d_1)I,\\
 \mathsf C_2={}&\sum_{\substack{\text{words of length }4\\
                         \text{with two }X\text{ and two }Y}}
                  \text{word}\\
       &-\tfrac C2X^2-B(XY+YX)-\tfrac A2Y^2-w_2X-w_1Y\\
       &+(AC/4+B^2/2-3d_2/2)I.
\end{align*}
Here $\mathsf C_1=\mathsf C_1(Y)$ in the directional notation below. These are the coefficients of $t^0,t^1,t^2$ in the characteristic
polynomial identity for $X+tY$, and hence all vanish.  Expanding traces
cyclically, the difference between the left and right sides of
\eqref{sr:n4-R8-classical} is exactly
\begin{align*}
 &6\operatorname{tr}(\mathsf C_0Y^2X^2)-18\operatorname{tr}(\mathsf C_0YX^2Y)
   +12\operatorname{tr}(\mathsf C_1YX^3)-6\operatorname{tr}(\mathsf C_2X^4)\\
 &\quad+3A\operatorname{tr}(\mathsf C_0Y^2)-3C\operatorname{tr}(\mathsf C_0X^2)
   +6w_1\operatorname{tr}(\mathsf C_0Y)-6w_2\operatorname{tr}(\mathsf C_0X).
\end{align*}
The corresponding classical relative primitive is
\begin{align*}
 U_8={}&-6\operatorname{tr}(bX^2MX^2)
          -3A\operatorname{tr}(bX^2M)-6A\operatorname{tr}(bXMX)\\
       &-6w_0\operatorname{tr}(bXM)
          +\tfrac12(d_0-A^2)\operatorname{tr}(bM).
\end{align*}
It is an invariant trace polynomial with no undifferentiated $c$.
Its classical BRST image is $R_8$. The complete quantum identity and
an algebraic verification of all three quantum lifts are given next.

For the rank-four cubic fields I use
\[
 w_i=\frac{(3-i)!}{3!}f_{(0)}^iP_3,\qquad
 B_i=\frac{(2-i)!}{2!}f_{(0)}^iB_3,\qquad
 C_i=\frac{(2-i)!}{2!}f_{(0)}^iC_3.
\]
The matrices in these definitions are now $4\times4$. Quartic bottoms
are $d_i=(4-i)!f_{(0)}^iP_4/4!$, so $d_0=d_{40}$,
$d_1=d_{31}$ and $d_2=d_{22}$.

\subsection{The case \texorpdfstring{$N=4$}{N=4}: the complete lowest-space algebra}
\label{sr:rank-four-zero}
Let $E$ be a finite-dimensional lowest space at weight $r>0$.
Write $e,H,f$ for the current zero modes, and
\[
 \Omega=ef+fe+\tfrac12H^2.
\]
All products below are compositions of operators on $E$, in the displayed
order.  Write $F_i=B_i$ or $F_i=C_i$ for the zero modes of the two cubic
fermion triplets, with
\[
 [e,F_i]=iF_{i-1},\quad [f,F_i]=(2-i)F_{i+1},\quad
 [H,F_i]=(2-2i)F_i.
\]
Let $D_i=o(d_{4-i,i})$, and let $t_i$ be the zero modes of the normalized
polarizations of $(\operatorname{tr}X^2)^2$.  Thus
\[
 t_0=4e^2,\quad t_1=-(He+eH),\quad
 t_2=\tfrac23(H^2-ef-fe),
\]
with the last two components determined by lowering.  It is convenient
to use
\[
 \mathsf R_i=D_i-\tfrac14t_i=\widehat W_i+\tfrac3{17}t_i,
\]
where $\widehat W_i$ are the zero modes of the harmonic quartic bottom with
coefficient $29/68$.  The $\mathsf R_i$ form a spin-two tensor.

\subsubsection{Direct mode-sum rules}
For an odd supercurrent $S$, define $\psi_S(v)=o(S_{(0)}v)$ when $v$
has half-integer conformal weight.  For a current $J$ and a cubic bottom
component $w$ of weight $3/2$, normal-mode splitting gives
\begin{align*}
 \psi_S(:Jw:)&=\psi_S(w)J-\psi_{S_{(0)}J}(w),\\
 \psi_S(:J\partial w:)&=-2\psi_S(:Jw:),\qquad
 \psi_S(:(\partial J)w:)=-\psi_S(:Jw:),\\
 \psi_S(:Tw:)&=(r+\tfrac52)\psi_S(w),\qquad
 \psi_S(\partial^2w)=6\psi_S(w).
\end{align*}
For a quartic bottom $d$ of weight two,
\[
 \psi_S(:wd:)=o(d)\psi_S(w)-\psi_S(w_{(0)}d).
\]
For completeness the rules use the following finite splitting identities.
If $a,b$ have weight two, then
\[
 o(:ab:)=(-1)^{|a||b|}o(b)o(a)-o(a_{(0)}b).
\]
If the first factor has weight $3/2$ and the product has integer weight,
then $o(:ab:)=-o(a_{(0)}b)$; if it has weight $5/2$, then
$o(:ab:)=-2o(a_{(0)}b)-o(a_{(1)}b)$.
These follow by writing
\[
 (:ab:)_{(n)}=\sum_{i<0}a_{(i)}b_{(n-1-i)}
 +(-1)^{|a||b|}\sum_{i\geq0}b_{(n-1-i)}a_{(i)}.
\]
The lowest-space condition kills the first sum and all but finitely many
terms of the second.  Commuting its remaining lowering modes to the
right gives the formulas.  In particular
$S_{(0)}T=\frac12\partial S$ contributes $\frac12\psi_S(w)$ to
$\psi_S(:Tw:)$, while $o(:T(S_{(0)}w):)=(r+2)\psi_S(w)$.
No parity-twisted Zhu convention is used.

\subsubsection{The two degree-seven descendant constraints}
The contraction formula \eqref{of:wick-formula} gives
\[
 w_{0(0)}d_{22}-2w_{1(0)}d_{31}+w_{2(0)}d_{40}
       =-6:ew_1:-3:Hw_0:+\tfrac{40}{3}\partial w_0.
\]
Applying the preceding mode rules to the two supercurrent descendants
of the degree-seven chain identity gives
\begin{align}
 0={}&2D_0F_1-2D_1F_0-2F_1e^2-F_0He-8F_0e,
 \label{sr:n4-R7-Xtop}\\
 0={}&-3D_2F_0+4D_1F_1-D_0F_2+F_2e^2+2F_1He
       +F_0(H^2+fe)\notag\\
 &\hspace{1cm}-2F_1e+5F_0H-(10r+12)F_0.
 \label{sr:n4-R7-Ytop}
\end{align}
In spin-two and spin-one highest components these become
\begin{align}
 \mathsf R_0F_1-\mathsf R_1F_0&=\tfrac32F_0e,
 \label{sr:n4-RF-spin2}\\
 \tfrac32(\mathsf R_2F_0-2\mathsf R_1F_1+\mathsf R_0F_2)
 &=F_0\Omega-\tfrac14(2F_1e+F_0H)-(10r+17)F_0.
 \label{sr:n4-RF-spin1}
\end{align}
For example, subtracting one quarter of the $f$-commutator of
\eqref{sr:n4-R7-Xtop} from \eqref{sr:n4-R7-Ytop} extracts the
spin-one component.  The barred and unbarred supercurrent calculations
give the same two formulas for $C$ and $B$, respectively.

\subsubsection{Odd anticommutators and the quartic--odd commutators}
In field identities below, $\mathsf R_i$ denotes the field
$d_{4-i,i}-t_i/4$ before taking its zero mode. For two weight-two fields set
\[
 \mathcal Z(a,b)=a_{(0)}b-\tfrac12\partial(a_{(1)}b).
\]
Its zero mode is the supercommutator of the two zero modes.
The following chain identities give
\[
 \{B_i,B_j\}=\{C_i,C_j\}=0,
 \qquad \{B_i,C_j\}=-\{B_j,C_i\}.
\]
Here the vanishing of the symmetric mixed scalar is justified by the
chain identity
\begin{align*}
 &\mathcal Z(B_0,C_2)-2\mathcal Z(B_1,C_1)+\mathcal Z(B_2,C_0)\\
 &\quad=-\tfrac32\bigl(:G_X\bar G_Y:-:G_Y\bar G_X:+\partial T\bigr)
              -\tfrac32 Q\operatorname{tr}(bbp).
\end{align*}
The expression in parentheses has zero mode zero on $E$.
The same-ghost scalar is
\[
 \mathcal Z(B_0,B_2)-\mathcal Z(B_1,B_1)
       =-\tfrac32:G_XG_Y:-\tfrac12Q\operatorname{tr}(bbb),
\]
whose zero mode also vanishes.  The spin-two highest components vanish
directly.  The barred scalar is, explicitly,
\[
 \mathcal Z(C_0,C_2)-\mathcal Z(C_1,C_1)
       =-\tfrac32:\bar G_X\bar G_Y:-\tfrac32Q\operatorname{tr}(bpp).
\]
Thus $C,C$ vanishing does not assume that a ghost rotation preserves
the BRST differential.

The antisymmetric mixed vector is the chain identity
\begin{align}\label{sr:mixed-vector-chain}
 \mathcal Z(B_0,C_1)-\mathcal Z(B_1,C_0)
 ={}&\tfrac43L_4+\tfrac1{12}\Delta_4
 -\tfrac{58}{51}:eT:+\tfrac7{51}:G_X\bar G_X:
 +\tfrac{11}{51}:H\partial e:\notag\\
 &-\tfrac{11}{51}:(\partial H)e:
 -\tfrac{19}{17}\partial^2e
 +\tfrac13Q\operatorname{tr}(bX\partial X)
 -\tfrac13Q\operatorname{tr}(b(\partial X)X).
\end{align}

Let $Z=\{B_0,C_1\}-\{B_1,C_0\}$.  Eliminating the quartic top with the
degree-six chain identity gives the particularly simple expression
\begin{equation}
 Z=\tfrac23K,\qquad
 K=\mathsf R_2e+\mathsf R_1H-\mathsf R_0f+\tfrac13(\Omega-r-3)e.
 \label{sr:n4-BC-top}
\end{equation}
All other mixed anticommutators follow by $\mathfrak{sl}_2$ covariance.
In particular $\{B_0,C_1\}=K/3$, $\{B_1,C_1\}=0$.
To check the constants in this elimination, if $\widehat H_4$ is the quartic top
used in the transition calculation, the intermediate formulas are
\begin{align*}
 Z&=\tfrac43o(\widehat H_4)-\tfrac{58r+107}{51}e,\\
 4o(\widehat H_4)&=2(D_2e+D_1H-D_0f)
                  +(2\Omega+\tfrac{140}{51}r+\tfrac{39}{17})e,\\
 t_2e+t_1H-t_0f&=(-\tfrac83\Omega+4)e.
\end{align*}

The quartic--odd commutators are completely fixed by
\begin{equation}
 [\mathsf R_0,F_0]=0,\qquad [\mathsf R_0,F_1]=2F_0e,\qquad
 [\mathsf R_0,F_2]=-2F_0H-2F_0,
 \label{sr:n4-RF-comm}
\end{equation}
and flavor covariance.  Explicitly, rows are indexed by $i=0,\ldots,4$
and columns by $j=0,1,2$:
\[
\begin{array}{c|ccc}
 [\mathsf R_i,F_j]&0&1&2\\\hline
0&0&2F_0e&-2F_0H-2F_0\\
1&-F_0e&F_1e+\tfrac12F_0&-F_1H-F_0f-F_1\\
2&-\tfrac43F_1e+\tfrac13F_0H-\tfrac13F_0
 &\tfrac13F_2e+\tfrac13F_0f+\tfrac23F_1
 &-\tfrac13F_2H-\tfrac43F_1f-\tfrac13F_2\\
3&-F_2e+F_1H-F_1&F_1f+\tfrac12F_2&-F_2f\\
4&2F_2H-2F_2&2F_2f&0
\end{array}
\]
The chain-level identities for the nontrivial first row are
\begin{align*}
 \mathcal Z(\mathsf R_0,B_1)&=2:eB_0:+\tfrac13:G_Xw_0:,\\
 \mathcal Z(\mathsf R_0,B_2)&=-2:HB_0:+2:G_Xw_1:
       -\tfrac43:G_Yw_0:-2\partial B_0+2Q\operatorname{tr}(bbX^2),\\
 \mathcal Z(\mathsf R_0,C_2)&=-2:HC_0:+2:\bar G_Xw_1:
       -\tfrac43:\bar G_Yw_0:-2\partial C_0\\
 &\hspace{2cm}+2Q\operatorname{tr}(bpX^2)-2Q\operatorname{tr}(bX^2p).
\end{align*}
The formula for $C_1$ is the same as the first line with bars and $C$.

\subsubsection{A consequence of odd Jacobi: the cubic odd zero modes vanish}
The mixed antisymmetry, same-ghost vanishing, and super Jacobi imply
$[K,F_1]=0$.  Substituting \eqref{sr:n4-RF-comm} and then eliminating
the two quartic combinations by
\eqref{sr:n4-RF-spin2}--\eqref{sr:n4-RF-spin1} gives
\begin{equation}
 F_2e^2+F_1He-F_0fe+3(r+2)F_0=0.
 \label{sr:n4-odd-Casimir}
\end{equation}
One check on this calculation is that, before using the degree-seven
relations,
\begin{align*}
 [K,F_1]={}&\mathsf R_2F_0-\mathsf R_0F_2+F_2e^2+F_1He-F_0fe
      +\tfrac43F_1e-\tfrac56F_0H\\
 &+\tfrac13F_0\Omega+\tfrac{1-r}{3}F_0.
\end{align*}

Decompose $E$ into finite-dimensional irreducible $\mathfrak{sl}_2$
modules.  A spin-one tensor between spins $j$ and $j'$ can have
$j'=j+1,j,j-1$.  On its raising and lowering transition components, the
first three terms of \eqref{sr:n4-odd-Casimir} vanish identically;
on its diagonal component they equal $-2j(j+1)F_0$.
These statements can be checked without a tensor-operator theorem:
take a highest vector $v$ with $e v=0$ and
$e f^a v=a(2j-a+1)f^{a-1}v$.  For the raising transition evaluate on
$v$; for the lowering transition evaluate on $f^2v$.  The diagonal
spin-one tensor is proportional to $(e,-H/2,-f)$, and substitution gives
$-\Omega e$.
For $r>0$, the off-diagonal transitions therefore vanish and the only
possible nonzero diagonal blocks obey
\[
 2j(j+1)=3(r+2),\qquad j\geq\tfrac32.
\]
There is at most one such spin.  On its isotypic block write
\[
 B=(e\beta,-H\beta/2,-f\beta),\qquad
 C=(e\gamma,-H\gamma/2,-f\gamma),
\]
where the odd multiplicity operators $\beta,\gamma$ commute with the
current action.  The diagonal spin-two block of $R$ has the form
\[
 \mathsf R_0=e^2\rho,\quad \mathsf R_1=-\tfrac12e(H+1)\rho,\quad
 \mathsf R_2=(H^2/4-\Omega/6)\rho
\]
for an even multiplicity operator $\rho$.
Projecting \eqref{sr:n4-RF-comm} to this block involves no off-diagonal
$R$ terms, because $B,C$ preserve spin.  Since $e^3\ne0$ for
$j\geq3/2$, the first two commutators give
\[
 [\rho,\beta]=[\rho,\gamma]=0,\qquad
 \rho\beta=\beta,\quad \rho\gamma=\gamma.
\]
The same-ghost and symmetric mixed anticommutators give
$\beta^2=\gamma^2=0$ and $\beta\gamma+\gamma\beta=0$.
On the other hand $Z=2e\beta\gamma$, while
\eqref{sr:n4-BC-top} and $\Omega=3(r+2)$ give
\[
 \beta\gamma=\frac{2r+3}{9}(1-3\rho).
\]
Right multiplication by $\beta$ has zero left side and right side
$-2(2r+3)\beta/9$.  Hence $\beta=0$.  The analogous multiplication
gives $\gamma=0$.  I conclude
\begin{equation}
 B_i|_E=C_i|_E=0\qquad(i=0,1,2),\qquad r>0.
 \label{sr:n4-odd-zero-vanish}
\end{equation}

\subsubsection{The even commutators}
After the odd operators vanish, \eqref{sr:n4-BC-top} becomes
$\mathsf R_2e+\mathsf R_1H-\mathsf R_0f+(\Omega-r-3)e/3=0$.
The highest quartic commutator is

\[
 [\mathsf R_0,\mathsf R_1]=2e\mathsf R_0,
\]
because its chain-level representative is
$2:e\mathsf R_0:+\frac13:w_0w_0:$ and the last term has zero mode zero.
The independent spin-one quartic commutator has the zero-mode consequence
\begin{align*}
 [\mathsf R_0,\mathsf R_3]-3[\mathsf R_1,\mathsf R_2]
 &=\tfrac73(\mathsf R_2e+\mathsf R_1H-\mathsf R_0f)
       +\left(\tfrac{25\Omega-160r}{9}-\tfrac{185}{6}\right)e\\
 &=\left(2\Omega-17r-\tfrac{57}{2}\right)e
       \qquad\text{after }B=C=0.
\end{align*}
Its chain-level representative is
\begin{align*}
 &6Jd+16J^3-\tfrac92QU_6-\tfrac{22}{3}L_4-\tfrac{11}{24}\Delta_4\\
 &\quad-\tfrac{650}{51}:eT:-\tfrac{115}{51}:G_X\bar G_X:
 +\tfrac{628}{51}:H\partial e:-\tfrac{1036}{51}:(\partial H)e:
 -\tfrac{91}{17}\partial^2e\\
 &\quad+\tfrac{299}{12}Q\operatorname{tr}(bX\partial X)
 +\tfrac{229}{12}Q\operatorname{tr}(b(\partial X)X)\\
 &\quad-7Q\operatorname{tr}(bXMX)-8Q\operatorname{tr}(bX^2M),
\end{align*}
where $Jd=:ed_{22}:+:Hd_{31}:-:fd_{40}:$ and
$J^3=:e(:ef:):+\frac14:e(:HH:):$.
Together with $[\mathsf R_0,\mathsf R_1]=2e\mathsf R_0$, flavor lowering determines all
quartic--quartic commutators.
\subsubsection{The symmetric degree-eight relation}
Let $\widehat W_i$ be the normalized flavor components of
$\widehat W_0=\widehat P_4$, so that $f_{(0)}\widehat W_i=(4-i)\widehat W_{i+1}$, and let
$B^{(4)}_0=-G_{Y,(0)}\widehat W_0/4$, $C^{(4)}_0=-\bar G_{Y,(0)}\widehat W_0/4$.
Use the explicit normal ordering
\begin{align*}
 R_8^{\rm n}={}&12:d_0d_2:-12:d_1d_1:
 -3:A(:Ad_2:):+6:A(:Bd_1:):-3:B(:Bd_0:):\\
 &+4:A(:w_0w_2:):-3:A(:w_1w_1:):
       +:C(:w_0w_0:):-2:B(:w_0w_1:):.
\end{align*}
The exact quantum identity is
\begin{equation}\label{sr:r8-quantum}
 R_8^{\rm n}-QU_8=\mathcal L+Q\mathcal V,
\end{equation}
where
\begin{align*}
\mathcal L={}&\tfrac{39}{2}\partial^2\widehat W_0
 +\tfrac{8257}{34}:e\partial^2e:
 +12:e\partial \widehat W_1:-\tfrac{1860}{17}:(\partial e)(\partial e):
 +144:(\partial e)\widehat W_1:\\
 &-39:H\partial \widehat W_0:+30:(\partial H)\widehat W_0:
 +24:G_XC^{(4)}_0:-24:\bar G_XB^{(4)}_0:+28:T\widehat W_0:\\
 &-5:w_0\partial w_1:+13:(\partial w_0)w_1:+12:B_0C_0:\\
 &-\tfrac{18}{17}:e(:e\partial H:):
 -\tfrac{480}{17}:e(:eT:):
 -\tfrac{2280}{17}:e(:(\partial e)H:):
 -\tfrac{120}{17}:e(:G_X\bar G_X:):,
\end{align*}
and, with products within traces taken in the elementary Wick sense,
\begin{align*}
\mathcal V={}&54\operatorname{tr}(b(\partial X)X^3)
 +38\operatorname{tr}(bX(\partial X)X^2)
 +26\operatorname{tr}(bX^2(\partial X)X)
 -38\operatorname{tr}(bX^3\partial X)\\
 &+\tfrac{128}{3}w_0\operatorname{tr}((\partial b)X)
       -44A\operatorname{tr}(b(\partial X)X).
\end{align*}
Every primitive is invariant and contains no undifferentiated $c$.
The complete algebraic Wick verification is given in
subsection~\ref{sr:quantum-verification}; all terms of $U_8$ and
$\mathcal V$ have been displayed.

Here is its direct zero-mode reduction on a positive lowest space, after
$B_i=C_i=0$.  Besides the earlier mode-sum rules, the required short OPEs
are
\begin{align*}
 G_{X,(0)}C^{(4)}_0&=\tfrac14\partial \widehat W_0,&
 \bar G_{X,(0)}B^{(4)}_0&=-\tfrac14\partial \widehat W_0,\\
 w_{0,(0)}w_1&=3\mathsf R_0^{\rm field},&
 B_{0,(1)}C_0&=\mathsf R_0^{\rm field},\qquad
 B_{0,(0)}C_0=\tfrac12\partial \mathsf R_0^{\rm field}.
\end{align*}
Here $\mathsf R_0^{\rm field}=d_0-\frac14A^2$ has zero mode $\mathsf R_0$.
Consequently
\begin{align*}
 o(\mathcal L)={}&(28r+197)\widehat W_0-168\widehat W_1e+48\widehat W_0H+21\mathsf R_0\\
 &+\tfrac{6277-480r}{17}e^2+\tfrac{2298}{17}He^2\\
 ={}&(28r+218)\mathsf R_0-168\mathsf R_1e+48\mathsf R_0H+(425-48r)e^2+42He^2.
\end{align*}
On the left, set $K_d=d_{0,(0)}d_2-d_{1,(0)}d_1$.  The required exact OPE identities are
\begin{align*}
K_d={}&8:ed_1:-2:Hd_0:+\tfrac23:w_0w_1:
       +4:e(:eH:):+\tfrac{37}{12}\partial d_0
       +\tfrac{173}{4}:e\partial e:,\\
w_{1,(0)}w_1={}&-6\partial e,\qquad
w_{0,(0)}w_2=6d_1+3:eH:+12\partial e.
\end{align*}
Using $o(:ab:)=o(b)o(a)-o(a_{(0)}b)$ for the weight-two bosons gives
\begin{align*}
o(R_8^{\rm n})={}&12(D_2D_0-D_1^2)-12D_2e^2-12D_1He-3D_0H^2
 -144D_1e+24D_0H\\
 &+74D_0+24\mathsf R_0-72He^2+651e^2-6\mathsf R_0H.
\end{align*}
Subtracting the preceding right side, substituting
$D_0=\mathsf R_0+e^2$, $D_1=\mathsf R_1-(He+eH)/4$,
$D_2=\mathsf R_2+H^2/4-\Omega/6$, and using
\[
 [\mathsf R_0,\mathsf R_2]=\tfrac13(8\mathsf R_1e-2\mathsf R_0H),\qquad
 [\Omega,\mathsf R_0]=8\mathsf R_1e+4\mathsf R_0H+12\mathsf R_0,
\]
yields the symmetric spin-two relation
\begin{equation}
12\left(\tfrac12\{\mathsf R_0,\mathsf R_2\}-\mathsf R_1^2\right)
 -\{\Omega,\mathsf R_0\}-(28r+66)\mathsf R_0+(48r+87)e^2=0.
\label{sr:n4-R8-top}
\end{equation}
The first commutator used here follows by one flavor lowering of
$[\mathsf R_0,\mathsf R_1]=2e\mathsf R_0$; the second follows directly from current covariance.
\subsubsection{Every finite positive lowest space is a current singlet}
I now use \eqref{sr:n4-R8-top} to exclude all nontrivial current spins;
including arbitrary multiplicities.
A rank-two tensor $R$ between irreducible spins $j,j'$ permits
$j'-j\in\{-2,-1,0,1,2\}$.  The vector relation $K=0$ kills the
$\pm1$ transitions.  For example, put $n=2j$ and use a highest vector
$v$, with the unnormalized powers $f^a v$ in this paragraph.
For an upward $j\to j+1$ transition put $\mathsf R_0fv=\alpha u$, where $u$ is
the target highest vector.  Then
\[
 (\mathsf R_2e+\mathsf R_1H-\mathsf R_0f)v=-\tfrac{n+4}{4}\alpha u.
\]
For a downward $j\to j-1$ transition put $\mathsf R_0f^3v=\alpha u$; evaluation
on $f^2v$ gives $-(n+2)\alpha u/12$.  The scalar-current part of $K$
has no off-diagonal block, so both coefficients force the transition
to vanish.  The usual edge cases in which the displayed source vector
does not exist also have no such tensor transition.

For $j\geq1$, the diagonal block must be
\[
 \mathsf R_0=e^2\rho_j,\quad \mathsf R_1=-e(H+1)\rho_j/2,\quad
 \mathsf R_2=(H^2/4-\Omega/6)\rho_j,
\]
with a multiplicity operator $\rho_j$.  The same vector relation gives
\[
 \rho_j=\frac{c_j-r-3}{2c_j-3}\,1,\qquad c_j=2j(j+1).
\]
For $j=1/2$ the diagonal rank-two tensor is absent; $K=0$ already forces
$r=-3/2$, so that spin cannot occur when $r>0$.

It remains to kill $j\leftrightarrow j+2$.  Use divided-power bases
$f v_k=(k+1)v_{k+1}$, and put $n=2j$.
If an upward highest block is $\mathsf R_0v_0=\alpha u_0$, covariance gives
\[
 \mathsf R_1v_0=\frac{\alpha}{n+4}u_1,\qquad
 \mathsf R_2v_0=\frac{2\alpha}{(n+4)(n+3)}u_2.
\]
Project \eqref{sr:n4-R8-top} from spin $j$ to spin $j+2$.
In its quadratic term only paths through these two spins can contribute,
because the $\pm1$ blocks were already killed.
Writing $c_n=n(n+2)/2$ and
$\rho_n=(c_n-r-3)/(2c_n-3)$, the coefficient of $\alpha u_0$ is
\begin{align*}
 &n(n-1)\rho_n+(n+6)(n+7)\rho_{n+4}
       -(c_n+c_{n+4})-28r-66\\
 &\hspace{2cm}=-30(r+2).
\end{align*}
For instance the first two terms simplify to
$n^2+6n+18-2r$, since $2c_n-3=(n-1)(n+3)$.
The formula includes $j=0$, whose diagonal tensor is zero.  Its only
singular case is $j=1/2$, already excluded.
For a downward block, normalize $\mathsf R_0u_{k+4}=\beta v_k$.
Lowering gives $\mathsf R_1u_{k+3}=-\beta v_k$ and
$\mathsf R_2u_{k+2}=\beta v_k$.  Evaluating \eqref{sr:n4-R8-top} on
$u_4$ gives the identical coefficient $-30(r+2)$.
This direct calculation uses no positivity or unitarity.  As $r>0$, all $\pm2$ blocks vanish.

Now $R$ preserves each spin.  For $j\geq3/2$, the commutator
$[\mathsf R_0,\mathsf R_1]=2e\mathsf R_0$ and $e^3\ne0$ give $\rho_j^2=\rho_j$.
The choice $\rho_j=1$ would give $r=-c_j\leq0$.
The choice $\rho_j=0$ gives $r=c_j-3$, and the spin-one $RR$ relation
then gives $c_j=3/2$, incompatible with $j\geq3/2$.
For the remaining spin $j=1$, $c_j=4$ and $\rho_j=(1-r)/5$.
On a diagonal rank-two tensor,
\[
12\left(\tfrac12\{\mathsf R_0,\mathsf R_2\}-\mathsf R_1^2\right)
       =(15-2c_j)\rho_j^2e^2.
\]
Thus the coefficient of $e^2$ in \eqref{sr:n4-R8-top} is
\[
 (15-2c_j)\rho_j^2-(2c_j+28r+66)\rho_j+48r+87
       =\frac{3}{25}(49r^2+472r+604)>0,
\]
a contradiction.  Only spin zero remains, on which the rank-two
operator also vanishes.  Consequently every finite positive lowest
space satisfying the displayed exact identities obeys
\[
 e|_E=H|_E=f|_E=0,\qquad \mathsf R_i|_E=B_i|_E=C_i|_E=0.
\]
The remaining scalar lowest space is excluded by the half-integer
mode calculation that follows.

\subsubsection{Exclusion of a positive scalar lowest space}
Let $v\in E$, and set $u=G_{Y,(0)}v$.  It has weight $r+1/2$ and lies
in the spin-$1/2$ isotypic component of that weight space.  Let $\pi$
be the current-equivariant projection onto that component.
All positive current and Virasoro modes kill $u$, since they lower
weight by at least one.  The cubic bottom modes $w_{i,(n)}$ for
$n\geq2$ also kill $u$ by weight.  For $n=1$ their output would lie
in the scalar space $E$, whereas spin $3/2$ tensored with spin $1/2$
contains only spins $1$ and $2$.  Hence these modes kill $u$ as well.
The quartic operators $D_i,\widehat W_i$ have spin two, so their zero modes have
no block from spin $1/2$ to spin $1/2$; inserting a current zero mode
before them does not change this conclusion.

Put $g=G_{X,(0)}v=e_{(0)}u$.  Normal-order mode splitting gives
\begin{align*}
 \pi o(R_6^{\rm n})u&=16g,&
 \pi o(:eT:)u&=(r+\tfrac12)g,\\
 \pi o(:G_X\bar G_X:)u&=(r+1)g,&
 \pi o(:H\partial e:)u&=\pi o(:(\partial H)e:)u=g,\\
 \pi o(\partial^2e)u&=2g,&
 \pi o(\Delta_4)u&=0.
\end{align*}
Here $R_6^{\rm n},\Delta_4$ use the normalization of the earlier exact
quantum degree-six relation.  The first equality can be checked
explicitly: the $D_i$ terms vanish after projection, while the remaining
operator is
$-3He+18e+4fe^2+H^2e$, equal to $16e$ on spin $1/2$.
For the fermion normal product, the two surviving summands give
\[
 o(:G_X\bar G_X:)u
 =e_{(0)}u+G_{X,(0)}\bar G_{X,(1)}u
               -\bar G_{X,(0)}G_{X,(1)}u=(r+1)g.
\]
Indeed $G_{X,(1)}G_{Y,(0)}v=0$, while
$\bar G_{X,(1)}G_{Y,(0)}v=rv$ follows from the small
$\mathcal N=4$ OPE and the scalar-bottom condition.
This extra $r$ term is essential; $u$ has not been claimed to be a
lowest vector for the entire vertex algebra.

Substituting these finite mode sums into the quantum $R_6$ identity
therefore gives
\begin{equation}
 4\pi \widehat H_{4,(2)}G_{Y,(0)}v=\tfrac{10}{3}(r+2)G_{X,(0)}v.
 \label{sr:n4-singlet-R6}
\end{equation}
There is an independent way to evaluate the left side.  Write
$\widetilde B^{(4)}_1=f_{(0)}B^{(4)}_0$ in this paragraph, without division by $3$.
The exact short-multiplet OPEs are
\[
 \widehat H_{4,(0)}G_Y=-\partial \widetilde B^{(4)}_1,\qquad
 \widehat H_{4,(1)}G_Y=-\tfrac54\widetilde B^{(4)}_1,\qquad
 \widehat H_{4,(2)}G_Y=0.
\]
They have a short direct verification.  In trace-orthonormal coordinates,
write $P=\widehat P_4$ and $P_a=\partial_aP$, $P_{ab}=\partial_a\partial_bP$.
The displayed definitions of the multiplet give the elementary Wick
expressions
\[
 \widehat H_4=\tfrac14P_{ab}b^ap^b-\tfrac3{16}P_a\partial Y^a
                  +\tfrac1{16}P_{ab}Y^a\partial X^b,
 \qquad \widetilde B^{(4)}_1=\tfrac14P_{ab}Y^ab^b.
\]
Direct contraction gives
$G_{Y,(0)}\widehat H_4=-\partial \widetilde B^{(4)}_1/4$ and
$G_{Y,(1)}\widehat H_4=-5\widetilde B^{(4)}_1/4$.
The possible third-order coefficient is proportional to
$\partial_a(\Delta_X P)b^a$ and vanishes:
\[
 \Delta_X\operatorname{tr}X^4=29\operatorname{tr}X^2,\qquad
 \Delta_X(\operatorname{tr}X^2)^2=68\operatorname{tr}X^2.
\]
These identities follow from
$\sum_aT^aMT_a=\operatorname{tr}(M)I-M/4$ and
$\sum_aT^aT_a=15I/4$.  Thus the coefficient $29/68$ in $\widehat P_4$ makes
it harmonic.  Vertex skew symmetry gives the stated $\widehat H_4,G_Y$ OPEs.
The operator $\widehat H_{4,(2)}$ kills $E$, since it has spin one and $E$ is
scalar.  Its commutator with $G_{Y,(0)}$ consequently gives
\[
 \widehat H_{4,(2)}G_{Y,(0)}v
 =\bigl(-\partial \widetilde B^{(4)}_1\bigr)_{(2)}v
                 -\tfrac52\widetilde B^{(4)}_{1,(1)}v
 =-\tfrac12\widetilde B^{(4)}_{1,(1)}v.
\]
The last vector lies in spin $3/2$, so its projection by $\pi$ is zero.
Equation \eqref{sr:n4-singlet-R6} and $r>0$ force
$G_{X,(0)}E=0$.  Finally the small $\mathcal N=4$ anticommutator is
\[
 \{G_{X,(0)},\bar G_{Y,(1)}\}=-L_0+\tfrac12H_{(0)}.
\]
Both operators on the left kill $E$: the first by the preceding result,
the second by its negative conformal degree.  The right side is $-r$
on $E$.  This contradicts $r>0$ and $E\ne0$.

Thus $\mathcal H_4/A_4=0$. Together with
Theorem~\ref{thm:strong-closure}, this proves strong generation by the
complete degree-two, degree-three and degree-four multiplets in every
ghost degree and every conformal weight.

\subsection{Algebraic verification of the rank-four quantum identities}
\label{sr:quantum-verification}

I give the complete finite contraction data for
\eqref{sr:n4-R6-quantum}, \eqref{sr:n4-R7-quantum} and
\eqref{sr:r8-quantum}. The calculation takes place in the space of
signed cyclic trace polynomials and uses Cayley--Hamilton only in the
explicit combinations displayed below. In particular, it does not
specialize to a representation of the zero-mode algebra.

For an adjoint-covariant matter polynomial $S=S^at_a$ with no
derivatives, $b^aS_a$ is gauge invariant. This applies to each $S_k$
below, whose matrix words have invariant trace coefficients.
The three possible residue types in the BRST current give
\begin{align}\label{sr:brst-residue-formula}
 Q(b^aS_a)={}&\mu^aS_a+\mathcal D(S),\qquad \mu=[X,Y],\notag\\
 \mathcal D(S)={}&f_{abc}\left(
       S_{a,Y^b}\partial Y^c-S_{a,X^c}\partial X^b\right)
       +f_{abc}b^dp^aS_{d,Y^bX^c},\notag\\
 &f_{abc}=\operatorname{tr}(t_a[t_b,t_c]).
\end{align}
The first term is the single $c$--$b$ contraction; its other
single-contraction gauge terms cancel. The two derivative terms
contract $c$ with $b$ and one matter pair, then take the first Taylor
coefficient of the surviving matter field. The last term contracts
two matter pairs and takes the first Taylor coefficient of $c$.
These are all possibilities; a triple contraction leaves no field at
the insertion point and has no residue. This proves the formula.

For two derivative-free matter polynomials, the full regular
coefficient is
\begin{equation}\label{sr:regular-residue-formula}
 :UV:=\left.\sum_{\ell\ge0}\frac{1}{(\ell!)^2}
   \partial_z^\ell\Omega^\ell(U(z)V(w))\right|_{z=w},\qquad
 \Omega=\sum_a(\partial_{X^a(z)}\partial_{Y^a(w)}
                    -\partial_{Y^a(z)}\partial_{X^a(w)}).
\end{equation}
One factorial counts the unordered matching and the other comes from
Taylor expansion. Differentiating propagators extends this formula to
jets and ghosts, with the Koszul signs of
\eqref{of:wick-formula}. For nested products it is applied successively
in the stated order; no reassociation is performed. Every monomial
in $R_k$ has two $Y$'s, so globally at most two matter contractions
can contribute to its regular correction.

Write $U_k=\operatorname{tr}(bS_k)$, retaining the scalar trace factors
in the displayed definitions of $U_k$, and let $R_k^{\rm PBW}$ denote
the corresponding simultaneous product. Define
\[
 \mathcal W_k=R_k^{\rm n}-R_k^{\rm PBW},\qquad
 \mathcal D_k=\mathcal D(S_k)\quad(k=6,7,8).
\]
The following three tables list every nonzero cyclic-trace coefficient
of these two expressions. To keep the entries compact, brackets mean
a simultaneous trace and
\[
 X_1=\partial X,\qquad Y_1=\partial Y,\qquad
 X_2=\tfrac12\partial^2X,\qquad p=\partial c.
\]
The $X_2$ convention includes its factor $1/2$; all other displayed
jets have order one. Products of brackets keep their displayed order,
including the signs of odd traces. A blank coefficient means zero.
Each row follows from \eqref{sr:brst-residue-formula} or
\eqref{sr:regular-residue-formula} and the completeness identity
\eqref{fnd:completeness}. For example, for $S=XMX$ the two derivatives are
\[
 S_{Y^a}=X[X,t_a]X,\qquad
 S_{X^a}=t_aMX+X[t_a,Y]X+XMt_a.
\]
Completeness gives their derivative contribution
$-2A\langle XY_1\rangle+2\langle XX_1\rangle\langle XY\rangle
 -4\langle M[X,X_1]\rangle$ and their ghost contribution
$-2\langle Xb\rangle\langle Xp\rangle
 -4\langle b[X,[X,p]]\rangle$.
One may obtain the latter directly from
\[
 \sum_a\bigl([p,t_a][X,t_a]X+X[[p,t_a],t_a]X
                  +X[X,t_a][p,t_a]\bigr)
 =-2\operatorname{tr}(Xp)X-4[X,[X,p]].
\]
This gives every entry contributed by the $XMX$ primitive. The other
primitive words use the same differentiation rule with the derivative
inserted at each $X$ position. Thus the tables include the scalar part
of the traceless projector and all split-trace terms.

\begingroup\small
\begin{longtable}{@{}lrr@{}}
\hline \phantomsection\label{sr:table-6}Trace monomial & $\mathcal W_6$ & $\mathcal D_6$\\ \hline
\endfirsthead
\multicolumn{3}{c}{\tablename~\thetable\ (continued)}\\
\hline Trace monomial & $\mathcal W_6$ & $\mathcal D_6$\\ \hline
\endhead
\hline
\multicolumn{3}{r}{\textit{Continued on the next page}}\\
\endfoot
\hline
\caption{Complete contraction coefficients for the degree-6 relation.}\\
\endlastfoot
$\langle X^{2}\rangle\langle XY_{1}\rangle$ & $1$ & $\frac{13}{3}$ \\
$\langle X^{2}\rangle\langle X_{1}Y\rangle$ & $-4$ & $-\frac{1}{3}$ \\
$\langle X^{2}\rangle\langle bp\rangle$ &  & $\frac{1}{3}$ \\
$\langle X^{3}Y_{1}\rangle$ & $-12$ & $-12$ \\
$\langle X^{2}X_{1}Y\rangle$ & $\frac{19}{3}$ & $-\frac{53}{6}$ \\
$\langle X^{2}YX_{1}\rangle$ & $\frac{19}{3}$ & $\frac{91}{6}$ \\
$\langle X^{2}bp\rangle$ &  & $\frac{53}{6}$ \\
$\langle X^{2}pb\rangle$ &  & $\frac{91}{6}$ \\
$\langle XX_{1}\rangle\langle XY\rangle$ & $-2$ & $-4$ \\
$\langle XX_{1}XY\rangle$ & $\frac{34}{3}$ & $\frac{17}{3}$ \\
$\langle XX_{2}\rangle$ & $16$ &  \\
$\langle Xb\rangle\langle Xp\rangle$ &  & $4$ \\
$\langle XbXp\rangle$ &  & $-\frac{17}{3}$ \\
$\langle X_{1}^{2}\rangle$ & $-5$ &  \\
\end{longtable}\endgroup

\begingroup\small
\begin{longtable}{@{}lrr@{}}
\hline \phantomsection\label{sr:table-7}Trace monomial & $\mathcal W_7$ & $\mathcal D_7$\\ \hline
\endfirsthead
\multicolumn{3}{c}{\tablename~\thetable\ (continued)}\\
\hline Trace monomial & $\mathcal W_7$ & $\mathcal D_7$\\ \hline
\endhead
\hline
\multicolumn{3}{r}{\textit{Continued on the next page}}\\
\endfoot
\hline
\caption{Complete contraction coefficients for the degree-7 relation.}\\
\endlastfoot
$\langle X^{2}\rangle\langle X^{2}Y_{1}\rangle$ &  & $-6$ \\
$\langle X^{2}\rangle\langle XX_{1}Y\rangle$ & $-\frac{3}{2}$ & $-4$ \\
$\langle X^{2}\rangle\langle XYX_{1}\rangle$ & $-\frac{3}{2}$ & $10$ \\
$\langle X^{2}\rangle\langle Xbp\rangle$ &  & $4$ \\
$\langle X^{2}\rangle\langle Xpb\rangle$ &  & $10$ \\
$\langle X^{3}\rangle\langle XY_{1}\rangle$ & $\frac{8}{3}$ & $-\frac{10}{3}$ \\
$\langle X^{3}\rangle\langle X_{1}Y\rangle$ & $\frac{11}{3}$ & $\frac{10}{3}$ \\
$\langle X^{3}\rangle\langle bp\rangle$ &  & $-\frac{10}{3}$ \\
$\langle X^{4}Y_{1}\rangle$ & $-12$ & $8$ \\
$\langle X^{3}X_{1}Y\rangle$ & $-2$ & $2$ \\
$\langle X^{3}YX_{1}\rangle$ & $-2$ & $-11$ \\
$\langle X^{3}bp\rangle$ &  & $-2$ \\
$\langle X^{3}pb\rangle$ &  & $-11$ \\
$\langle X^{2}X_{1}\rangle\langle XY\rangle$ & $3$ &  \\
$\langle X^{2}X_{1}XY\rangle$ & $8$ & $1$ \\
$\langle X^{2}X_{2}\rangle$ & $24$ &  \\
$\langle X^{2}Y\rangle\langle XX_{1}\rangle$ & $-5$ &  \\
$\langle X^{2}YXX_{1}\rangle$ & $8$ &  \\
$\langle X^{2}bXp\rangle$ &  & $-1$ \\
$\langle XX_{1}^{2}\rangle$ & $-6$ &  \\
\end{longtable}\endgroup

\begingroup\small
\begin{longtable}{@{}lrr@{}}
\hline \phantomsection\label{sr:table-8}Trace monomial & $\mathcal W_8$ & $\mathcal D_8$\\ \hline
\endfirsthead
\multicolumn{3}{c}{\tablename~\thetable\ (continued)}\\
\hline Trace monomial & $\mathcal W_8$ & $\mathcal D_8$\\ \hline
\endhead
\hline
\multicolumn{3}{r}{\textit{Continued on the next page}}\\
\endfoot
\hline
\caption{Complete contraction coefficients for the degree-8 relation.}\\
\endlastfoot
$\langle X^{2}\rangle\langle X^{2}\rangle\langle XY_{1}\rangle$ & $\frac{3}{2}$ & $5$ \\
$\langle X^{2}\rangle\langle X^{2}\rangle\langle X_{1}Y\rangle$ & $\frac{3}{2}$ & $7$ \\
$\langle X^{2}\rangle\langle X^{2}\rangle\langle bp\rangle$ &  & $-7$ \\
$\langle X^{2}\rangle\langle X^{3}Y_{1}\rangle$ &  & $-24$ \\
$\langle X^{2}\rangle\langle X^{2}X_{1}Y\rangle$ & $-2$ & $-16$ \\
$\langle X^{2}\rangle\langle X^{2}YX_{1}\rangle$ & $-2$ & $8$ \\
$\langle X^{2}\rangle\langle X^{2}bp\rangle$ &  & $16$ \\
$\langle X^{2}\rangle\langle X^{2}pb\rangle$ &  & $8$ \\
$\langle X^{2}\rangle\langle XX_{1}\rangle\langle XY\rangle$ & $-12$ & $-12$ \\
$\langle X^{2}\rangle\langle XX_{1}XY\rangle$ & $16$ & $32$ \\
$\langle X^{2}\rangle\langle XX_{2}\rangle$ & $\frac{507}{2}$ &  \\
$\langle X^{2}\rangle\langle Xb\rangle\langle Xp\rangle$ &  & $12$ \\
$\langle X^{2}\rangle\langle XbXp\rangle$ &  & $-32$ \\
$\langle X^{2}\rangle\langle X_{1}^{2}\rangle$ & $\frac{287}{4}$ &  \\
$\langle X^{3}\rangle\langle X^{2}Y_{1}\rangle$ & $-5$ & $-12$ \\
$\langle X^{3}\rangle\langle XX_{1}Y\rangle$ & $19$ & $-24$ \\
$\langle X^{3}\rangle\langle XYX_{1}\rangle$ & $19$ & $48$ \\
$\langle X^{3}\rangle\langle Xbp\rangle$ &  & $24$ \\
$\langle X^{3}\rangle\langle Xpb\rangle$ &  & $48$ \\
$\langle X^{4}\rangle\langle XY_{1}\rangle$ & $-3$ & $4$ \\
$\langle X^{4}\rangle\langle X_{1}Y\rangle$ & $3$ & $-4$ \\
$\langle X^{4}\rangle\langle bp\rangle$ &  & $4$ \\
$\langle X^{5}Y_{1}\rangle$ & $-36$ &  \\
$\langle X^{4}X_{1}Y\rangle$ & $20$ & $-2$ \\
$\langle X^{4}YX_{1}\rangle$ & $20$ & $22$ \\
$\langle X^{4}bp\rangle$ &  & $2$ \\
$\langle X^{4}pb\rangle$ &  & $22$ \\
$\langle X^{3}X_{1}\rangle\langle XY\rangle$ & $24$ &  \\
$\langle X^{3}X_{1}XY\rangle$ & $76$ & $8$ \\
$\langle X^{3}X_{2}\rangle$ & $196$ &  \\
$\langle X^{3}Y\rangle\langle XX_{1}\rangle$ & $12$ &  \\
$\langle X^{3}YXX_{1}\rangle$ & $76$ & $-40$ \\
$\langle X^{3}bXp\rangle$ &  & $-8$ \\
$\langle X^{3}pXb\rangle$ &  & $-40$ \\
$\langle X^{2}X_{1}\rangle\langle X^{2}Y\rangle$ & $-105$ & $-12$ \\
$\langle X^{2}X_{1}X^{2}Y\rangle$ & $132$ & $12$ \\
$\langle X^{2}X_{1}^{2}\rangle$ & $106$ &  \\
$\langle X^{2}b\rangle\langle X^{2}p\rangle$ &  & $12$ \\
$\langle X^{2}bX^{2}p\rangle$ &  & $-12$ \\
$\langle XX_{1}\rangle\langle XX_{1}\rangle$ & $-8$ &  \\
$\langle XX_{1}XX_{1}\rangle$ & $40$ &  \\
\end{longtable}\endgroup

Let $\mathcal E_6,\mathcal E_7,\mathcal E_8$ denote the complete
right-hand sides, including their explicit BRST terms, of
\eqref{sr:n4-R6-quantum}, \eqref{sr:n4-R7-quantum}, and
\eqref{sr:r8-quantum}. The remaining comparison is a short
Cayley--Hamilton calculation. Define, for any traceless matrix letter
$Z$ of either parity,
\begin{align}\label{sr:ch-directional}
 \mathsf C_0={}&X^4-\tfrac12AX^2-\tfrac13DX
                      +(A^2/8-d_0/4)I,\notag\\
 \mathsf C_1(Z)={}&\sum_{i=0}^3X^iZX^{3-i}
  -\operatorname{tr}(XZ)X^2-\tfrac A2(XZ+ZX)
  -\operatorname{tr}(X^2Z)X-\tfrac D3Z\notag\\
  &+\bigl(\tfrac A2\operatorname{tr}(XZ)-\operatorname{tr}(X^3Z)\bigr)I.
\end{align}
Here $D=\operatorname{tr}X^3=w_0$. Cayley--Hamilton gives
$\mathsf C_0=0$ and its directional derivative gives
$\mathsf C_1(Z)=0$. For odd $Z$ introduce an auxiliary odd parameter
$\varepsilon$, apply the ordinary identity to the even matrix
$X+\varepsilon Z$, and extract its linear coefficient. This proves
exactly the same directional identity with its graded signs.

Expanding the normal products and BRST terms in $\mathcal E_k$ by
\eqref{sr:brst-residue-formula} and
\eqref{sr:regular-residue-formula}, Tables~\ref{sr:table-6}--\ref{sr:table-8}
give the following complete remainders:
\begin{align}
 \mathcal W_6-\mathcal D_6-\mathcal E_6={}&0,
 \label{sr:q6-final}\\
 \mathcal W_7-\mathcal D_7-\mathcal E_7={}&
 -20\operatorname{tr}(\mathsf C_0\partial Y)
 +5\operatorname{tr}(\mathsf C_1(\partial X)Y)\notag\\
 &-2\operatorname{tr}(\mathsf C_1(b)p),
 \label{sr:q7-final}\\
 \mathcal W_8-\mathcal D_8-\mathcal E_8={}&
 -36\operatorname{tr}(\mathsf C_0X\partial Y)
 -16\operatorname{tr}(\mathsf C_0(\partial X)Y)
 -188\operatorname{tr}(\mathsf C_0Y\partial X)\notag\\*
 &-56\operatorname{tr}(\mathsf C_0bp)
 +40\operatorname{tr}(\mathsf C_0pb)
 +132\operatorname{tr}(\mathsf C_1(\partial X)XY)
 +24\operatorname{tr}(\mathsf C_1(b)Xp).
 \label{sr:q8-final}
\end{align}
These equalities are identities in the free signed cyclic-word space:
expanding \eqref{sr:ch-directional} on their right gives each
coefficient on their left. For example, the three terms containing
$\partial Y$ in \eqref{sr:q7-final} are
$-20\langle X^4Y_1\rangle+10A\langle X^2Y_1\rangle
 +(20/3)D\langle XY_1\rangle$, and its odd terms are
\begin{align*}
 &-2\langle X^3bp\rangle+2\langle X^3pb\rangle
 -2\langle X^2bXp\rangle+2\langle X^2pXb\rangle\\
 &\quad+A\langle Xbp\rangle-A\langle Xpb\rangle
 +\tfrac23D\langle bp\rangle\\
 &\quad+2\langle X^2b\rangle\langle Xp\rangle
 -2\langle X^2p\rangle\langle Xb\rangle.
\end{align*}
This explicitly fixes the cyclic signs in the directional identity.
All right-hand sides of \eqref{sr:q6-final}--\eqref{sr:q8-final}
vanish for $N=4$. Their classical leading terms vanish by the
three displayed Cayley--Hamilton decompositions of $R_k$.
Thus $R_k^{\rm n}=QU_k+\mathcal E_k$, proving the complete quantum
identities, with their relative primitives and with every normal-order
correction retained.

The ancillary OPE identities used in the rank-four module calculation
are obtained from the same contraction rule. Their completeness can
also be seen from the polynomial degrees: the displayed $w$--$d$ and $\mathcal Z(R,F)$ components have at
most two matter contractions, and the spin-one $\mathcal Z(R,R)$
component has at most three. For $\mathcal Z(B,C)$ there are at most
two matter contractions and one ghost contraction; the completely
contracted term contributes only to the fourth pole, and hence does
not enter $\mathcal Z$. The symmetric highest components are zero
because $B_{0,(1)}C_0=\mathsf R_0^{\rm field}$ and
$B_{0,(0)}C_0=\partial \mathsf R_0^{\rm field}/2$; the same-ghost highest
components have no contraction at all. Their scalar and antisymmetric
components, including all relative primitives, have been displayed
in the lowest-space calculation. Substitution into
\eqref{of:wick-formula}, cyclic rotation, the completeness identity
\eqref{fnd:completeness}, and the Cayley--Hamilton reductions below
give the stated coefficients. No higher
contraction or independent tensor component is left over, because
$V_1\otimes V_1=V_2\oplus V_1\oplus V_0$ and
$\bigwedge^2V_2=V_3\oplus V_1$.

For clarity, the finite-rank matrix reduction in these subsidiary
identities can also be recorded explicitly. Let
$\mathscr E_{\mathsf R_0,F_i}$ be the exact Wick value of
$\mathcal Z(\mathsf R_0,F_i)$ minus its displayed field expansion in
subsection~\ref{sr:rank-four-zero}, with
$(F_i,g)=(B_i,b)$ or $(C_i,p)$.
Let $\mathscr E_{RR,3}$ and $\mathscr E_{RR,1}$ be the corresponding
remainders for $\mathcal Z(\mathsf R_0,\mathsf R_1)$ and
$\mathcal Z(\mathsf R_0,\mathsf R_3)-3\mathcal Z(\mathsf R_1,\mathsf R_2)$,
respectively,
and let $\mathscr E_{K_d}$ be the remainder in the displayed expansion
of $K_d=d_{0,(0)}d_2-d_{1,(0)}d_1$.
All these are field identities, before taking zero modes.
Before imposing Cayley--Hamilton, the contraction formulas give
\begin{equation}
\label{sr:ancillary-ch-remainders}
\begin{aligned}
 \mathscr E_{\mathsf R_0,F_1}
    &=4\operatorname{tr}(\mathsf C_0g),&
 \mathscr E_{\mathsf R_0,F_2}
    &=2\operatorname{tr}(\mathsf C_1(Y)g),\\
 \mathscr E_{RR,3}
    &=4\operatorname{tr}(\mathsf C_0X^2),&
 \mathscr E_{K_d}
    &=8\operatorname{tr}(\mathsf C_0XY),\\
 \mathscr E_{RR,1}
    &=18\operatorname{tr}(\mathsf C_0Y^2)
       -\tfrac92\operatorname{tr}(\mathsf C_1(Y)XY).
\end{aligned}
\end{equation}
For example, the first remainder is
$4\operatorname{tr}(X^4g)-2A\operatorname{tr}(X^2g)
-\frac43D\operatorname{tr}(Xg)$; the constant term of
$\mathsf C_0$ drops out because $g$ is traceless.
The other remainders in \eqref{sr:ancillary-ch-remainders} are obtained
by inserting the displayed matrix polynomials
\eqref{sr:ch-directional}, with the order of every matrix factor
retained. All the remaining subsidiary field identities have zero
remainder already in the signed cyclic-word calculation.
Thus \eqref{sr:ch-directional} completes their finite-rank reduction,
including both odd families, without an additional relation or an
external calculation.

\bibliographystyle{JHEP}
\clearpage
\phantomsection
\addcontentsline{toc}{section}{References}
\bibliography{references}
\end{document}